# Solitons and Quantum Behavior

Richard A. Pakula

email: rpakula1@gmail.com

April 9, 2017

## Abstract

In applied physics and in engineering intuitive understanding is a boost for creativity and almost a necessity for efficient product improvement. The existence of soliton solutions to the quantum equations in the presence of self-interactions allows us to draw an intuitive picture of quantum mechanics. The purpose of this work is to compile a collection of models, some of which are simple, some are obvious, some have already appeared in papers and even in textbooks, and some are new. However this is the first time they are presented (to our knowledge) in a common place attempting to provide an intuitive physical description for one of the basic particles in nature: the electron. The soliton model applied to the electrons can be extended to the electromagnetic field providing an unambiguous description for the photon. In so doing a clear image of quantum mechanics emerges, including quantum optics, QED and field quantization. To our belief this description is of highly pedagogical value. We start with a traditional description of the old quantum mechanics, first quantization and second quantization, showing the tendency to renounce to 'visualizability', as proposed by Heisenberg for the quantum theory. We describe an intuitive description of first and second quantization based on the concept of solitons, pioneered by the 'double solution' of de Broglie in 1927. Many aspects of first and second quantization are clarified and visualized intuitively, including the possible achievement of ergodicity by the so called 'vacuum fluctuations'.

# 1. Chapter: Quantum Mechanics

## 1.1. Classical physics

In classical physics, including classical particle mechanics, fluid mechanics or electromagnetism, the description of the system under consideration is given in terms of physical objects, like solids, fluids (liquids, gases, plasmas), different types of fields (scalar fields, vector fields, tensor fields), following well defined motion-laws described by equations in real space-time. In a classical description, physical intuition or visualization is fundamental. The physicist start with well-defined objects and systems in space and time that can be visualized without difficulty, and try to find the equations of motion that best describe the observations he made with the help of measurements performed with well-known and controllable instruments. In this process the physicist is free to modify the mathematical equations until finding the most adequate to the observations. In this sense the primary source of the theory is the observation and intuition gained through experience, whereas the mathematics is relegated to a description tool or language.

## 1.2. Quantum mechanics.

### 1.2.1. Old Quantum Theory

The equations of classical physics known at the beginning of the XX century were unable to explain the evolution of atomic systems. At the end of the XIX century, Rutherford discover the 'orbital model' for the atom: negatively charged electrons orbited around the positively charged nucleus, much as our solar system. However when trying to explain this model with the available equations of classical electrodynamics, wrong frequencies and intensities were found: the most studied model was the hydrogen atom constituted by a single proton and a single electron orbiting around. Classically the electron should slow down due to radiation reaction and collapse against the nucleus after a few nanoseconds. The hydrogen atom like most others is however stable. Additionally the spectrum of light emitted by the 'flames' generated by heated materials (the typical one was common salt or sodium Na) consisted of a discrete set of lines following certain laws that were a puzzle at the start of the century. These set of lines corresponded to the so called Balmer series.

Another problem was related to pure radiation fields: classical electrodynamics predicted a divergence for the energy content of blackbody radiation at increasing frequencies. The opposite was however observed at higher temperatures in the UV region, the intensity decreased monotonically. The first correct modeling of the blackbody spectrum was done by Max Planck in 1900-1901, where he was able to derive the blackbody spectrum using classical electrodynamic wave theory together with the extraordinary assumption that the energy content of the electromagnetic field at each frequency $f$ could not have an arbitrary value as allowed by all rational assumptions, but should be a multiple of the quantity $hf$, where $h$ was a new discovered constant, later universally known as Planck constant. The physical model behind Planck's calculations was then that the energy contents of the electromagnetic field was quantized by the condition $E = nhf$. Moreover, in 1905 Einstein showed that the photoelectric effect consisting on the spontaneous emission of electrons from metallic surfaces when irradiated by UV radiation, that also could not be explained by classical electrodynamics, could however be 'most naturally' explained by assuming that light was not a wave as believed since the time of Huygens, but was a flux of particles, each one with an energy $E$ equal to $hf$. These particles became later to be known as 'photons'. In 1909, Einstein showed that the statistics of the blackbody radiation at lower frequencies corresponded to a wave-like system, while at high frequencies approached the statistics of a gas of free particles. Fifteen years later Bose and Einstein were able to derive the correct blackbody spectrum based on a fully particle model following Einstein's energy quantization condition, instead of the wave model used by Planck originally. The key point in their model was the introduction of the concept of so called 'indistinguishable particles', at the time of 'counting the number of states'. This counting operation used in the statistical derivation of the energy spectrum was traditionally based on the intuitive visualization that 'each particle has a given energy'. The new theory was based on the principle that the energy was a property of the entire system, without any reference, even in principle, to the individuality of the particles. Particles following this principle were later called 'indistinguishable'. The statistics derived from this model was called 'boson statistics' and its principal application was for photons and particles with integer spin, such as helium atoms. Any explanation for the existence of this new principle was totally unknown to classical physics.

The first partial explanation for the atomic stability and spectral emission was given by Niels Bohr on his work on the hydrogen atom in 1913. There he postulated that the electrons orbiting according to the Rutherford model could do it only at a set of preselcted orbits characterized by the condition that its orbital momentum was 'quantized' to multiples of the Planck constant $\hbar$, like blackbody radiation. These orbits were called 'atomic levels' and had an energy that was also quantized. Bohr's theory was able to find a derivation for the hydrogen Balmer lines. Additionally Frank and Hertz performed in 1914 an experiment where by applying a potential difference across a vapor of mercury atoms, they could verify that the non-uniformities in the variation of current with voltage could be explained by assuming that the atomic energy states were indeed quantized. This theory was later known as 'old quantum theory'. Successful for the hydrogen atom, Bohr's theory could not explain more complicated atoms such as Helium with two electrons orbiting around the nucleus. In Bohr's theory, the electron could be found in any of that orbits, but only the lowest energy orbit was assumed to be stable. All others 'excited states' were assumed unstable and eventually decayed through finite transitions or 'quantum jumps' to the 'ground state'. The frequency of the radiation emitted during the quantum jump was not given by the orbital frequency of motion of the electron but was obtained through Einstein's equation $\Delta E = hf$ , where $\Delta E$ was the energy difference between the two atomic levels participating in the transition. Obvious question: where is the source for this radiation oscillating at that frequency?

Soon after the work of Bose-Einstein, Pauli developed another statistics to be applied in the architecture of multi-electron atomic systems, where he postulated that not 'two electrons could ever occupy the same state'. This principle helped in creating stability conditions for multi-electron atoms. When adding a new electron in the building of an atom, these new electron could go only to some of the upper, unoccupied levels; in this way it was possible to explain the chemical properties of the elements as classified in the periodic table of the elements. The statistics derived from this principle was developed by Dirac and Fermi around 1926 and was applied to particles of half-integer spin, most notably to electrons. These particles were called fermions and its statistics Fermi-Dirac statistics for indistinguishable quantum particles.

### 1.2.2. First quantization

In 1916, after the development of the Bohr's atom, Einstein started using the terminology of probabilities when referring to the processes of absorption and emission of radiation by transitions between the stationary states of atomic systems, and created the statistical concepts for spontaneous and induced emission of radiation (lasing). In his model the equations were not describing actual events occurring in space and time, but were just statistical models calculating probabilities for the transitions to occur, in this case the probability for the emission or absorption of a photon particle by the atomic system. In this work the probability to find a photon particle in space was assumed to be proportional to the electromagnetic energy density given by the square of the field magnitude. In 1921 Ladenburg apparently incorporated for the first time these transition probabilities in the description of radiation scattered by atomic systems settled in the Bohr's orbits. In so doing, he was providing the first steps in the construction of the probabilistic model that was going to be used in the new quantum theory still under development. Evidently Einstein had a primary role in the creation of the basic new ideas later used by quantum mechanics, but he didn't take a principal part in the actual development of the mathematical machinery behind quantum mechanics, and he was always against the indeterministic, counterintuitive philosophy maintained by the Copenhagen interpretation.

The first attempt to create a complete theory able to explain the characteristics of Bohr's atomic model was developed by Werner Heisenberg around 1925. Classical mechanics obtains the equation of motion for a given dynamical variable as partial derivatives of the Hamiltonian expression with respect to the corresponding complementary variable.

Heisenberg created a new theory where the equations where obtained through the 'commutator' of the Hamiltonian with the complementary variable. Classically physical variables are described by real numbers, and therefore they always commute. In Heisenberg's model $x$ and $p_x$ for example, shouldn't commute. Heisenberg proceeded to describe these physical variables not by real numbers, but by matrices of complex numbers that are non-commuting mathematical objects. He further postulated the fundamental commutation relation $[x, p_x] = i\hbar$ where $\hbar$ is Planck's constant. In this way he was able to derive equations of motion replicating Bohr's model.

A series of new various principles had to be incorporated into the quantum theory in order to explain the newly discovered behavior of atomic and subatomic systems. Among them the principle of identical particles, exclusion, uncertainty, wave particle duality, correspondence. Eventually we will see that all those multiple principles can be explained by just one: the existence of a new force in nature that can be called the quantum force.

The philosophy behind the works of most of the creators of quantum mechanics was contrary to the basic ideas behind classical mechanics, namely the concepts of motion and forces, first defined by Newton in his famous equations. Instead of reconciling the equations of motion with the experiments by the incorporation of still unknown forces, they believed that the basic ideas of motion and forces were fundamentally wrong: they tried to create a new indeterministic model for this world, based on probabilities instead of objective realities. We will see in this work that this was an excessively radical position: it was not necessary to change the philosophy behind Newton's law, it was only needed to incorporate the newly discovered forces and to generalize the scope of these forces to include possible non-local interactions.

In his work on matrix quantum mechanics, Heisenberg was gradually renouncing to any possibility for a realistic and intuitive description of the physical 'quantum systems' but followed an opposite philosophy, trusting in abstract mathematical formulations and equations, using matrices and commutation relations. Around the same time Paul Dirac was creating a sort of (mystical ?) description of quantum systems in terms of new objects that were not following traditional, 'classical laws' as the objects we see in our everyday life, but were following a set of different 'quantum laws'. The classical system was said to be based on the classical 'c numbers', while the quantum system were depending on the quantum 'q numbers'. So 'q numbers' were providing a sort of description for physical systems with properties different to those of classical systems. This dichotomy between natural, classical objects and special, quantum objects survived up to the most sophisticated models of canonical quantization by Heisenberg and Pauli in 1929. In this last theory, physical fields which classically were following evolution equations in real space-time, were replaced by abstract field operators evolving in the abstract Hilbert space, residence of the quantum states of the system.

In 1926 Schrödinger, following an original idea of de Broglie, was able to show that the quantization of energy levels of Bohr could be explained as a wave phenomenon much as a musical instrument at resonance. In 1927 Davisson and Germer were able to verify experimentally the existence of electron waves by observing electron diffraction at the surface of a crystal of nickel. Schrödinger created his famous equation by replacing the momentum variable in the expression for the classical Hamiltonian by the gradient or 'nabla' operator. So, in Schrödinger's model, $p_x$ was replaced by $-i\hbar \frac{d}{dx}$. Dirac was able to show soon after, that this replacement together with the use of wave functions to describe a quantum system was equivalent to the abstract and enigmatic Heisenberg's 'matrix mechanics' based on the proposed commutation relations between position and momentum. In fact, differential operators like nabla or gradient don't commute with multiplication by their respective coordinate: $x \frac{d}{dx} f(x)$ is not equal to $\frac{d}{dx} x f(x)$ as can be easily verified. By using a differential equation acting over a wave function instead of a dynamical equation over a q-number variable, the first mystification of quantum mechanics was resolved. Commutation relations among position and momentum, were just impersonating the existence of differential equations providing the evolution of wave functions as descriptions of the system. Dirac and also Jordan and Heisenberg postulated that both descriptions: that from Heisenberg in terms of matrices and that from Schrödinger in terms of differential equations for wave functions were fully 'equivalent' representations of the same theory. This was the basis of the so called 'transformation theory' generalizing the concept of variables as operators and of functions as states of the system. The 'transformation theory' was one of the basic pillars used in the philosophical interpretation in terms of probabilities, by claiming that space and space- coordinates were nothing more fundamental than any other group of variables used in the description of the state of the system. In other words, events don't need to exist or take place necessarily in our usual space, for the theory this is only a simple 'representation' detail. We don't agree with that point of view. From our perspective Schrödinger's equation is a much more fundamental theory than Heisenberg's matrix mechanics. What Heisenberg's equations were describing was merely the evolution of the average values of the dynamical variables over the wave functions, and could be completely derived from Schrödinger's model. These equations are known today as the Heisenberg equations of motion and can be routinely derived using Schrodinger equation.

This was however a brief success for the intuition and the reason. This step raised immediately the following question: Why the differential equation for the wave function could be obtained by replacing the momentum variable with the gradient operator in the Hamiltonian expression? This was just the beginning of wave-quantum mechanics and many additional puzzles were awaiting to appear and remain apparently unsolved for a long time to come. However an answer presents by itself: Is not why quantum mechanics derives in that particular way from classical mechanics, but how classical mechanics derives from quantum mechanics. If a rational, intuitive modeling can be given to the quantum wave theory, and if it can be shown that classical theory is the limiting behavior of the quantum system under some conditions, then the answer would be self-evident. This limiting process couldn't be deduced systematically by traditional quantum mechanics and had to be postulated as a new principle: the principle of correspondence which stated more or less that for high 'quantum numbers' or higher energy states, the evolution of the system could be approximated by the classical equations of motion.

The original idea of Schrödinger after creation of his quantum equation was to provide a description for the electron as a fluid field in space. However he introduced complex numbers in the definition of the electron wave function, moving away from a classical physical field that would be given in terms of real numbers. Nevertheless, the single particle Schrödinger equation was still assumed to be an equation defined in real space-time. The traditionally accepted connection of these wave functions with measurement was given by Max Born in 1926 in terms of 'probability densities' proportional to the magnitude squared of the wave function. In so doing he was following the original 1916 idea of Einstein providing a link between the existence of corpuscles of light and the electromagnetic fields.

Step by step, the original idea of Heisenberg for abandoning any attempt at an 'intuitive visualization or understanding' of microscopic processes was growing in favor of abstract mathematical formalisms, as long as the final theoretical prediction agreed with observed measurements. The equations of quantum mechanics and its solutions were losing any physical reality and any intuitive description in terms of real physical objects. The relativistic generalization of the Schrödinger equation by Klein and Gordon in 1926-1927 provided more puzzles to the quantum problem: the appearance of negative energies and negative probability densities in the relativistic current-density four-vector. For a very lucid description of the problems associated with relativistic quantum equations see the work by H. Nikolic[1]. Pauli incorporated the electron spin description in 1927 in terms of complex spinors, defined as a set of two complex numbers, and generalized the equation of motion incorporating the 'Pauli matrices', a set of 2x2 square matrices of complex numbers. These matrices were associated with the internal or intrinsic angular momentum of the electron. Finally Dirac provided the relativistic version of the Pauli equation in terms of 4-spinors and the set of 'Dirac matrices' which are 4x4 versions of the original Pauli matrices. Neither the spinors nor the matrices had a straightforward interpretation in terms of classical space-time vectors or tensors, and were initially assumed to provide a more basic (although very counterintuitive) description of physics. However these equations provided better and better agreement with the measured spectra of hydrogen atomic systems under different experimental situations.

After 1927, two more problems still remained to be solved: multi-particle systems and quantization of the electromagnetic fields. Many-electron systems could be described in terms of the Schrödinger, Pauli and Dirac equations, which are first-order differential equations in time. However the multi-particle wave function solution of those equations was not defined in real space-time, but in the abstract configuration space, where every particle was associated with a unique, independent axis of coordinates. In this way a system of 3 electrons in the multi-particle Schrödinger equation is described by a wave function defined in a 9-dimensional space, 3 additional dimensions for each electron. Additionally, the statistics of indistinguishable quantum particles was achieved by requesting that the wave function describing the state of the system of fermions be provided by the determinant of the products of the single particle states $\left(f_1(r_1)f_2(r_2) - f_2(r_1)f_1(r_2)\right)$, while wave function for bosons should be given by the corresponding 'permanent' (same structure as determinant but with all terms adding together) $\left(f_1(r_1)f_2(r_2) + f_2(r_1)f_1(r_2)\right)$. These functions were said to be eigenstates of the permutation operator.

What meaning should be attributed to this description? The answer provided by Heisenberg's logic was as simple as disappointing: 'that question is meaningless'.

Around 1960, Feynman described one of the most challenging concepts of quantum mechanics given by the wave-particle duality, with the phrase "the only mystery of quantum mechanics is the interference of particles". This was one of the properties that mostly puzzled the creators of quantum mechanics: how is it possible that particles in vacuum, in the absence of any interaction or external fields appear to follow non-straight trajectories leading to interference patterns? The answer provided by traditional quantum mechanics appears somehow unclear if not controversial to an independent observer: instead of trying to understand and find the laws followed by the trajectories of the particles, they refuted the very existence of trajectories at all.

### 1.2.3. Second quantization

The introduction of electromagnetism inside the quantum theory was achieved by the creation of the so-called 'Canonical or Second Quantization'.

The Schrödinger equation was able to describe in 1926 the architecture of the hydrogen atom through a resonance phenomenon of a 'wavefunction' obtained by the calculation of eigenfunctions. The eigenvalues of the equation provided the frequencies of resonance that were associated with the quantized energy levels through the relation *E=hf*. This relation was postulated for the electron by de Broglie in 1923, and really was a generalization of the same relationship first enunciated by Einstein in 1905 for photons. In this way it was achieved the 'quantization of the energy levels' allowed for the hydrogen atom, or equivalently, available for the electron particle.

After the development of the Schrodinger equation, people continued investigating if a similar theory could explain the quantization of the energy levels allowed for the electromagnetic fields *E=nhf* as postulated originally by Planck in 1900. By comparing the wave equations provided by the electron, like the Klein Gordon or the Schrödinger equations with the Maxwell's equations, the first idea that comes to our mind is that the Maxwell's equation could be just interpreted as the 'photon wave equation', namely the quantum equation for the assumed photon particle envisaged by Einstein. In fact the only main difference between the Maxwell and Klein Gordon equations was the presence of a source term in the Maxwell's equations. However in those regions of space where no charges are present the equations can be written in a totally equivalent form. In both cases the boundary conditions were responsible for the geometry of the wave and the quantization of the frequencies allowed at resonance.

What was different however was the prevailing philosophy behind the interpretation of those equations. It is a fact in physics that in many cases some equations allow solutions that are not observed experimentally. In those cases it is the interpretation who establishes rules to define what solutions should be allowed and what solutions should be rejected. In fact, the solutions that were allowed for those equations were different. While the classical solutions to the Maxwell's equation could have any amplitude, the only allowed quantum solutions to the electron equations were wavefunctions normalized to unity.

Some formal problems associated with the mathematics of the system, appeared when trying to associate the electromagnetic solutions with a probabilistic interpretation. First, the existence of a source term in the Maxwell's equations prevented the achievement of a constant value in the norm of the wave. This problem could be avoided however in regions where the source term was null. A second problem was strictly related to the formal equation defining the value of the probability density for finding the photon particle at some point in space. While in the electron case this probability density could be defined exclusively as a function of the square of the magnitude of the wave function, being a fully local quantity, in the electromagnetic case it appeared that the 'probability density' would also be a function of derivatives of the wave, generating some sort of 'nonlocality' in the definition. A final word on this question appears to be founded lately by Margaret Hawton[2] and coworkers in favor of the wave function probability interpretation. This would represent an example where a technical difficulty associated with a mathematical tool was confused with a physical law and prevented the development of the concept of photon as a particle. This problem and the non-constancy of the norm of the electromagnetic waves induced that for many years the studies in terms of 'photon wavefunctions' were totally avoided.

A deep difference appeared on the assignment of energy content to the wave. In the electron case, the energy was defined by the frequency through the de Broglie relation and was obviously independent of the wave amplitude, always normalized to unity. On the contrary, on the electromagnetic case, the energy was classically given as an integral over space of a quantity proportional to the square of the amplitude of the wave and independent of its frequency. What the new quantum electrodynamic theory was attempting to achieve was readily quantizing the value of that integral, and therefore indirectly the amplitude of the wave.

This was the link followed by Dirac in his works on quantization of the electromagnetic fields. Dirac, following a previous work by Jordan et al. noticed that the wave equation and the Hamiltonian for transverse electromagnetic waves in the absence of sources was formally identical to the respective Schrödinger equation and Hamiltonian for a mechanical harmonic oscillator. The formal equivalence could be achieved when the identification of the position coordinate for the mechanical oscillator was identified with the amplitude of the electromagnetic wave. Then Dirac 'borrowed' the quantization results from the mechanical harmonic oscillator and applied them to the electromagnetic field. This procedure was the source for 'Field Quantization'. The allowed energy levels of the quantum harmonic oscillator are quantized by the condition $E=nhf$ . In this way the energy content of the electromagnetic field was quantized at the levels $E=nhf$, that was nothing else than Planck's original assumption. We can notice that the description of the electromagnetic system was still done in terms of quantized waves, and not in terms of particles, as Einstein first assumed in 1905. The entire eigenstate of the Maxwell's equations was now a field whose energy content was quantized. The 'photon' was identified with these eigenstates, and somehow associated with Einstein's particle concept, by arguing that when the field has an energy $E=nhf$, it has $n$ 'photons'. A primordial role in Dirac's theory was played by the equivalent to the creation and destruction operators in the mechanical oscillator, that were directly associated with the electric and magnetic fields.

Two different but equivalent representations of second quantization were created. The first one is the one previously described, created by Dirac, Heisenberg and Pauli, called 'Field quantization' or 'Canonical quantization', where the system description was given in terms of 'field operators' evolving in the Hilbert space of the states of the system. For bosons, commutation relations were defined for the field creation and destruction operators in analogy with the similar operators for the quantum harmonic oscillator. Anticommutation relations were defined for fermions. The equation of motion was defined by the commutator of the field operator with the field Hamiltonian, similarly to the original Heisenberg's matrix theory, but now applied to fields instead of particles. An independent second representation was based on wave functions in Fock space, defined as the direct sum of configuration spaces with different number of dimensions. The different models of second quantization had its roots on the original quantum statistics and works of Planck, Bose, Einstein and Dirac. In fact, Bosons Canonical Quantization, originated in Dirac's work on the quantization of the electromagnetic fields, followed the wave model developed by Planck on waves with a prescribed energy content. The representation in Fock's space followed the particle's description originally developed by Bose and Einstein. The first description is mainly used in quantum optics and quantum electrodynamics, while the second one mostly in solid state physics. Some aspects from the theory of the quantum harmonic oscillator were introduced into boson's theory and also indirectly into the fermions quantization, like zero point fluctuations and the form of the Hamiltonian operator.

The commutation relation prescribed for the operations of creation and destruction operators in the Dirac representation was the same as that existing for the operators in the quantization of the harmonic oscillators. Fock was able to show that this commutation relation when expressed in terms of operators in Fock space, guaranteed that the wavefunctions were built as the permanent of the product of single-particle states. In the second quantization for fermions, the anticommutation relation prescribed for the operations of creation and destruction operators had no equivalent in the quantization of the harmonic oscillator, but they guaranteed that in Fock space, the wavefunctions were built as the determinant of the product of single-particle states.

The dynamics of the system was defined by the commutator of the Hamiltonian with the field operators. As we see this last description is very far from any 'intuitive visualization and understanding' of the set of particles composing the system and its evolution. Even the object of evolution is a totally abstract concept: the 'quantum field operator'. The present day link of this mathematical formalism and the experimental results is still in terms of probabilities. The mathematics can provide probabilistic distributions for all variables of interest that can be verified with experimentally measured statistical distributions. They agree to an excellent level of accuracy. This agreement is used by followers of the so called Copenhagen interpretation as a demonstration that everything is in order and that we shouldn't be asking for more. Even the more radical followers of this interpretation may argue that our universe is nothing more than a set of probabilities turned into realities at the time measurements are performed over the quantum systems.

## 1.3. Intuitive interpretation

Despite this huge abstract machinery, an inquisitive student can ask himself: is it possible to find an interpretation that is not renouncing to our desire of understanding the world in which we are living, and to provide a realistic description for it? The feeling of many of the founders of the quantum theory was that they had found the 'true description' of nature, and that attempts to find alternative interpretations were meaningless and worthless. It is possible that this feeling prevented the development of alternative interpretations other than the traditional one. Over the years however, many people attempted to find such an interpretation. We will attempt in the following to provide a description of those attempts.

We start by noticing that even when there are solutions to the quantum equations, which can only reproduce experimental measurements on a statistical sense, there exists other set of solutions that can describe measured quantum events on an individual basis. The first one to study this type of solutions was de Broglie in 1927, and he called his model “double solution”.

### 1.3.1. First quantization

Let's start from the beginning, the definition of the Schrödinger wave function in terms of complex numbers. The natural and simpler description for the electron should be in terms of real scalars or vectors fields. It is clear that a scalar description is not enough, we need to provide a vector-field description and define a vector for each point in space. It is well known that complex numbers are a description of vectors in a two dimensional plane. However real space has three dimensions, if we are providing a description in terms of vectors inside a plane, what is the orientation of that plane in real space? The Schrödinger equation cannot provide the answer, but by looking at the Pauli equation, the needed orientation is provided unambiguously: is the spin direction. The plane providing the basis for the complex notation is perpendicular to the spin orientation. This was shown by Hestenes using the techniques of geometrical algebra. Therefore in the Pauli matricial and spinor formalism, one is readily using a deceptive description of vectors in space.

Dirac himself provided the meaning of his equation as a generalization of the relativistic Klein Gordon second-order partial differential equation in time, in the presence of spin. Two of the spinor members can be understood in terms of the description of the spin direction, what are the other two spinor members? They can be understood as a mathematical technique used to reduce a third-order differential equation in time to a first order equation. And this can be mathematically proved. Similarly, one can reduce an equation of third order in the energy to the expression provided by Dirac's equation. The reason for the third order in energy is the relativistic variation of mass with velocity. By using a particular gauge the equation can also be written as a fourth order equation[3]. Additionally the 4x4 Dirac matrices when sandwiched between the 4-spinors provide descriptions for the interaction of a 'fluid' that at rest possess an internal magnetic dipole. The relativistic law of transformation for the magnetic dipole provides a term proportional to the fluid velocity with the form of an electric dipole interaction. In summary, the Dirac equation describes a relativistic charged vectorial field with an internal rotational motion, which is described in the dipole approximation when it is in interaction with external electromagnetic fields. And all of this can be rigorously proved.

Moreover it can be shown that a chain model exists that can provide a great level of detail for this realistic fluid field. The evolution of this fluid is expressed by equations of motion containing the different types of forces acting on every element of the fluid. This model postulate the existence of a spring-like force, trying to restore the equilibrium position and proportional to the extension of the deviation from the equilibrium position. Next a similar force proportional to the relative deviation between neighboring elements, playing the role of a tension force. These two forces are enough to give rise to the Klein Gordon equation in the continuum limit. A third force producing a similar effect to the Biot-Savart force for a magnetic field is assumed and can be shown to be the origin for the electron spin. This last force can be described by a vector potential field given by the cross product between the distance between neighbor chain elements and the particular direction in space defined by the spin-vector. As shown by Hestenes this force generates a rotation of the fluid in the spin plane perpendicular to the spin vector $\vec{s}$.

The previous description provides a solution for the presence of complex numbers and spinors in the wave functions for a single electron. Once we have found this model, the uncertainty principle reduces to an application of Fourier analysis to the wave functions. But what about the wave-particle duality, the configuration space and Fock space descriptions and the presence of field operators?

The wave-particle duality and the resolution to the Feynman mystery due to ‘particle interference’ was however resolved very early in time by de Broglie as soon as 1926 after the discovery of the Klein-Gordon equation. His proposal however was totally ignored, neglected and forgotten by traditional quantum mechanics after the Solvay Conference of 1927. De Broglie called his model ‘double solution’ because provided the existence of an additional type of solutions to the quantum equation. De Broglie simply showed that a valid solution to the Klein Gordon existed, showing a sort of protuberance or protrusion that was moving following the gradient of the imaginary phase of the wave, therefore drawing interference patterns. De Broglie was able to find the equation of motion for this object, which was confirmed in 1952 by David Bohm who showed that de Broglie’s ‘guidance condition’ could be derived from the existence of a force derived from an effective potential energy that Bohm called the ‘quantum potential’. The physical origin of this quantum force is just the ‘chain force’ propagating through the body of the fluid, with full respect for all classical laws of energy and momentum conservation. In opposition to all other known wave solutions that tend to disperse and disaggregate with time, this one remained self-confined. This was his original model for a “particle”. Two problems presented these solutions to be considered seriously as particle models: i) they are not falling down sufficiently fast to zero from the center, and ii) they diverge at the center of symmetry. One would like the solutions to be finite everywhere, to have a finite energy content, and to decay sufficiently fast to infinity, in order to be said that they are “localized”, one would desire they to decay exponentially with distance. With time these solutions gave rise to the study of ‘solitons’ which appear as solutions to nonlinear wave equations. The introduction of nonlinearities in the quantum equations is to a certain degree arbitrary and seems to be artificial. Moreover, when considering nonlinearities with orders of magnitude comparable with the linear terms of the Schrödinger equation they lead to energy levels for the hydrogen atom that not reproduce experimental measurements. This implies that this nonlinearities, if existing, should appear at electromagnetic energy levels much higher than typical atomic values. The solution to this puzzle is provided by quantum electrodynamics itself. It can be rigorously shown that natural self-interactions of the electrons and photons with themselves lead genuinely to the appearance of effective nonlinear terms in the otherwise linear quantum differential equations. The solutions to these equations are solitons with the desired characteristics. Let us remark that the consideration of self-interactions is not something extraneous to canonical quantization, but it readily provides the basic elements for the theory of renormalization. Self-interactions can be incorporated for the electron by solving simultaneously the Dirac and the Poisson equations for the field generated by the electron itself. In the case of photons, the self-interaction appears in first order perturbation theory, as additional terms in the Maxwell equations generated by the Uehling and Heisenberg-Euler vacuum polarization terms present in QED. The great news is that solitons following the quantum potential are the most natural resolution of the wave-particle dichotomy: they are localized and can follow interference patterns.

In our interpretation, probability densities, even when available in some cases, are not required as a 'must do' of the theory. The real objects of our interpretation are the fields and the appearance of negative probabilities represent no real problem at all.

Now we are in a position to understand the use of spinors in the Pauli and Dirac equations. The reason for their introduction is because the quantum potential is responsible not only for providing guidance to the soliton motion, but also to produce a torque on the spin vector. The spinorial representation provides the mathematical vehicle to encode the full quantum force. The equilibrium orientation for the spin, as provided by the quantum potential, can only be parallel or antiparallel to external magnetic fields. This explains the splitting of energy levels for atomic systems in magnetic fields observed in the Zeeman Effect and the discontinuous splitting of the trajectories followed by magnetic particles traversing magnetic fields in the Stern Gerlach experiment.

### 1.3.2. Second quantization

Some of the commutation properties of quantum operations like rotation or excitation and de-excitation of an atomic system readily have their roots in classical systems. Classical rotations have been described in some text books and shown that they follow the same commutation properties as those defined in their quantum mechanical counterparts. For example a 90° around the x axis, followed by a 90° around the y axis produce a final rotation different than that obtained by permuting the order of the two rotations, as can easily be verified experimentally. Similarly a position measurement performed before or after applying a translation gives different values. When we realize that the typical technique to measure the speed is by measuring a position translation, and to measure an angular momentum we need to measure a finite rotation process, this properties lead directly to the commutation properties of measurements of position and momentum on one side and angle and angular momentum on the other side. Also measuring the number of particles before and after the creation of a particle implies a change in value of 1. Quantum mechanics formalized these relationships and used them as basic properties for the definition of the different variables, making use of the similar properties of the differentiation and multiplication operations in calculus.

Note that all operations having the same properties can be described by the same mathematical formalism. What we wanted to remark here is that the commutation properties are not something that belong exclusively to a quantum regime, but they find their place also in classical systems, and moreover, they can be explained also in classical terms.

In first quantization for single and multiple particles, solitons could be found as solutions of nonlinear equations induced by electromagnetic self-interactions. However an apparent inconsistency appears when trying to interpret those solutions in the frame of Canonical Quantization. It is traditional in classical quantum optics textbooks to interpret nonlinear terms in the Maxwell's equation as well as in the corresponding Hamiltonian, as responsible for photon frequency conversions consistent with the annihilation and creations of multiple photons, but not as a straightforward source for a coherent nonlinear index of refraction in the presence of single photons. It is easy to verify that this is the case however. It can be shown through the proper application of commutation rules to the field operators, that both interpretations are fulfilled simultaneously. Nonlinear terms are responsible for multiple-photon frequency conversions and nonlinear changes in single photon index of refraction. The same can be applied to other particles.

In the single particle case, a soliton can be found following the law of motion predicted by the quantum potential, which in turn is given as a function of the wave function. In the case of many particles, solutions to the wave equation can be found in Fock space that present a number of solitons, one per particle. Each soliton follows the quantum potential which in this case is now a function of the actual position of all the particles composing the system: electrons and photons. The Fock space in canonical quantization offers a mathematical tool for the representation of these solitons and their mutual interactions. The difference between the first quantization multi-particle description and canonical quantization is that in the first case the system consists only of electron wavefunctions solutions of the Schrödinger or Dirac equations with a fix number of particle coordinates, while the electromagnetic fields are given as external fields. For second quantization, the system is composed of electrons, positrons and photons described by relativistic (or non-relativistic) equations, whose solutions consist of functions with variable number of particle coordinates, and no external fields are required.

The solitons provide reality of presence for the particles while the quantum potential represent interactions between the particles. In fact, the solution to the wave equation describes simultaneously the position of the particles and its mutual interaction energy. The wavefunction, and therefore also the quantum potential, depends on the actual position of many particles, implying nonlocal interactions among them.

The quantum force does not depend exclusively on the amplitude of the field. This is radically different from classical systems, where the forces are proportional to the field amplitude (or some power of them). The reason why the quantum force can comply with energy and momentum conservation is because the force is not between the particles and the wavefunction, but between the particles among themselves or with the physical system composing the boundary conditions. For example, it is the wall where the slits are drilled, who receives the energy and recoil in a two slit experiment, before the interfering particle-waves reach the final screen. The wavefunction serves only as an (instantaneous, nonlocal) intermediary.

In the same way as the Lagrangian and Hamiltonian expressions can be considered mathematical tools to derive the equations of motion for the trajectories of real particles or physical objects in space, Field Quantization can be considered a mathematical tool that can be used to derive the wave equations followed by particle- and photon- wave functions. It can be argued that the Field Quantization Schrödinger equation for Field Operators has only a formal meaning. The real application of the theory is to offer a template in the form of an operator expression that when sandwiched between initial ('ket') and final ('bra') states provides effective Hamiltonians or Lagrangians. From these effective Lagrangians one can extract the actual wave equations in the presence of many particle-fields. In other words, the purpose of second quantization is to write down the "effective" equations to solve for traditional, numerical field variables. Then one can solve those using standard analytical techniques as in first quantization. Attempting to solve the problem directly using operator variables in second quantization is similar to the pioneering works of Pauli[4] and Dirac[5], where they were able to find the hydrogen spectrum using exclusively operator properties, instead of solving the Schrödinger equation. Similarly Born and Jordan[6] solved the harmonic oscillator quantization problem based on matrix commutation properties in 1925 before the discovery of the Schrödinger equation. A drawback of this technique is that it allows to find energy levels but makes very difficult to find any dynamical behavior.

It has been shown by Dirac that the value of the commutators between the field operators with the Hamiltonian as defined in Canonical Quantization are equal to the Poisson brackets utilized in classical physics. This property is particularly evident in QED[7]. As a consequence the equations followed by the field operators appear equal to the equations followed by the classical fields. The main difference however between the two descriptions is that while the classical field equations are settled equations with a fixed form relating fields and its sources, their quantum counterparts are only formal dynamical equations depending on the environmental conditions, and they don't acquire their final form until after being sandwiched between initial and final states. Basically, they depend on the number of field-particles present in the system definition. In this sense they behave as templates for the equations to be followed by the real fields.

The most impressive property of these wave equations is that they include source and sink terms, in a similar way as the classical Maxwell equations. This is in parallel with Feynman propagator formalism, but instead of working with Huygens principle, one is working directly with the wave equations and the fields themselves. The wave functions derived from second quantization can be interpreted as real electron and electromagnetic particle-fields evolving and interacting in real space-time.

It is noticeable that the equation for the electromagnetic field operators is almost identical with the classical Maxwell's equation. However this similarity is only formal and holds only in the single photon case. In the multiple photons case, the equation followed by the photons can evolve into a very complicated system, as can be seen in the laser rate equations.

One of the most peculiar properties of the 'effective wave equations' is that they incorporate sources as sinks, in a similar case as in electrodynamics. For example the product of a positron field and an electromagnetic field appears as a source for the electron field, and vice versa. This point appears to have been overlocked by the traditional interpretation, where the particles seem to appear in vacuum 'spontaneously'. To our knowledge Julian Schwinger proposed after the development of quantum electrodynamics the existence for sources and sinks for the quantum fields, but was totally confronted by his peers. In our model the soliton-particles are born from a source and die, or are absorbed by a sink, which are regions of high field intensity.

One of the virtues of second quantization was to show that the existence of negative energies as eigenvalues of the Klein-Gordon and Dirac equations were just mathematical artifacts related to the higher order time derivatives. Nevertheless, based on the relativistic symmetry properties of the equations, they could be related to the existence of antiparticles.

The different quantum states in quantum optics, coherent, thermal, n-photon number, all define different inter-photon potentials that are responsible for the quantum optics statistics. The functions and states in Fock space define the multi photon functions, and with them, the potentials.

We should mention that Canonical quantization in addition to providing a generalization to many particle quantum systems also handles self-interactions and self-mass. These last effects are however not provided through the natural description of classical physics but through the effects associated with the so called vacuum fluctuations or zero energy fields. Many of the properties and effects found in this way were attributed initially to second quantization, but it was later shown that they can also be described in a first quantization theory. In fact, it can be argued that the so called vacuum fluctuations effects are generated by the combination of self-interactions and the quantum potential. In Feynman's diagrams the vacuum fluctuations look like self-interactions, and they readily should be interpreted as such. One of the peculiarities of the self-interactions and the quantum potential is that when both are considered together, they create dynamic resonant states that generate a sort of wandering of the particle across and along the fields. This might be the reason behind the so called 'zero field fluctuations' responsible for the achievement of a statistical ergodic distribution, even in the presence of a single particle.

As a final point we would like to mention that in many practical cases in the traditional textbook presentations of QED, it is difficult to distinguish mathematical tools allowing to increase the performance of calculations from real physical principles. An example is the decomposition of the electron states in QED in terms of plane waves. It appears as if plane waves would play a primordial role in the physics of electrodynamics, when in reality they are just an application of Fourier analysis to simplify calculations. In the following chapters we are going in more detail over these points.

## 1.4. Conclusion

In this work we have discussed how the wave-particle paradox can be resolved starting from a field description, by the "double solution" hypothesis of De Broglie in 1927, complemented with the calculation of quantum self-field solitons originated in the Dirac-Poisson equations. Equivalently we saw that interference phenomena can be understood as a manifestation of a new force in nature derived from the so called "Quantum Potential", best described by David Bohm in 1952. We mentioned that complex numbers, operators in Hilbert space (field operators), matrices and spinors are mathematical tools to simplify notation, but that obscure the understanding of the physical reality. In fact, it was known from long time ago that the Dirac equation is a mathematical formalism able to represent a relativistic tensorial equation in a highly compact way. The existence of solitons as solutions to the quantum equations in the presence of self-interactions provides a realistic interpretation for quantum mechanics, reproducing individual events in space-time.

In many cases the equation that was discovered first historically and was able to reproduce experimental results corresponded to a shortest notation in terms of complex numbers and spinors, but was highly counterintuitive and prevented a straightforward discovery of its physical interpretation.

Next we want to make some remarks on notation. The importance of a 'user friendly notation' is fundamental at the time to perform calculations and understanding the meaning of the theory. This is evident when we consider a 'simple' sum operation when expressed by normal numbers and by roman numbers:

$$\begin{array}{r} 375 \\ \underline{+\,258} \\ 633 \end{array} \qquad\qquad \begin{array}{r} CCCLXXV \\ \underline{+\,CCLVIII} \\ DCXXXIII \end{array}$$

User friendly notation — Not user friendly notation

As a second remark on notation let's note that there is a tendency in physics to shorten and abbreviate the notation used to express equations. In many cases this tendency generates expressions that are less intuitive and clear than the original, more explicit equation. As a first example consider the continuity equation expressing the conservation of mass, where in space-time notation is $\frac{\partial \rho}{\partial t} + \vec{\nabla} \cdot \vec{j} = 0$, while in relativistic notation is $\partial_\mu j^\mu = 0$. The space-time notation can be intuitively interpreted as indicating that temporal changes in density are associated with spatial changes in the current density or velocity, literally, that the time rate of change in density is proportional to the divergence of the velocity, and can be visualized by the graphics in Figure 1:

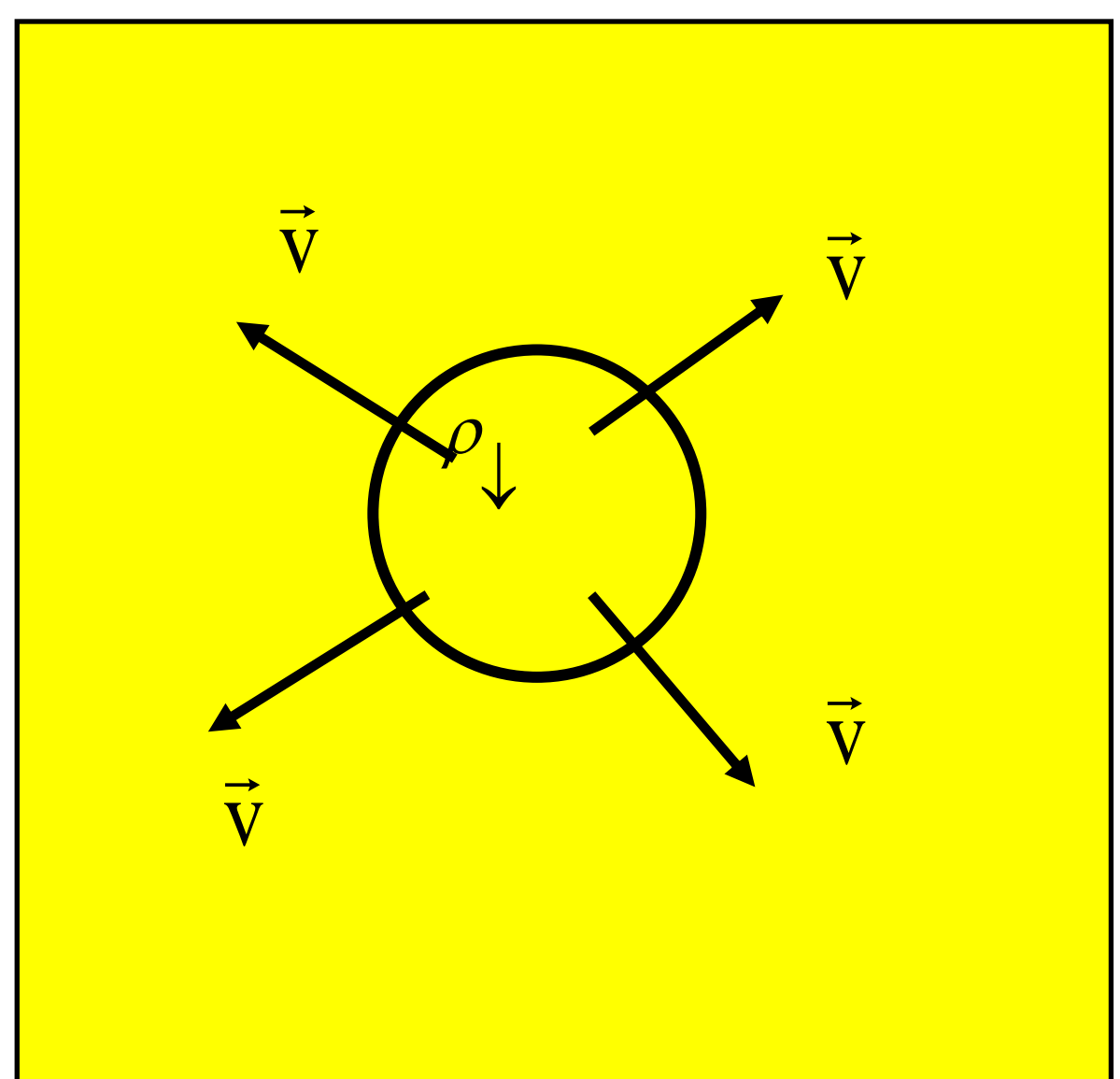

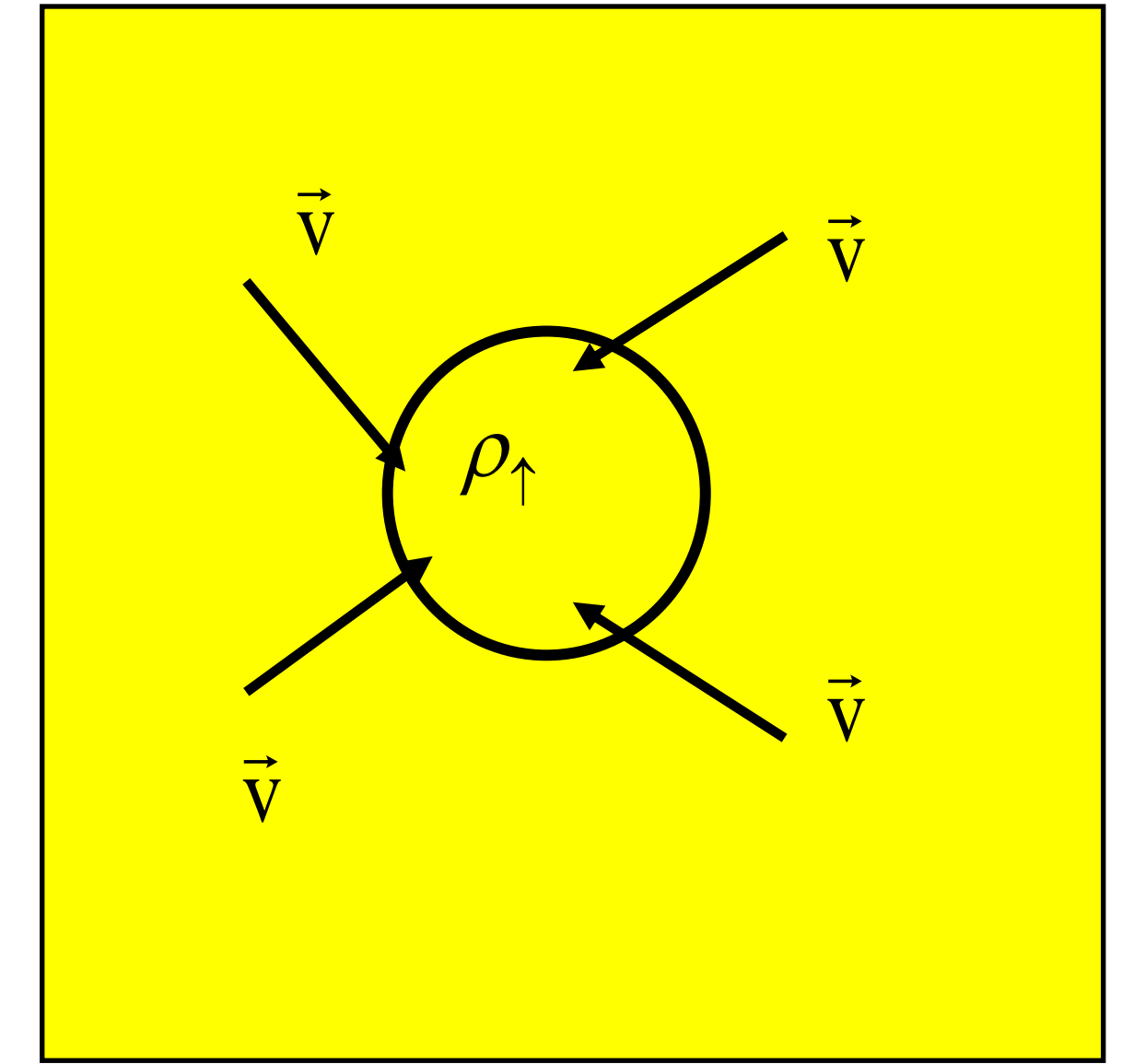

**Figure 1**

The relativistic notation is more compact and express the relativistic symmetry between spatial and temporal components, but this last quality can be appreciated only for a specialist in relativity, obscuring the meaning of the equation for a more general reader. Similar comments can be said about Feynman slash notation $\not{A} \equiv \gamma^{\mu} A_{\mu}$ and about the short hand notations $a^{[\mu}b^{\nu]} \equiv \frac{1}{2}(a^{\mu}b^{\nu} - a^{\nu}b^{\mu})$ and $a^{(\mu}b^{\nu)} \equiv \frac{1}{2}(a^{\mu}b^{\nu} + a^{\nu}b^{\mu})$ used in classical electromagnetic radiation theory, which lead very easily to confusion and misunderstanding. This tendency can be seen also in the use of complex notation. In some cases complex notation was the first historical choice at the time to write equations describing quantum mechanics. In many cases this notation was counterintuitive and prevented a straightforward understanding of its physical meaning. Another comment on notation is about the multiplicity in representing physical variables, depending on the particular section of physics where the variables are described. This multiplicity of notation in general creates some sense of uneasy trying to unify the different points of view available from different situations. This is particularly notable with electromagnetic fields and potentials, when comparing classical and quantum theories. Also the so called “natural units” where the speed of light and Planck constant are made dimensionless constants equal to one $c = 1, \hbar = 1$ may sound as a nice interpretation realization, but only bring problems and inconsistencies at the time to rewrite the equations into usual dimensional variables, able to be compared with experimental values.

A note is in place at this time, in the course of these sections we will recall many well-known models and expressions from textbooks. This will help to put our results in perspective and make clear the similarities and differences between the different models. The reader familiar with those models can skip the corresponding sections.

# 2. Chapter: Dirac-Poisson Soliton

## 2.1. The Quantum Equations. Traditional Presentation.

### 2.1.1. Schrödinger equation

#### 2.1.1.1. Equation

In 1926 Erwin Schrödinger[8] was able to show that the hydrogen atom could be qualitatively and quantitatively described by a resonance wave model, similar to musical instruments vibrating at their proper resonances. The Schrödinger equation[8] admits only explicit complex solutions. In the absence of magnetic fields reads:

$$i\hbar\frac{\partial\psi}{\partial t} = -\frac{1}{2m}\hbar^2\nabla^2\psi + eV\psi$$

In order to introduce the interaction between the electron fields and magnetic fields the replacement

$$i\hbar\frac{\partial}{\partial x} \;\rightarrow\; \left(i\hbar\frac{\partial}{\partial x} - \frac{e}{c}A_x\right)$$

**Eq 1**

has to be done on the electron equations. By applying the rule one can find the equation

$$i\hbar\frac{\partial\psi}{\partial t} = \frac{1}{2m}\left(-i\hbar\vec{\nabla} + \frac{e}{c}\vec{A}\right)^2\psi + eV\psi$$

for the interaction with electromagnetic fields.

#### 2.1.1.2. Hamiltonian

In a similar way as done in classical particle physics, Lagrangian and Hamiltonian functions can be defined for fluids, fields, continuous media in general, from which the equations of motion can be derived. A difference between the energy expressions for particles and for fluids is that for a continuous medium, the potential energy may depend not only on the amplitude, but also on the gradient of the fluid density or field. As a result, forces derived from these potentials can involve higher space derivatives than in the case of discrete particle systems. The spatial and temporal properties of the Lagrangian and Hamiltonian expressions provide the confirmation for the compliance of desired properties such as conservation laws and relativity transformations for the equations of motion. In particular, quantities fulfilling a continuity equation can be found, that serve as candidates for charge densities and currents. These last ones are given by the partial derivative of the Hamiltonian respect to the electric potential field.

In the absence of external fields the Hamiltonian for the Schrödinger equation reads:

$$H = \pi\frac{\partial\psi}{\partial t} - L = \frac{\hbar^2}{2m}\vec{\nabla}\psi^*\cdot\vec{\nabla}\psi$$

And with fields:

$$H = \frac{1}{2m}\left[\left(i\hbar\vec{\nabla} - \frac{e}{c}\vec{A}\right)\psi^*\right]\cdot\left[\left(-i\hbar\vec{\nabla} - \frac{e}{c}\vec{A}\right)\psi\right] + eV\psi\psi^*$$

#### 2.1.1.3. Charge density

From the expression of the Hamiltonian the expression for the charge density and charge current can be obtained. This density can also be interpreted as a probability density and in the absence of external fields is given by:

$$\rho_e(r) = e\psi^*\psi(r)$$

and the charge-current density is given by:

$$\vec{j} = \frac{ie\hbar}{2m}\left(\psi^*\vec{\nabla}\psi - \psi\vec{\nabla}\psi^*\right)$$

**Eq 2**

The first work trying to provide an alternative "physical" interpretation for the Schrödinger equation was Madelung[9], who in 1926 consider the equation as describing a "physical" fluid following a current density given by Eq 2.

### 2.1.2. Pauli equation:

Despite the success of Schrödinger wave equation, there were some details that were still missing. There was experimental evidence that the electron has an internal magnetic moment. This was apparent in the presence of external magnetic fields, where the spectra of hydrogen presented a splitting consistent with the existence of a dipolar interaction, known as the Zeeman effect[10]. The particular splitting of a beam of silver atoms in a strong magnetic field as observed in the experiment by Stern and Gerlach in 1921, provided also an indisputable demonstration that the magnetic moment of the electron appeared to be 'spatially quantized'[11].

Pauli provided a new equation[12] in 1927 that included the magnetic dipole, was able to duplicate the Zeeman splitting and explained the Stern Gerlach experiment. However, his model incorporated complex spinors and matrices in a very counter-intuitive way. The Pauli equation can be considered an effective formalism from a more detailed theory, where the electron mass has not been considered.

#### 2.1.2.1. Spinors

A complex spinor is a set of two complex numbers arranged in a column. They provide a way to encode 4 numbers:

$$\chi(r,t) = \begin{pmatrix} f(r,t) \\ g(r,t) \end{pmatrix} = \begin{cases} \begin{pmatrix} a+ib \\ c+id \end{pmatrix} & in\ cartesian\ representation \\ \begin{pmatrix} b_1 e^{i\varphi_1} \\ b_2 e^{i\varphi_2} \end{pmatrix} & in\ polar\quad representation \end{cases}$$

In particular, a spinor can be used to store vector coordinates, or a given direction in space. The choice of encoding scheme is defined by the particular user of the spinor. In a particular type of encoding, that can be called 'Pauli encoding', the Pauli matrices are a particular key to decode the value of those vectors or angles. The Pauli matrices $\vec{\boldsymbol{\sigma}}$ are defined by the set of three 2x2 matrices

$$\boldsymbol{\sigma_x} = \begin{pmatrix} 0 & 1 \\ 1 & 0 \end{pmatrix} \quad \boldsymbol{\sigma_y} = \begin{pmatrix} 0 & -i \\ i & 0 \end{pmatrix} \quad \boldsymbol{\sigma_z} = \begin{pmatrix} 1 & 0 \\ 0 & -1 \end{pmatrix}$$

The same property is expressed in mathematically terminology as follows (Wikipedia: Pauli matrices article):

"The Pauli matrices (after multiplication by $i$ to make them anti-Hermitian), also generate transformations in the sense of Lie algebras: the matrices $i\sigma_1$, $i\sigma_2$, $i\sigma_3$ form a basis for $\mathbf{su}(2)$, which exponentiates to the **special unitary group** SU(2). The algebra generated by the three matrices $\sigma_1$, $\sigma_2$, $\sigma_3$ is isomorphic to the Clifford algebra of $\mathbb{R}^3$, called the algebra of physical space."

A mathematical problem is to find how to transform the two components of the spinor in front of rotations, translations, in such a way that they keep representing the same vector in all frames of reference. The solution to this problem is the algebra-representation-model and is shown in textbooks on quantum mechanics.

To show how the Pauli matrices can be used to decode the information stored in the spinor, first we note that any spinor can be written without loss of generality in the form

$$\chi = \sqrt{b}e^{i\alpha}\begin{pmatrix} \cos\frac{\theta}{2} \\ \sin\frac{\theta}{2}e^{i\delta} \end{pmatrix}$$

And its adjoint:

$$\bar{\chi} = \sqrt{b}e^{-i\alpha}\begin{pmatrix} \cos\frac{\theta}{2} & \sin\frac{\theta}{2}e^{-i\delta} \end{pmatrix}$$

Where we have encoded the values of the components of the vector $\vec{s}$

$$\vec{s} = (b\sin\theta\cos\delta\,, \quad b\sin\theta\sin\delta\,, \quad b\cos\theta)$$

These components encoded in the spinor $\chi$ can be recovered by use of the following bilinear expressions:

$$\vec{s} = \bar{\chi}\vec{\boldsymbol{\sigma}}\chi$$

Or in parts:

$$\bar{\chi}\boldsymbol{\sigma_z}\chi = b\left(\cos^2\frac{\theta}{2} - \sin^2\frac{\theta}{2}\right) = b\cos\theta = s_z$$

$$\bar{\chi}\boldsymbol{\sigma_x}\chi = b\left(\cos\frac{\theta}{2}\sin\frac{\theta}{2}e^{i\delta} + \cos\frac{\theta}{2}\sin\frac{\theta}{2}e^{-i\delta}\right) = b\sin\theta\frac{\left(e^{i\delta} + e^{-i\delta}\right)}{2} = b\sin\theta\cos\delta = s_x$$

$$\bar{\chi}\boldsymbol{\sigma_y}\chi = ib\left(-\cos\frac{\theta}{2}\sin\frac{\theta}{2}e^{i\delta} + \cos\frac{\theta}{2}\sin\frac{\theta}{2}e^{-i\delta}\right) = -b\sin\theta\frac{i\left(e^{i\delta} - e^{-i\delta}\right)}{2} = b\sin\theta\sin\delta = s_y$$

The angles $\theta(r,t)$ and $\delta(r,t)$ and the magnitude $a(r,t)$ could be function of space and time that are recovered once the spinor is given as an explicit space-time function:

$$\vec{s}(r,t) = \bar{\chi}(r,t)\vec{\boldsymbol{\sigma}}\chi(r,t)$$

#### 2.1.2.2. Equation

The Pauli equation is a nonrelativistic quantum equation incorporating internal rotational and magnetic effects. In the absence of fields it reads:

$$i\hbar\frac{\partial\chi}{\partial t} = \left[\frac{1}{2m}\left(\vec{\boldsymbol{\sigma}}\cdot\hat{\vec{p}}\right)^2 + qV\right]\chi$$

where $\vec{\boldsymbol{\sigma}}$ is a matrix called 'electron spin operator'. In the presence of electromagnetic fields, the Pauli equation is:

$$i\hbar\frac{\partial\chi}{\partial t} = \left[\frac{1}{2m}\left(\vec{\boldsymbol{\sigma}}\cdot\left(\hat{\vec{p}} - q\vec{A}\right)\right)^2 + qV\right]\chi$$

By using the relation

$$(\vec{a}\cdot\vec{\boldsymbol{\sigma}})\left(\vec{b}\cdot\vec{\boldsymbol{\sigma}}\right) = \left(\vec{a}\cdot\vec{b}\right)\hat{I} + i\left(\vec{a}\times\vec{b}\right)\cdot\vec{\boldsymbol{\sigma}}$$

**Eq 3**

it can be found that the equation is equivalent to

$$\underbrace{i\hbar\frac{\partial\chi}{\partial t} = \left[\frac{1}{2m}\left(\hat{\vec{p}} - q\vec{A}\right)^2 + qV\right]\chi}_{Schrodinger\ equation} - \underbrace{\frac{q\hbar}{2m}\vec{\boldsymbol{\sigma}}\cdot\vec{B}\chi}_{\substack{Stern\ Gerlach \\ term}}$$

**Eq 4**

which is the basis to define the physical meaning of the matrix $\vec{\boldsymbol{\sigma}}$: as an operator representing an 'internal magnetic dipole moment or angular momentum' for the electron. The upper element in the spinor represents a particle with 'spin up', and the lower element represents a particle with 'spin down'. Moreover, the expression

$$\vec{s}(r,t) = \bar{\chi}(r,t)\vec{\boldsymbol{\sigma}}\chi(r,t)$$

provides a "classical" vector field in space and time representing this dipole moment.

We see that the matrix and spinorial form of the Pauli equation is just encoding a 'classical' differential equation in space involving a vector field. There is a difference however with a normal differential equation: the use of a 'spinor', or two-component column to describe the particle allows for the existence of two states that are orthogonal to each other in all space and for all times, one with 'spin up' corresponding to the solution

$$\begin{pmatrix} f(r,t) \\ 0 \end{pmatrix}$$

And the other with 'spin down' corresponding to the solution

$$\begin{pmatrix} 0 \\ g(r,t) \end{pmatrix}$$

#### 2.1.2.3. Charge density

With the solution of the Pauli equation given by the spinor with components *f, g* as follows:

$$\chi(r,t) = \begin{pmatrix} f(r,t) \\ g(r,t) \end{pmatrix},$$

the charge density and current density for a Pauli particle are given by

$$\rho_e(r) = e(f^*f + g^*g)$$

$$\vec{j}_e(r) = \frac{ie\hbar}{2m}\left(\left(f^*\vec{\nabla}f - f\vec{\nabla}f^*\right) + \left(g^*\vec{\nabla}g - g\vec{\nabla}g^*\right)\right)$$

The important point is that the two components don't interfere with each other, as if they were two independent objects. In fact, by the use of spinors to define/describe the state of the particle, a spin vector field can be defined in space, with the particularity that the state possess only two degrees of freedom, instead of the three that would naturally be possessed by a typical spatial vector.

#### 2.1.2.4. Alternative interpretation, Dipole approximation

In traditional quantum mechanics, the Pauli equation is considered as a basic equation, describing a point-like non-relativistic electron, with no other possible interpretation. In this sense it is considered that the 'spin' is a fundamental property, impossible to be derived from other models.

On the other hand, one can consider the Pauli equation as the dipole approximation for a sizeable electron with a finite radius and physical internal rotational motion generating a magnetic dipole moment, which in the limit of zero size can be described by Eq 4. However the kinetic energy associated with this internal motion is absent from the Pauli equation. This however is not a serious drawback, it can be argued that an additional negative potential energy term exactly cancels the positive term associated with the kinetic energy. This additional potential energy would provide a centripetal force that will overcome the centrifugal force generated by the internal rotational motion. It would be something like 'filling the square' technique:

$$\left(\vec{\mathrm{p}} - \vec{s} - q\vec{A}\right)^2 - \vec{s}^2 = \left(\vec{\mathrm{p}} - q\vec{A}\right)^2 - \vec{s}\cdot q\vec{A} - \vec{\mathrm{p}}\cdot\vec{s} \underset{\lim r\to 0}{\underbrace{\to}} \left(\vec{\mathrm{p}} - q\vec{A}\right)^2 - \vec{s}\cdot q\vec{B}$$

**Eq 5**

## 2.1.3. Klein Gordon equation

### 2.1.3.1. Equation

The Klein Gordon equation is the typical and most intuitive equation in relativistic quantum mechanics. It admits real as well as complex solutions. In the absence of external electric and magnetic fields it reads:

$$-\frac{1}{c^2}\hbar^2\frac{\partial^2\psi}{\partial t^2} = m^2c^2\psi - \hbar^2\nabla^2\psi$$

**Eq 6**

With fields, The Klein Gordon for a charged particle[13] is

$$\frac{1}{c^2}\left(i\hbar\frac{\partial}{\partial t} - eV\right)^2\psi = m^2c^2\psi + \left(i\hbar\vec{\nabla} + \frac{e}{c}\vec{A}\right)^2$$

Explicitly it reads:

$$\frac{1}{c^2}\left(-\hbar^2\frac{\partial^2\psi}{\partial t^2} + e^2V^2\psi \cancel{- i\hbar\left(\frac{\partial V}{\partial t}\right)\psi} - 2i\hbar eV\frac{\partial\psi}{\partial t}\right) = m^2c^2\psi + \left(-\hbar^2\nabla^2\psi + \left(\frac{e}{c}\right)^2 A^2\psi + \cancel{i\hbar(\vec{\nabla}\cdot\vec{A})\psi} + 2i\hbar e\vec{A}\cdot\vec{\nabla}\psi\right)$$

**Eq 7**

Where we have made

$$\frac{\partial V}{\partial t} = 0; \qquad \vec{\nabla}_{r_1}\cdot\vec{A} = 0$$

which is valid in the Lorenz gauge or a stationary problem in the Coulomb gauge. Then we get

$$\frac{-\hbar^2}{c^2}\frac{\partial^2\psi}{\partial t^2} = \underbrace{-\hbar^2\nabla^2\psi}_{F_{tR}} + m^2c^2\psi - e^2\left(\frac{V^2}{c^2} - A^2\right)\psi + 2i\hbar e\left(\frac{V}{c^2}\frac{\partial\psi}{\partial t} + \vec{A}\cdot\vec{\nabla}\psi\right)$$

**Eq 8**

which in relativistic four-vector notation reads:

$$\frac{-\hbar^2}{c^2}\frac{\partial^2\psi}{\partial t^2} = -\hbar^2\nabla^2\psi + \left(m^2c^2 - e^2A^\mu A_\mu\right)\psi + 2i\hbar eA^\mu\nabla_\mu\psi$$

**Eq 9**

The interaction adds a term proportional to the 4-dimensional relativistic magnitude of the electromagnetic potential $A^\mu A_\mu$:

$$A^\mu\nabla_\mu\psi = V\frac{\partial\psi}{c^2\partial t} + \vec{A}\cdot\vec{\nabla}\psi$$

Also adds the 4-dimensional relativistic dot product between the four-vectors $A^\mu$ and $\nabla_\mu\psi$ . This last term can be considered the relativistic generalization of the dot product between $\vec{A}$ and $\vec{\nabla}\psi$.

### 2.1.3.2. Hamiltonian

In the absence of fields the Hamiltonian reads:

$$T_0^0 = \frac{\hbar^2}{2m}\left[\frac{1}{c^2}\frac{\partial\psi^*}{\partial t}\frac{\partial\psi}{\partial t} + (\vec{\nabla}\psi^*)\cdot(\vec{\nabla}\psi) + \frac{m^2c^2}{\hbar^2}\psi^*\psi\right]$$

**Eq 10**

And with fields it reads (see Greiner):

$$\mathcal{H} = \mathcal{H}_0^{KG} + \mathcal{H}_0^{em} + \mathcal{H}_I$$

where the interaction term is given by

$$\mathcal{H}_I = ie\psi^*\overleftrightarrow{\partial}_k\psi A^k + ie(\pi^*\psi^* - \pi\psi)A^k - e^2\psi^*\psi A_\mu A^\mu + e^2\psi^*\psi(A^0)^2$$

The momentum energy tensor is given by:

$$T^\mu{}_\nu = \frac{1}{4}\delta_\nu^\mu F_{\sigma\rho}F^{\sigma\rho} - \partial_\nu A_\rho F^{\rho\mu} + (T^\mu{}_\nu)_\psi$$

where

$$(T^\mu{}_\nu)_\psi = \frac{1}{2m}\Big[\left(i\hbar\partial^\mu - \frac{e}{c}A^\mu\right)\psi^*(-i\hbar\partial_\nu\psi) + (i\hbar\partial_\nu\psi^*)\left(-i\hbar\partial^\mu - \frac{e}{c}A^\mu\right) - \delta_\nu^\mu\left(i\hbar\partial_\rho - \frac{e}{c}A_\rho\right)\psi^*\left(-i\hbar\partial^\rho - \frac{e}{c}A^\rho\right)\psi + \delta_\nu^\mu m^2c^2\psi^*\psi\Big]$$

#### 2.1.3.3. Charge and current density:

From the Hamiltonian equations, the charge current density is obtained as:

$$\rho_e = \frac{ie\hbar}{2mc^2}\left(\psi^*\frac{\partial\psi}{\partial t} - \psi\frac{\partial\psi^*}{\partial t}\right) - \frac{e^2}{mc^2}A_0\psi\psi^*$$

and the charge-current density is given by:

$$\vec{j} = \frac{ie\hbar}{2m}\left(\psi^*\vec{\nabla}\psi - \psi\vec{\nabla}\psi^*\right) - \frac{e^2}{mc}\vec{A}\psi\psi^*$$

**Eq 11**

From here one can directly see two very important properties:

a) A connection between sign of frequency (or energy) and sign of charge. In the absence of electromagnetic fields, solutions with positive time derivative or frequency have positive charge. Solutions with negative time derivative or frequency have negative charge, really describing two different particles. This is possible because this equation possess 2 degrees of freedom as expected for a second degree differential equation. If that frequency is interpreted as the energy of the particle, then the sign of the charge will be provided by the sign of the energy of the solution.
b) The value of the charge and current densities depend on the presence of electromagnetic fields at the location of the Klein-Gordon fields.

### 2.1.4. Klein Gordon equation in Schrödinger form

#### 2.1.4.1. Linearization of energy:

The relativistic expression for the kinetic energy of a particle is given by

$$E^2 = (mc^2)^2 + c^2p^2$$

**Eq 12**

This is an expression quadratic in the energy. One possible (obvious) way to obtain the value of the energy is just taking the square root of the previous expression:

$$E = \sqrt{(mc^2)^2 + c^2p^2}$$

Another way is by using some algebraic steps as the following

$$E^2 - (mc^2)^2 = c^2p^2 \;\; \rightarrow \;\; (E + mc^2)(E - mc^2) = c^2p^2 \;\; \rightarrow \;\; (E - mc^2) = \frac{c^2p^2}{(E + mc^2)}$$

or

$$E = mc^2 + \frac{c^2p^2}{(E + mc^2)}$$

**Eq 13**

or even by solving the system of linear equations

$$\begin{cases} E = mc^2 + \dfrac{c^2p^2}{\alpha} \\ E = -mc^2 + \alpha \end{cases}$$

It is important to see how the definition of the Hamiltonian taken from the different previous expressions can modify the definition of the interaction of the field with electric fields. In the case of an electric potential $V$ the Hamiltonian corresponding to Eq 12 goes over:

$$H = \frac{(E - qV)^2 + (mc^2)^2 + c^2p^2}{2mc^2}$$

From here the charge is given by

$$q' = \frac{\partial H}{\partial V} = q\,\frac{(E - V)}{mc^2}$$

And from Eq 13:

$$H = mc^2 + qV + \frac{c^2p^2}{(E - qV + mc^2)}$$

Giving a charge:

$$q' = \frac{\partial H}{\partial V} = q + \frac{qc^2p^2}{(E - qV + mc^2)^2} = q\left(1 + \frac{c^2p^2}{(E - qV + mc^2)^2}\right)$$

We see here explicitly that the definition of the charge depends on the explicit form in which the Hamiltonian is written.

In a similar way we can verify that the second order differential equation

$$\frac{d^2}{dt^2}\varphi = a^2\varphi$$

can be written as the system of first order equations

$$\begin{cases} \frac{d}{dt}\varphi = a\chi \\ \frac{d}{dt}\chi = a\varphi \end{cases}$$

**Eq 14**

which can be verified by differentiating the first equation again:

$$\frac{d^2}{dt^2}\varphi = a\frac{d}{dt}\chi = a^2\varphi$$

Now, the subsystem of equations Eq 14 can be written as the spinor first order equation:

$$\frac{d}{dt}\begin{pmatrix}\varphi \\ \chi\end{pmatrix} = \begin{pmatrix}0 & a \\ a & 0\end{pmatrix}\begin{pmatrix}\varphi \\ \chi\end{pmatrix} = a\begin{pmatrix}\chi \\ \varphi\end{pmatrix}$$

Here we can identify the existence of 2 degrees of freedom provided by the upper and down components. If we take the traditional Cartesian definition for the scalar product, we would see that the two functions wouldn't generate interference because are orthogonal degrees of freedom!

So, we can see that relativity which originally can be associated with a quadratic expression in energy and give rise to a second order differential equation, can also be expressed as a system of two linear equations for the energy or equivalently as a first order differential equation in terms of a two-component spinor.

Additionally, this is the practical way how a higher order differential equation can be reduced in order to be solved by a computer algorithm that can handle only systems of first order differential equations such as Runga-Kutta.

#### 2.1.4.2. Linear Equation

The so called Klein Gordon equation in Schrödinger form is a first order differential equation that provides the same eigenenergy as the Klein Gordon equation through the Eigensolutions of the following equation:

$$i\hbar\frac{\partial}{\partial t}\Psi = \hat{H}_f\Psi \qquad \hat{H}_f = (\boldsymbol{\tau_3} + i\boldsymbol{\tau_2})\frac{\hat{p}^2}{2m} + \boldsymbol{\tau_3}mc^2$$

**Eq 15**

where

$$(\boldsymbol{\tau_3} + i\boldsymbol{\tau_2}) = \begin{pmatrix}1 & 1 \\ -1 & -1\end{pmatrix}; \qquad \boldsymbol{\tau_3} = \begin{pmatrix}1 & 0 \\ 0 & -1\end{pmatrix}$$

are the Pauli matrices and $\Psi$ is a 2 members column vector similar to the Pauli spinors:

$$\Psi = \begin{pmatrix}\varphi \\ \chi\end{pmatrix}$$

In terms of these functions the equation reads:

$$-i\hbar\frac{\partial}{\partial t}\begin{pmatrix}\varphi \\ \chi\end{pmatrix} = (\boldsymbol{\tau_3} + i\boldsymbol{\tau_2})\frac{\hat{p}^2}{2m}\begin{pmatrix}\varphi \\ \chi\end{pmatrix} + \boldsymbol{\tau_3}mc^2\begin{pmatrix}\varphi \\ \chi\end{pmatrix}$$

Now the two functions are defined in terms of the original Klein Gordon solution $\psi$ as

$$\varphi = \frac{\psi}{2} + \frac{i\hbar}{2mc^2}\frac{\partial\psi}{\partial t}$$
$$\chi = \frac{\psi}{2} - \frac{i\hbar}{2mc^2}\frac{\partial\psi}{\partial t}$$

The new functions are defined as a linear transformation of the original function and its time derivative. The Klein Gordon wavefunction and time derivative are recovered as

$$\psi = \varphi + \chi \qquad i\hbar\frac{\partial}{\partial t}\psi = mc^2(\varphi - \chi)$$

which is easily seen directly by the definition: the sum gives $\psi$ the difference gives $\frac{\partial\psi}{\partial t}$.

In order to recover the original Klein Gordon charge density and current Eq 11 one cannot use the Cartesian definition of scalar product, but it has to be replaced by

$$\varrho = e(\varphi^*\varphi - \chi^*\chi) = e\Psi^\dagger \boldsymbol{\tau}_3 \Psi$$

**Eq 16**

and

$$\vec{j} = \frac{e\hbar}{i2m}\left[\Psi^\dagger \boldsymbol{\tau}_3(\boldsymbol{\tau}_3 + i\boldsymbol{\tau}_2)\vec{\nabla}\Psi - (\vec{\nabla}\Psi^\dagger)\boldsymbol{\tau}_3(\boldsymbol{\tau}_3 + i\boldsymbol{\tau}_2)\Psi\right]$$

With these definitions, the original and the new representations provide the same densities in space. The two degrees of freedom associated with the second degree of the Klein Gordon differential equation are now encoded in the two members of the spinor used to represent the solution of this first order differential equation. In the presence of an electric potential the equation goes over

$$\left(-i\hbar\frac{\partial}{\partial t} - eV\right)\begin{pmatrix}\varphi\\ \chi\end{pmatrix} = \begin{pmatrix}1 & 1\\ -1 & -1\end{pmatrix}\frac{\hat{p}^2}{2m}\begin{pmatrix}\varphi\\ \chi\end{pmatrix} + \begin{pmatrix}1 & 0\\ 0 & -1\end{pmatrix}mc^2\begin{pmatrix}\varphi\\ \chi\end{pmatrix}$$

From here, because the equation is linear in the electric potential *V*, the charge density is given by Eq 16 which is independent from the applied electromagnetic field. This is a different expression than the one for the Klein Gordon equation. Then, even when in the absence of fields the solutions are the same, in the presence of fields are different. We can see that the fact that the density depends on the fields is not a strictly relativistic property, but it depends on the way in which the Hamiltonian is written. We see here two conflicting possible definitions for the charge and current densities. The traditional viewpoint is to recover the original Klein Gordon charge and current densities from Eq. 11. We note however, that it is possible to adopt the linear Hamiltonian from Eq 13 and derive from there the definition for the charge and current densities. The charge and current densities definition for the Dirac equation follows this last option. Only experiment can decide for the right expression.

Even when the two expressions in the absence of fields allow to calculate the same energy and wavefunction $\psi$, the properties associated with the system are not the same. In fact the linear and original Klein Gordon descriptions are not identical. They don't accept the same functions as eigenstates. This is related to the fact that the original Klein Gordon is a second order while the other is a first order differential equation. For example the original equation accepts as eigenstates

a) $\sin(kx - \omega t)$
b) $\cos(kx - \omega t)$
c) $e^{i(kx-\omega t)}$
d) $e^{-i(kx-\omega t)}$

While the linear version accepts only the last two.

In general spinors of the form $\begin{pmatrix} a \\ b \end{pmatrix}$ and $\begin{pmatrix} b \\ a \end{pmatrix}$ are mutually perpendicular when following Eq 16 as definition of dot product, because it gives *ab-ba=0*, independently of any external phase or function. For example the positive energy and negative energy wavefunctions for free particles with the same absolute value of the energy are:

$$\begin{pmatrix} A_+\varphi_0^+ \\ A_+\chi_0^+ \end{pmatrix} = \frac{1}{\sqrt{L^3}} \begin{pmatrix} (mc^2 + E_p)/\sqrt{E_p 4mc^2} \\ (mc^2 - E_p)/\sqrt{E_p 4mc^2} \end{pmatrix}$$

and

$$\begin{pmatrix} A_-\varphi_0^- \\ A_-\chi_0^- \end{pmatrix} = \frac{1}{\sqrt{L^3}} \begin{pmatrix} (mc^2 - E_p)/\sqrt{E_p 4mc^2} \\ (mc^2 + E_p)/\sqrt{E_p 4mc^2} \end{pmatrix}$$

And because they are of the form $\begin{pmatrix} a \\ b \end{pmatrix}$ and $\begin{pmatrix} b \\ a \end{pmatrix}$ they are orthogonal and will not generate interference.

This is however not true if the magnitude of the energy of the positive and negative eigensolutions is different. Negative and positive energy eigenfunctions with different eigenvalue will produce interference, however not as intense as the original Klein Gordon equation.

#### 2.1.4.3. The Feshbach-Villars representation

It is possible to perform a canonical transformation on the Hamiltonian appearing in Eq 15 that allows for fully separation of the solutions with positive energy from the solutions with negative energy. In this representation, called the Feshbach-Villars representation, it is possible to have all positive energy eigensolutions represented always by the upper component of the spinor, and negative energy always by the lower component. This has as a very important effect for the interference properties of these solutions: if the Cartesian definition of scalar product or the one provided by Eq 16 are used, positive energy solutions are always orthogonal to negative energy solutions in all space and for all times. They will never interfere with each other, even when their respective value for the magnitude of the energy be different. Therefore a new definition for the scalar product and charge/current density is required in this representation. This re-definition of charge density also comes with a reinterpretation of all operators appearing in the quantum equations, even leading to non-local expressions for the position operator.

Similar considerations can be applied to the Dirac equation, which can be considered as a combination of the Pauli spinor and the previous spinor. This explains the non-localities found in the Foldy Wouthuyssen representation.

It can be argue that the correct representation should be the one where the charge density in the absence of fields (equivalent to the norm and scalar product) is independent of time and space derivatives. The source for the Maxwell equations in the presence of fields should be taken from the expression for the Hamiltonian in this particular representation. This would take the Klein Gordon equation in Schrödinger form as the correct expression for the spin-0 particles, and not the original Klein Gordon equation. In so doing, a soliton model for pions would be possible.

### 2.1.5. Dirac Equation.

#### 2.1.5.1. Traditional

##### *2.1.5.1.1. Equation*

The Dirac equation is the most fundamental equation describing a relativistic electron[14]. In the absence of fields the equation is

$$\left(i\hbar\frac{\partial}{\partial t} - i\hbar c\vec{\boldsymbol{\alpha}}\cdot\vec{\nabla} + \boldsymbol{\beta} mc^2\right)\psi = 0$$

Where $\psi$ is a 4-members column vector called 4-spinor

$$\psi(\vec{r},t) = \begin{pmatrix} \psi_1(\vec{r},t) \\ \psi_2(\vec{r},t) \\ \psi_3(\vec{r},t) \\ \psi_4(\vec{r},t) \end{pmatrix}$$

And $\vec{\boldsymbol{\alpha}}$ , $\boldsymbol{\beta}$ are typically taken to be the 4x4 matrices called Dirac matrices, generalizations of the Pauli matrices:

$$\vec{\boldsymbol{\alpha}} = \begin{pmatrix} \begin{matrix} 0 & 0 \\ 0 & 0 \end{matrix} & \vec{\boldsymbol{\sigma}} \\ \vec{\boldsymbol{\sigma}} & \begin{matrix} 0 & 0 \\ 0 & 0 \end{matrix} \end{pmatrix}; \qquad \boldsymbol{\beta} = \begin{pmatrix} \mathbf{1} & \begin{matrix} 0 & 0 \\ 0 & 0 \end{matrix} \\ \begin{matrix} 0 & 0 \\ 0 & 0 \end{matrix} & \mathbf{-1} \end{pmatrix}$$

All what can be said at this point is that the 4-spinor is a way to encode up to 8 numbers.

With fields the equation is:

$$i\hbar\frac{\partial\psi}{\partial t} = \left(c\vec{\boldsymbol{\alpha}}\cdot\left(\vec{p} - \frac{e}{c}\vec{A}\right) + \boldsymbol{\beta} mc^2 + eV\right)\psi$$

*2.1.5.1.2. Hamiltonian*

For the Dirac equation the eigenenergy obtained as a solution of the equation is

$$E\psi = \left(c\vec{\boldsymbol{\alpha}}\cdot\left(\hat{\vec{p}} - \frac{e}{c}\vec{A}\right) + \boldsymbol{\beta} mc^2 + eV\right)\psi$$

**Eq 17**

Is the same as the energy density tensor obtained from

$$T^0{}_0 = \psi^\dagger\left(c\vec{\boldsymbol{\alpha}}\cdot\left(\hat{\vec{p}} - e\vec{A}\right) + \boldsymbol{\beta} mc^2 + eV\right)\psi$$

This is a property unique to the Dirac equation, and depends on the correct normalization of the wavefunctions to one. However simple and attractive, this equation presents practically no hint about the physics it represents!

*2.1.5.1.3. Probability density*

The probability or charge density is:

$$\rho = e\bar{\psi}\gamma_0\psi = e\psi^\dagger\psi$$

**Eq 18**

From this definition it appears that the charge density is independent from the presence of electromagnetic fields, which is wrong. Here again what is the physical meaning of the 4-spinors? Eq 18 has the consequence that for energy eigenstates, the density is proportional to the field energy:

$$\rho = \psi^\dagger\psi = \frac{1}{E}\psi^\dagger E\psi = \frac{1}{E}\psi^\dagger\hat{H}\psi = \frac{1}{E}\mathcal{E}_{Field} = \frac{1}{E}T_0^0$$

Where $\mathcal{E}_{Field} = T_0^0$ is the field energy density.

By using the Heisenberg equation of motion to get the velocity of the particle operator, Schrödinger[15] found

$$\frac{d\vec{r}}{dt} = \left[\vec{r}, \hat{H}\right] = c\vec{\boldsymbol{\alpha}}$$

which is used to provide the physical meaning of the $\vec{\boldsymbol{\alpha}}$ matrices: they represent the 'velocity' operator normalized to the velocity of light.

The best description of the current density is provided by the bilinear form $\psi^\dagger\vec{\boldsymbol{\alpha}}\psi$ as given by the Gordon[13] decomposition:

$$c\bar{\psi}\boldsymbol{\gamma^\mu}\psi = \frac{1}{2m_0}[\bar{\psi}\hat{p}^\mu\psi - (\hat{p}^\mu\bar{\psi})\psi] - \frac{i}{2m_0}\hat{p}^\nu(\bar{\psi}\hat{\boldsymbol{\sigma}}^{\boldsymbol{\mu}}{}_{\boldsymbol{\nu}}\psi)$$

where the 1-component is equivalent to

$$\vec{j}_{eD_1}(r) = 2m_0c\bar{\psi}\boldsymbol{\beta}\vec{\alpha}_1\psi = [\bar{\psi}\hat{p}^1\psi - (\hat{p}^1\bar{\psi})\psi] + \hbar\left[\vec{\nabla} \times (\bar{\psi}\,\vec{\boldsymbol{\sigma}}\psi)\right]_1 - \frac{i\hbar}{c}\frac{\partial}{\partial t}\left(\bar{\psi}\hat{\boldsymbol{\alpha}}^{\mathbf{1}}\psi\right)$$

By replacement, starting with a function in the first element of the 4-spinor we find:

$$\vec{\boldsymbol{\alpha}}_{\mathbf{1}} = 2m_0c\frac{\hat{p}^1 + \frac{\hbar}{2m_0}\left[\vec{\nabla} \times \vec{\boldsymbol{\sigma}}\right]_1 + \frac{i\hbar}{2m_0c}\frac{\partial}{\partial t}\hat{\boldsymbol{\alpha}}^{\mathbf{1}}}{2m_0c + \left(i\hbar\frac{\partial\psi}{\psi\partial t} - eV\right)}$$

**Eq 19**

Solutions of Eq 19 have been found by Barut[14].

In the traditional interpretation, the term $\frac{1}{2m_0}[\bar{\psi}\,\hat{p}^\mu\psi - (\hat{p}^\mu\bar{\psi})\psi]$ is called convection current density while $-\frac{i}{2m_0}\hat{p}^\nu(\bar{\psi}\,\hat{\boldsymbol{\sigma}}^{\boldsymbol{\mu}}{}_{\boldsymbol{\nu}}\psi)$ is the spin-current density. In the non-relativistic limit the Gordon current density reduces to

$$\vec{j} = \frac{ie\hbar}{2m}\left(\psi^\dagger\vec{\nabla}\psi - \psi\vec{\nabla}\psi^\dagger\right) - \frac{e\vec{A}}{m}\psi^\dagger\psi + \frac{1}{m}\vec{\nabla} \times \left(\psi^\dagger\vec{\boldsymbol{\sigma}}\psi\right) \equiv \rho\vec{\mathrm{v}} = \rho\frac{\vec{p} - e\vec{A}}{m} + \frac{\vec{\nabla} \times (\rho\vec{s})}{m}$$

Work performed in this direction can be found in reference 31.

The expression in the presence of electromagnetic fields can be obtained directly by the usual substitution.

#### 2.1.5.2. Decoding the traditional Dirac equation

For a long time after the discovery of the Dirac equation, a lot of arguments were developed trying to understand the physical meaning of the 4-spinors and the Dirac matrices. Even it was argued that spinors and matrices were a most fundamental way of expressing physical relationships, over space-time classical tensorial equations used in classical relativistic electromagnetism. It took many years, and the work of many people, to finally understand that the big advantage of this equation lies in its compactness and mathematical simplicity. Anyway it can be fully rewritten in terms of physical tensorial quantities[16]. The big drawback of this equation is that the price one has to pay for compactness is a deep loss in physical and intuitive understanding.

The first attempts to provide a tensorial form for the Dirac equation were developed by the disciples from the Broglie and by Takabayashi mostly in the 1950's [17] . However their results were not susceptible of simple interpretation as can be seen in the next set of 12 equations, provided by Takabayashi[18] as equivalent to the Dirac equation:

- a continuity equation for the current $S_\mu$

$$\partial_\mu S_\mu = 0$$

- a continuity-like equation for the spin density $\hat{S}_\mu$

$$\partial_\mu \hat{S}_\mu + 2\kappa\hat{\Omega} = 0$$

- the Euler equation for the flow

$$S_\nu \partial_\nu S_\mu - \hat{S}_\nu \partial_\nu \hat{S}_\mu = -P\partial_\mu P - P^{-2}\left(S_\mu \hat{S}_\nu - S_\nu \hat{S}_\mu\right) Z_\nu + iP^{-2}\varepsilon_{\mu\nu\chi\lambda} S_\nu \hat{S}_\chi K_\lambda$$

- the equation of motion for the spin distribution

$$S_\nu \partial_\nu \hat{S}_\mu - \hat{S}_\nu \partial_\nu S_\mu = -Z_\nu - P^{-1}\partial_\nu P\left(S_\mu \hat{S}_\nu - S_\nu \hat{S}_\mu\right) + iP^{-2}\varepsilon_{\mu\nu\chi\lambda} S_\nu \hat{S}_\chi Q_\lambda$$

- and two additional scalar equations

$$\sum_\mu T_{\mu\mu} + 2\kappa\Omega = 0$$

$$\sum_\mu \hat{T}_{\mu\mu} = P^{-2}\left(\hat{S}_\mu K_\mu + S_\mu Q_\mu + i\varepsilon_{\chi\lambda\mu\nu} S_\chi \hat{S}_\lambda \partial_\mu \hat{S}_\nu\right)$$

The importance of these attempts however was that they removed the idea that the equations were representing a fundamentally different way to express physical quantities and equations as that provided by tensor analysis.

##### *2.1.5.2.1. Relativistic particle Hamiltonian in the presence of a dipole interaction.*

Similarly as the Pauli equation, and by the way it was created, just following mathematical properties, the Dirac equation can be clearly considered as an effective formalism that could possibly be based on a more detailed and intuitive theory.

In order to find a physical interpretation to the Dirac equation we will consider the relativistic classical case first. In relativistic mechanics magnetic and electric dipoles cannot exist independently, both of them must coexist simultaneously, because they transform into each other by Lorentz transformations, in the same way as electric and magnetic fields do. In fact the magnetic-electric dipole moment can be described by a tensor exactly in the same way as electric and magnetic fields.

In the dipole approximation, the internal energy contents of the dipole particles is neglected. The relativistic energy equation in the dipole approximation can be written as:

$$(E - qV)^2 = mc^2 + \left(c\vec{p} - q\vec{A}\right)^2 + \vec{\mathrm{s}} \cdot \vec{B} + \vec{\mathrm{d}} \cdot \vec{E}$$

**Eq 20**

Where $\vec{\mathrm{s}}$ is the dipolar magnetic moment and $\vec{\mathrm{d}}$ the dipolar electric moment. The quantity

$$\vec{\mathrm{s}} \cdot \vec{B} + \vec{\mathrm{d}} \cdot \vec{E}$$

Is a relativistic invariant, given by the dot product between the electromagnetic field tensor and the relativistic dipole tensor as follows:

$$\frac{1}{2}\begin{pmatrix} 0 & d_x & d_y & d_z \\ -d_x & 0 & s_z & -s_y \\ -d_y & -s_z & 0 & s_x \\ -d_z & s_y & -s_x & 0 \end{pmatrix}\begin{pmatrix} 0 & E_x & E_y & E_z \\ -E_x & 0 & B_z & -B_y \\ -E_y & -B_z & 0 & B_x \\ -E_z & B_y & -B_x & 0 \end{pmatrix} = \vec{d}\cdot\vec{E} + \vec{\mathrm{s}}\cdot\vec{\mathrm{B}}$$

**Eq 21**

In the case that the particle has no electric dipole moment at rest, then the electric dipole relativistically transformed from the magnetic moment is proportional to the particle velocity:

$$\vec{\mathrm{d}} = k\vec{\mathrm{v}} = \frac{k\vec{\mathrm{p}}}{m_1}$$

**Eq 22**

and then the energy takes the form

$$(E-qV)^2 = m_0c^2 + \left(c\vec{p}-q\vec{A}\right)^2 + \vec{\mathrm{s}}\cdot\vec{H} + \frac{k\vec{\mathrm{p}}}{m_1}\cdot\vec{E}$$

The mass appearing in the denominator of the energy density is not a constant, but proportional to the energy and in first approximation would be given by

$$m_1 = m_a + \frac{E-qV}{c^2} = \frac{m_ac^2+E-qV}{c^2}$$

**Eq 23**

where $m_a$ is a constant with units of mass. Using this model we get

$$(E-qV)^2 = m_0c^2 + \left(c\vec{p}-q\vec{A}\right)^2 + \vec{\mathrm{s}}\cdot\vec{H} + \frac{c^2k\vec{\mathrm{v}}\cdot\vec{E}}{m_ac^2+E-qV}$$

**Eq 24**

Or

$$(m_ac^2+E-qV)(E-qV)^2 = (m_ac^2+E-qV)\left[m_0c^2 + \left(c\vec{p}-q\vec{A}\right)^2 + \vec{\mathrm{s}}\cdot\vec{H}\right] + k\vec{\mathrm{v}}\cdot\vec{E}$$

**Eq 25**

Eq 25 shows that this last quantity is a polinom of third degree in the energy. As a consequence the wave equation following this energy would be a third degree differential equation in time. That means that a term $\frac{\partial^3\psi}{\partial t^3}$ would appears in this equation. We will see that the Dirac equation really is a 3-order differential equation in time.

#### *2.1.5.2.2. Alternative expression for the Dirac equation*

By using the property that

$$(\vec{\boldsymbol{\alpha}}\cdot\vec{A})(\vec{\boldsymbol{\alpha}}\cdot\vec{B}) = (\vec{A}\cdot\vec{B}) + i\boldsymbol{\Sigma}\cdot(\vec{A}\times\vec{B})$$

And the commutation properties of the Dirac matrices, Dirac was able to show[19] that his equation was equivalent to

$$\left[(E-eV)^2 - \left(c\hat{\vec{p}}-e\vec{A}\right)^2 - m^2c^4 + e\hbar c\vec{\boldsymbol{\sigma}}'\cdot\vec{B} + \boldsymbol{i}e\hbar c\vec{\boldsymbol{\alpha}}\cdot\vec{E}\right]\psi = 0$$

**Eq 26**

where $\psi$ is the Dirac 4-spinor, the fields are given by

$$\vec{E} = \left(\vec{\nabla}V + \frac{\partial\vec{A}}{\partial t}\right), \qquad \vec{B} = \left(\vec{\nabla}\times\vec{A}\right)$$

and the new matrix is

$$\overrightarrow{\boldsymbol{\sigma}'} = \begin{pmatrix} \vec{\boldsymbol{\sigma}} & \begin{matrix} 0 & 0 \\ 0 & 0 \end{matrix} \\ \begin{matrix} 0 & 0 \\ 0 & 0 \end{matrix} & \vec{\boldsymbol{\sigma}} \end{pmatrix}$$

This is a 4x4 square matrix composed by the $\vec{\boldsymbol{\sigma}}$ 2-dimensional Pauli matrices. $\overrightarrow{\boldsymbol{\sigma}'}$ is a generalization of the Pauli matrices, from the definition it can be seen that the bilinear forms of this operator provide just the sum or average of the value of the Pauli matrices when calculated over the 2 upper spinor members and over the 2 lower spinor members. It is clear the equivalence between Eq 26 and Eq 20: The Dirac equation is a relativistic Hamiltonian in the dipole approximation, where the electric dipole is proportional to the particle velocity as will be explained later.

The presence of the imaginary unit in the expression Eq 26 is not a problem. It has the same meaning as the Schrödinger interaction with magnetic fields, where the product

$$\hat{\vec{p}}\cdot\vec{A} \equiv i\hbar\vec{\nabla}\cdot\vec{A}$$

also appears. The answer is the same: the eigenfunction is not a real function, but a complex function, in such a way that

$$i\hbar\vec{\nabla}\cdot\vec{A}\psi$$

Provides a real number multiplied by $\psi$. Like

$$\psi = e^{iS(r,t)}, \quad i\hbar\vec{A}\cdot\vec{\nabla}\psi = -\hbar\vec{A}\cdot\vec{\nabla}S(r,t)\psi$$

If Eq 26 is inserted in a bilinear form with the 4-spinor $\psi$, $\bar{\psi}\vec{\boldsymbol{\Sigma}}\psi$ and $\bar{\psi}i\vec{\boldsymbol{\alpha}}\psi$ represent real vectorial functions in space. The apparent complex value of the term multiplying the electric field is reduced to a real number, when the 4-spinors represent a solution for a real state. In that case it represents an electric dipole interaction with the relativistic transform of the spin $\vec{\boldsymbol{\sigma}}$ due to motion (remember that alpha is the velocity in Schrödinger interpretation of the Dirac equation).

Obviously Eq 26 provides a more physical expression than Eq 17. One can define the interaction between electromagnetic fields and the electron directly by the previous equation Eq 26, just by adding to the Klein Gordon equation, the tensorial dot product between the dipoles and the fields, as an effective additional interaction between the electron and the electromagnetic field. However this description looks very deceptive and we would like to obtain a more intuitive understanding of it. In fact we can think that this equation is not giving the full story, it may be just the dipole approximation for some internal motion of a sizeable charged electron field. We left for the present time as open the question about the interpretation of the Compton radius based on the fact that most (if not all) calculations performed in the full second quantization of quantum optics[20] take the electron field operator as the effective source for the radiation field, and on the fact that Compton[21] was able to explain the properties of radiation scattering at wavelengths of the order of the Compton wavelength by assuming a flexible-ring-like electron with a radius of the order of the Compton wavelength.

#### *2.1.5.2.3. Alternative expression for the Charge density*

When working with the Schrödinger and the Klein Gordon equation one works with a single function as the solution. All quantities of interest can be expressed in terms of this single function. When one works with spinorial equations, the quantities of interest are expressions involving spinors, namely a set of functions and not a single function as before. This raise the question: How can we compare the energy density, for example, for the solutions to the Klein Gordon and the Dirac equations?

In general the different functions building the spinor are not independent but related to each other. If for example one takes the Dirac equation and set the z axis in the direction of the spin one is not losing generality. The spinor simplifies in that the second component is null now. The third and fourth components can then be fully expressed in terms of the first one. In this way the other physical quantities such as the energy density can be written in terms of the first component alone. This allows us to perform meaningful comparisons among physical quantities like charge density and current as important examples.

In the case of energy eigenstates of the Dirac equation we have,

$$\hat{H}_D\Psi = E\Psi$$

The energy density is equal to the probability density times the eigenenergy value:

$$\Psi^{\dagger}\hat{H}_D\Psi = \Psi^{\dagger}E\Psi = E\Psi^{\dagger}\Psi = E\rho_{eD}(r)$$

And the same happens with the charge density:

$$Q\Psi^{\dagger}\Psi = Q\rho_{eD}(r)$$

Then we can conclude that in energy eigenstates the charge density is proportional to the energy density and to the probability density. In the following we are going to find an expression for these quantities that is much more intuitive than Eq 18. To this end we define the wavefunction as

$$\Psi = \begin{bmatrix}\varphi(r,t)\\ \chi(r,t)\end{bmatrix}\psi(r,t),$$

**Eq 27**

where $\varphi$ and $\chi$ is a pair of 2-component Pauli spinors. In Appendix 2.1: Charge Density for the Dirac equation we find that under the approximations that $= \begin{pmatrix}1\\0\end{pmatrix}$, $\frac{\partial\chi}{\partial t} = 0$ and the state is an energy eigenstate $\psi = \psi_{sp}(\vec{r})e^{-i\frac{E}{\hbar}t}$ the electron charge density for a state associated with the Dirac equation can be expressed as:

$$\rho_{eD} = \frac{\mathrm{e}}{E^2 + (mc^2 - e\mathrm{V})^2}\left\{\left[E^2 + \left(mc^2 - e\mathrm{V}\right)^2 + \left(e\vec{A}\right)^2 + 2\vec{s}\cdot\left(\vec{\nabla}\times\vec{A}\right)\right]\psi^*\psi + \left(\hbar c\vec{\nabla}\psi\right)^2\right\}$$

where $\vec{s} = \varphi^*\vec{\boldsymbol{\sigma}}\varphi$

Of course, this is not the most general expression, but it shows the complexity of the charge density. We are going to compare this expression for the Dirac charge density with the expression for the Klein Gordon case. In the absence of fields, and replacing $E\psi$ by $i\hbar\frac{\partial\psi}{\partial t}$ in the square bracket, the Dirac charge density can be written as:

$$\rho_{eD} = \frac{\hbar^2 ec^2}{E^2 + (mc^2)^2}\left[\frac{1}{c^2}\frac{\partial\psi^*}{\partial t}\frac{\partial\psi}{\partial t} + \left(\vec{\nabla}\psi^*\right)\cdot\left(\vec{\nabla}\psi\right) + \frac{m^2c^2}{\hbar^2}\psi^*\psi\right]$$

while the Klein Gordon charge density is:

$$\rho_{eKG} = \frac{ie\hbar}{2mc^2}\left(\psi^*\frac{\partial\psi}{\partial t} - \psi\frac{\partial\psi^*}{\partial t}\right)$$

we observe that both expressions are totally different. They refer to different physical objects. However, when we compare the Dirac energy density with the Klein Gordon **field energy density** we find:

$$\rho_{ED} = \frac{\hbar^2 E c^2}{E^2 + (mc^2)^2}\left[\frac{1}{c^2}\frac{\partial\psi^*}{\partial t}\frac{\partial\psi}{\partial t} + (\vec{\nabla}\psi^*)\cdot(\vec{\nabla}\psi) + \frac{m^2c^2}{\hbar^2}\psi^*\psi\right]$$

and

$$T^0_{0\mathrm{KG}} = \frac{\hbar^2}{2m}\left[\frac{1}{c^2}\frac{\partial\psi^*}{\partial t}\frac{\partial\psi}{\partial t} + (\vec{\nabla}\psi^*)\cdot(\vec{\nabla}\psi) + \frac{m^2c^2}{\hbar^2}\psi^*\psi\right]$$

and we can observe that the parts inside the square bracket are identical. This solves our question: in reality the probability, the charge, and the energy densities associated with the Dirac equation correspond all three to the **field energy density** associated with the Klein Gordon equation. Moreover in the presence of the scalar potential $V$ we get for the charge densities:

$$\rho_{eD} = \frac{e}{E^2 + (mc^2 - e\mathrm{V})^2}\left[\left(\hbar\frac{\partial\psi^*}{\partial t}\right)\left(\hbar\frac{\partial\psi}{\partial t}\right) + \left(mc^2 - eV\right)^2\psi^*\psi + \left(\hbar c\vec{\nabla}\psi\right)^2\right]$$

and

$$\rho_{eKG} = \frac{ie\hbar}{2mc^2}\left(\psi^*\frac{\partial\psi}{\partial t} - \psi\frac{\partial\psi^*}{\partial t}\right) - \frac{e^2}{mc}V\psi\psi^*$$

We see that the sign of the charge in the Klein Gordon equation can be reversed in the presence of very strong external fields, but that is impossible in the Dirac equation, where the potential field appears always inside a squared expression. This property we will see is fundamental in obtaining soliton-solutions in the presence of self-interaction: while the Dirac equation allows for the existence of solitons, the Klein Gordon equation prevents them completely.

However one apparent inconsistency of the Dirac equation is the following: if the current density implies a circular, spin related motion, where is its kinetic energy? In the Dirac expression for the energy density we find a term related to the interaction of this current with magnetic fields but not a term corresponding to its kinetic energy. The solution to this paradox would be the relativistic generalization to the case of the Pauli equation in Eq 5. In the same way as the Pauli equation can be seen as an effective dipole approximation for a more fundamental motion, the same can be said about the Dirac equation.

Similarly to the Feshbach-Villar transformation, the Foldy Wouthuyssen transformation allows for a representation of the Dirac equation where positive energy solutions (electrons) are represented by the two upper spinor components, while the negative energy solutions (positrons) are represented by the two lower spinor components. This property is also maintained in second quantization under the field operators' formalism.

We can summarize the properties of using Dirac 4-spinors as follows:

1. Allow to write a higher order differential equation as a first order differential equation
2. Allow for the encoding of data in the members of the 4-spinor
3. Commutation properties: it is associated with cross product or curl
4. Creation of a 4-dimensional state space
5. Modify the definition of the charge and current densities respect to a second order equation.

Now a question appears: The Dirac and the Pauli equations really represent a charged particle in the dipole approximation to a more complete equation, or are they exact expressions for a wave equation that we still don't understand completely? We will find an answer to this question later, when we analyze the quantum potential for these equations.

### 2.1.6. Maxwell equations

For completeness we introduce here the Maxwell equations.

#### 2.1.6.1. Equations

These equations can be written in mixed 4-vector notation as

$$\nabla^2 \mathrm{A}_\mu - \frac{1}{c^2}\frac{\partial^2 \mathrm{A}_\mu}{\partial t^2} = -\mu_\mu \mathrm{J}_\mu$$

In the presence of linear polarization

$$-\mu_0 \mathrm{J}_\mu = m\mathrm{A}_\mu$$

It becomes a vectorial version of the Klein Gordon equation. Despite the fundamental similarity between the Maxwell and Klein Gordon equations, these similarity is broken by second quantization where photons are considered bosons while electrons are treated as fermions, moreover the Dirac and Klein Gordon equations are traditionally interpreted as describing the electron as a point particle, while the photon is considered to have no possible space localization at all.

#### 2.1.6.2. Hamiltonian

$$H = \frac{1}{2}\sum_{k=1}^{3}\left[\left(\dot{A}^k\right)^2 + \left(\nabla A^k\right)^2\right] - \frac{1}{2}\left[\left(\dot{A}^0\right)^2 + (\nabla A^0)^2\right]$$

$$\Theta^{00} = \frac{1}{2}\left(\partial^0 \vec{A} + \vec{\nabla} A^0\right)^2 + \frac{1}{2}\left(\vec{\nabla}\times\vec{A}\right)^2 = \frac{1}{2}\left(\vec{E}^2 + \vec{B}^2\right)$$

From Jackson, the electrostatic energy is

$$W = \frac{\epsilon_0}{2}\int \left|\vec{\nabla}\mathrm{V}\right|^2 d^3x = \frac{\epsilon_0}{2}\int \left|\vec{\mathrm{E}}\right|^2 d^3x$$

## 2.2. The Chain Model.

At this point we would like to attempt to find a physical model for the equations described above.

### 2.2.1. Klein Gordon Equation

Our starting point is to consider quantum mechanics as a fluid or field theory. The evolution of this fluid or field is expressed by equations of motion containing different types of forces acting on every element of the fluid.

We start our analysis with the one dimensional chain model. This model is used to identify the forces acting between different elements of the fluid or field. Consider a physical field[1] where $\vec{\eta}$ represents the field strength. We call the distance between the chain elements $a$ as shown in Figure 2.

[1] Could also be considered a sort of vectorial fluid. In this case $\vec{\eta}$ would represent the deviation from equilibrium

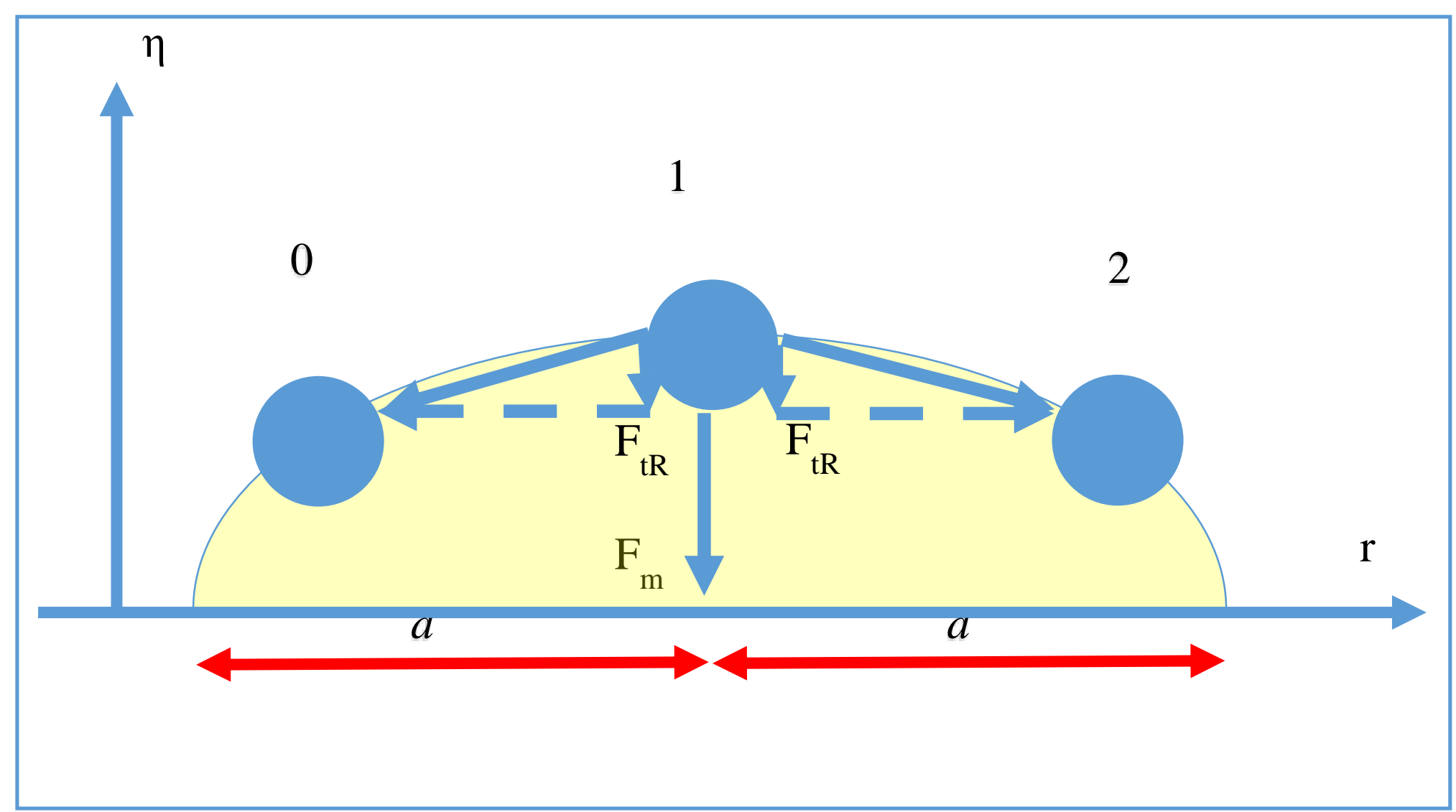

**Figure 2**

The ‘acceleration’ of the $\vec{\eta}$ coordinate of particle 1 is given by

$$\frac{\partial^2 \eta_1}{\partial t^2}$$

We will assume the existence of two forces in the $\vec{\eta}$ direction, the first one a spring-like force, trying to restore the equilibrium position and proportional to the extension of the deviation itself, given by

$-m\eta_1$ restoring force

Next a similar force proportional to the relative deviation between neighboring elements, playing the role of a tension force and given by

$k(\eta_2 - \eta_1)$ tension force

The horizontal tension is assumed to cancel by symmetry. The equation of motion for the element $\eta_1$ located at $\vec{x}_1$is

$$m\frac{\partial^2 \eta_1}{\partial t^2} = F_1 = [k(\eta_2 - \eta_1) - k(\eta_1 - \eta_0)] - m\eta_1 = \left[k\big(\eta(\vec{x}_2) - \eta(\vec{x}_1)\big) - k\big(\eta(\vec{x}_1) - \eta(\vec{x}_0)\big)\right] - m\eta(\vec{x}_1)$$

In the limit for the distance a separating the chain elements $a$ going to zero, the equation of motion takes the form

$$m\frac{\partial^2 \eta_1}{\partial t^2} = -ka^2\frac{\big(\eta(\vec{x}_1 - a) - 2\eta(\vec{x}_1) + \eta(\vec{x}_1 + a)\big)}{a^2} - m\eta(\vec{x}_1) \underbrace{\rightarrow}_{lim\, a\to 0} - ka^2\frac{d^2\eta}{dx^2} - m\eta(x)$$

**Eq 28**

Where $-ka^2\frac{d^2\eta}{dx^2}$ is the tension force, while $m\eta(x)$ is the restoring force. The equation per unit distance can be defined as

$$\frac{m}{a}\frac{\partial^2 \eta}{\partial t^2} = ka\frac{d^2\eta}{dx^2} - \frac{m}{a}\eta(\vec{x})$$

**Eq 29**

The potential energy from which the restoring force derives, is like an internal harmonic force given by

$$V_m = \frac{1}{2}m\big(\eta(\vec{x}_1)\big)^2 \underbrace{\rightarrow}_{lim\, a\to 0} \frac{1}{2}m\eta^2$$

**Eq 30**

the energy from where $m\vec{\eta}_1$ can be derived. Similarly we can find that the potential energy for the tension force is:

$$V_i = \frac{1}{2}ka^2\frac{\big(\eta(\vec{x}_2)-\eta(\vec{x}_1)\big)^2}{a^2} \underbrace{\rightarrow}_{lim\, a\to 0} \frac{1}{2}ka^2\left(\frac{\partial\eta}{\partial x}\right)^2$$

**Eq 31**

The kinetic energy density per unit length is given by:

$$T = \frac{m}{a}\left(\frac{\partial\eta}{\partial t}\right)^2$$

**Eq 32**

In the absence of external forces the Lagrangian and Hamiltonian from which the equation of motion can be derived is:

$$L = \frac{m}{a}\left(\frac{\partial\eta}{\partial t}\right)^2 - \frac{1}{2}ka^2\left(\frac{\partial\eta}{\partial x}\right)^2 - \frac{1}{2}m\eta^2$$

**Eq 33**

and

$$H = \frac{m}{a}\left(\frac{\partial\eta}{\partial t}\right)^2 + \frac{1}{2}ka^2\left(\frac{\partial\eta}{\partial x}\right)^2 + \frac{1}{2}m\eta^2$$

**Eq 34**

Providing the equation of motion:

$$\frac{m}{a}\frac{\partial^2\eta}{\partial t^2} = ka\frac{d^2\eta}{dx^2} - \frac{m}{a}\eta(\vec{x})$$

**Eq 35**

This linear model can be extended to three dimensions in space. In this case the equation is generalized to

$$\frac{m}{a}\frac{\partial^2\eta}{\partial t^2} = ka\nabla^2\eta - \frac{m}{a}\eta(\vec{x})$$

Which is nothing else than the Klein Gordon equation Eq 6!

The real solutions to this wave equation are typically waves describing propagating oscillations of the field. For harmonic waves is given by the product of the amplitude A and the phase $\varphi$ in the following form:

$$\psi = A\sin\varphi$$

A monochromatic solution is given by

$$\psi = A\sin(\omega t - kx)$$

The field speed (no the propagation speed) is given by

$$\dot{\psi} = \omega A \cos(\omega t - kx)$$

Obviously it is a time dependent quantity. The local wave amplitude is modulated by the phase and does not remain constant. Similarly the kinetic energy follows a periodic pattern characteristic of an oscillatory motion. These waves describe propagating oscillations.

This simple model in most of textbooks is provided as an example of a similar model to the Klein Gordon equation. But as we will see, it may provide something more than just an example, it may provide a real physical model for interference in quantum mechanics. Can it be considered the model for the actual **electron substance**?

It is clear that the tension force we have considered can transmit a force from the boundary conditions to any point in the wave. Also it is clear that this system in contact with boundary conditions will comply with energy and momentum conservation, as seen by the existence of a valid Hamiltonian independent of space and time.

So far we restricted the direction of the field to a single direction in space. In this case it can be seen that no property associated with the wavefunction can be found, that could be considered as a fluid or current density.

The case is different when we allow the direction of the field to rotate in a plane, like passing from linear to elliptical or circular polarization. The direction of this plane in space will be defined by an axis perpendicular to the plane. We will call this axis the 'spin axis', and the plane 'spin plane'. The solutions to the wave equation in this case can describe different types of propagating waves with different polarization. Circular polarization will be described by propagating solutions with constant amplitude and phase rotating in the spin plane. The direction of propagation doesn't need to be the same as the spin axis. It is clear that a rotating field in this plane has two projections on two perpendicular directions which follows each the one-dimensional model we had described before. In the case of circular polarization, each of the two projections oscillate 90 degrees out of phase respect to the other.

Instead of calling these two perpendicular axes as the *x* and *y* axis, we can use well known properties of complex numbers as a representation of the plane, and associate those axes with the real and imaginary axes of the complex plane. Indeed, it can be shown that in this case, the same Klein Gordon equation is able to describe the propagation of rotating waves, by just allowing complex solutions for the equations. Who called attention to the fact that complex solutions to the quantum equations were describing motion on a plane was Hestenes[22]. He proposed that the complex plane should be identified with the plane perpendicular to the spin direction. He rewrote the equations of quantum mechanics in the formalism of 'geometric algebra'. A representation of the chain, able to rotate as well as oscillate in the complex plane is shown in Figure 3.

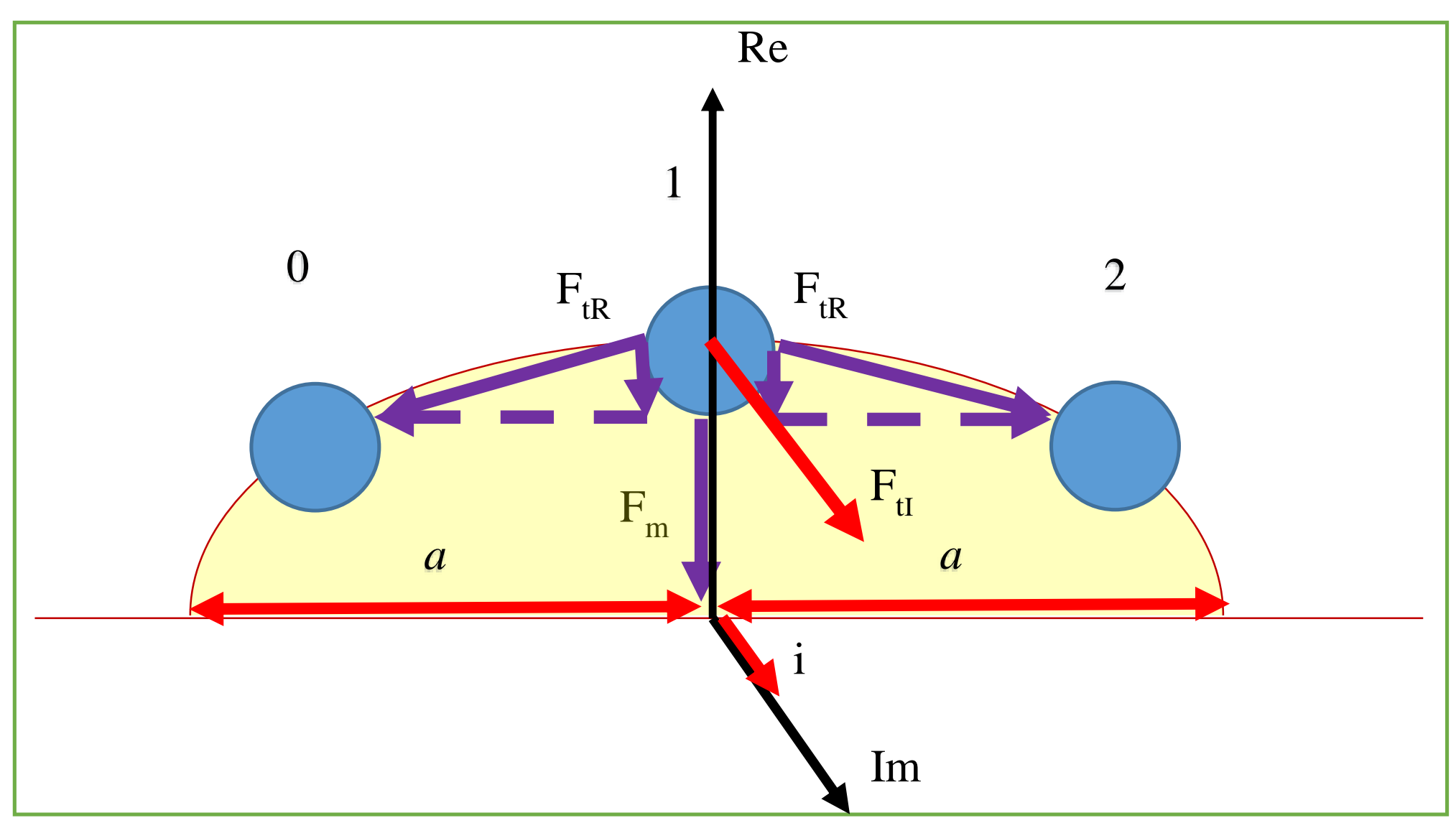

**Figure 3**

Contrary to the case of linear motion described by a single set of real numbers, on a plane the local amplitude and the phase can remain as two independent variables. The phase represents the rotation of the field on the plane while the amplitude can remain constant. This type of motion is represented in complex notation as

$$\psi = Ae^{i\varphi}$$

In a plane circular motion we have in contrast

$$\psi = Ae^{i(\omega t - kx)}$$

The field speed is given by

$$\dot{\psi} = i\omega Ae^{i(\omega t - kx)}$$

which has constant magnitude. Similarly the kinetic energy can be constant in time. These waves describe propagating rotations.

By definition, the product of two complex numbers can be represented in polar coordinates in the complex plane. If $Z_c$ is the product of the complex numbers $Z_a$ and $Z_b$,

$$Z_c = Z_a Z_b$$

the magnitude of $Z_c$ is the product of the magnitudes of $Z_a$ and $Z_b$. The angle that $Z_c$ makes in the complex plane respect to the real axis is given by sum of the angles $Z_a$ and $Z_b$ make with respect to that real axis.

An important property derived from the algebra and geometrical interpretation of complex numbers is that the product of a complex number $Z_a$ by a pure imaginary number $Z_b = ib$ (b a real number) reproduces the cross product

$$\vec{c} = \vec{b} \times \vec{a}$$

when the first vector is identified with $Z_a$ and the second vector has a magnitude equal to $b$ and is localized on the axis perpendicular to the complex plane:

$Z_c = Z_a Z_b \equiv \vec{c} = \vec{b} \times \vec{a}$

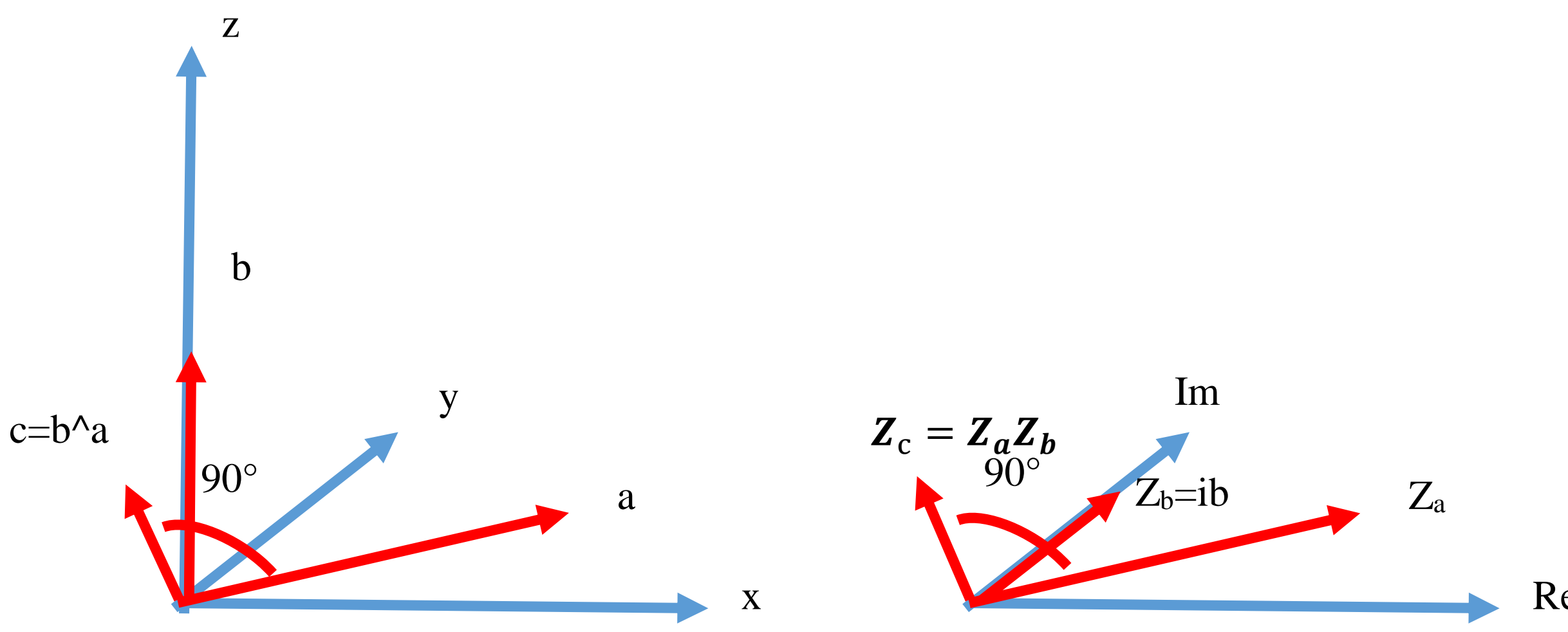

**Figure 4**

Multiplying by a pure imaginary number is equivalent to the cross product by a perpendicular vector as shown in Figure 4. This identification of the vector cross product and complex multiplication allowed Hestenes to propose that the complex wavefunction in the quantum equations really represents a vector rotating in a plane perpendicular to the direction of the “spin”. In this sense we can say that the use of complex numbers in the quantum equations represents a compact and sophisticated way to represent mathematical operations that otherwise in usual vectorial notation would be more complicated. The price for this compactness in notation is the loss of physical intuition and understanding.

For completeness, we can mention that the vector cross product also can be expressed as the product of a skew-symmetric matrix and a vector:

$$\vec{a} \times \vec{b} = [\vec{a}]_{\times} \vec{b} = \begin{bmatrix} 0 & -a_3 & a_2 \\ a_3 & 0 & -a_1 \\ -a_2 & a_1 & 0 \end{bmatrix} \begin{bmatrix} b_1 \\ b_2 \\ b_3 \end{bmatrix}$$

**Eq 36**

This form of the cross product can be generalized to relativity as we will see in the discussion about the Dirac equation.

Similarly, the use of the spinors and Pauli matrixes provides a fancy way to express the cross and dot product between vectors. We have mentioned already in Eq 3 the equivalent notation for the dot product and cross product in terms of the Pauli matrices:

$$(\vec{a} \cdot \vec{\boldsymbol{\sigma}})(\vec{b} \cdot \vec{\boldsymbol{\sigma}}) = (\vec{a} \cdot \vec{b})\hat{I} + i(\vec{a} \times \vec{b}) \cdot \vec{\boldsymbol{\sigma}}$$

When considering complex solutions for the Klein Gordon equation, the phase of the wavefunction is not more associated with the local field amplitude but with the direction of the field in the plane perpendicular to spin, and the frequency is not more a frequency of oscillation but a frequency of rotation. We see that a totally different physical picture appears when one is working with real numbers or with complex numbers, in terms of plane oscillations and rotations respectively. The equation however remains the same:

$$-\frac{1}{c^2}\hbar^2 \frac{\partial^2 \psi}{\partial t^2} = m^2 c^2 \psi - \hbar^2 \nabla^2 \psi$$

We can mention here that the interpretation we have provided here for the $-m\eta_1$ term appearing in the particle wave equation as a restoring force in the chain model, would provide an alternative to the Higgs boson as source for the electron mass.

## 2.3. De Broglie's Guiding Condition and Bohm's Quantum Potential

### 2.3.1. Introduction

#### 2.3.1.1. Effects of Boundary conditions in the absence of external fields, De Broglie

Already in 1927, by the time the Schrödinger equation just appeared and the probabilistic interpretation started being developed by the Copenhagen group, de Broglie[23] was working with wave equations and started visualizing what he called "the double solution". The double solution model can be considered a natural answer provided by quantum mechanics itself, to one of the most challenging questions of interpretation: the wave-particle duality. However 30 years later Feynman was referring to this paradox saying that interference of particles remained one of the mysteries of quantum mechanics. The reason for that was because de Broglie's ideas were crushed down in the 1927 Solvay Conference and quantum physics had to wait more than 20 years for a revival of those ideas in the works of David Bohm and De Broglie's disciples.

De Broglie considered a set of two solutions for the Klein Gordon Eq 37. First was the stationary solution:

$$u(x_0, y_0, z_0, t_0) = \frac{const.}{r_0} e^{\frac{2\pi i}{\hbar} mc^2 t_0} \qquad \left( r_0 = \sqrt{x_0^2 + y_0^2 + z_0^2} \right)$$

**Eq 38**

followed by the solution moving with speed v:

$$\psi_r(r,t) = u(x,y,z,t) = \frac{const.}{\sqrt{x^2 + y^2 + \frac{(z - \mathrm{v}t)^2}{1-\beta^2}}} e^{\frac{2\pi i}{\hbar}(Wt - pz)}$$

**Eq 39**

This second solution is nothing else than the first one moving along the z axis with speed equal to v. These solutions, especially the second one have two very important properties:

a) Its shape remains constant over time.
b) the velocity of motion of the constant shape is directly proportional to the gradient of the imaginary phase.

Respect to the property a) it is to remark that in opposition to all other known wave solutions that tend to disperse and disaggregate with time, this one remained self-confined. This was his original model for a "particle". With time these solutions gave rise to the study of 'solitons' which appear as solutions to nonlinear wave equations. Two problems presented these solutions to be considered seriously as particle models: i) they are not falling down sufficiently fast to zero from the center, and ii) they diverge at the center of symmetry. One would like the solutions to be finite everywhere, to have a finite energy content, and in order to be said that they are "localized", one would desire they to decay exponentially with distance.

Respect to the b) property one can ask if it is an indication of a direct relationship with the Hamilton Jacoby equation from classical physics? Readily, the answer to this question should be for the affirmative because Schrödinger was inspired precisely by the Hamilton Jacoby theory when devoping his quantum equation. To find if this property represents a general property of the equations and not a special case associated with this solution, we will continue working now with the Schrödinger equation for simplicity:

$$i\hbar\frac{\partial\psi}{\partial t} = -\frac{\hbar^2}{2m}\nabla^2\psi$$

**Eq 40**

We will try as solution for the Schrödinger equation the product of an internal function with fixed shape and decaying very fast to zero, multiplied by an external solution to the Schrödinger equation fulfilling the conditions imposed by the presence of external boundary conditions. It will be further assumed that the spatial dimension $\ell_i$of the fixed-shape internal solution is much smaller than the typical dimension $\ell_e$ imposed by the boundary conditions to the external solution.

$$\psi = \psi_{int}(r,t)\psi_{ext}(r,t)$$

**Eq 41**

The reasons how a fixed-shape function can have real existence will be described later. By replacing Eq 41 in Eq 40 we get:

$$i\hbar\frac{\partial\psi_{int}}{\partial t}\psi_{ext} + i\hbar\psi_{int}\frac{\partial\psi_{ext}}{\partial t} = -\frac{\hbar^2}{2m}\left(\vec{\nabla}^2\psi_{int}\psi_{ext} + \psi_{int}\vec{\nabla}^2\psi_{ext} + 2\vec{\nabla}\psi_{int}\cdot\vec{\nabla}\psi_{ext}\right)$$

**Eq 42**

We describe the internal function playing the role of a “particle” by:

$$\psi_{int}(r,t) = f(\vec{r_1}) \quad ; \quad \vec{r_1} = \vec{r} - \vec{r_0}(t); \quad \frac{d\vec{r_0}(t)}{dt} = \vec{\mathrm{v}}$$

**Eq 43**

It can be seen that the local partial time derivative of this function is given by the equation:

$$\frac{\partial\psi_{int}(r,t)}{\partial t} = -\vec{\nabla}_{r_1}f(\vec{r_1})\cdot\vec{\mathrm{v}} \equiv -\vec{\nabla}f\cdot\vec{\mathrm{v}}$$

**Eq 44**

Calculating both sides of Eq 42 we find:

$$i\hbar\frac{\partial\psi}{\partial t} = -i\hbar\vec{\nabla}f\cdot\vec{\mathrm{v}}\psi_{ext} + i\hbar\psi_{int}\frac{\partial\psi_{ext}}{\partial t}$$

**Eq 45**

$$-\frac{\hbar^2}{2m}\vec{\nabla}^2\psi = -\frac{\hbar^2}{2m}\left((\vec{\nabla}^2 f)\psi_{ext} + 2\vec{\nabla}f\cdot\vec{\nabla}\psi_\rho e^{i\left(\frac{S(\vec{r})}{\hbar}\cdot-\omega t\right)} + \frac{i}{\hbar}2\vec{\nabla}f\cdot\vec{\nabla}S\psi_{ext} + \psi_{int}\vec{\nabla}^2\psi_{ext}\right)$$

**Eq 46**

The green part is equal to the initial Schrodinger equation

$$i\hbar\frac{\partial\psi_{ext}}{\partial t} = -\frac{\hbar^2}{2m}\nabla^2\psi_{ext}$$

**Eq 47**

The fact that both functions $\psi$ and $\psi_{ext}$ are solutions of the same wave equation give the name of 'double solution' to this model. Let's take for $\psi_{ext}$ an energy eigenfunction subject to the imposed boundary conditions:

$$\psi_{ext}(r,t) = \psi_\rho(r) e^{i\left(\frac{S(\vec{r})}{\hbar} - \omega t\right)}$$

**Eq 48**

Where $S(\vec{r})$ is a function only of coordinates with units of action and $\omega$ is the Eigenfrequency of the solution. Following with our interpretation of the complex solutions, we can say that in the polar coordinates representation of the wave function in the complex plane $\psi_\rho$ and $\left(\frac{S(\vec{r})}{\hbar} - \omega t\right)$ represent amplitude and phase of the wavefunction in the spin plane.

The grey part can be written as[2]

$$\vec{\nabla}^2 f = -2\vec{\nabla} f \cdot \frac{\vec{\nabla}|\psi_{ext}|}{|\psi_{ext}|}$$

**Eq 49**

And represents a new equation that $f(\vec{r_1})$ must fulfill. This condition can be seen to be a small perturbation of order $\ell_i/\ell_e$ that can be neglected in first approximation. In order of magnitude, the gradient of the amplitude of the external function in terms of characteristic values is

$$\vec{\nabla}|\psi_{ext}| \approx \frac{|\psi_{ext}|}{l_e}$$

or

$$\frac{\vec{\nabla}|\psi_{ext}|}{|\psi_{ext}|} \approx \frac{1}{l_e}$$

Similarly the gradient of the profile of the internal function in terms of characteristic values is

$$\vec{\nabla} f \approx \frac{f}{l_i}$$

and the Laplacian is given in order of magnitude by Eq 49 as

$$\vec{\nabla}^2 f \approx \frac{\vec{\nabla} f}{l_e} \approx \frac{f}{l_i l_e}$$

The deviations in the value of $f$ (the soliton profile) from the solution with zero laplacian can be found to be

$$\delta f \approx \left(\vec{\nabla}^2 f\right) l_i^2 \approx \frac{f}{l_i l_e} l_i^2 \approx f \frac{l_i}{l_e}$$

Then in first approximation they can be neglected in the case

$$\frac{l_i}{l_e} \ll 1$$

making simply

---

[2] The author would like to acknowledge Andrew Laidlaw's private communication.

$\vec{\nabla}^2 f \cong 0.$

At this point, we should point out that in the following sections below, we will obtain solitons by solving the self-interaction problem for the electron. The self-interaction problem has the property that the solutions cannot be renormalized at will. The normalization is provided by the solution of the differential equations and cannot be modified. In that case what solves the soliton equation is not the partial function $\psi_{int}$ but the total function $\psi = \psi_{int}\psi_{ext}$. Because in general $\psi_{ext}$is not constant in magnitude, the question arises if the total function still fulfills the self-field equation after traversing positions where $\psi_{ext}$ has changed in magnitude. The total function $\psi$ cannot change shape and/or volume appreciably. We have verified that this is the case in the calculation of the guiding condition for the Klein Gordon equation, and in general is guaranteed by the fact that the quantum equations imply a continuity equation, that as long as the shape of the soliton remains relatively constant by the internal, Poisson-stress-like forces, the volume and magnitude of the total function will remain also constant, guaranteeing that still fulfills the soliton solution.

The two previous parts Eq 47 and Eq 49 are real, the only imaginary part is the yellow one, which reads

$$i\hbar\vec{\nabla}f \cdot \vec{\mathrm{v}}\psi_{ext} = \frac{\hbar^2}{2m}\frac{i}{\hbar}2\vec{\nabla}f \cdot \vec{\nabla}S\psi_{ext}$$

**Eq 50**

Or, in the direction parallel to $\vec{\nabla}f$

$$\vec{\mathrm{v}}(r,t) = \frac{\vec{\nabla}S(r,t)}{m}$$

**Eq 51**

This is a fundamental equation, which provides the basic relationship between the propagation of the soliton and the phase of the external wavefunction: the local velocity with which the front of the wave moves, is given by the gradient of the phase!! This is a fundamental equation!!

Eq 51 is an exact expression in the case when $\vec{\nabla}S = constant$ for the straight motion described in Eq 39 . However this is the value of the soliton velocity in the direction of propagation of the external field. The guidance condition says nothing about the possibility of a component perpendicular to the direction of propagation. In fact, the Dirac equation predicts a current density with a component perpendicular to that direction, given by the Gordon decomposition and interpreted as the 'spin motion'.

In the approximation that the dimensions of the 'particle' are much smaller than the geometry of the boundary conditions, the velocity of propagation of the particle is given by the gradient of the phase at the position of the particle. In the case of the hydrogen atom, $\ell_e$ would be the atomic size. In the case of a two-slit interference experiment, $\ell_e$ would be given by the size of the slits. If the dimensions of the particle are smaller than these quantities, de Broglie's model will work.

In Appendix 2.1: Charge Density for the Dirac equation we provide a derivation of the guiding condition for the Klein Gordon equation.

Eq 51 means that the phase $S(\vec{r})$ works as a velocity potential field, de Broglie called this relation the "guiding condition" for the particle. This is exactly the same role the action potential plays in classical particle mechanics for the particle velocity.

For non-trivial cases, one can just think in a double slit interference experiment without any external field applied. Here the guiding condition does not follow a straight line but the interference pattern. The solution and interference are determined exclusively by the boundary conditions.

De Broglie proposed that in the absence of certainty about the initial conditions, the external function allows a statistical interpretation. We recover the traditional view. But as it can be seen in this model the statistical interpretation is only half of the story…

#### 2.3.1.2. External fields, Bohm

The next step is to understand the action of external fields in the complex plane chain model. We can identify three types of modifications of the field $\eta$. We have seen already two: modification of the magnitude of the field, achieved by multiplying the field by a real number, and rotation of the field in the complex plane achieved by multiplication by a pure imaginary field. We are going to consider now translations of the field in space.

Suppose we have the soliton from de Broglie moving in space, and it is charged, and an electric field is present in space.

Our purpose is to write an equation that will describe what we see, namely the trajectory of a charged particle in the presence of a field. In other words, the question we would like to answer is: What equation should be followed now by the wavefunction so that the soliton follows the appropriate trajectory of the charged particle in the electric field? We will consider the situation shown in Figure 5.

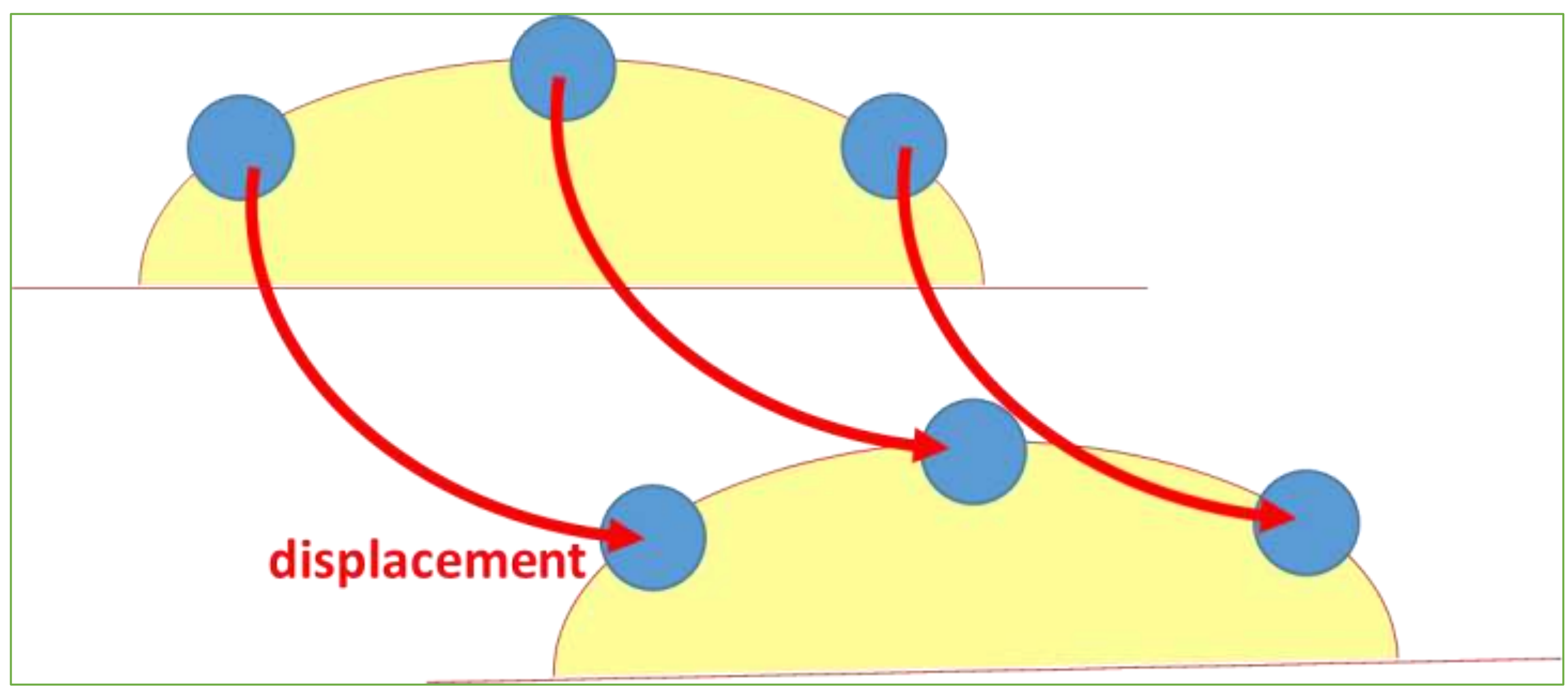

**Figure 5**

We have seen that the phase of the wavefunction plays the same role for the motion of the soliton as the action field plays for a classical particle. In both cases the 'particle' trajectory follows the gradient of an effective velocity potential or action field. Therefore we can guide us by the equations of classical mechanics in order to see how we should include the external fields in the equation for the imaginary phase in such a way that the soliton trajectory be modified by the presence of the field.

First we are going to see how we can express the equation for a velocity potential function for a classical particle following Newton's law. In the work by A. Granik[24] it is shown how the equation for the action *S* can be derived directly from Newton's equation in the presence of fields. For a non-relativistic particle in the presence of an electric potential field the equation can be derived as follows:

Newton's equation reads

$$\frac{d\vec{p}}{dt} = \frac{\partial\vec{p}}{\partial t} + \frac{1}{m}(\vec{p}\cdot\vec{\nabla})\vec{p} = -\vec{\nabla}U$$

Assuming irrotational flow in Euler picture implies

$$\vec{p} = \vec{\nabla}S$$

Replacing back

$$\vec{\nabla}\left\{\frac{\partial S}{\partial t} + \frac{1}{2m}\left(\vec{\nabla}S\right)^2 + U\right\} = 0$$

Implies

$$\frac{\partial S}{\partial t} + \frac{1}{2m}\left(\vec{\nabla}S\right)^2 + U = f(t)$$

Defines now

$$S' = S - \int f(t)dt$$

And finally gets

$$\frac{\partial S'}{\partial t} + \frac{1}{2m}\left(\vec{\nabla}S'\right)^2 + U = 0$$

**Eq 52**

Which is known as the Hamilton-Jacobi equation. It can be seen that the derivation of this equation is relatively straightforward, however in traditional textbooks on classical mechanics, this item is covered at a very high theoretical level after reading about %80 of the entire book.

By performing a Fourier decomposition of the solution for the function S in Eq 52 and applying a well-known theorem of Fourier analysis it can be shown that $\Delta(\nabla S) \times \Delta x \geq S_0$ remembering that $\nabla S = p$ we also get $\Delta p \times \Delta x \geq S_0$. Is this the uncertainty principle demonstrated in classical physics? Not yet, but we are close.

In the presence of electric and magnetic fields the Hamilton-Jacobi equation reads:

$$\left(\frac{\partial S}{\partial t} - V\right) = \frac{1}{2m}\left(\vec{\nabla}S + \frac{e}{c}\vec{A}\right)^2$$

**Eq 53**

The Hamilton-Jacoby equation can be seen as an alternative way for solving mechanical problems. Instead of starting with the forces in the Newton equation, and from there find the acceleration, integrate and find the velocity, integrate again and find the solution to the motion, one can solve this differential equation for the action *S* in the presence of the scalar and vector electromagnetic potentials, and from there find the velocity by direct calculation of the gradient of the function *S*. In problems where the total energy *E* is conserved one can replace the energy *E* for $\frac{\partial S}{\partial t}$. This is indeed an alternative way in classical physics to solve mechanical problems by knowing the potential fields, without considering the force descriptions at all.

These results show us the type of equation that the phase of the complex wave function should fulfill in the presence of potential energy fields, in a way that the resulting motion of the de Broglie "soliton" would respond to the given forces.

Because we have seen that the phase really describes the direction of the field vector in the complex plane, a visual representation of the action of the electromagnetic potentials, would be that they generate a rotation in the complex plane where the wavefunction field is located. From Eq 53 we see that a scalar potential *V* will tend to modify the rotation velocity of the field, while the vector potential *A* would tend to "bend" the direction of the phase gradient in a similar way as a torsion force. In the Hamilton Jacobi of classical particle physics one has to make the replacements

$$\frac{\partial S}{\partial t} \rightarrow \left(\frac{\partial S}{\partial t} - V\right)$$

$$\vec{\nabla} S \rightarrow \left(\vec{\nabla} S + \frac{e}{c}\vec{A}\right)$$

**Eq 54**

This short deduction could be seen as one of the most compelling reasons behind the postulation of the gauge condition in quantum physics: to obtain the desired bulk motion in the presence of external fields just make the replacements provided by Eq 54 in the quantum wave equation[3]. Now we have to answer the question: How can we achieve these replacements in the Schrödinger/Klein Gordon equation?

We have seen that the wavefunction in polar coordinates is given by

$$\psi(\vec{r},t) = \psi_\rho(\vec{r},t)e^{i\frac{S(\vec{r},t)}{\hbar}}$$

At this point is convenient to remember that

$$\rho = \frac{ie\hbar}{2mc^2}\left(\psi^*\frac{\partial \psi}{\partial t} - \psi\frac{\partial \psi^*}{\partial t}\right)$$

and

$$\vec{j} = \frac{ie\hbar}{2m}\left(\psi^*\vec{\nabla}\psi - \psi\vec{\nabla}\psi^*\right)$$

are just alternative expressions for $\psi\frac{\partial S}{\partial t}$ and $\psi\vec{\nabla}S$ respectively. We find

$$\frac{\partial \psi}{\partial t} = \frac{\partial \psi_\rho}{\partial t}e^{i\frac{S}{\hbar}} + \frac{i}{\hbar}\psi_\rho\frac{\partial S}{\partial t}e^{i\frac{S}{\hbar}}$$

$$\vec{\nabla}\psi = \vec{\nabla}\psi_\rho e^{i\frac{S}{\hbar}} + \frac{i}{\hbar}\psi_\rho\vec{\nabla}S e^{i\frac{S}{\hbar}}$$

Then we have

$$\vec{\nabla}S = Im\left(\frac{\hbar}{\psi}\vec{\nabla}\psi\right)$$

$$\frac{\partial S}{\partial t} = Im\left(\frac{\hbar}{\psi}\frac{\partial \psi}{\partial t}\right)$$

If we try the replacement

$$i\vec{\nabla}\psi \rightarrow i\vec{\nabla}\psi - \vec{A}\psi = \left(i\vec{\nabla} - \vec{A}\right)\psi$$

---

[3] Indeed considering that in order to preserve the stability of mass in front of relativistic transformation, all other interactions other than electromagnetic should follow the same receipt. This can be considered also a valid argument to provide a logical basis why all interactions have to follow equivalent transformations laws as electromagnetism, and in this way provide a basis for the universality of the Lorentz transformations without the need to postulate the speed of light in vacuum as a maximal upper limit for the transfer of energy or any other type of interaction.

$$i\frac{\partial\psi}{\partial t} \to i\frac{\partial\psi}{\partial t} - V\psi = \left(i\frac{\partial}{\partial t} - V\right)\psi$$

**Eq 55**

we see that we get the right equation for the phase or action S in the imaginary part of the equation. However we find that additional terms appear in the real part of the equation, modifying the magnitude of the field $\psi_\rho$ as well.

### 2.3.2. Schrödinger Equation

In order to understand the modifications provided by the previous changes we need to look first at the work of David Bohm[25] in 1952, who rediscovered de Broglie's guiding condition. He wrote the solution to the Schrödinger equation

$$i\hbar\frac{\partial\psi}{\partial t} = \left(-\frac{\hbar^2}{2m}\nabla^2 + V\right)\psi$$

**Eq 56**

In the polar form

$$\psi = Re^{i\frac{S}{\hbar}}$$

**Eq 57**

Then he found that the equation decomposes in an imaginary part

$$\frac{\partial R}{\partial t} = -\frac{1}{2m}\left(R\nabla^2 S + 2\vec{\nabla}R \cdot \vec{\nabla}S\right)\psi$$

**Eq 58**

and a real part

$$\frac{\partial S}{\partial t} = -\left[\frac{\left(\vec{\nabla}S\right)^2}{2m} + V + Q\right]$$

**Eq 59**

Then he argued that if the particle density is defined as

$$\rho = R^2$$

**Eq 60**

and its velocity given by

$$\vec{\mathrm{v}} = \frac{\vec{\nabla}S}{m}$$

**Eq 61**

recovering equation Eq 51 of deBroglie, then the first equation can be rewritten as a continuity equation

$$\frac{\partial\rho}{\partial t} + \vec{\nabla}\cdot(\rho\vec{v}) = 0$$

**Eq 62**

While the second equation is just the Hamilton Jacoby Eq 52 for a classical particle subject to an additional effective potential energy field given by

$$Q = -\frac{\hbar^2}{2m}\frac{\nabla^2 R}{R}$$

**Eq 63**

He called this potential energy the 'Quantum Potential'[26]. Bohm arrived at the conclusion that the Schrödinger equation is nothing else than the composition of the continuity equation and the Hamilton-Jacobi equation for a classical particle (or fluid) with the addition of an effective quantum potential energy. Immediately we can recognize that the phase of the wavefunction already follows basically the equation we were looking for. The only difference is the appearance of this new potential field from which a "quantum force" could be extracted. Bohm gave however no explanation for this potential energy, and no justification for the assumption that the particles of the fluid where following $\frac{\vec{\nabla} S}{m}$ as de Broglie did 25 years ago.

By comparison of Eq 1 and Eq 54 we see then that the replacement proposed by Eq 1 , known as the "gauge condition" is just the receipt we needed: how to modify the Schrödinger equation to include the effect of external fields.

$$i\vec{\nabla}\psi \rightarrow i\vec{\nabla}\psi - \vec{A}\psi = \left(i\vec{\nabla} - \vec{A}\right)\psi$$

$$i\frac{\partial \psi}{\partial t} \rightarrow i\frac{\partial \psi}{\partial t} - V\psi = \left(i\frac{\partial}{\partial t} - V\right)\psi$$

In the presence of the vector potential the Schrödinger equation reads:

$$\left(i\hbar\frac{\partial}{\partial t} - V\right)\psi = \frac{1}{2m}\left(-i\hbar\vec{\nabla} + \frac{e}{c}\vec{A}\right)^2\psi$$

or

$$i\hbar\frac{\partial \psi}{\partial t} = \frac{1}{2m}\left(-i\hbar\vec{\nabla} + \frac{e}{c}\vec{A}\right)^2\psi + V\psi$$

And now we know it will reproduce the classical trajectories for solitons in those cases where the quantum potential can be neglected.

At this point the quantum potential and the force that can be derived from it seems to assume a 'non physical' character, and it could not be considered a real force because it doesn't fulfill basic physics principles such as energy or momentum conservation. This point of view is wrong. We can see that the origin of this force is the tension-like force of our pristine chain model, and that force obviously was derived from a potential energy and it was possible at that time to write an energy density. In fact, the tension force from our original chain model, can be understood as the origin of the quantum potential, and we have seen that it can propagate forces from the boundary conditions to the body of the chain, and now we see that it cannot only create oscillations, but also move the chain as a whole.

Bohm recovered the guidance condition from De Broglie, but at the same time provided another interpretation for the magnitude part of the wavefunction: its squared can be interpreted as a probability function or fluid density fulfilling the continuity equation. Actually this interpretation was obtained originally by Madelung in 1927, shortly after the discovery of the Schrödinger equation. In the chain model the internal tension forces, generate an effective quantum potential field that directs the soliton to follow quantum trajectories. At the same time the magnitude of the field squared follows the continuity equation. Everything seems to work cooperatively in the Schrödinger equation!!

### 2.3.3. Klein Gordon equation

By repeating the calculation leading to Eq 51, in the presence of a magnetic field (vector potential A) de Broglie[27] found that now the guidance condition for the Klein Gordon equation was:

$$\vec{\mathrm{v}}(x, y, z, t) = -c^2 \frac{\vec{\nabla}\varphi + \frac{e}{c}\vec{\mathrm{A}}}{\frac{\partial \varphi}{\partial t} - eV}$$

**Eq 64**

We see that now the 'particle' velocity in the presence of a vector potential A is not given just by $\vec{\nabla}S$ but now by

$$\vec{\mathrm{v}} = \vec{\nabla}S - \vec{A}\,.$$

**Eq 65**

In a similar way as that followed by Bohm, Schiller[28] showed in 1962 that also the Pauli, Klein Gordon[29] and Dirac equations can be reduced to a continuity equation for a fluid and an action equation. What Schiller did was readily to rewrite classical mechanics as a wave function similar to the quantum equations for point and spinning particles with and without electromagnetic dipole moments in the non-relativistic and relativistic regime. In this way he was able to compare the differences between the classical and quantum theories and obtain explicit expressions for the quantum potential. Apart from the presence of this quantum potential, the papers by Schiller really showed that what Quantum Mechanics brought was not new physical phenomena or new mathematics, but a change in the philosophy inducing a reinterpretation of the physical phenomena and associated mathematical expressions. The action function, which classically was a mathematical sophisticated tool describing particles moving in space, in quantum mechanics was interpreted as a probability density. Our intuition is that the solution to the quantum wave equations is a real object in space, and that particles are like divergences or bumps in that wave, similar to Einstein's ideas.

The relativistic generalization of the Hamilton-Jacoby as shown by Schiller[30] is given by:

$$\left(\frac{\partial S}{\partial t} + V\right)^2 - m^2c^4 + c^2 \sum_{i,k} g^{ik} \left(\frac{\partial S}{\partial q_i} - \frac{e}{c}A_i\right)\left(\frac{\partial S}{\partial q^k} - \frac{e}{c}A_k\right) = 0$$

If we now rewrite the continuity equation in terms of this expression for the velocity, we find that the modifications in the magnitude of $\psi_\rho$ have the effect to provide the correct continuity equation!! Everything seems to work again! The replacements given in Eq 55 have the exact desired effect to modify the phase equation in the form given in Eq 54 and additionally provide the correct continuity equation!!

In classical particle physics the canonical momentum $\vec{\mathrm{P}}$ of a particle is defined by the equation

$$.\vec{\mathrm{v}} = \vec{\mathrm{P}} - \vec{A}$$

But this is just the same as Eq 65 above if we identify $\vec{\nabla}S$ with $\vec{\mathrm{P}}$. This expression justifies the identification of $\vec{\nabla}S$ with the particle canonical momentum in quantum mechanics. By extending the reasoning to $\frac{\partial S}{\partial t}$, being by considering relativity arguments involving the space-time symmetry, or just by considering the Hamilton Jacobi classical equation where $\frac{\partial S}{\partial t}$ is equal to the particle energy, it can be said that $\frac{\partial S}{\partial t}$ is the energy of the "soliton". In reality, in our model $\frac{\partial S}{\partial t}$ is the angular velocity of the field rotation in the spin-plane, but it plays also the role of total energy for the soliton particle that is moving in space.

At this point a clear understanding can be given to De Broglie's double solution function: the continuity equation together with the Hamilton Jacobi equation for the phase of the function, implies that if we have a packet-like function, the body of the function will move with a velocity given by some averaging of the gradient of the phase over the body of the wavefunction. If some internal force would play the role of the Poincare stresses and somehow keep the particle stable and free from dispersing, the theory would work. He and his disciples restarted in 1952 this direction of research working with nonlinear wave equations, which are known to generate stable soliton-solutions that would behave as particles.

In Appendix 2.2: Double Solution for the Klein Gordon self-field equation we derive the double solution model for the Klein Gordon equation in the absence of magnetic potentials, and found that given the Klein Gordon equation

$$\left(i\hbar\frac{\partial}{\partial t}-V\right)^2\psi=-c^2\hbar^2\nabla^2\psi+(mc^2)^2\psi$$

and with the assumptions

$$V=V_{int}+V_{ext}$$

$$m=m_{int}+m_{ext}$$

$$\psi_{ext}=\psi_{\rho}^{ext}(r,t)e^{i\left(\frac{S_{ext}(\vec{r})}{\hbar}-\frac{E_{ext}}{\hbar}t\right)}$$

$$\psi_{int}=\psi_{\rho}^{int}(\vec{r}_1,t)e^{-i\frac{E_{int}}{\hbar}t}$$

$$\vec{r}_1=\vec{r}-\vec{R}_0(t);\qquad \frac{d\vec{r_1}(t)}{dt}=-\frac{d\vec{R}_0(t)}{dt}=-\vec{\mathrm{v}}$$

the Klein Gordon equation splits in the following terms:

**A.** $\left(i\hbar\frac{\partial}{\partial t}-V_{ext}\right)^2\psi_{ext}=-c^2\hbar^2\nabla^2\psi_{ext}+(mc^2)^2\psi_{ext}$

which is the traditional equation for the external function with the external fields,

**B.** $(E-V_{int})^2\psi_{int}=-c^2\hbar^2\vec{\nabla}_{r_1}^2\psi_{\rho}^{int}+\left(m_{int}^2\right)c^4\psi_{int}-c^2\hbar^2 2\left(\vec{\nabla}_{r_1}\psi_{\rho}^{int}\right)\cdot\frac{\vec{\nabla}\psi_{\rho}^{ext}}{\psi_{ext}}$

which is a modified equation for the internal field due to the internal self-field and the addition of the effective extra potential due to the motion :

$$-c^2\hbar^2 2\left(\vec{\nabla}_{r_1}\psi_{\rho}^{int}\right)\cdot\frac{\vec{\nabla}\psi_{\rho}^{ext}}{\psi_{ext}}$$

and the guiding condition

**C.** $\vec{\mathrm{v}}=-c^2\frac{2\vec{\nabla}_{r_1}S_{ext}}{(2E_{int}-V_{int})}$

If the electron is at rest, the equations reduced to

$$(E-V_{int})^2\psi_{int}=-c^2\hbar^2\vec{\nabla}_{r_1}^2\psi_{\rho}^{int}+\left(m_{int}^2\right)c^4\psi_{int}$$

where $V_{int}(r)$ is the self-field given as solution of the equation

$$\nabla^2 V_{int}(r)=-\frac{1}{4\pi\epsilon_0}e\rho(r)$$

We have solved this last problem for the Dirac equation and present our results in the chapter on the electron soliton. Once the soliton has been found, the equations **A.** and **C.** can be used to solve for the motion of the electron soliton. The guiding condition is completely equivalent to the postulation of the existence of an effective 'quantum potential' Q of the form:

$$Q = -\frac{\hbar^2}{8\pi^2 m}\left(\frac{\Box|\psi_{ext}|}{|\psi_{ext}|}\right)$$

The existence of the internal function, soliton like, following the guiding condition imposed by the external wave function can resolve most of the so called quantum paradoxes, and in particular resolves unambiguously the wave-particle dichotomy. In traditional quantum mechanics only equation **A.** is solved, but we can see that in the double solution model that term represents only a partial solution to the problem.

### 2.3.4. Pauli Equation

Two different interpretation for the spin have been developed. The first one is based on the Gordon decomposition of the electron current density in terms of the convection and spin-current densities as we have already described in Section 2.1.5.1.3. In this interpretation the spin-current density in the non-relativistic dipole approximation reduces to the Stern-Gerlach term of Eq 4. We can identify two concerns with this interpretation:

a) By creating a new current density, different from $\vec{\nabla}S$ , one is moving away from a classical description of the system where $S$ can be directly identified with the classical action function.
b) There are no experimental evidence supporting the existence of contributions to the electron internal energy from the kinetic energy associated with this spin-current, or its dot product with the convection term. Why the dipole approximation works so well, not only in the non-relativistic Pauli equation, but also in the full relativistic Dirac equation?

The second interpretation, as shown by Peter Holland[32] considers the spin tensor appearing in the Gordon spin-current density not as a kinetic term, but as a new 'vector potential field', playing a similar role to the magnetic vector potential $\vec{A}$. Moreover in this interpretation the quantum potential has two parts that behave in the same way as the scalar and vectorial electromagnetic potentials $V$ an d $\vec{A}$. The quantum forces acting on the electron fluid now have an electric component and a magnetic component, following the same relativistic transformations as the electric and magnetic fields from electromagnetism. The possibility for having these two contrasting interpretations is because relativistic electric- magnetic dipoles follow the same transformation laws as the electric and magnetic fields as can be easily seen in its relativistic tensorial representation. The large advantage of this new interpretation in terms of quantum scalar and vector potential fields is that by a judicious definition of the quantum potentials, the 'missing' energy terms quadratic in the spin can be fully accounted for by counteracting terms in the potential field. The two concerns mentioned above are solved in this model: first the fluid velocity is still given by $\vec{\nabla}S$ as in a classical model, and second the equation of motion and energy expressions can be understood as exact, not merely 'dipole approximations' from a deeper model.

Now we are in a position that we can answer the question: Why the electron is best described by a spinor?

In the presence of a magnetic field the symmetry of space is locally cylindrical, and not spherical as it is in free space. The quantum potential acquires this cylindrical symmetry and imposes it to the possible electron states. As a result there are just two steady states: parallel and anti-parallel to the magnetic field. If one chose the eigenstates as the basis for the Hilbert's space, one has then two states in the absence of translations: up and down. The system can show clean resonances only at the Eigen frequencies of the Hamiltonian operator. This is what happens with the Zeeman Effect, there are only two eigenstates: parallel and anti-parallel to the magnetic field. The spin corresponds to the description of the electron on this basis. This is the reason to write the states as spinors.

#### 2.3.4.1. Pauli Quantum Potential

Defining the wave function as

$$\Psi(\vec{x},t) = \psi(\vec{x},t)\chi(\vec{x},t)$$

with a complex scalar function of position and time $\Psi(\vec{x},t)$, in polar decomposition given by

$$\psi = \sqrt{\rho} e^{i\frac{S}{\hbar}} \qquad \rho = \psi^* \psi$$

**Eq 66**

and a normalized spinor field $\chi$ where

$$\chi^\dagger \chi = 1$$

defining the spin vector field $\vec{s}(\vec{x},t)$ by

$$\vec{s} = \frac{\hbar}{2} \chi^\dagger \vec{\boldsymbol{\sigma}} \chi$$

**Eq 67**

It can be shown that the original Pauli equation

$$i\hbar \frac{\partial \Psi}{\partial t} = \left[ \frac{1}{2m} \left( \vec{\boldsymbol{\sigma}} \cdot (\hat{\vec{p}} - q\vec{A}) \right)^2 + qV \right] \Psi$$

is equivalent to

$$i\hbar \frac{\partial \Psi}{\partial t} = \left[ \frac{1}{2m} \left( -i\hbar\vec{\nabla} - q\vec{A} \right)^2 + qV \right] \Psi - \frac{q\hbar}{2m} \vec{\boldsymbol{\sigma}} \cdot \vec{B} \Psi$$

which has the same form as a Schrödinger equation for a magnetic body in the dipole approximation. In turn, as shown by Holland in 2003[32], this last equation is equal to the equation

$$i\hbar \frac{\partial \Psi}{\partial t} = \frac{1}{2m} \left( -i\hbar\vec{\nabla} + \vec{A}_Q + \frac{e}{c}\vec{A} \right)^2 \Psi + \left( V + V_Q \right) \Psi$$

**Eq 68**

when the term $\vec{A}_Q$ is given by

$$\vec{A}_Q = -\vec{\nabla} \log \rho \times \vec{s}$$

and the term $V_Q$ by

$$V_Q = Q + \frac{1}{m} \nabla S \cdot \vec{A}_Q - \frac{1}{2m} \vec{A}_Q^{\;2}$$

where $\rho$ and $\vec{s}$ are given by Eq 66 and Eq 67. In turn Eq 68 can be split into a continuity equation and a Hamilton Jacoby-like equation:

$$\frac{\partial S}{\partial t} + \frac{1}{2m}\left(\vec{\nabla}S - \vec{A}_Q\right)^2 + V_Q + V = 0$$

where allows us to interpret the expressions $\vec{A}_Q$ and $V_Q$ as a quantum vector and scalar potential respectively. This is the interpretation we were looking for! This really means that the quantum potential has not only an electric component $V_Q$, but also a magnetic component as well $\vec{A}_Q$ whose formal manifestation is the presence of the enigmatic 'spin'. The practical meaning of this formalism is that in the event a soliton grows in the fluid, it will follow trajectories determined by the equation of motion:

$$m\ddot{\vec{x}} = \vec{E}_Q + \dot{\vec{x}} \times \vec{B}_Q - \nabla V$$

where the quantum electric field is

$$\vec{E}_Q = -\nabla V_Q - \frac{\partial \vec{A}_Q}{\partial t},$$

and the quantum magnetic field by

$$\vec{B}_Q = \nabla \times \vec{A}_Q$$

$$\vec{B}_Q = -\vec{\nabla}\left(\vec{\nabla}\log\rho \cdot \vec{s}\right) + \vec{s}\nabla^2\log\rho$$

Following Schiller in 1962[30] showed that this equation can be written also as the equation of motion for a magnetic dipole in the point dipole approximation:

$$m\frac{d\vec{\mathrm{v}}}{dt} = -e\left(\vec{E} + \vec{\mathrm{v}} \times \vec{\mathrm{B}}\right) - \frac{e}{mc}\left(\vec{\nabla}\,\vec{\mathrm{B}}\right)\cdot\vec{s}$$

and in addition the spin vector is a second dynamical variable following the equation

$$\frac{d\vec{s}}{dt} = -\frac{e}{mc}\vec{s} \times \vec{\mathrm{B}} - \vec{\mathrm{F}}_Q^{Pauli}$$

**Eq 69**

Where $\vec{\mathrm{F}}_Q^{Pauli}$ is the force derived from the Pauli quantum potential:

$$\frac{\hbar^2}{8m}\vec{\nabla}\left(\Psi^\dagger\vec{\boldsymbol{\sigma}}\Psi\right):\vec{\nabla}\left(\Psi^\dagger\vec{\boldsymbol{\sigma}}\Psi\right)\frac{1}{\Psi^\dagger\Psi}$$

**Eq 70**

Which includes torques modifying the direction of the spin vector.

The equation of motion for the spin vector includes terms depending on both vector potentials, the external electromagnetic $\vec{A}$ and the quantum vector potential $\vec{A}_Q$. This behavior is shared also by the Dirac equation.

In this interpretation the expression for the energy and the equation of motion are exact, they are not a sort of 'dipole approximation', but the apparent missing terms are included in the definition of the scalar quantum potential.

#### 2.3.4.2. Chain model for the Spin Vector Potential Field

In the following we are going to follow Holland's interpretation for the Pauli quantum potential and consider an additional force in the chain model. This force will produce a similar effect to the Biot-Savart force for a magnetic field. Let's consider a vector potential field given by the cross product between the distance between neighbor chain elements and the spin-vector. We define the 'spin vector potential field' by:

$$\vec{A}_{sp} = -\frac{\vec{s}}{2} \times \hat{e}_{12}\big(\eta(\vec{x}_1) - \eta(\vec{x}_2)\big) - \frac{\vec{s}}{2} \times \hat{e}_{10}\big(\eta(\vec{x}_1) - \eta(\vec{x}_0)\big)$$
$$= -a\frac{\vec{s}}{2} \times \hat{e}_{12}\frac{\eta(\vec{x}_1) - \eta(\vec{x}_1 + a)}{a} - a\frac{\vec{s}}{2} \times \hat{e}_{10}\frac{\eta(\vec{x}_1) - \eta(\vec{x}_1 - a)}{a} \underbrace{\rightarrow}_{\lim a \to 0} - a\vec{s} \times \vec{\nabla}|\eta|$$

**Eq 71**

where

$$\hat{e}_{12} = \frac{\vec{x}_1 - \vec{x}_2}{|\vec{x}_1 - \vec{x}_2|} \qquad \hat{e}_{10} = \frac{\vec{x}_1 - \vec{x}_0}{|\vec{x}_1 - \vec{x}_0|}$$

**Eq 72**

And $\theta$ is the angle between $\vec{\nabla}\eta$ and $\vec{s}$. We call this field "spin vector potential". This vector potential would generate a rotation of the entire field in space perpendicular to the spin direction. This motion should not be confused with the rotation of the direction of the field generated by the first part of the Klein Gordon equation. Even when both motions take place in the same plane, the last one is related to the field body as a whole, while the first one is related to the rotation of the direction of the field alone. It is similar to a gyrating arrow mounted on a rotating frame of reference (like an old turntable). However both rotations typically follow different frequencies.

In the presence of the spin vector potential the Schrödinger equation would read:

$$i\hbar\frac{\partial\psi}{\partial t} = \frac{1}{2m}\left(-i\hbar\vec{\nabla} + \vec{A'}_{sp} + \frac{e}{c}\vec{A}\right)^2 \psi + V\psi$$

Where

$$\vec{A'}_{sp} = -a\vec{s} \times \vec{\nabla}$$

**Eq 73**

This force will generate a rotation of the fluid in the spin plane perpendicular to the spin vector $\vec{s}$ (Hestenes)[31]. We will add an additional term[32] in the equation that will provide the centripetal force to counteract the centrifugal force generated in the spin circular motion. Under these assumptions, the Schrödinger equation in the presence of an external vector potential A can be written as

$$-i\frac{\hbar}{c}\frac{\partial}{\partial t}\psi = \left(-\hbar i\vec{\nabla} - \vec{A'}_{sp} - e\vec{A}\right)^2 \psi + V\,\psi - \left(\vec{A'}_{sp}\right)^2 \psi$$

**Eq 74**

where

$$\left(\vec{A'}_{sp}\right)^2 \psi = \left[\left(\vec{\Sigma} \times \hbar\vec{\nabla}\right) \cdot \left(\vec{\Sigma} \times \hbar\vec{\nabla}\right)\right]\psi = \hbar^2 \sin^2\theta \left|\vec{\Sigma}\right|^2 \nabla^2\psi$$

$\theta$ is the angle between $\vec{\Sigma}$ and $\vec{\nabla}\psi$, can be incorporated into the definition of the electron mass.

In Appendix 2.3: Double solution for the Pauli equation, we find that this expression is equivalent to the classical Pauli equation:

$$\underbrace{i\hbar\frac{\partial\phi}{\partial t}=\left[\frac{1}{2m}\left(-i\hbar\vec{\nabla}-q\vec{A}\right)^{2}+qV\right]\phi}_{Schrodinger\ equation}-\underbrace{\frac{q\hbar}{2m}\vec{\boldsymbol{\sigma}}\cdot\vec{B}\phi}_{\substack{Stern\ Gerlach\\ term}}$$

We have seen that the spin vector provides the direction of the plane of rotation of the electron field due to the internal magnetic-type force. Two electrons having parallel axes of rotation can 'rotate in phase' and their field can have a certain degree of coherence. Two electrons with the spin axis antiparallel, rotating in opposite directions have not a common rotating phase and therefore their fields cannot interfere coherently. Also, due to the quantum potential, the only stable states are those where the spin direction is parallel or antiparallel to externally applied magnetic fields. This has the consequence that the energy spectra are split in pairs of levels, one with spin up and one with spin down. This behavior can be simulated by assuming they belong to two orthogonal states and can be the reason of why the electron has to be described by a spinor in the Pauli representation instead of the most natural spatial three-dimensional representation.

### 2.3.5. Dirac equation.

The Dirac equation is a 4-spinor first order differential equation in time that is the combination of the Klein-Gordon-Schrödinger equation and the Pauli equation. Despite its apparent mystical appearance defying any realistic interpretation, it allows for a real physical description as the relativistic generalization of the Pauli equation and its intrinsic spin force generating internal rotations.

In fact as we have already mentioned from the work of Schrödinger that the velocity operator is given by

$$\frac{dx}{dt}=\vec{\alpha}$$

**Eq 75**

Writing the Gordon decomposition for an individual component one finds:

$$2m_0c\bar{\psi}\beta\vec{\alpha}_1\psi+\frac{i\hbar}{c}\frac{\partial}{\partial t}(\bar{\psi}\hat{\alpha}^1\psi)-2m_0\bar{\psi}\hat{\alpha}^1eV\psi=[\bar{\psi}\hat{p}^1\psi-(\hat{p}^1\bar{\psi})\psi]+\hbar\left[\vec{\nabla}\times(\bar{\psi}\vec{\sigma}\psi)\right]_1$$

**Eq 76**

In Eq 76 can be clearly seen the spin rotation of the fluid given by the last term proportional to

$$\vec{\nabla}\times\vec{\sigma}$$

**Eq 77**

However its kinetic energy doesn't appear in the equation of motion or the Hamiltonian. Something similar we have seen in the Pauli equation, some apparent circular motion seems to interact with magnetic fields in the dipole approximation, but no kinetic energy or effect on the equation of motion is present. Holland provided an alternative interpretation for the Pauli equation solving these problems. In the following we attempt to develop a relativistic generalization of the Holland's model of 2003[32]. We propose to use the relativistic generalization of the cross product to find the relativistic quantum potential. We consider a continuous fluid model. Call $\sigma_{\mu\nu}$ the relativistic generalization of the axial vector $\vec{s}$ and has the following form

$$\vec{s} \;\rightarrow\; \sigma_{\mu\nu} = \begin{pmatrix} 0 & i\alpha_1 & i\alpha_2 & i\alpha_3 \\ -i\alpha_1 & 0 & \vec{\Sigma}_3 & -\vec{\Sigma}_2 \\ -i\alpha_2 & -\vec{\Sigma}_3 & 0 & \vec{\Sigma}_1 \\ -i\alpha_3 & \vec{\Sigma}_2 & -\vec{\Sigma}_1 & 0 \end{pmatrix}$$

**Eq 78**

The relativistic version of the cross product Eq 73 then takes the form, derived from Eq 36

$$\times \qquad \begin{pmatrix} \frac{\hbar}{c}\frac{\partial}{\partial t} \\ \hbar\vec{\nabla}_1 \\ \hbar\vec{\nabla}_2 \\ \hbar\vec{\nabla}_3 \end{pmatrix}\eta$$

$$\begin{pmatrix} 0 & i\alpha_1 & i\alpha_2 & i\alpha_3 \\ -i\alpha_1 & 0 & \vec{\Sigma}_3 & -\vec{\Sigma}_2 \\ -i\alpha_2 & -\vec{\Sigma}_3 & 0 & \vec{\Sigma}_1 \\ -i\alpha_3 & \vec{\Sigma}_2 & -\vec{\Sigma}_1 & 0 \end{pmatrix} \boxed{= \begin{pmatrix} i\vec{\alpha} \cdot \hbar\vec{\nabla} \\ -i\alpha_1 \frac{\hbar}{c}\frac{\partial}{\partial t} + \left(\vec{\Sigma} \times \hbar\vec{\nabla}\right)_1 \\ -i\alpha_2 \frac{\hbar}{c}\frac{\partial}{\partial t} + \left(\vec{\Sigma} \times \hbar\vec{\nabla}\right)_2 \\ -i\alpha_3 \frac{\hbar}{c}\frac{\partial}{\partial t} + \left(\vec{\Sigma} \times \hbar\vec{\nabla}\right)_3 \end{pmatrix}\eta}$$

**Eq 79**

and the 4-spin-potential generalization of Eq 73 is given by

$$A^{\mu}_{sp}\eta = \left( \underbrace{i\vec{\alpha} \cdot \hbar\vec{\nabla}\eta,}_{V_{sp}} \quad \underbrace{\left(-i\vec{\alpha}\frac{\hbar}{c}\frac{\partial}{\partial t} + \vec{\Sigma} \times \hbar\vec{\nabla}\right)\eta}_{\vec{A}_{sp}} \right)$$

**Eq 80**

It can be verified that by construction the relativistic dot product $A^{\mu}_{sp}\nabla_{\mu}\psi$ is null, both 4-vectors are relativistically perpendicular. Following the treatment of electromagnetic fields, the equation of motion for the continuous chain would be

$$\left(-i\frac{\hbar}{c}\frac{\partial}{\partial t} - \frac{V_{sp}}{c}\right)^2 \psi = mc^2\psi + \left(-\hbar i\vec{\nabla} - e\vec{A}_{sp}\right)^2 \psi$$

**Eq 81**

or

$$\frac{-\hbar^2}{c^2}\frac{\partial^2\psi}{\partial t^2} = -\hbar^2\nabla^2\psi + \left(m^2c^2 + e^2A^{\mu}_{sp}A_{\mu_{sp}}\right)\psi + 2i\hbar e A^{\mu}_{sp}\nabla_{\mu}\psi$$

**Eq 82**

Due to the perpendicular character $A^{\mu}_{sp}\nabla_{\mu}\psi = 0$ this is equal to

$$\frac{-\hbar^2}{c^2}\frac{\partial^2\psi}{\partial t^2} = -\hbar^2\nabla^2\psi + \left(m^2c^2 + e^2 A^{\mu}_{sp}A_{\mu_{sp}}\right)\psi$$

$$= -\hbar^2\nabla^2\psi + \left(m^2c^2 + e^2\left(\mp\left(\vec{\alpha}\cdot\hbar\vec{\nabla}\right)^2 - \left(\vec{\alpha}\frac{\hbar}{c}\frac{\partial}{\partial t}\right)^2 + \left(\vec{\Sigma}\times\hbar\vec{\nabla}\right)^2 - 2i\vec{\alpha}\frac{\hbar}{c^2}\frac{\partial}{\partial t}\cdot\left(\vec{\Sigma}\times\hbar\vec{\nabla}\right)\right)\right)\psi$$

**Eq 83**

Some authors have proposed to define an effective mass $m_e$ by the relation

$$m_e{}^2c^2 = \left(m^2c^2 + e^2 A^{\mu}_{sp}A_{\mu_{sp}}\right)$$

and that even the full electron mass is the result of the energy stored in the quantum potential $A^{\mu}_{sp}$ . This is the so called Zitterbewegung interpretation of quantum mechanics[22].

In the presence of electromagnetic fields the equation goes over:

$$\left(-i\frac{\hbar}{c}\frac{\partial}{\partial t} - \frac{V_{sp}}{c} - \frac{V}{c}\right)^2\psi = mc^2\psi + \left(-\hbar i\vec{\nabla} - e\vec{A}_{sp} - e\vec{A}\right)^2\psi$$

If the quantum four potential is defined in this case by

$$\begin{pmatrix} 0 & i\alpha_1 & i\alpha_2 & i\alpha_3 \\ -i\alpha_1 & 0 & \vec{\Sigma}_3 & -\vec{\Sigma}_2 \\ -i\alpha_2 & -\vec{\Sigma}_3 & 0 & \vec{\Sigma}_1 \\ -i\alpha_3 & \vec{\Sigma}_2 & -\vec{\Sigma}_1 & 0 \end{pmatrix} \times \begin{pmatrix} \frac{\hbar}{c}\frac{\partial}{\partial t} - eV \\ \hbar\vec{\nabla}_1 - eA_1 \\ \hbar\vec{\nabla}_2 - eA_2 \\ \hbar\vec{\nabla}_3 - eA_3 \end{pmatrix}\eta = \boxed{\begin{pmatrix} i\vec{\alpha}\cdot\hbar\vec{\nabla} \\ -i\alpha_1\left(\frac{\hbar}{c}\frac{\partial}{\partial t} - eV\right) + \left(\vec{\Sigma}\times\left(\hbar\vec{\nabla} - e\vec{A}\right)\right)_1 \\ -i\alpha_2\left(\frac{\hbar}{c}\frac{\partial}{\partial t} - eV\right) + \left(\vec{\Sigma}\times\left(\hbar\vec{\nabla} - e\vec{A}\right)\right)_2 \\ -i\alpha_3\left(\frac{\hbar}{c}\frac{\partial}{\partial t} - eV\right) + \left(\vec{\Sigma}\times\left(\hbar\vec{\nabla} - e\vec{A}\right)\right)_3 \end{pmatrix}\eta}$$

the quantum potential and the ‘canonical momentum’ remain perpendicular by definition, their dot product being zero and the equation reduces to

$$\left(-i\frac{\hbar}{c}\frac{\partial}{\partial t} - \frac{V}{c}\right)^2\psi = m_e c^2\psi + \left(-\hbar i\vec{\nabla} - e\vec{A}\right)^2\psi + 2e^2\psi A^{\mu}_{sp}A_{\mu_{sp}}$$

$$A^{\mu}_{sp}\eta = \left(\underbrace{i\vec{\alpha}\cdot\hbar\vec{\nabla}\eta}_{V_{sp}},\quad \underbrace{\left(-i\vec{\alpha}\frac{\hbar}{c}\frac{\partial}{\partial t} + \vec{\Sigma}\times\hbar\vec{\nabla}\right)\eta}_{\vec{A}_{sp}}\right)$$

$$\left(-i\frac{\hbar}{c}\frac{\partial}{\partial t} - \frac{V}{c}\right)^2\psi = m_e c^2\psi + \left(-\hbar i\vec{\nabla} - e\vec{A}\right)^2\psi + 2e^2\psi\left(i\vec{\alpha}\cdot\hbar\vec{\nabla}V + \left(-i\vec{\alpha}\frac{\hbar}{c}\frac{\partial}{\partial t} + \vec{\Sigma}\times\hbar\vec{\nabla}\right)\vec{A}\right)$$

which reduces to Eq 26:

$$\left[(E - e\phi)^2 - \left(c\hat{\vec{p}} - e\vec{A}\right)^2 - m^2c^4 + e\hbar c\vec{\boldsymbol{\sigma}} \cdot \vec{B} + \boldsymbol{i}e\hbar c\vec{\boldsymbol{\alpha}} \cdot \vec{E}\right]\psi = 0$$

The expressions for the $V_{sp}$ and $\vec{A}_{sp}$ can be obtained from the theory of tensors in the Dirac equation and the identification of $\vec{\Sigma}$ and $\vec{\alpha}$ with the vector fields defined by the decoding of the 4-spinors by those matrices. See Schiller[33]. In the 4-spinor, the direction of two spin vectors is encoded: one for the negative charge component and one for the positive charge component of the particle. The total spin is the average of the spin of both components. Encoded can be found also the value of $\vec{\alpha}$. In the same way as $\vec{\Sigma}$ can be interpreted as a magnetic dipole moment, $\vec{\alpha}$ can be interpreted as proportional to an electric dipole moment. The electron possess a magnetic dipole at rest, but not an electric dipole. The electric dipole $b\vec{\alpha}$ can be interpreted as the relativistic transform of the magnetic dipole, and therefore is proportional to the particle velocity, which is the primary interpretation of $\vec{\alpha}$ .

At this point we have found a classical field that following the chain model can provide a basic realistic interpretation for the Dirac electron. Our next step is to find under what conditions a soliton could be generated.

Moreover, the existence of a quantum 4-potential depending on the density, or shape of the fluid, makes us wonder if it were possible to engineer the shape of the object to generate a desired force? Can we model the shape of the wavefunction with an external electromagnetic field? Can we remove the external fields faster than the reaction time of the wavefunction, creating a self-propulsion shape!!

## 2.4. Self-Field Soliton

### 2.4.1. Nonlinear Equations

We have already seen that the solution used by de Broglie didn't fulfilled the desired requirements as a particle model. Let's see if it is possible to find an electron soliton model and the Poincare stresses that keep it stable!!!

There have been proposed nonlinear versions of the Schrödinger, Klein Gordon and Dirac equations. But there is a very important problem with the solutions of these equations, even when they are everywhere finite[34], stable and localized[35,36]. When one calculates the energy levels of the hydrogen atom using nonlinear terms of the same order of magnitude as the linear terms, they are not able to reproduce the experimental values. The only equations that reproduce the experimental energy levels to the best degree of precision are the original linear equations.

See discussion on nonlinear optics[37] about solitons for the needs on the differential equation to get solitons.

An important attempt to find localized solutions to the linear quantum equations was by considering the self-interaction through the electric and magnetic fields originated at the charge and current of the particle itself. Indeed the theory of mass renormalization in QED has to do exactly with attempts to remove and/or ignore the infinite energy associated with the electron self-energies. And here again the first order solution[38] (Weisskopf model) for the self-interaction diverges at the origin[39].

It can be shown that self-field interactions are equivalent to adding nonlinear terms to the equation. Let's consider a self-interaction in the nonrelativistic limit. The Schrödinger equation reads:

$$i\hbar\frac{\partial\psi}{\partial t} = \frac{1}{2m}\left(-i\hbar\vec{\nabla} + \frac{e}{c}\vec{A}\right)^2\psi + eV\psi$$

**Eq 84**

For an energy eigenstate in the absence of magnetic fields reduces to

$$\frac{\hbar^2}{2m}\nabla^2\psi = -(E - eV)\psi$$

**Eq 85**

The charge density is given by

$$\rho_e(r) = e\psi^*\psi(r)$$

**Eq 86**

Which can be inserted into the Poisson's equation to get the electrostatic field

$$\nabla^2 V(r) = -\frac{1}{4\pi\epsilon_0} e\psi^*\psi(r)$$

**Eq 87**

By taking the Laplacian of Eq 85 and replacing the electrostatic field by Eq 87 we get the following nonlinear equation:

$$\frac{\hbar^2}{2m}\nabla^2(\nabla^2\psi) = \cdots + \frac{\hbar^2}{2m}e(\nabla^2 V)\psi + \cdots = \cdots + \frac{\hbar^2}{2m}e\frac{1}{4\pi\epsilon_0}e(\psi^*\psi)\psi + \cdots$$

**Eq 88**

In this expression we can verify that the solution to the self-interaction problem really is equivalent to the addition of non-linear terms to the electron equation. It is known that this type of equations admits soliton's type solutions. Therefore it is expected that the full solution of the original equations Eq 84 and Eq 87 will provide also soliton-like solutions.

However nonlinear equations have never been seen with good eyes because quantum mechanical systems have always been considered as following a unitary evolution.

### 2.4.2. Boundary conditions

A misconception is that the 'quantum potential' is always some sort of repulsive force, like in the case of the hydrogen atom where it prevents the electron from collapsing into the central nucleus. This is wrong, the quantum potential can also generate an attractive force, all depends on the direction of $\vec{\nabla}\left(\nabla^2\left|\hat{\psi}\right|\right)$. The problem however in the generation of acceptable soliton models is that the solitons generally obtained are divergent functions.

We are going to consider further these attempts. As mentioned previously the solutions must remain finite, their energy content must be finite, and should remain stable. Let's look first at some simple cases:

The solutions to these equations are

1. $\hat{\nabla}^2\hat{\psi}(\rho) = \mathrm{a}^2\hat{\psi}(\rho) \quad \rightarrow \quad \frac{e^{\pm ar}}{ar}$ localized, diverges at 0, is normalizable over infinity

**Eq 89**

2. $\hat{\nabla}^2\hat{\psi}(\rho) = -\mathrm{a}^2\hat{\psi}(\rho) \quad \rightarrow \quad \frac{\begin{cases}\sin ar\\ \cos ar\end{cases}}{ar}$ finite at 0 but is not normalizable over infinity

**Eq 90**

3. $\hat{\nabla}^2\hat{\psi}(\rho) = 0 \cdot \hat{\psi}(\rho) \quad \rightarrow \quad \frac{1}{r}$ diverges at 0, and is not normalizable over infinity

**Eq 91**

We need that our equation behaves as case 1. Eq 89 at long distances, but behaves as case 2. Eq 90 at short distances.

### 2.4.3. Charge density

Let's consider now different cases:

### 2.4.3.1. Non-relativistic Schrödinger

$\rho_e(r) = e\psi^*\psi$

By using the Schrödinger equations described in equations Eq 84 to Eq 87, it is found that it is not possible to fulfill conditions Eq 89 - Eq 90 simultaneously, therefore they are discarded.

### 2.4.3.2. Relativistic Klein Gordon

Our equation is

$$c^2\hbar^2\nabla^2\psi = -\left[\left(E - \frac{eV}{2}\right)^2 - (mc^2)^2\right]\psi$$

**Eq 92**

So we need

$$-\left[\left(E - \frac{eV}{2}\right)^2 - (mc^2)^2\right] > 0, \qquad r \gg 1 \quad \rightarrow \left(E - \frac{eV}{2}\right)^2 < (mc^2)^2 \quad \rightarrow |E| < mc^2$$

**Eq 93**

and

$$-\left[\left(E - \frac{eV}{2}\right)^2 - (mc^2)^2\right] < 0, \qquad r \ll 1 \quad \rightarrow \left(E - \frac{eV}{2}\right)^2 > (mc^2)^2$$

**Eq 94**

We see that there is no reason why both conditions cannot be fulfilled, at infinity and at the origin! In the case of energy eigenfunctions the equations we have to solve are

$$\boxed{\begin{gathered} c^2\hbar^2\nabla^2\psi(r) = -\overbrace{\left[\left(E - \frac{eV(r)}{2}\right)^2 - (mc^2)^2\right]}^{effective\ potential\ energy}\psi(r) \\ \nabla^2 V(r) = -\frac{1}{4\pi\epsilon_0} e\left(\frac{E - eV(r)}{mc^2}\right)\psi^*\psi(r) \end{gathered}}$$

**Eq 95**

Where we made use of the expression for the charge density for a Klein Gordon field:

$$\rho_{eKG}(r) = e\left(\frac{E - eV(r)}{mc^2}\right)\psi^*\psi(r)$$

**Eq 96**

We see that the charge density depends explicitly on the electric potential field. A change in the sign of the difference between total energy and potential energy $\left(E - eV(r)\right)$, has the effect to invert the sign of the charge described by Eq 96.

Here we provide a numerical calculation where we are showing the value of the "effective" potential seen by a Klein Gordon particle in the presence of a 1/r potential field:

$$m = 1 \qquad V = \frac{1}{r} \qquad \psi = const. \qquad charge = \rho = -\frac{(E-V)}{1.49}\psi^*\psi$$

$$Dirac\ energy\colon\ m^2 - (E-V)^2$$

$$Klein\ Gordon\ energy\colon\ m^2 - sign(E-V)\cdot(E-V)^2$$

**Eq 97**

For total energy E=0.50 we get for an assumed constant wavefunction independent of position:

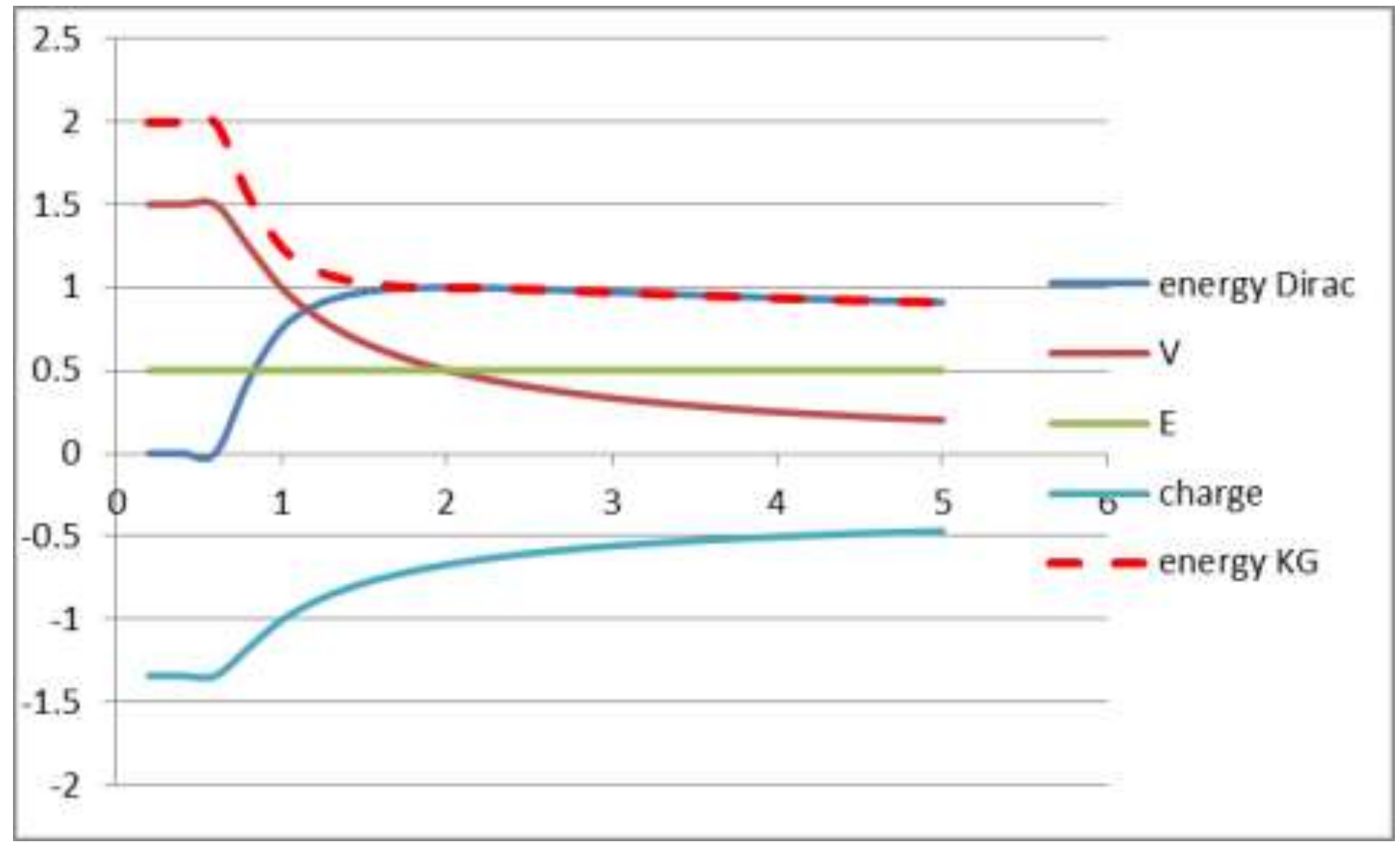

**Figure 6**

curve E shows the value of the eigenenergy, curve V shows the shape assumed for the real potential field. The line named “energy Dirac” shows the effective potential energy the Laplacian would see if the charge density were independent of the real potential field. The line named “energy KG” shows the effective potential energy the Laplacian actually see for the constant wavefunction due to the variation of the charge density with value of the real potential field. The line named charge corresponds to the charge density given by Eq 96.

We can see that if the charge density would have been independent of the value of V, the effective potential would have a minimum at the center and a localized soliton-solution would be possible, similar to the hydrogen atom solution. However in the case of the Klein Gordon equation, due to the potential-dependence of the charge density, the effective potential has not a minimum but a maximum and a confined solution is not possible. This is the reason why the initial attempts to find solitons based on the self-interaction with the Klein Gordon and Poisson equation by Rosen et al.[34] gave unsuccessful results.

It is possible that the Klein Gordon in Schrödinger form, being a first order equation in time, may imply a charge density definition, having a weak dependency on external electromagnetic fields. In this way, it may lead to the construction to solitons in a similar way as the Dirac equation. However, we have not verified this possibility so far.

## 2.5. Dirac soliton

The problems we have seen with the Klein Gordon equations to generate self-field solitons due to the external fields-dependent charge density, might be overcome by the Dirac equation. In fact in first inspection the charge density in the Dirac solutions appears to be independent of the existence of external fields:

$$\rho = e\bar{\psi}\gamma_0\psi = e\psi^{\dagger}\psi$$

**Eq 98**

We have seen however that some dependence on the fields remain. In the absence of magnetic fields:

$$\rho_{eD} = e\psi^*\psi + e\frac{\left(\hbar\vec{\nabla}\psi\right)^2}{E^2 + (mc^2 - e\mathrm{V})^2}$$

The most secure way to find if solitons are possible is to solve numerically the traditional Maxwell-Dirac equations. In the case we are going to consider we are neglecting all magnetic fields. In this approximation, if a soliton is found, it will be independent from the spin term, and will be possible by the combined action of the quantum potential and the spin-orbit self-interaction. These two effects are hidden inside the definition of the $\vec{\boldsymbol{\alpha}}$ matrix in the Dirac equation and the electron mass.

### 2.5.1. Dimensional analysis

We start by deriving the dimensionless versions of the Dirac equation by applying dimensional analysis. We define the 4-spinors we are going to use in terms of spinorial harmonics:

$$\psi_{\frac{1}{2}\frac{1}{2}} = \begin{pmatrix} ig(r)\Omega_{\frac{1}{2},0,\frac{1}{2}} \\ -f(r)\Omega_{\frac{1}{2},1,\frac{1}{2}} \end{pmatrix} = \begin{pmatrix} ig(r)\dfrac{1}{\sqrt{4\pi}}\begin{pmatrix}1\\0\end{pmatrix} \\ f(r)\sqrt{\dfrac{1}{4\pi}}\begin{pmatrix}\cos\theta \\ \sin\theta\, e^{i\phi}\end{pmatrix} \end{pmatrix} = \frac{1}{\sqrt{4\pi}}\begin{pmatrix} ig(r)\begin{pmatrix}1\\0\end{pmatrix} \\ f(r)\begin{pmatrix}\cos\theta \\ \sin\theta\, e^{i\phi}\end{pmatrix} \end{pmatrix}$$

In terms of the spinorial functions, the equations to solve in spherical coordinates are

$$c\hbar\frac{dg(r)}{dr} + (1+\kappa)c\hbar\frac{g(r)}{r} - (E - V(r) + mc^2)f(r) = 0$$

$$c\hbar\frac{df(r)}{dr} + (1-\kappa)c\hbar\frac{f(r)}{r} - (E - V(r) - mc^2)g(r) = 0$$

for an s state $j = \frac{1}{2}$; $\kappa = -1$

$$\boxed{\begin{aligned} c\hbar\frac{dg(r)}{dr} &= \left(E - \frac{eV(r)}{2} + mc^2\right)f(r) \\ c\hbar\frac{df(r)}{dr} &= c\hbar\frac{2}{r}f(r) - \left(E - \frac{eV(r)}{2} - mc^2\right)g(r) \\ \nabla^2 V(r) &= -\frac{1}{4\pi\epsilon_0}e\left(f^*f(r) + g^*g(r)\right) \end{aligned}}$$

**Eq 99**

Now we define the dimensional quantities

$$\beta = \frac{|E|}{mc^2}; \qquad \alpha = \frac{e^2}{4\pi\epsilon_0\hbar c} = \frac{1}{137};$$

$$r_0 = \frac{\hbar c}{mc^2}; \quad \rho = \frac{r}{r_0}; \quad r = \rho r_0 = \rho \frac{\hbar c}{mc^2}$$

$$V(r) = \frac{e_0}{4\pi\epsilon_0 r_0}\hat{V}(\rho); \quad \psi(r) = \frac{1}{r_0^{3/2}}\hat{\psi}(\rho)$$

And the equations reduce to

$$\boxed{\begin{aligned} \frac{\partial \hat{g}(\rho)}{\partial \rho} &= \left(1 + \left(\beta - sign(eE)\frac{\alpha}{2}\hat{V}(\rho)\right)\right)\hat{f}(\rho) \\ \frac{\partial \hat{f}(\rho)}{\partial \rho} &= \frac{2}{\rho}\hat{f}(\rho) + \left(1 - \left(\beta - sign(eE)\frac{\alpha}{2}\hat{V}(\rho)\right)\right)\hat{g}(\rho) \\ \frac{\partial^2 \hat{V}(\rho)}{\partial \rho^2} &= -\frac{2}{\rho}\frac{\partial \hat{V}(\rho)}{\partial \rho} - sign(eE)\left(\hat{f}^*\hat{f}(\rho) + \hat{g}^*\hat{g}(\rho)\right) \end{aligned}} \quad Dirac$$

It is interesting to compare at this point with the dimensionless Klein Gordon equations in spherical coordinates:

$$\boxed{\begin{aligned} \frac{\partial \hat{g}(\rho)}{\partial \rho} &= \hat{f}_1(\rho) \\ \frac{\partial \hat{f}_1(\rho)}{\partial \rho} &= -\frac{2}{\rho}\hat{f}_1(\rho) + \left[1 - \left(|\beta| - sign(eE)\frac{\alpha\hat{V}(\rho)}{2}\right)^2\right]\hat{g}(\rho) \\ \frac{\partial^2 \hat{V}(\rho)}{\partial \rho^2} &= -\frac{2}{\rho}\frac{\partial \hat{V}(\rho)}{\partial \rho} - sign(eE)\left(|\beta| - sign(eE)\alpha\hat{V}(\rho)\right)\hat{\psi}^*\hat{\psi}(\rho) \end{aligned}} \quad Klein\ Gordon$$

We notice two main differences:

a) The Dirac equation is more symmetric in the derivatives of $f$ and $g$
b) The Klein Gordon equation provides a charge density directly dependent on the presence of electric fields.

Numerical simulations show that in the case of self-fields the difference is generated not in the electron fields, but in the electric field, with a change in the sign of the charge density given by the term $|\beta| - sign(eE)\alpha\hat{V}(\rho)$.

The Hamiltonian, which also corresponds to the total energy content inside the Dirac field is given by

$$H = \int d^3x\, \psi^\dagger(\vec{x},t)\left(-i\vec{\boldsymbol{\alpha}} \cdot \vec{\nabla} + \boldsymbol{\beta} m\right)\psi(\vec{x},t)$$

The energy density in terms of our dimensional functions reads

$$H = mc^2\left[\frac{c\hbar}{mc^2}\left(-g\frac{df}{dr} + f\frac{dg}{dr} + \frac{2\kappa}{r}gf\right) + (g^2 + f^2)\right] = \frac{mc^2}{{r_0}^3}\left[\left(-\hat{g}\frac{d\hat{f}}{d\rho} + \hat{f}\frac{d\hat{g}}{d\rho} + \frac{2\kappa}{\rho}\hat{g}\hat{f}\right) + \left(\hat{g}^2 + \hat{f}^2\right)\right]$$

Integrated over the volume gives the total energy:

$$H = mc^2\int\left(-\hat{g}\frac{d\hat{f}}{d\rho} + \hat{f}\frac{d\hat{g}}{d\rho} + \frac{2\kappa}{\rho}\hat{g}\hat{f}\right)4\pi\rho^2 d\rho + mc^2\int\left(\hat{g}^2 + \hat{f}^2\right)4\pi\rho^2 d\rho$$

### 2.5.2. Boundary conditions:

Other quantities of interest given in terms of the functions are:

### 2.5.2.1. Spin:

$$\vec{S} = mc\int \langle\psi^{\dagger}|\vec{r}\times\vec{\alpha}|\psi\rangle dr^{3} = mc\int \langle\psi^{\dagger}|\vec{r}\times\vec{\alpha}|\psi\rangle r^{2}dr\sin\theta\, d\theta d\varphi$$

$$\vec{S}_z = mc\int \langle\psi^{\dagger}|\vec{r}\times\vec{\alpha}|_z|\psi\rangle r^{2}dr\sin\theta\, d\theta d\varphi = \hat{k}2mc\int fgr^{3}sin^{3}(\theta)drd\theta d\varphi = \frac{\hat{k}16\pi\hbar}{3}\int f(\rho)g(\rho)\rho^{3}d\rho$$

$$\vec{S} = \hat{k}\frac{\hbar}{2}; \qquad spin = J_d = \int 4\pi f(\rho)g(\rho)\rho^{3}d\rho = \frac{3}{8}$$

### 2.5.2.2. Charge:

$$r_e = \frac{e^{2}}{4\pi\epsilon_0 mc^{2}} \qquad ƛ_C = \frac{\hbar c}{mc^{2}} \qquad \boldsymbol{\alpha} = \frac{\boldsymbol{e^{2}}}{4\pi\epsilon_0\hbar c} \qquad \rho = \frac{r}{ƛ_C}$$

$$q = e\int 4\pi r^{2}(g^{*}g + f^{*}f)dr = e\int\left(\hat{g}^{2} + \hat{f}^{2}\right)4\pi\rho^{2}d\rho = e(m_d + g_d)$$

$$\alpha_{out} = \frac{q^{2}}{4\pi\epsilon_0\hbar c} = \frac{e^{2}(m_d + g_d)^{2}}{4\pi\epsilon_0\hbar c} = \alpha_{in}\left(m_d(\alpha_{in}) + g_d(\alpha_{in})\right)^{2}$$

### 2.5.2.3. Mass-energy

$$\mathrm{m_0c^{2}} = \beta mc^{2} = mc^{2}\int\left(-\hat{g}\frac{d\hat{f}}{d\rho} + \hat{f}\frac{d\hat{g}}{d\rho} + \frac{2\kappa}{\rho}\hat{g}\hat{f}\right)4\pi\rho^{2}d\rho + mc^{2}\int\left(\hat{g}^{2} + \hat{f}^{2}\right)4\pi\rho^{2}d\rho + 4\pi\epsilon_0\int 4\pi r^{2}\frac{\overbrace{\vec{\nabla}V(r)}^{-\vec{E}}{}^{2}}{2}dr$$

Here we add to the electron mass the electrostatic energy. This is consistent with the idea that the non-radiation fields are part of the electron, and are 'attached' to it.

### 2.5.2.4. Similarity transformations:

We can find scaling laws for the solutions when the parameter $\alpha$ is modified. The equations are:

$$\hat{\nabla}^{2}\hat{\psi}(\rho) = \left[1 - \left(|\beta| - sign(eE)\frac{\alpha\hat{V}(\rho)}{2}\right)^{2}\right]\hat{\psi}(\rho)$$

$$\hat{\nabla}^{2}\hat{V}(\rho) = -sign(eE)\left(|\beta| - sign(eE)\alpha\hat{V}(\rho)\right)\hat{\psi}^{*}\hat{\psi}(\rho) \qquad \text{for Klein Gordon charge density}$$

$$\hat{\nabla}^{2}\hat{V}(\rho) = -sign(eE)|\beta|\hat{\psi}^{*}\hat{\psi}(\rho) \qquad \text{for Dirac charge density}$$

Define the new variables

$$\alpha' = \tau\alpha$$

$$\hat{V}' = \frac{\hat{V}}{\tau}$$

$$\hat{\psi}' = \frac{\hat{\psi}}{\sqrt{\tau}}$$

Now replace the old variables in function of the new variables:

$$\sqrt{\tau}\hat{\nabla}^2\hat{\psi}'(\rho) = \left[1 - \left(|\beta| - sign(eE)\frac{\alpha'\tau\hat{V}'(\rho)}{2\tau}\right)^2\right]\sqrt{\tau}\hat{\psi}'(\rho)$$

$$\tau\hat{\nabla}^2\hat{V}'(\rho) = -sign(eE)\left(|\beta| - sign(eE)\frac{\alpha'\tau\hat{V}'(\rho)}{\tau}\right)\tau\hat{\psi}^{*\prime}\hat{\psi}'(\rho)$$ for Klein Gordon charge density

$$\tau\hat{\nabla}^2\hat{V}'(\rho) = -sign(eE)|\beta|\tau\hat{\psi}^{*\prime}\hat{\psi}'(\rho)$$ for Dirac charge density

Which simplify to:

$$\hat{\nabla}^2\hat{\psi}'(\rho) = \left[1 - \left(|\beta| - sign(eE)\frac{\alpha'\hat{V}'(\rho)}{2}\right)^2\right]\hat{\psi}'(\rho)$$

$$\hat{\nabla}^2\hat{V}'(\rho) = -sign(eE)\left(|\beta| - sign(eE)\alpha'\hat{V}'(\rho)\right)\hat{\psi}^{*\prime}\hat{\psi}'(\rho)$$ for Klein Gordon charge density

$$\hat{\nabla}^2\hat{V}'(\rho) = -sign(eE)|\beta|\hat{\psi}^{*\prime}\hat{\psi}'(\rho)$$ for Dirac charge density

And we see that the new variables fulfill exactly the same equation than the old ones. This means that if we had a solution in the old variables, the same solution still holds in the new variables as well. In other words when the $\alpha$ parameter is multiplied by a factor $\tau$, the new solutions can be found by dividing the electric potential by $\tau$ and the wavefunction by $\sqrt{\tau}$. The shape of the new functions is the same as of the old functions. This means that we need to find the solutions for a fix value of $\alpha$, and then we can find the solutions for other values by using the founded scaling laws.

We were not able to find a simple relationship for the modifications of the solutions when we modify the $\beta$ parameter. It can be found only by explicit numerical integration of the equations.

### 2.5.3. Numerical solution

The asymptotic solution for the wave function at infinity in spherical coordinates is

$$f = \frac{Ae^{-kr}}{r}$$

the asymptotic solution for the electric potential is

$$V = \frac{Be}{r}$$

Where *A* and *B* are amplitudes to be determined. We take $\alpha$ and $\beta$ as parameters. We start our integration at infinity using the asymptotic solutions with arbitrary values of *A* and *B* and go to the center at $\rightarrow 0$ . Typically one finds that for our starting initial conditions the solutions diverge. Then we vary the amplitudes *A* and *B* until the solution at the origin converges to a finite value or zero for all functions of interest. Next step we vary $\alpha$ and $\beta$ and repeat the procedure. In this way we find that for all $\beta$ in the range $0 < \beta < 1$, and for a huge range of $\alpha$ from $\alpha \ll 1$ to $\alpha \gg 1$, as far as we can say solutions can be found. Then the "real" electron solution will be the one where

1. the energy eigenvalue $E = \hbar\omega$ of the Dirac equation equals the expression for the field Hamiltonian energy Eq 15 and simultaneously is equal to the experimentally measured electron mass,
2. the total charge is equal to the measured electron charge and
3. the angular momentum of the system is $\hbar/2$

The condition 1. on the energy above is really the same as the quantization of the electromagnetic field, the total integral of the energy density of the field in space $E$ is made equal to $E = \hbar\omega$, the oscillation frequency times Planck constant.

We solved it for $\kappa = -1$ and found the following values: $\alpha = 0.101 \cong {}^{1}/_{9.9}$ (about ten times the measured value of $\alpha$), and $\beta = 0.35$ . Figure 7 shows the variation of the total energy-mass, charge and other integrated quantities for the range $0 < \beta < 1$, keeping the spin value constant equal to $\frac{3}{8} = 0.375$ .

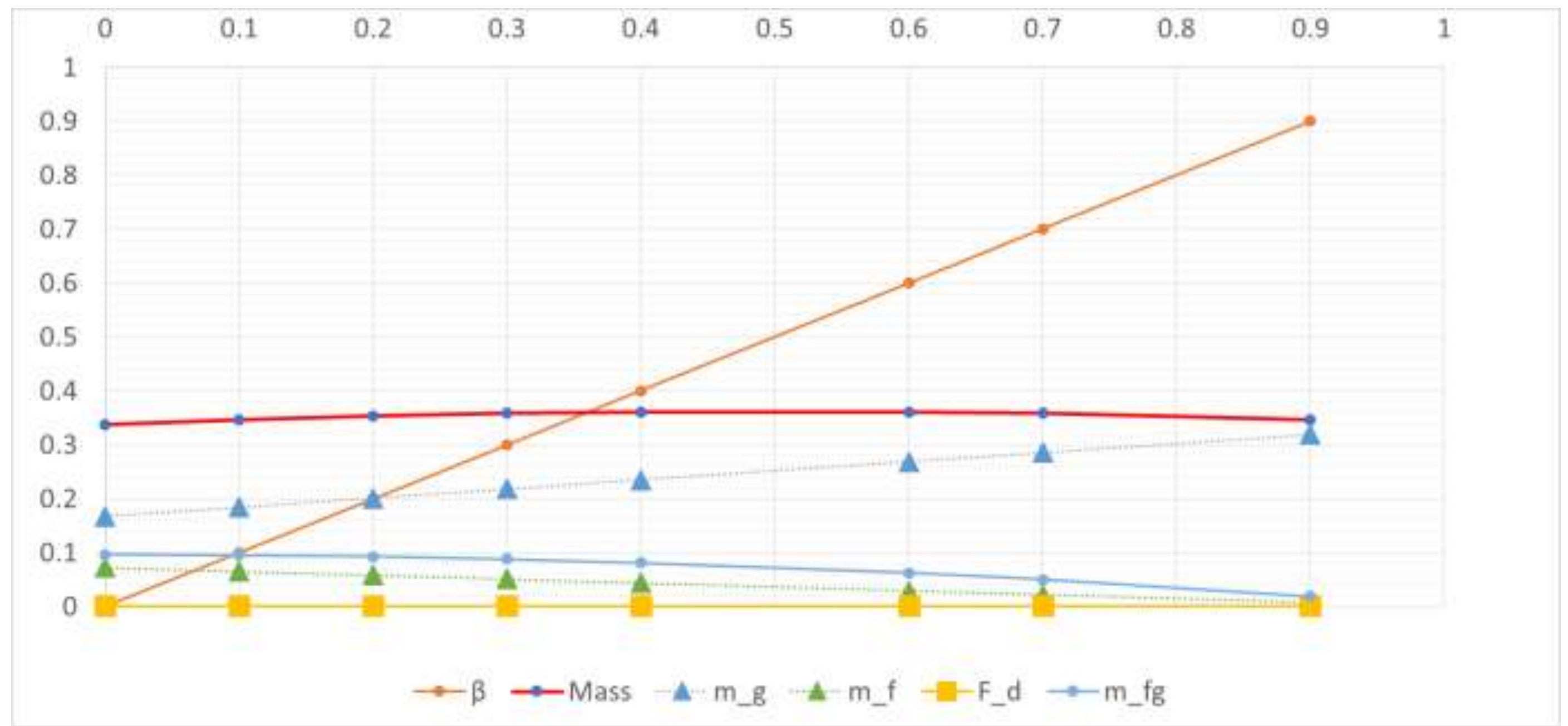

**Figure 7**

where

$$m_g = -\int 4\pi r^2 g^* g(r) dr$$

$$m_f = -\int 4\pi r^2 f^* f(r) dr$$

$$m_{fg} = -\int \rho_{fg}\, 4\pi r^2 dr$$

$$\rho_{fg} = \hat{g}\frac{d\hat{f}}{dr} - \hat{f}\frac{d\hat{g}}{dr} + \frac{2}{r}\hat{g}\hat{f}$$

$$F_d = -\frac{1}{4\pi}\int 4\pi r^2\, \vec{\nabla}\varphi(r)^2 dr$$

$$V_d = -\frac{1}{4\pi}\int 4\pi r^2 \varphi(r)\psi^*\psi(r) dr$$

The last two integrals represent the electrostatic energy and should be equal numerically.

$$Mass = m_g + m_f + m_{fg} + F_d$$

$$spin = J_d = \int 4\pi f(\rho)g(\rho)\rho^3 d\rho = \frac{3}{8}$$

From the last graph we can see that $\beta$ and the field energy are equal for $= 0.35$ , and simultaneously the value of the integrated charge equals the experimental charge when $\alpha = 0.101$ . At all times in these graphs the spin integrated over the particle has the experimental value of $\hbar/_2$. This is possible only when the dimensionless integral is to $\frac{3}{8} = 0.375$ as chosen.

In the Figure 8 and Figure 9 we can observe values of the profiles of different quantities of interest for the solution fulfilling the right boundary conditions.

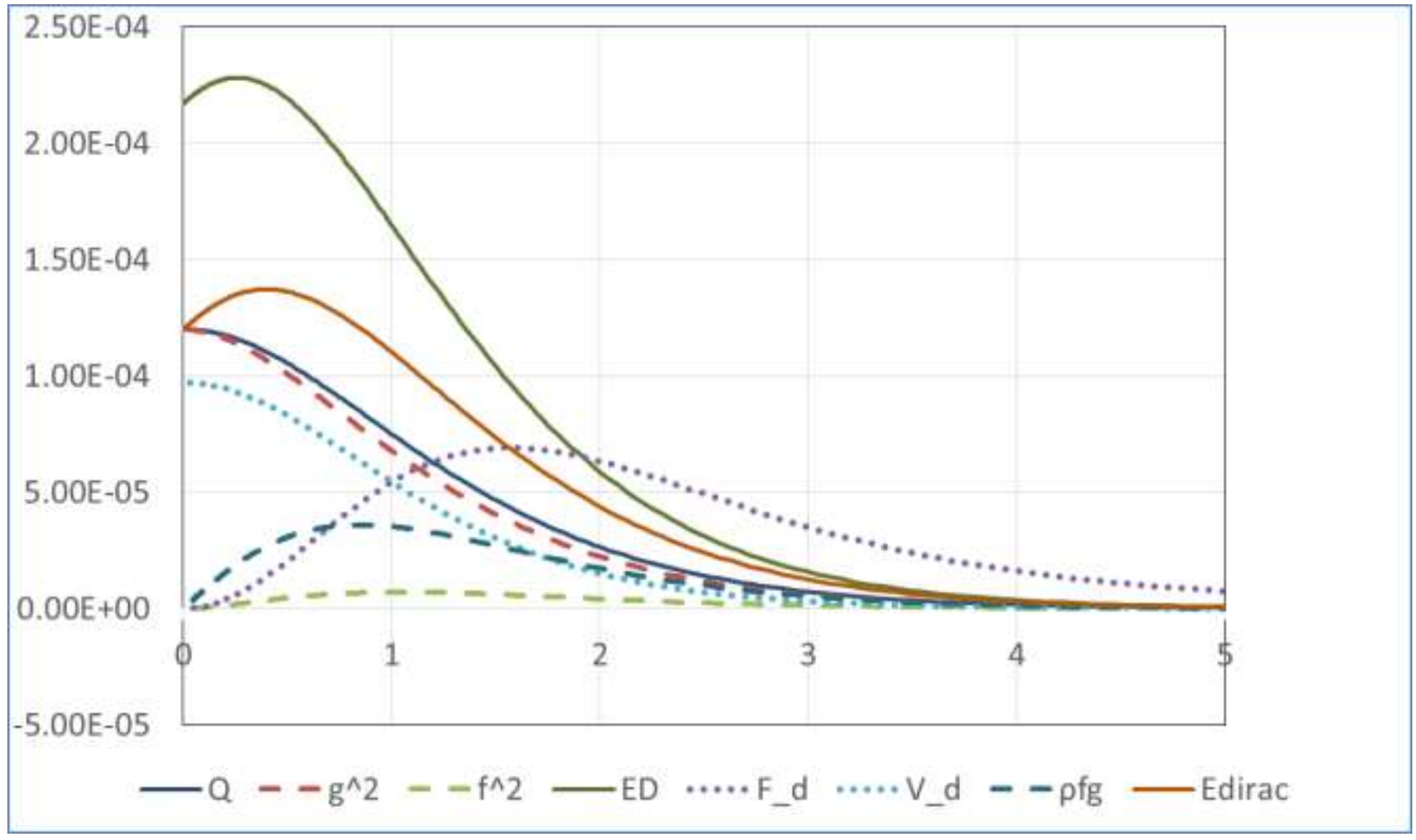

**Figure 8**

where

$$Q = g^2 + f^2$$

$$ED = g^2 + f^2 + \rho_{fg} + V_d$$

$$E_{Dirac} = g^2 + f^2 + \rho_{fg}$$

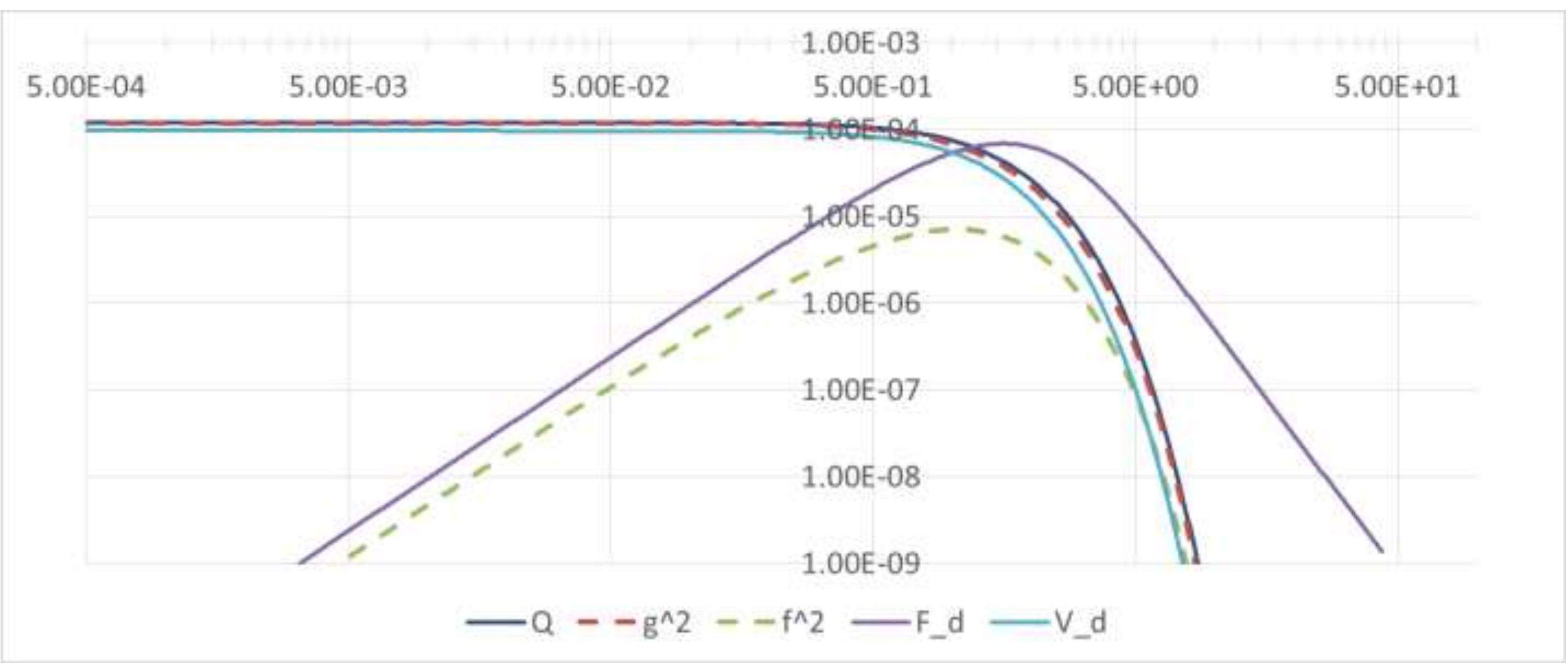

**Figure 9**

We have repeated the calculations for $\kappa = 1$ and obtained the values $\alpha = 0.0685 \cong {}^{1}/_{14.6}$ and $\beta = 0.20$. The corresponding plots are shown in Figure 10- Figure 12.

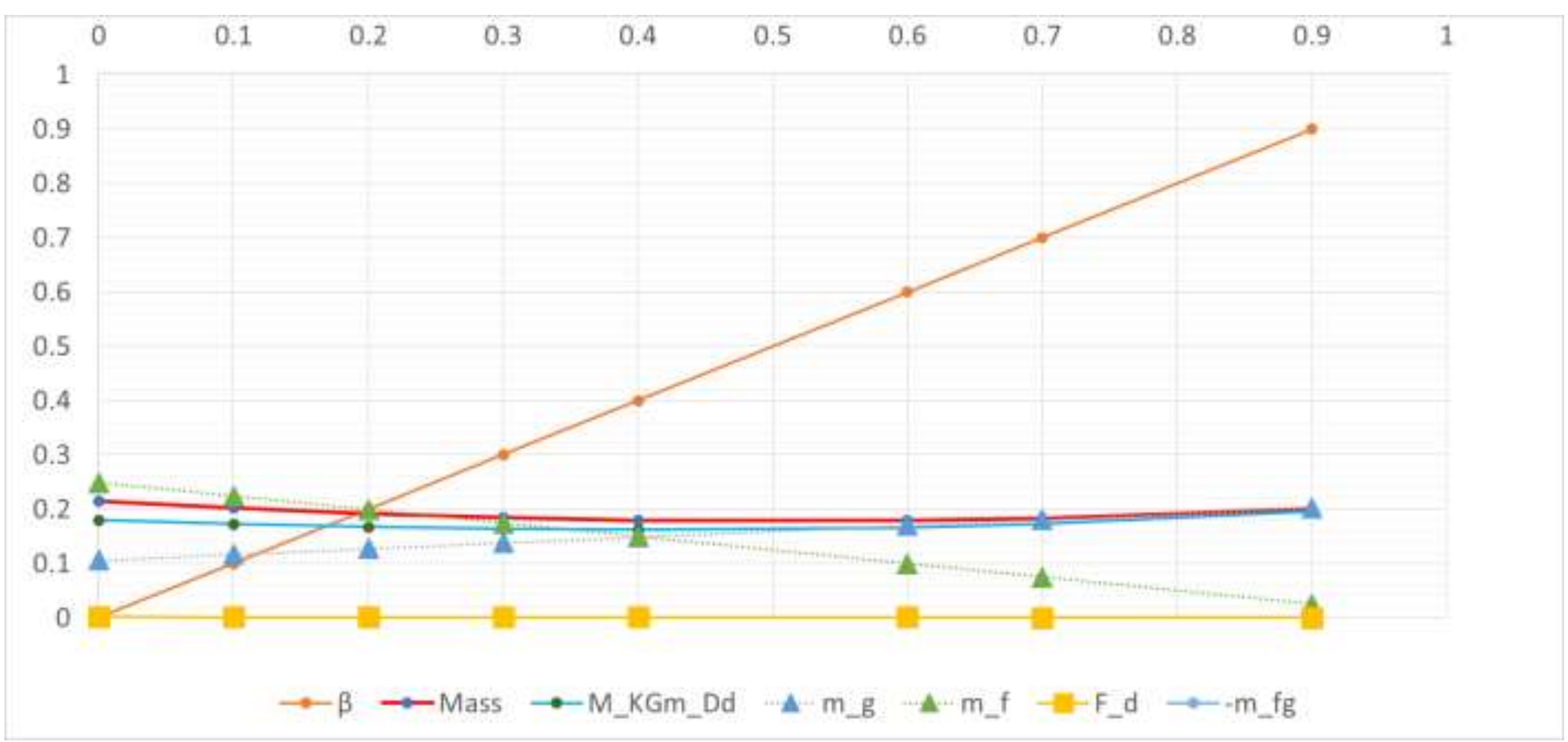

**Figure 10**

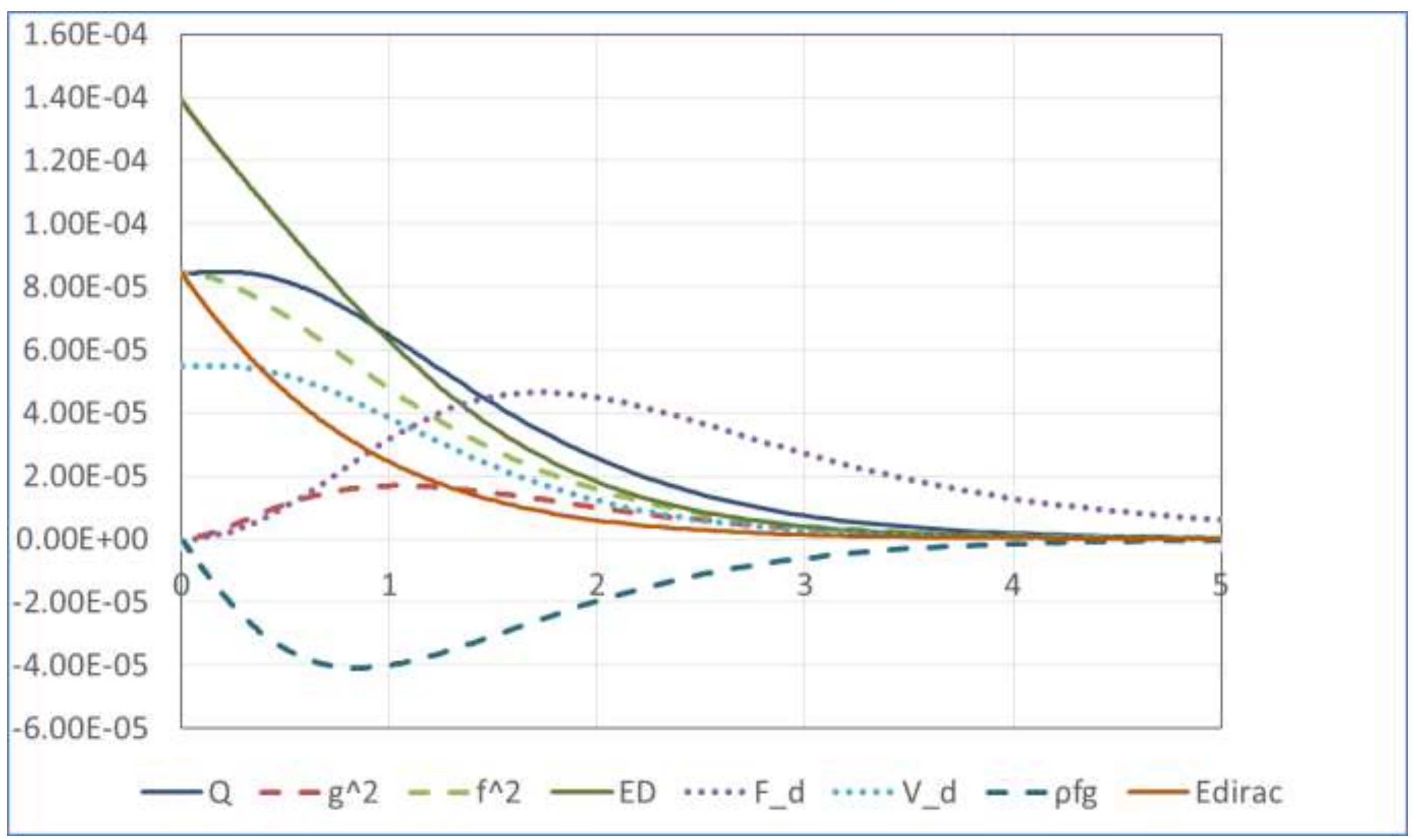

**Figure 11**

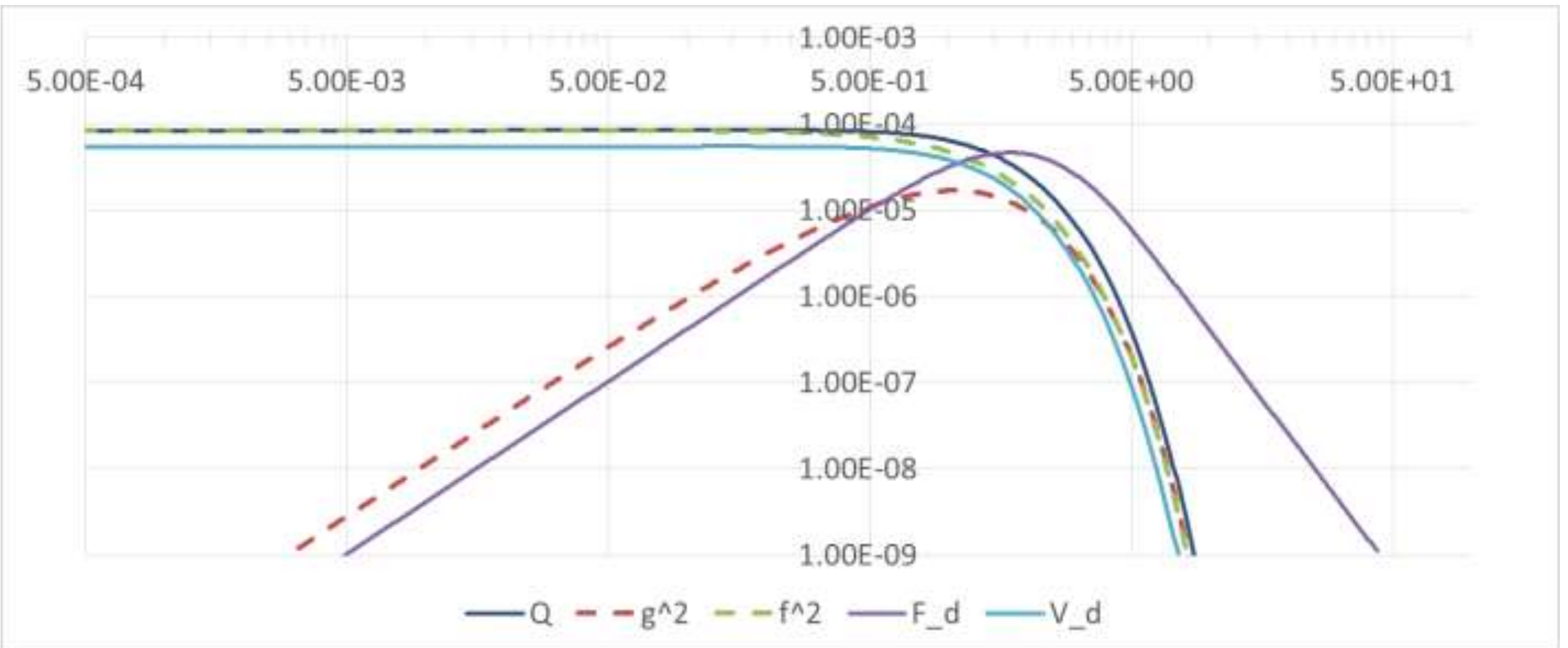

**Figure 12**

The physics behind the electron soliton confinement is the following: The repulsive electrostatic Coulomb potential and the centrifugal force is counter acted by the attractive quantum potential and the spin-orbit interaction.

### 2.5.3.1. Feranchuk model

Feranchuk[40] calculated a soliton with a mixture of a positive and negative energy solutions which are orthogonal and normalized to 1.

$\langle \psi^c | \psi \rangle = 0$

The values for alpha and electron mass he calculates are

$\alpha_0 \cong 174, \;\; m_0 \cong 222m \cong 113\; Mev, \quad r_e^0 \cong 0.416 r_e$

In his model the energy density as well as the charge density is given by the difference in energies between the electron and positron solutions. However he is considering solutions normalized to 1 without checking that the total internal angular momentum be equal to $\hbar/_2$. This model in some sense is similar to ours in that a mixture of negative and positive solutions gives as net result the same as a thinner wavefunction. The same holds for the energy.

We can adapt our results to his model. The equations are

$$\frac{\partial \hat{g}(\rho)}{\partial \rho} = \left(1 + \left( \beta - sign(eE)\frac{\alpha}{2}\widehat{V}(\rho)\right)\right)\hat{f}(\rho)$$
$$\frac{\partial \hat{f}(\rho)}{\partial \rho} = \frac{2}{\rho}\hat{f}(\rho) + \left( 1 - \left(\beta - sign(eE)\frac{\alpha}{2}\widehat{V}(\rho)\right)\right)\hat{g}(\rho)$$
$$\frac{\partial^2 \widehat{V}(\rho)}{\partial \rho^2} = -\frac{2}{\rho}\frac{\partial \widehat{V}(\rho)}{\partial \rho} - sign(eE)\left(\hat{f}^*\hat{f}(\rho) + \hat{g}^*\hat{g}(\rho)\right)$$

Take for the electron and positron functions

$G_e = AG_n, \qquad F_e = AF_n$

$G_p = BG_n, \qquad F_p = BF_n$

Where $F_n$ and $G_n$ are normalized to 1. We have seen that our solution is not normalized to 1 but to a smaller value N. With F and G being our solutions we have

$G = CG_n, \qquad F = CF_n$

The first two equations are linear in G and F and therefore are fulfilled automatically as long as $V$ keeps its value at all positions. Feranchuk showed that the electron and positron functions can be found that are orthogonal and follow the same differential equations. The Poisson equation will provide the same value for $V$ if

$C^2(F^*F + G^*G) = [A^2(F^*F + G^*G) - B^2(F^*F + G^*G)]$

or

$C^2 = (A^2 - B^2)$

**Eq 100**

On the other hand the field energy for the system is

$$H_{e,p} = \pm H_{part} \equiv mc^2 \int \left(-\hat{g}\frac{d\hat{f}}{d\rho} + \hat{f}\frac{d\hat{g}}{d\rho} + \frac{2\kappa}{\rho}\hat{g}\hat{f}\right) 4\pi\rho^2 d\rho + mc^2 \int \left(\hat{g}^2 + \hat{f}^2\right) 4\pi\rho^2 d\rho$$

$$H_{ph} = 4\pi\epsilon_0 \int 4\pi r^2 \,\frac{\vec{E}^2}{2}\, dr$$

The value of the total field energy will be given by

$H = A^2 H_{part} - B^2 H_{part} + H_{ph}$

As long as the value of the electric potential/field remains constant, $H_{ph}$ will remain the same while Eq 100 insures that the other terms remain constant too.

$$H = A^2 H_{part} - B^2 H_{part} + H_{ph} = C^2 H_{part} + H_{ph}$$

The total field energy is the same, the electric field is the same, and the eigenenergy is the same. Therefore it describes the same state.

#### 2.5.3.2. Relationship to QED

Weisskopf, between the years 1934-1939 applied non-relativistic perturbation theory to the problem of self-energy and mass of the electron using the electrostatic interaction. Basically, our equations are the first quantization version of his problem, this means without vacuum fluctuations or vacuum polarization, etc. As an intuitive result of his calculations, he provided a 'density profile' decaying exponentially with the Compton wavelength as parameter for r>>1, and like $1/r^{\frac{5}{2}}$ for r<<1. The exponential decay is the same we see in our calculations. In the center, our calculation represents the exact solution, not just as a first order perturbation approximation in a first quantization model. The relativistic version of the spin related circulation is called Zitterbewegung by many authors. The dynamic part of Weisskopf calculations due to vacuum fluctuations and responsible for the anomalous electron magnetic moment can be interpreted as the self-field interaction of the Zitterbewegung and its own (Biot-Savart) magnetostatic field. In present-day QED, Weisskopf calculations are a part of the self-energy Feynman diagram.

**At this point we may claim that a mechanical model for the electron exists and is self-consistent: the Dirac equation allows for solitons with a radius of the order of the Compton wavelength. We may think that the argumentation about 'point-like particles' in QED is based on a particular interpretation for the mathematical results.**

## 2.6. Conclusion

We have seen that starting from a mechanical field model, the combined action of self-field nonlinear interactions and the 'quantum force' derived from the internal tension forces of the chain model can provide an intuitive picture for the electron, including the wave-particle duality, interference, spin, vacuum fluctuations among others.

We presented the chain model, a classical model of particles linked by inter-particle forces that can reproduce all the properties of the Klein Gordon and Dirac equations in the continuum limit. In fact the quantum potential and the electron mass can be found to be manifestations of the forces and energy content of the chain model. This model is fully compatible with all classical conservation laws and provides a realistic model for what we can call the 'electron fluid', essence of the electron. The quantum force in turn exerts the Poincare stresses needed to maintain the electron fluid together. The energy content of the chain model serves as the source of mass for the electron. In this sense it can be considered as a possible substitute for the Higgs boson as potential generator of mass.

The model described here fulfills all basic conservation principles of classical physics, and with the exception of non-local interactions and non-local entanglement, the full model can be understood in terms of classical local physics. Taking as a working premise that physics is a human description of nature, we leave as an open issue exceeding the limits of this work the following question: is it possible that nature can be better described and understood by us, human beings, in terms of a classical non-local theory?

## 2.7. Appendix 2.1: Charge Density for the Dirac equation

Define the wavefunction as

$$\Psi = \begin{bmatrix} \varphi(r,t) \\ \chi(r,t) \end{bmatrix} \psi(r,t),$$

**Eq 101**

where $\varphi$ and $\chi$ is a pair of 2-component Pauli spinors, the Dirac equation in terms of these variables is

$$i\hbar \frac{\partial}{\partial t}\left( \begin{bmatrix} \varphi \\ \chi \end{bmatrix} \psi \right) = c\vec{\boldsymbol{\sigma}} \cdot \vec{\pi} \begin{bmatrix} \chi \\ \varphi \end{bmatrix} \psi + e\mathrm{V} \begin{bmatrix} \varphi \\ \chi \end{bmatrix} \psi + mc^2 \begin{bmatrix} \varphi \\ -\chi \end{bmatrix} \psi$$

where

$$\frac{\vec{\sigma} \cdot \vec{\pi}}{2mc} = \frac{\sigma^1 \pi_x + \sigma^2 \pi_y + \sigma^3 \pi_z}{2mc} = \frac{1}{2mc} \begin{pmatrix} \pi_z & \pi_x - i\pi_y \\ \pi_x + i\pi_y & -\pi_z \end{pmatrix} = \frac{1}{2mc} \begin{pmatrix} \frac{\hbar}{i}\vec{\nabla}_z - (e/c)\vec{A}_z & \pi_x - i\pi_y \\ \pi_x + i\pi_y & -\left( \frac{\hbar}{i}\vec{\nabla}_z - (e/c)\vec{A}_z \right) \end{pmatrix}$$

Separating for the spinor components and solving for $\chi$ we find:

$$\chi = \frac{c\vec{\boldsymbol{\sigma}} \cdot \vec{\pi}\varphi - i\hbar \frac{\partial \chi}{\partial t}}{\left( i\hbar \frac{\partial \psi}{\psi \partial t} - e\mathrm{V} + mc^2 \right)}$$

The effect of the last term in the numerator, $i\hbar \frac{\partial \chi}{\partial t} \psi$ plays a rol analogous to radiation reaction in electromagnetism. Its effects have been analyzed by Hestenes, Barut and others. It might be responsible for an additional zitterbewegung generating the typical spin algebra. Reinserting $\chi$ in the Dirac spinor we get

$$\begin{bmatrix} \varphi \\ \chi \end{bmatrix} \psi = \begin{bmatrix} \varphi \\ \dfrac{c\vec{\boldsymbol{\sigma}} \cdot \vec{\pi}\varphi - i\hbar \frac{\partial \chi}{\partial t}}{\left( i\hbar \frac{\partial \psi}{\psi \partial t} - e\mathrm{V} + mc^2 \right)} \end{bmatrix} \psi$$

**Eq 102**

By using Eq 102 we can find an approximated expression for the density:

$$\rho = \begin{pmatrix} & \begin{bmatrix} \varphi \\ \chi \end{bmatrix} \psi \\ [\varphi^* \quad \chi^*]\psi^* & \end{pmatrix} = (\varphi^*\varphi + \chi^*\chi)\psi^*\psi = \varphi^*\varphi\psi^*\psi + \left( \frac{c\vec{\boldsymbol{\sigma}} \cdot \vec{\pi}\varphi - i\hbar \frac{\partial \chi}{\partial t}}{i\hbar \frac{\partial \psi}{\psi \partial t} - e\mathrm{V} + mc^2} \right)^* \left( \frac{c\vec{\boldsymbol{\sigma}} \cdot \vec{\pi}\varphi - i\hbar \frac{\partial \chi}{\partial t}}{i\hbar \frac{\partial \psi}{\psi \partial t} - e\mathrm{V} + mc^2} \right) \psi^*\psi$$

If we now simplify the problem by specializing the electron state by the case when $\varphi = \begin{pmatrix} 1 \\ 0 \end{pmatrix}$ and $\frac{\partial \chi}{\partial t} = 0$ we get

$$\vec{\boldsymbol{\sigma}} \cdot \vec{\pi}\varphi\psi = \begin{pmatrix} \frac{\hbar \mathrm{c}}{i}\nabla_z \psi - eA_z \psi \\ \frac{\hbar \mathrm{c}}{i}\nabla_x \psi - eA_x \psi + i\left( \frac{\hbar \mathrm{c}}{i}\nabla_y \psi - eA_y \psi \right) \end{pmatrix}$$

and for the density

$$\rho_{eD} = (\varphi^*\varphi + \chi^*\chi)\psi^*\psi = \varphi^*\varphi\psi^*\psi + \frac{\left( \hbar \mathrm{c}\vec{\nabla}(\varphi\psi) \right)^2 + \left( e\vec{A} \right)^2 \varphi^*\varphi\psi^*\psi + \varphi^*\psi^* 2\vec{s} \cdot \left( \vec{\nabla} \times \vec{A} \right)\varphi\psi}{\left( i\hbar \frac{\partial \psi}{\psi \partial t} - e\mathrm{V} + mc^2 \right)\left( \left( i\hbar \frac{\partial \psi}{\psi \partial t} \right)^* - e\mathrm{V} + mc^2 \right)}$$

To simplify further assume we have an energy eigenstate, so that

$$\psi = \psi_{sp}(\vec{r})e^{-i\frac{E}{\hbar}t}, \quad \frac{\partial \psi}{\partial t} = -i\frac{E}{\hbar}\psi, \qquad \frac{\partial \psi^*}{\partial t} = -i\frac{E}{\hbar}\psi^*$$

And for an energy eigenstate it reduces to:

$$\rho_{eD} = \varphi^*\varphi\psi^*\psi + \frac{\left(\hbar c\vec{\nabla}(\varphi\psi)\right)^2 + \left(e\vec{A}\right)^2\varphi^*\varphi\psi^*\psi + \varphi^*\psi^* 2\vec{s}\cdot\left(\vec{\nabla}\times\vec{A}\right)\varphi\psi}{E^2 + (mc^2 - e\mathrm{V})^2}$$
$$= \frac{1}{E^2 + (mc^2 - e\mathrm{V})^2}\left[E^2 + \left(mc^2 - e\mathrm{V}\right)^2\psi^*\psi + \left(e\vec{A}\right)^2\psi^*\psi + 2\vec{s}\cdot\left(\vec{\nabla}\times\vec{A}\right)\psi^*\psi + \left(\hbar c\vec{\nabla}\psi\right)^2\right]$$

with $\vec{s} = \varphi^*\vec{\boldsymbol{\sigma}}\varphi$

Of course, this is not the general expression, but it shows us the complexity of the charge density.

## 2.8. Appendix 2.2: Double Solution for the Klein Gordon self-field equation

### Klein Gordon equation for a point particle

Suppose we have a 'point' electron in the presence of external fields. The Klein Gordon equation reads:

$$\left(i\hbar\frac{\partial}{\partial t} - \mathrm{e}V_{ext}\right)^2\psi_e = -c^2\hbar^2\nabla^2\psi_e + (m_e c^2)^2\psi_e$$

**Eq 103**

In the absence of external fields it obviously reads

$$-\hbar^2\frac{\partial^2\psi_e}{\partial t^2} = -c^2\hbar^2\nabla^2\psi_e + (m_0 c^2)^2\psi_e$$

**Eq 104**

### Self-field localized Klein Gordon solution at rest

Suppose that it were possible to find a real, localized, bounded soliton eigensolution to the problem of the self-field Klein Gordon equation at rest in the absence of external fields. Namely a solution for the electron and electromagnetic self-field following the equations

$$Klein\ Gordon: \quad \left(i\hbar\frac{\partial}{\partial t} - \mathrm{e}V_{sf}\right)^2\psi_{sf} = -c^2\hbar^2\nabla^2\psi_{sf} + (m_0 c^2)^2\psi_{sf}$$

$$Poisson: \qquad \nabla^2 V_{sf}(r) = -\rho(r) = \frac{ie\hbar}{2m_0c^2}\left(\psi_{sf}^*\frac{\partial\psi_{sf}}{\partial t} - \psi_{sf}\frac{\partial\psi_{sf}^*}{\partial t}\right) - \frac{e^2}{m_0c^2}V_{sf}\psi_{sf}^*\psi_{sf}$$

**Eq 105**

Where we have not included a vector potential field $\vec{A}_{sf}$ because we are not considering magnetic effects. It is clear that the electric potential field is quadratic in the Klein Gordon electron fields.

We note here that a similar treatment can be given to the soliton solution to the self-field Dirac equation found later in this work, when neglecting spin effects. In this case the charge density should be replaced by the corresponding Dirac expression which in terms of the spinor $\psi$ is defined as:

$\rho = e\bar{\psi}\gamma_0\psi = e\psi^\dagger\psi$ which is also a quadratic form in the wave function.

To find an eigensolution of Eq 105 we take a function with the form

$$\psi_{sf}(r,t) = f(r)e^{-i\frac{m_{sf}c^2}{\hbar}t}$$

where we define the quantity $m_{sf}c^2$ has units of energy and represents the energy eigenvalue for the eigensolution. The equation reads:

$$\left(\left(m_{sf}c^2\right)^2 + e^2V_{sf}^2 - 2m_{sf}c^2eV_{sf}\right)\psi_{sf} = -c^2\hbar^2\vec{\nabla}^2\psi_{sf} + (m_0c^2)^2\psi_{sf}$$

that can be written as

$$c^2\hbar^2\vec{\nabla}^2 f(r) = -\left(m_{sf}c^2\right)^2 f(r) + (m_0c^2)^2 f(r) + \left(-e^2V_{sf}^2 + 2m_{sf}c^2eV_{sf}\right)f(r)$$

or

$$\left(m_{sf}c^2\right)^2 f(r) = -c^2\hbar^2\vec{\nabla}^2 f(r) + (m_0c^2)^2 f(r) + \left(-e^2V_{sf}^2 + 2m_{sf}c^2eV_{sf}\right)f(r)$$

**Eq 106**

which is a second order differential equation in r. This equation has to be solved together with the solution to the Poisson equation in Eq 105. We note that we find a self-field solution to the Dirac-Poisson equations later in this work.

## Self-field localized solution moving with a constant velocity, $\vec{\nabla}|\psi_{ext}| = 0$

Suppose one can find a solution $\psi_{sf}$ to Eq 105 and Eq 106 for a field at rest. Next we can ask for the problem of such a soliton moving with constant velocity $\vec{v}$ respect to the unprimed $(r,t)$ frame. On one hand, we know such a solution $\psi'(r,t)$ exists because it can be simply found as the Lorentz transformed of the self-field solution $\psi_{sf}(\vec{r}') = f(\vec{r}')$ from Eq 106, at rest in the primed frame $(\vec{r}', t')$. Alternatively we can try to solve the problem directly from the Klein Gordon equation in the $(r,t)$ frame, in that case the equation reads:

$$\left(i\hbar\frac{\partial}{\partial t} - \mathrm{e}V_{sf}'\right)^2\psi'(\vec{r},t) = c^2\left(i\hbar\vec{\nabla} + e\vec{A}_{sf}'\right)^2\psi'(\vec{r},t) + (m_0c^2)^2\psi'(\vec{r},t)$$

**Eq 107**

The internal electric potential field $V_{sf}'(\vec{r},t)$ will be given by the Lorentz transformed from the self-field at rest in the primed frame $V_{sf}(\vec{r}')$ :

$$V_{sf}'(\vec{r},t) = \gamma V_{sf}(\vec{r}')$$

Additionally an electric potential vector field $\vec{A}_{sf}(\vec{r},t)$ will appear in the unprimed frame given by

$$\vec{A}'_{sf}(\vec{r},t) = V_{sf}'(\vec{r},t)\frac{\vec{v}}{c} = \gamma V_{sf}(\vec{r}')\frac{\vec{v}}{c}.$$

We use the Lorentz gauge so that $\frac{\partial}{\partial t}V_{sf} = \vec{\nabla}\cdot\vec{A}_{sf}$ in all frames of reference. We take a solution of the form

$$\psi'(r,t) = \psi_{ext}(r,t)f'(\vec{r},t) = \psi_{ext}(r,t)f(\vec{r}')$$

Explicitly Eq 107 reads

$$-\hbar^2\frac{\partial^2\psi'}{\partial t^2} - 2i\hbar \mathrm{e}V'\frac{\partial\psi'}{\partial t} + \left(\mathrm{e}V_{sf}'\right)^2\psi' = -c^2\hbar^2\nabla^2\psi' + 2ic^2e\vec{A}_{sf}'\cdot\vec{\nabla}\psi' + c^2\left(e\vec{A}_{sf}'\right)^2\psi' + (m_0c^2)^2\psi'$$

## Transformation properties

$$\vec{r} = (x, y, z)$$

$$\vec{r}' = (x', y', z') = \big(\gamma\big(x - x_0(t)\big), y, z\big) = \vec{r}'(x, y, z, t)$$

$$\frac{\partial x_0(t)}{\partial t} = v, \quad \frac{\partial^2 x_0(t)}{\partial t^2} = \frac{\partial v(t)}{\partial t} = a$$

$$t' = \gamma\left(t - \frac{vx}{c^2}\right)$$

where

$$\gamma = \frac{1}{\sqrt{1 - \frac{\mathrm{v}^2}{c^2}}}$$

$$\vec{\mathrm{v}} = \mathrm{v}\hat{n}$$

$$\gamma(x - vt) = x + (\gamma - 1)x - \gamma \mathrm{v}t$$

$$\vec{r}'(\vec{r}, t) = \vec{r} + (\gamma - 1)(\hat{r} \cdot \hat{\mathrm{v}})\vec{r} - \gamma\vec{\mathrm{v}}t = \big(1 + (\gamma - 1)(\hat{r} \cdot \hat{\mathrm{v}})\big)\vec{r} - \gamma\vec{\mathrm{v}}t$$

$$\frac{\partial \vec{r}'}{\partial t} = -\gamma\vec{\mathrm{v}}, \quad \frac{\partial^2 \vec{r}'}{\partial t^2} = -\gamma\frac{\partial \vec{\mathrm{v}}(t)}{\partial t} = -\gamma\vec{a}$$

where $\vec{a}$ is the possible acceleration of the moving system.

$$t'(\vec{r}, t) = \gamma\left(t - \frac{\vec{\mathrm{v}} \cdot \vec{r}}{c^2}\right)$$

$$\vec{\nabla}_{r'} f(\vec{r}') = \hat{n}'\big|\vec{\nabla}_{r'} f(\vec{r}')\big|$$

$$\vec{\nabla}_r f'(\vec{r}) = \frac{\partial \vec{r}'}{\partial \vec{r}}\vec{\nabla}_{r'} f(\vec{r}') + \cancel{\frac{\partial r'}{\partial t}\frac{\partial f(\vec{r}')}{\partial t'}} = \big(1 + (\gamma - 1)(\hat{n}' \cdot \hat{\mathrm{v}})\big)\vec{\nabla}_{r'} f(\vec{r}')$$

$$\begin{aligned}\nabla_r^2 f'(\vec{r}) = \vec{\nabla}_r \cdot \vec{\nabla}_r f(\vec{r}') &= \frac{\partial \vec{r}'}{\partial \vec{r}}\vec{\nabla}_{r'}\left(\frac{\partial \vec{r}'}{\partial \vec{r}}\vec{\nabla}_{r'} f(\vec{r}')\right) = \big(1 + (\gamma - 1)(\hat{n}' \cdot \hat{\mathrm{v}})\big)\vec{\nabla}_{r'}\Big(\big(1 + (\gamma - 1)(\hat{n}' \cdot \hat{\mathrm{v}})\big)\vec{\nabla}_{r'} f(\vec{r}')\Big) \\ &= \big(1 + (\gamma - 1)(\hat{n}' \cdot \hat{\mathrm{v}})\big)^2 \nabla_{r'}^2 f(\vec{r}')\end{aligned}$$

$$\frac{\partial f'(\vec{r})}{\partial t} = \frac{\partial \vec{r}'}{\partial t} \cdot \vec{\nabla}_{r'} f(\vec{r}') + \cancel{\frac{\partial t'}{\partial t}\frac{\partial f(\vec{r}')}{\partial t'}} = -\gamma\vec{\mathrm{v}} \cdot \vec{\nabla}_{r'} f(\vec{r}') = -\gamma\vec{\mathrm{v}} \cdot \hat{n}'\big|\vec{\nabla}_{r'} f(\vec{r}')\big|$$

$$\begin{aligned}\frac{\partial^2 f'(\vec{r})}{\partial t^2} &= \frac{\partial \vec{r}'}{\partial t} \cdot \vec{\nabla}_{r'}\left(-\gamma\vec{\mathrm{v}} \cdot \vec{\nabla}_{r'} f(\vec{r}')\right) + \frac{\partial^2 \vec{r}'}{\partial t^2} \cdot \vec{\nabla}_{r'} f(\vec{r}') = \gamma\vec{\mathrm{v}} \cdot \vec{\nabla}_{r'}\left(\gamma\vec{\mathrm{v}} \cdot \vec{\nabla}_{r'} f(\vec{r}')\right) - \gamma\vec{a} \cdot \vec{\nabla}_{r'} f(\vec{r}') \\ &= \gamma^2\vec{\mathrm{v}} \cdot \vec{\nabla}_{r'}\left(\vec{\mathrm{v}} \cdot \vec{\nabla}_{r'} f(\vec{r}')\right) - \gamma\vec{a} \cdot \vec{\nabla}_{r'} f(\vec{r}') = \gamma^2 \mathrm{v}^2\big(\hat{\mathrm{v}} \cdot \vec{\nabla}_{r'}\big)\big(\hat{\mathrm{v}} \cdot \vec{\nabla}_{r'}\big) f(\vec{r}') - \gamma\vec{a} \cdot \vec{\nabla}_{r'} f(\vec{r}') \\ &= \gamma^2 \mathrm{v}^2\big(\hat{\mathrm{v}} \cdot \vec{\nabla}_{r'}\big)^2 f(\vec{r}') - \gamma\vec{a} \cdot \vec{\nabla}_{r'} f(\vec{r}') = \gamma^2 \mathrm{v}^2(\hat{\mathrm{v}} \cdot \hat{n}')^2 \vec{\nabla}_{r'}^{\,2} f(\vec{r}') - \gamma\vec{a} \cdot \vec{\nabla}_{r'} f(\vec{r}')\end{aligned}$$

Take the velocity constant in the x direction: $\vec{\mathrm{v}} = \mathrm{v}\hat{\imath}$; then

$$\frac{\partial^2 f'(\vec{r})}{\partial x^2} = \big(1 + (\gamma - 1)\big)^2 \frac{\partial^2 f(\vec{r}')}{\partial x'^2} = \big(1 + (\gamma - 1)\big)^2 \frac{\partial^2 f(\vec{r}')}{\partial x'^2} = \gamma^2 \frac{\partial^2 f(\vec{r}')}{\partial x'^2}$$

$$\frac{\partial^2 f'(\vec{r})}{\partial t^2} = \gamma^2 \mathrm{v}^2 (\hat{\mathrm{v}} \cdot \hat{n}')^2 \vec{\nabla}_{r'}^{\,2} f(\vec{r}') = \gamma^2 \mathrm{v}^2 \frac{\partial^2 f(\vec{r}')}{\partial x'^2}$$

$$\gamma^2 = \frac{1}{1 - \frac{\mathrm{v}^2}{c^2}}$$

$$\boxed{\frac{\partial^2 f'(\vec{r})}{\partial x^2} - \frac{\partial^2 f'(\vec{r})}{c^2 \partial t^2} = \gamma^2 \left(1 - \frac{\mathrm{v}^2}{c^2}\right) \frac{\partial^2 f(\vec{r}')}{\partial x'^2} = \frac{\partial^2 f(\vec{r}')}{\partial x'^2}}$$

As it is expected because the D'Alembertian is a relativistic invariant : $\Box_{r'} f(r') = \Box_r f'(r)$

The different terms in the Klein Gordon equation read:

$$\frac{\partial \psi'}{\partial t} = \frac{\partial f'(\vec{r}, t)}{\partial t} \psi_{ext} + \frac{\partial \psi_{ext}}{\partial t} f'(\vec{r}, t) = -\gamma \vec{\mathrm{v}} \cdot \vec{\nabla}_{r'} f(\vec{r}') \psi_{ext} + \frac{\partial \psi_{ext}}{\partial t} f'(\vec{r}, t)$$

$$\begin{aligned} \frac{\partial^2 \psi'}{\partial t^2} &= \frac{\partial^2 f'(\vec{r}, t)}{\partial t^2} \psi_{ext} + \frac{\partial^2 \psi_{ext}}{\partial t^2} f'(\vec{r}, t) + 2 \frac{\partial f'(\vec{r}, t)}{\partial t} \frac{\partial \psi_{ext}}{\partial t} \\ &= \gamma^2 \mathrm{v}^2 \left(\hat{\mathrm{v}} \cdot \vec{\nabla}_{r'}\right)^2 f(\vec{r}') \psi_{ext} - \gamma \vec{a} \cdot \vec{\nabla}_{r'} f(\vec{r}') \psi_{ext} + \frac{\partial^2 \psi_{ext}}{\partial t^2} f'(\vec{r}, t) - 2\gamma \vec{\mathrm{v}} \cdot \vec{\nabla}_{r'} f(\vec{r}') \frac{\partial \psi_{ext}}{\partial t} \end{aligned}$$

$$\vec{\nabla} \psi' = \vec{\nabla} f' \psi_{ext} + \vec{\nabla} \psi_{ext} f' = \left(1 + (\gamma - 1)(\hat{r} \cdot \hat{\mathrm{v}})\right) \vec{\nabla}_{r'} f(\vec{r}') \psi_{ext} + \vec{\nabla} \psi_{ext} f'$$

$$\begin{aligned} \nabla^2 \psi' &= \nabla^2 f' \psi_{ext} + \nabla^2 \psi_{ext} f' + 2 \vec{\nabla} \psi_{ext} \cdot \vec{\nabla} f' \\ &= \left(1 + (\gamma - 1)(\hat{r} \cdot \hat{\mathrm{v}})\right)^2 \nabla_{r'}^2 f(\vec{r}') \psi_{ext} + \nabla^2 \psi_{ext} f' + 2 \vec{\nabla} \psi_{ext} \cdot \left(1 + (\gamma - 1)(\hat{r} \cdot \hat{\mathrm{v}})\right) \vec{\nabla}_{r'} f(\vec{r}') \end{aligned}$$

---

Then the equation Eq 107 reads:

$$\begin{aligned} -\hbar^2 \left( \frac{\partial^2 f'(\vec{r}, t)}{\partial t^2} \psi_{ext} + \frac{\partial^2 \psi_{ext}}{\partial t^2} f'(\vec{r}, t) + 2 \frac{\partial f'(\vec{r}, t)}{\partial t} \frac{\partial \psi_{ext}}{\partial t} \right) - 2i\hbar \mathrm{e} V' \left( \frac{\partial f'(\vec{r}, t)}{\partial t} \psi_{ext} + \frac{\partial \psi_{ext}}{\partial t} f'(\vec{r}, t) \right) + \left(\mathrm{e} V'_{sf}\right)^2 \psi' \\ = -c^2 \hbar^2 \left( \nabla^2 f' \psi_{ext} + \nabla^2 \psi_{ext} f' + 2 \vec{\nabla} \psi_{ext} \cdot \vec{\nabla} f' \right) + 2ic^2 e \vec{A}'_{sf} \cdot \left( \vec{\nabla} f' \psi_{ext} + \vec{\nabla} \psi_{ext} f' \right) + c^2 \left( e \vec{A}'_{sf} \right)^2 \psi' \\ + (m_0 c^2)^2 \psi' \end{aligned}$$

Take

$$\psi_{ext} = |\psi_{ext}(r)| e^{i\left(\frac{S_{ext}(\vec{r})}{\hbar} - \frac{E}{\hbar} t\right)}$$

$$\vec{\nabla} \psi_{ext} = \vec{\nabla} |\psi_{ext}| e^{i\left(\frac{S_{ext}(\vec{r})}{\hbar} - \frac{E}{\hbar} t\right)} + \frac{i}{\hbar} \vec{\nabla} S_{ext} \psi_{ext}$$

Then the equation Eq 107 reads:

$$
\begin{aligned}
&-\hbar^2\left(\frac{\partial^2 f'(\vec{r},t)}{\partial t^2}\psi_{ext} - 2\gamma\vec{\mathrm{v}}\cdot\vec{\nabla}_{r'}f(\vec{r}')\frac{\partial\psi_{ext}}{\partial t} + \frac{\partial^2\psi_{ext}(r,t)}{\partial t^2}f'^{(\vec{r})}\right) - 2i\hbar \mathrm{e}V'_{sf}\left(-\gamma\vec{\mathrm{v}}\cdot\vec{\nabla}_{r'}f(\vec{r}')\psi_{ext} + \boxed{\frac{\partial\psi_{ext}}{\partial t}f'(\vec{r},t)}\right)\\
&\quad + \left(\mathrm{e}V'_{sf}\right)^2\psi\\
&\quad = -c^2\hbar^2\left(\nabla^2 f'\psi_{ext} + \nabla^2\psi_{ext}f + 2\left(\cancel{\vec{\nabla}|\psi_{ext}|e^{i\left(\frac{S_{ext}(\vec{r})}{\hbar}-\frac{E}{\hbar}t\right)}} + \frac{i}{\hbar}\vec{\nabla}S_{ext}\psi_{ext}\right)\right.\\
&\quad \left.\cdot\left(1+(\gamma-1)(\hat{n}'\cdot\hat{\mathrm{v}})\right)\vec{\nabla}_{r'}f(\vec{r}')\right) + 2ic^2 e\vec{A}'_{sf}\\
&\quad \cdot\left(\cancel{\vec{\nabla}|\psi_{ext}|e^{i\left(\frac{S_{ext}(\vec{r})}{\hbar}-\frac{E}{\hbar}t\right)}f'} + \boxed{\frac{i}{\hbar}\vec{\nabla}S_{ext}\psi_{ext}f'} + \left(1+(\gamma-1)(\hat{n}'\cdot\hat{\mathrm{v}})\right)\vec{\nabla}_{r'}f(\vec{r}')\psi_{ext}\right) + c^2\left(e\vec{A}'_{sf}\right)^2\psi'\\
&\quad + (m_0c^2)^2\psi'
\end{aligned}
$$

where we have assumed $\vec{\nabla}|\psi_{ext}| = 0$ and the acceleration $\vec{a} = 0$ . The red terms are

$$\left(1+(\gamma-1)(\hat{n}'\cdot\hat{\mathrm{v}})\right)\vec{\mathrm{v}}\cdot\vec{\nabla}_{r'}f(\vec{r}') = \left(1+(\gamma-1)\right)\vec{\mathrm{v}}\cdot\vec{\nabla}_{r'}f(\vec{r}') = \gamma\vec{\mathrm{v}}\cdot\vec{\nabla}_{r'}f(\vec{r}')$$

$$2i\hbar \mathrm{e}V'\vec{\nabla}f'(\vec{r})\cdot\vec{\mathrm{v}}\psi_{ext} = 2ic^2 e\vec{A}'_{sf}\cdot\vec{\nabla}f'(\vec{r})\psi_{ext}$$

$$2i\hbar \mathrm{e}V'\vec{\nabla}f'(\vec{r})\cdot\vec{\mathrm{v}}\psi_{ext} = 2ic^2 eV'\vec{\mathrm{v}}\cdot\vec{\nabla}f'(\vec{r})\psi_{ext}$$

which is automatically fulfilled and can be omitted from the equation. The green terms are:

$$\left(\mathrm{e}V'_{sf}\right)^2\psi' = c^2\left(e\vec{A}'_{sf}\right)^2\psi'$$

$$\left(\mathrm{e}V'_{sf}\right)^2\psi' = c^2\left(eV'_{sf}\vec{\mathrm{v}}\right)^2\psi'$$

which can be taken into account with the replacement in the equation:

$$\left(\mathrm{e}V'_{sf}(\vec{r})\right)^2\psi' \quad\Rightarrow\quad \left(1-\frac{v^2}{c^2}\right)\left(\mathrm{e}V'_{sf}(\vec{r})\right)^2\psi' = \left(\mathrm{e}\frac{V'_{sf}(\vec{r})}{\gamma}\right)^2\psi' = \left(\mathrm{e}V_{sf}(\vec{r}')\right)^2\psi'$$

The equality of the two purple terms can be fulfilled when:

$$-2\hbar^2 i\frac{m'c^2}{\hbar}\vec{\nabla}f'\cdot\vec{\mathrm{v}} = -c^2\hbar^2 i2\vec{\nabla}f'\cdot\frac{\vec{\nabla}S_{ext}}{\hbar}$$

which is achieved when the velocity is defined by

$$\boldsymbol{\vec{\mathrm{v}}} = \boldsymbol{\frac{\vec{\nabla}S_{ext}}{m'}}$$

We see that this last equation, known as the 'de Broglie guidance condition' appears as the explicit law of motion followed by the self-field soliton, without the need for any ad-hoc assumption about the properties of the soliton trajectory. The terms in the squared part are:

$$\hbar^2 2i\hbar \mathrm{e}V'_{sf}(\vec{r})\frac{\partial\psi_{ext}}{\partial t}f' = 2ic^2 e\vec{A}'_{sf}(\vec{r})\cdot\frac{i}{\hbar}\vec{\nabla}S_{ext}\psi_{ext}f'$$

$$\hbar^2 2i\hbar \mathrm{e}V'_{sf}(\vec{r})i\frac{m'c^2}{\hbar} = 2ic^2 eV'_{sf}(\vec{r})\vec{\mathrm{v}}\cdot i\frac{m'\vec{\mathrm{v}}}{\hbar}\psi_{ext}$$

which can be taken into account with the replacement in the equation:

$$\hbar^2 2i\hbar eV'_{sf}(\vec{r})i\frac{m'c^2}{\hbar} \quad \Longrightarrow \quad \left(1-\frac{v^2}{c^2}\right)\hbar^2 2i\hbar eV'_{sf}(\vec{r})i\frac{m'c^2}{\hbar} = \hbar^2 2i\hbar eV_{sf}(\vec{r}')i\frac{mc^2}{\hbar}$$

with $m'c^2 = \gamma mc^2$. Now take $\vec{\nabla}|\psi_{ext}| = 0$, and the equation reduces to:

$$-\hbar^2\frac{\partial^2 f'(\vec{r},t)}{\partial t^2}\psi_{ext} - \hbar^2\frac{\partial^2\psi_{ext}(r,t)}{\partial t^2}f'(\vec{r}) - \hbar^2 2i\hbar eV_{sf}(\vec{r}')i\frac{E}{\gamma\hbar}\psi_{ext}(r,t)f'(\vec{r}) + \left(eV_{sf}(\vec{r}')\right)^2\psi_{ext}(r,t)f'(\vec{r})$$
$$= -c^2\hbar^2\left(\nabla^2 f'(\vec{r})\psi_{ext} + \nabla^2\psi_{ext}f'(\vec{r})\right) + (m_0c^2)^2\psi_{ext}(r,t)f'(\vec{r})$$

if we identify $E = \gamma m_{sf}c^2$ and take $f'(\vec{r}) = Bf(\vec{r}')$ the Klein Gordon equation reduces to

$$\hbar^2\frac{\partial^2\psi_{ext}(r,t)}{\partial t^2}f'(\vec{r}) - \hbar^2 2i\hbar eV_{sf}(\vec{r}')i\frac{E}{\gamma\hbar}\psi_{ext}(r,t)f'(\vec{r}) + \left(eV_{sf}(\vec{r}')\right)^2\psi_{ext}(r,t)f(\vec{r}')$$
$$= -c^2\hbar^2\left(\nabla^2_{r'}f'(\vec{r})\psi_{ext} + \nabla^2\psi_{ext}f(\vec{r}')\right) + (m_0c^2)^2\psi_{ext}(r,t)f(\vec{r}')$$

here, the red part can be written as

$$\left(m_{sf}c^2\right)^2 f(\vec{r}')\psi_{ext} = -c^2\hbar^2\nabla^2_{r'}f(\vec{r}')\psi_{ext} + (m_0c^2)^2 f(\vec{r}')\psi_{ext} + \left(-e^2V^2_{sf}(\vec{r}') + 2m_{sf}c^2 eV_{sf}(\vec{r}')\right)f(\vec{r}')\psi_{ext}$$

which is identical with Eq 106 in the moving frame of reference. Replacing this result in the Klein Gordon equation we get:

$$-\hbar^2\left(\frac{\partial^2\psi_{ext}}{\partial t^2}\right)f = -c^2\hbar^2(\nabla^2\psi_{ext})f + \left(m_{sf}c^2\right)^2\psi_{ext}f$$

recovering Eq 104, which shows that the energy content of the self-field soliton can be consistently considered as the mass of an effective particle in a Klein Gordon equation for the $\psi_{ext}$ wave function.

## Self-field localized solution moving with a constant velocity, $\vec{\nabla}|\psi_{ext}| \neq 0$

Now consider the case in the absence of external fields, but where $\vec{\nabla}\psi_{ext} \neq 0$ as is the case for a free particle in spherical or cylindrical geometry, or the case of interference. We take as a solution

$$\psi(\vec{r},t) = \psi_{ext}(r)A(t)f'(\vec{r},t)$$

**Eq 108**

where $f'(\vec{r},t)$ is related to the stationary solution we have described before in a moving frame

$f'(\vec{r},t) = f(\vec{r}')$. We further assume $\frac{\partial|\psi_{ext}(r)|}{\partial t} = 0, \frac{\partial S_{ext}(\vec{r})}{\partial t} = 0, \frac{\partial f(\vec{r}')}{\partial t'} = 0,$ while $\psi_{ext}(r)$ is a stationary eigen function of the energy to be determined later. As the soliton $\psi_{int}$ moves in space, it will find that the external wavefunction $\psi_{ext}$ changes in magnitude. In order to keep the soliton solution $f(\vec{r}')$ from disentangling, we need that the intensity of the internal electric fields $V_{sf}(\vec{r}')$ should remain relatively constant. Because the fields are quadratic in the magnitude of the total electron fields, the magnitude of the product $|\psi_{ext}(r)A(t)f'(\vec{r},t)|$ should remain constant at the position of the soliton, as it moves in space, or equivalently (remember $f'(\vec{r},t)$ is real)

$$\frac{d\left(|\psi_{ext}|A(t)f'(\vec{r},t)\right)}{dt} = 0$$

We expect this equation to appear as part of the solution. Now

$$\frac{d\left(|\psi_{ext}|\mathrm{A}(t)f'(\vec{r},t)\right)}{dt} = |\psi_{ext}|\frac{d\left(\mathrm{A}(t)f'(\vec{r},t)\right)}{dt} + \mathrm{A}(t)f'(\vec{r},t)\frac{d(|\psi_{ext}|)}{dt} = 0$$

or

$$\frac{d\left(\mathrm{A}(t)f'(\vec{r},t)\right)}{dt} = \frac{\partial \mathrm{A}}{\partial t}f'(\vec{r},t)$$

We know that

$$\frac{d\left(\left|\psi_{ext}(\vec{\mathrm{r}} = \vec{r_0}(t))\right|\right)}{dt} = \vec{\mathrm{v}}\cdot\vec{\nabla}|\psi_{ext}|$$

and therefore we get

$$|\psi_{ext}|\frac{\partial \mathrm{A}}{\partial t}f'(\vec{r},t) + \vec{\mathrm{v}}\cdot\vec{\nabla}|\psi_{ext}|\mathrm{A}f'(\vec{r},t) = 0$$

$$\frac{1}{A}\frac{\partial \mathrm{A}}{\partial t} = -\frac{\vec{\mathrm{v}}\cdot\vec{\nabla}|\psi_{ext}|}{|\psi_{ext}|}$$

**Eq 109**

On the other hand using Eq 108 we get the following expressions for the derivatives:

$$\frac{\partial\psi}{\partial t} = \psi_{ext}(r)\frac{\partial \mathrm{A}(t)}{\partial t}f'(\vec{r}) + \psi_{ext}(r)\mathrm{A}(t)\frac{\partial f'(\vec{r})}{\partial t} + \frac{\partial\psi_{ext}}{\partial t}\mathrm{A}(t)f'(\vec{r})$$

$$\frac{\partial^2\psi}{\partial t^2} = \psi_{ext}(r)\frac{\partial^2 \mathrm{A}}{\partial t^2}f'(\vec{r}) + 2\psi_{ext}(r)\frac{\partial \mathrm{A}(t)}{\partial t}\frac{\partial f'(\vec{r})}{\partial t} + 2\frac{\partial\psi_{ext}(r)}{\partial t}\frac{\partial \mathrm{A}(t)}{\partial t}f'(\vec{r}) + \psi_{ext}(r)\mathrm{A}(t)\frac{\partial^2 f'(\vec{r})}{\partial t^2} + 2\frac{\partial\psi_{ext}(r)}{\partial t}\mathrm{A}(t)\frac{\partial f'(\vec{r})}{\partial t} + \frac{\partial^2\psi_{ext}(r)}{\partial t^2}\mathrm{A}(t)f'(\vec{r})$$

$$\vec{\nabla}\psi = \mathrm{A}(t)\vec{\nabla}f'\psi_{ext} + \mathrm{A}(t)\vec{\nabla}\psi_{ext}f'$$

$$\nabla^2\psi = \mathrm{A}(t)\nabla^2 f'\psi_{ext} + \mathrm{A}(t)\nabla^2\psi_{ext}f' + 2\mathrm{A}(t)\vec{\nabla}\psi_{ext}\cdot\vec{\nabla}f'$$

Taking now

$$\psi_{ext} = |\psi_{ext}(r)|e^{i\left(\frac{S_{ext}(\vec{r})}{\hbar}-\frac{E}{\hbar}t\right)}$$

we get

$$\vec{\nabla}\psi_{ext} = \vec{\nabla}|\psi_{ext}|e^{i\left(\frac{S_{ext}(\vec{r})}{\hbar}-\frac{E}{\hbar}t\right)} + \frac{i}{\hbar}\vec{\nabla}S_{ext}\psi_{ext}$$

$$\vec{\nabla}\psi = \mathrm{A}(t)\vec{\nabla}f'\psi_{ext} + \mathrm{A}(t)\vec{\nabla}|\psi_{ext}|e^{i\left(\frac{S_{ext}(\vec{r})}{\hbar}-\frac{E}{\hbar}t\right)}f' + \frac{i}{\hbar}\mathrm{A}(t)\vec{\nabla}S_{ext}\psi_{ext}f'$$

$$\nabla^2\psi = \mathrm{A}(t)\nabla^2 f'\psi_{ext} + \mathrm{A}(t)\nabla^2\psi_{ext}f' + 2\mathrm{A}(t)\vec{\nabla}|\psi_{ext}|e^{i\left(\frac{S_{ext}(\vec{r})}{\hbar}-\frac{E}{\hbar}t\right)}\cdot\vec{\nabla}f' + 2\frac{i}{\hbar}\mathrm{A}(t)\vec{\nabla}S_{ext}\psi_{ext}\cdot\vec{\nabla}f'$$

Replacing these values into the Klein Gordon equation Eq 107, we get

$$
\begin{aligned}
&-\hbar^2\left(\psi_{ext}(r)\frac{\partial^2 \mathrm{A}}{\partial t^2}f'(\vec{r})+2\psi_{ext}(r)\frac{\partial \mathrm{A}(t)}{\partial t}\frac{\partial f'(\vec{r})}{\partial t}+2\frac{\partial \psi_{ext}(r)}{\partial t}\frac{\partial \mathrm{A}(t)}{\partial t}f'(\vec{r})+\psi_{ext}(r)\mathrm{A}(t)\frac{\partial^2 f'(\vec{r})}{\partial t^2}\right.\\
&\qquad\left.+2\frac{\partial \psi_{ext}(r)}{\partial t}\mathrm{A}(t)\frac{\partial f'(\vec{r})}{\partial t}+\frac{\partial^2 \psi_{ext}(r)}{\partial t^2}\mathrm{A}(t)f'(\vec{r})\right)\\
&\qquad-2i\hbar \mathrm{e}V'_{sf}\left(\boxed{\psi_{ext}(r)\frac{\partial \mathrm{A}(t)}{\partial t}f'(\vec{r})}+\psi_{ext}(r)\mathrm{A}(t)\frac{\partial f'(\vec{r})}{\partial t}+\frac{\partial \psi_{ext}}{\partial t}\mathrm{A}(t)f'(\vec{r})\right)+\left(\mathrm{e}V'_{sf}\right)^2\psi'\\
&\qquad=-c^2\hbar^2\left(\mathrm{A}(t)\nabla^2 f'\psi_{ext}-\gamma \mathrm{A}(t)\frac{\vec{a}}{c^2}\cdot\vec{\nabla}f'\psi_{ext}+\mathrm{A}(t)\nabla^2\psi_{ext}f'+2\mathrm{A}(t)\vec{\nabla}|\psi_{ext}|e^{i\left(\frac{S_{ext}(\vec{r})}{\hbar}-\frac{E}{\hbar}t\right)}\cdot\vec{\nabla}f'\right.\\
&\qquad\left.+2\frac{i}{\hbar}\mathrm{A}(t)\vec{\nabla}S_{ext}\psi_{ext}\cdot\vec{\nabla}f'\right)+2ic^2e\vec{A}'_{sf}\\
&\qquad\cdot\left(\mathrm{A}(t)\vec{\nabla}f'\psi_{ext}+\boxed{\mathrm{A}(t)\vec{\nabla}|\psi_{ext}|e^{i\left(\frac{S_{ext}(\vec{r})}{\hbar}-\frac{E}{\hbar}t\right)}f'}+\frac{i}{\hbar}\mathrm{A}(t)\vec{\nabla}S_{ext}\psi_{ext}f'\right)+c^2\left(e\vec{A}'_{sf}\right)^2\psi'+(m_0c^2)^2\psi'
\end{aligned}
$$

Here we have in red terms involving $\psi_{ext}$ alone. The new terms multiplying $V'_{sf}$ and $\vec{A}'_{sf}$ can be solved exactly:

$$
\boxed{2i\hbar eV'_{sf}\psi_{ext}(r)\frac{\partial \mathrm{A}(t)}{\partial t}f'(\vec{r})}=\boxed{2ic^2e\vec{A}'_{sf}\cdot \mathrm{A}(t)\vec{\nabla}|\psi_{ext}|e^{i\left(\frac{S_{ext}(\vec{r})}{\hbar}-\frac{E}{\hbar}t\right)}f'}
$$

or $\frac{\partial \mathrm{A}(t)}{A\partial t}=-\frac{\vec{\mathrm{v}}\cdot\vec{\nabla}|\psi_{ext}|}{|\psi_{ext}|}$, verifying Eq 109. The rest of the new terms can be re-arranged as follows:

$$
\begin{aligned}
&-\hbar^2\left(\psi_{ext}(r)\mathrm{A}(t)\frac{\partial^2 f'(\vec{r})}{\partial t^2}\left(1+2\frac{\partial \mathrm{A}(t)}{A\partial t}\frac{\partial f'(\vec{r})}{\partial t}\Big/\frac{\partial^2 f'(\vec{r})}{\partial t^2}+\frac{\partial^2 \mathrm{A}}{\partial t^2}\Big/\frac{\partial^2 f'(\vec{r})}{\partial t^2}\right)+2\frac{\partial \psi_{ext}(r)}{\partial t}\mathrm{A}(t)\frac{\partial f'(\vec{r})}{\partial t}\left(1+\frac{\partial \mathrm{A}(t)}{A\partial t}\right)\right.\\
&\qquad\left.+\frac{\partial^2 \psi_{ext}(r)}{\partial t^2}\mathrm{A}(t)f'(\vec{r})\right)-2i\hbar \mathrm{e}V'_{sf}\left(\psi_{ext}(r)\mathrm{A}(t)\frac{\partial f'(\vec{r})}{\partial t}+\frac{\partial \psi_{ext}}{\partial t}\mathrm{A}(t)f'(\vec{r})\right)+\left(\mathrm{e}V'_{sf}\right)^2\psi'\\
&\qquad=-c^2\hbar^2\left(\mathrm{A}(t)\nabla^2 f'\psi_{ext}\left(1+\frac{2\vec{\nabla}|\psi_{ext}|\cdot\vec{\nabla}f'}{|\psi_{ext}|\nabla^2 f'}-\frac{\gamma\vec{a}\cdot\vec{\nabla}f'}{c^2\nabla^2 f'}\right)+\mathrm{A}(t)\nabla^2\psi_{ext}f'+2\frac{i}{\hbar}\mathrm{A}(t)\vec{\nabla}S_{ext}\psi_{ext}\cdot\vec{\nabla}f'\right)\\
&\qquad+2ic^2e\vec{A}'_{sf}\cdot\left(\mathrm{A}(t)\vec{\nabla}f'\psi_{ext}+\frac{i}{\hbar}\mathrm{A}(t)\vec{\nabla}S_{ext}\psi_{ext}f'\right)+c^2\left(e\vec{A}'_{sf}\right)^2\psi'+(m_0c^2)^2\psi'
\end{aligned}
$$

One can verify that all new terms $\frac{\partial \mathrm{A}(t)}{A\partial t}$, $\left(\frac{\partial \mathrm{A}(t)}{A\partial t}\frac{\partial f'(\vec{r})}{\partial t}\Big/\frac{\partial^2 f'(\vec{r})}{\partial t^2}\right)$, $\frac{2\vec{\nabla}|\psi_{ext}|\cdot\vec{\nabla}f'}{|\psi_{ext}|\nabla^2 f'}$, $\frac{\gamma\vec{a}\cdot\vec{\nabla}f'}{c^2\nabla^2 f'}$ are at least of order $\mathcal{O}\left(\frac{\ell_i}{\ell_e}\right)\ll 1$ while $\left(\frac{\partial^2 \mathrm{A}}{\partial t^2}\Big/\frac{\partial^2 \psi_{ext}(r)}{\partial t^2}\right)$ is of order $\mathcal{O}\left(\frac{\ell_i}{\ell_e}\right)^2$ and can be neglected in a first order approximation. We have seen previously that the remaining of the original equation can be written as:

$$
\begin{aligned}
&\left(m_{sf}c^2\right)^2 f_e(\vec{r}')\psi_{ext}\\
&\qquad=-c^2\hbar^2\nabla^2_{r'}f_e(\vec{r}')\psi_{ext}\left(1+\mathcal{O}\left(\frac{\ell_i}{\ell_e}\right)+\mathcal{O}\left(\frac{\ell_i}{\ell_e}\right)^2\right)+(m_0c^2)^2 f_e(\vec{r}')\psi_{ext}\\
&\qquad+\left(-e^2V^2_{sf}(\vec{r}')+2m_{sf}c^2eV_{sf}(\vec{r}')\right)f_e(\vec{r}')\psi_{ext}
\end{aligned}
$$

in the primed frame of reference. We can recognize that the main effect of the new terms of order $\mathcal{O}\left(\frac{\ell_i}{\ell_e}\right)$ is to remove the spherical symmetry present in Eq 106, in the direction of $\vec{\nabla}|\psi_{ext}|$, implying that the new solution $f_e(\vec{r}')$ will have deformations of order $\mathcal{O}\left(\frac{\ell_i}{\ell_e}\right)$ respect to the original spherically symmetric solution $f(\vec{r}')$. However, this equation can still be solved for the same value of $m_{sf}$, preserving the internal energy of the self-field 'particle'. In this way we recover Eq 104 for $\psi_{ext}$ and the interpretation of the constant energy content of the internal function as the particle mass.

### Self-field localized solution in the presence of an external field $V_{ext}$

In the presence of a time independent external electric field, Eq 107 is replaced by

$$\left(i\hbar\frac{\partial}{\partial t} - \mathrm{e}V'_{sf} - eV_{ext}\right)^2 \psi' = c^2\left(i\hbar\vec{\nabla} + e\vec{A}'_{sf}\right)^2\psi' + (m_0c^2)^2\psi'$$

We observe that this inclusion is equivalent to perform the replacement $V'_{sf} \Longrightarrow V'_{sf} + V_{ext}$. Neglecting all terms of order $\left(\frac{\ell_i}{\ell_e}\right)$, the last equation can be written as

$$\begin{aligned}
-\hbar^2&\left(\psi_{ext}(r)\mathrm{A}(t)\frac{\partial^2 f'(\vec{r})}{\partial t^2} + 2\frac{\partial\psi_{ext}(r)}{\partial t}\mathrm{A}(t)\frac{\partial f'(\vec{r})}{\partial t} + \frac{\partial^2\psi_{ext}(r)}{\partial t^2}\mathrm{A}(t)f'(\vec{r})\right)\\
&- 2i\hbar \mathrm{e}V'_{sf}\left(\psi_{ext}(r)\mathrm{A}(t)\frac{\partial f'(\vec{r})}{\partial t} + \frac{\partial\psi_{ext}}{\partial t}\mathrm{A}(t)f'(\vec{r})\right)\\
&- 2i\hbar \mathrm{e}V_{ext}\left(\psi_{ext}(r)\mathrm{A}(t)\frac{\partial f'(\vec{r})}{\partial t} + \frac{\partial\psi_{ext}}{\partial t}\mathrm{A}(t)f'(\vec{r})\right) + \left(\mathrm{e}V'_{sf}\right)^2\psi' + (\mathrm{e}V_{ext})^2\psi' + 2eV_{ext}\mathrm{e}V'_{sf}\psi'\\
&= -c^2\hbar^2\left(\mathrm{A}(t)\nabla^2 f'\psi_{ext} + \mathrm{A}(t)\nabla^2\psi_{ext}f' + 2\frac{i}{\hbar}\mathrm{A}(t)\vec{\nabla}S_{ext}\psi_{ext}\cdot\vec{\nabla}f'\right) + 2ic^2e\vec{A}'_{sf}\\
&\cdot\left(\mathrm{A}(t)\vec{\nabla}f'\psi_{ext} + \frac{i}{\hbar}\mathrm{A}(t)\vec{\nabla}S_{ext}\psi_{ext}f'\right) + c^2\left(e\vec{A}'_{sf}\right)^2\psi' + (m_0c^2)^2\psi'
\end{aligned}$$

Removing the red terms, the rest can be re-written as

$$\begin{aligned}
-\hbar^2&\left(\psi_{ext}(r)\mathrm{A}(t)\frac{\partial^2 f'(\vec{r})}{\partial t^2} + 2\mathrm{A}(t)\frac{\partial f'(\vec{r})}{\partial t}\left(\frac{\partial\psi_{ext}(r)}{\partial t} + i\hbar \mathrm{e}V_{ext}\psi_{ext}(r)\right)\right)\\
&- 2i\hbar \mathrm{e}V'_{sf}\left(\psi_{ext}(r)\mathrm{A}(t)\frac{\partial f'(\vec{r})}{\partial t} + \mathrm{A}(t)f'(\vec{r})\left(\frac{\partial\psi_{ext}}{\partial t} + i\hbar \mathrm{e}V_{ext}\psi_{ext}(r)\right)\right) + \left(\mathrm{e}V'_{sf}\right)^2\psi'\\
&= -c^2\hbar^2\left(\mathrm{A}(t)\nabla^2 f'\psi_{ext} + 2\frac{i}{\hbar}\mathrm{A}(t)\vec{\nabla}S_{ext}\psi_{ext}\cdot\vec{\nabla}f'\right) + 2ic^2e\vec{A}'_{sf}\\
&\cdot\left(\mathrm{A}(t)\vec{\nabla}f'\psi_{ext} + \frac{i}{\hbar}\mathrm{A}(t)\vec{\nabla}S_{ext}\psi_{ext}f'\right) + c^2\left(e\vec{A}'_{sf}\right)^2\psi' + (m_0c^2)^2\psi'
\end{aligned}$$

while the blue terms can be incorporated into the definition of the total energy appearing in the internal equation by doing the replacement $E \Longrightarrow E' \equiv E - V_{ext}$, in the equation $\frac{\partial\psi_{ext}}{\partial t} = i\hbar E\psi_{ext}(r)$, recovering the internal wave equation:

$$\left(m_{sf}c^2\right)^2 f(\vec{r}')\psi_{ext} = -c^2\hbar^2\nabla^2_{\vec{r}'}f(\vec{r}')\psi_{ext} + (m_0c^2)^2 f(\vec{r}')\psi_{ext} + \left(-e^2V^2_{sf}(\vec{r}') + 2m_{sf}c^2eV_{sf}(\vec{r}')\right)f(\vec{r}')\psi_{ext}$$

but where now the mass represents the energy $E' = \gamma m_{sf}c^2$ equivalent to

$$E - V_{ext} = \gamma m_{sf}c^2$$

which is the expected relationship, as seen from the classical relativistic equation:

$$(E - V_{ext})^2 = (pc)^2 + \left(m_{sf}c^2\right)^2 = \left(\gamma m_{sf}c^2\right)^2$$

In this way the full equation can now be written as

$$\left(i\hbar\frac{\partial}{\partial t} - eV_{ext}\right)^2 \psi_{ext}f = -c^2\hbar^2(\nabla^2\psi_{ext})f + \left(m_{sf}c^2\right)^2 f(\vec{r}')\psi_{ext}$$

recovering Eq 103 as a Klein Gordon equation for an effective 'point' charged particle.

## 2.9. Appendix 2.3: Double solution for the Pauli equation

As shown by Holland:

$$\frac{\partial S}{\partial t} + \frac{1}{2m}(\nabla S)^2 + Q + V = 0$$

$$m\ddot{\vec{x}} = -\nabla(Q + V)$$

$$\frac{\partial S}{\partial t} + \frac{1}{2m}\left(\vec{\nabla}S - \vec{A}\right)^2 + Q' + V = 0 \qquad A = -\vec{\nabla}\log\rho \times \vec{s}$$

$$Q' = Q + \frac{1}{m}\nabla S \cdot A - \frac{1}{2m}A^2$$

$$m\ddot{\vec{x}} = E + \dot{\vec{x}} \times B - \nabla V$$

$$E = -\nabla Q' - \frac{\partial A}{\partial t}, \quad B = \nabla \times A$$

$$B = -\vec{\nabla}\left(\vec{\nabla}\log\rho \cdot \vec{s}\right) + \vec{s}\nabla^2\log\rho$$

We can try to obtain an inter-particle model for this force in the chain model, by adding a force proportional to the magnitude of

Under these assumptions, the Schrödinger equation in the presence of an external vector potential A can be written as

$$-i\frac{\hbar}{c}\frac{\partial}{\partial t}\psi = \left(-\hbar i\vec{\nabla} - \vec{A}_{sp} - e\vec{A}\right)^2\psi + Q'\,\psi + V\,\psi$$

With

$$\vec{A}_{sp} = \vec{\Sigma} \times \hbar\vec{\nabla}$$

$$Q' = Q + \frac{1}{m}\vec{\nabla}S \cdot \vec{A}_{sp} - \frac{1}{2m}\left(\vec{A}_{sp}\right)^2$$

Let's now apply de Broglie's double solution to this equation:

Pauli classical:

$$i\hbar\frac{\partial\psi}{\partial t} = \frac{\left(-i\hbar\vec{\nabla} - e\vec{A}\right)^2}{2m}\psi + V\,\psi$$

$$\left(-i\hbar\vec{\nabla} - e\vec{A}\right)^2\psi = \left(-\hbar^2\nabla^2 + ie\hbar\vec{\nabla}\cdot\vec{A} + ie\hbar\vec{A}\cdot\vec{\nabla} + e^2A^2\right)\psi$$

With $\vec{\nabla}\cdot\vec{A} = 0$, $\vec{\nabla}\times\vec{A} = 0$ because grad not acting on A

$$\left(-i\hbar\vec{\nabla} - e\vec{A}\right)^2\psi = \left(-\hbar^2\nabla^2 + 2ie\hbar\vec{A}\cdot\vec{\nabla} + e^2A^2\right)\psi$$

$$i\hbar\frac{\partial\psi}{\partial t} = \frac{\left(-i\hbar\vec{\nabla}+\vec{\Sigma}\times\hbar\vec{\nabla}-e\vec{A}\right)^2}{2m}\psi + V\,\psi$$

$$\begin{aligned}\left(-i\hbar\vec{\nabla}+\vec{\Sigma}\times\hbar\vec{\nabla}-e\vec{A}\right)^2\psi &= \Big(-\hbar^2\nabla^2 + ie\hbar\vec{\nabla}\cdot\vec{A} + 2ie\hbar\vec{A}\cdot\vec{\nabla} + e^2A^2 - i\hbar\vec{\nabla}\cdot\left(\vec{\Sigma}\times\hbar\vec{\nabla}\right) - i\hbar\left(\vec{\Sigma}\times\hbar\vec{\nabla}\right)\cdot\vec{\nabla} - e\vec{A}\cdot\left(\vec{\Sigma}\times\hbar\vec{\nabla}\right) \\ &\quad - \left(\vec{\Sigma}\times\hbar\vec{\nabla}\right)\cdot e\vec{A}\Big)\psi = \left(-\hbar^2\nabla^2 + ie\hbar\vec{A}\cdot\vec{\nabla} + e^2A^2 - e\vec{A}\cdot\left(\vec{\Sigma}\times\hbar\vec{\nabla}\right) - \left(\vec{\Sigma}\times\hbar\vec{\nabla}\right)\cdot e\vec{A}\right)\psi \\ &= \left(-i\hbar\vec{\nabla} - e\vec{A}\right)^2\psi - 2e\vec{A}\cdot\left(\vec{\Sigma}\times\hbar\vec{\nabla}\psi\right) - \overbrace{\hbar\left(\vec{\Sigma}\cdot\left(\vec{\nabla}\times\vec{A}\right)\right)}^{0\text{ because grad doesn't work on A}}\;\psi\end{aligned}$$

$$\begin{aligned}i2e\hbar\vec{A}\cdot\vec{\nabla}\psi + 2e\vec{A}\cdot\left(\vec{\Sigma}\times\hbar\vec{\nabla}\psi\right) - \hbar\left(\vec{\Sigma}\cdot\left(\vec{\nabla}\times\vec{A}\right)\right)\psi &= i2e\hbar\vec{A}\cdot\vec{\nabla}\psi + 2e\vec{A}\cdot\left(\vec{\Sigma}\times\hbar\vec{\nabla}\left(\psi^{int}\psi^{ext}\right)\right) - \hbar\left(\vec{\Sigma}\cdot\left(\vec{\nabla}\times\vec{A}\right)\right)\psi^{int}\psi^{ext} \\ &= i2e\hbar\vec{A}\cdot\left(\psi^{ext}\vec{\nabla}\psi^{int} + \psi^{int}\vec{\nabla}\psi^{ext}\right) + 2\hbar e\vec{A}\cdot\left(\vec{\Sigma}\times\left(\psi^{ext}\vec{\nabla}\psi^{int} + \psi^{int}\vec{\nabla}\psi^{ext}\right)\right) \\ &\quad - \overbrace{\hbar\left(\vec{\Sigma}\cdot\vec{B}\right)\psi^{int}\psi^{ext}}^{0\text{ because grad doesn't work on A}}\end{aligned}$$

In order to get an equation for the external wavefunction alone, Lets integrate over the internal wavefunction.

$$\int\psi^{int*}\left[i2e\hbar\vec{A}\cdot\left(\psi^{ext}\vec{\nabla}\psi^{int} + \psi^{int}\vec{\nabla}\psi^{ext}\right) + 2\hbar e\vec{A}\cdot\left(\vec{\Sigma}\times\left(\psi^{ext}\vec{\nabla}\psi^{int} + \psi^{int}\vec{\nabla}\psi^{ext}\right)\right) - \hbar\left(\vec{\Sigma}\cdot\vec{B}\right)\psi^{int}\psi^{ext}\right]dV =$$

$$\begin{aligned}&= i2\hbar e\psi^{ext}\overbrace{\int\psi^{int*}\vec{A}\cdot\vec{\nabla}\psi^{int}dV}^{B\cdot L} + 2\hbar e\vec{A}\cdot\vec{\nabla}\psi^{ext}\underbrace{\int\psi^{int*}\psi^{int}dV}_{1} + 2\hbar e\psi^{ext}\overbrace{\int\psi^{int*}\vec{A}\cdot\left(\vec{\Sigma}\times\vec{\nabla}\psi^{int}\right)dV}^{B\cdot L_S} + 2\hbar e\vec{A} \\ &\quad\cdot\left(\vec{\Sigma}\times\vec{\nabla}\psi^{ext}\right)\int\psi^{int*}\psi^{int}dV + \overbrace{\hbar\left(\vec{\Sigma}\cdot\vec{B}\right)\psi^{ext}\underbrace{\int\psi^{int*}\psi^{int}dV}_{1}}^{0\text{ because grad doesn't work on A}}\end{aligned}$$

$$\vec{A} = \vec{r}\times\vec{B}$$

$$\begin{aligned}\int\psi^{int*}\left(\vec{\Sigma}\times\vec{\nabla}\psi^{int}\right)\cdot\vec{A}dV &= \int\psi^{int*}\left(\vec{\Sigma}\times\vec{\nabla}\psi^{int}\right)\cdot\left(\vec{r}\times\vec{B}\right)dV = \vec{B}\cdot\int\psi^{int*}\vec{r}\times\left(\vec{\Sigma}\times\vec{\nabla}\psi^{int}\right)dV \\ &= \vec{B}\cdot\int\psi^{int*}\vec{\Sigma}\psi^{int}dV = \vec{B}\cdot\vec{\Sigma}\end{aligned}$$

$$\int\psi^{int*}\vec{\nabla}\psi^{int}\cdot\vec{A}dV = \int\psi^{int*}\vec{\nabla}\psi^{int}\cdot\left(\vec{r}\times\vec{B}\right)dV = \vec{B}\cdot\int\psi^{int*}\vec{r}\times\vec{\nabla}\psi^{int}dV = \vec{B}\cdot\int\psi^{int*}\vec{L}\psi^{int}dV = \vec{B}\cdot\vec{L}$$

$$\int \psi^{int^*}\vec{A}\cdot\left(\vec{\Sigma}\times\vec{\nabla}\psi^{ext}\right)\psi^{int}dV = \left(\vec{\Sigma}\times\vec{\nabla}\psi^{ext}\right)\cdot\int \psi^{int^*}\vec{A}\psi^{int}dV = \left(\vec{\Sigma}\times\vec{\nabla}\psi^{ext}\right)\cdot\int\psi^{int^*}\vec{r}\times\vec{B}\psi^{int}dV =$$

$$= -\left(\vec{\Sigma}\times\vec{\nabla}\psi^{ext}\right)\cdot\left(\vec{B}\times\overbrace{\int\psi^{int^*}\vec{r}\psi^{int}dV}^{=\mathbf{0}}\right) = 0$$

$$\int \psi^{int^*}\vec{A}\cdot\vec{\nabla}\psi^{ext}\psi^{int}dV = \vec{\nabla}\psi^{ext}\cdot\int \psi^{int^*}\vec{A}\psi^{int}dV = \vec{\nabla}\psi^{ext}\cdot\int\psi^{int^*}\vec{r}\times\vec{B}\psi^{int}dV =$$

$$= -\vec{\nabla}\psi^{ext}\cdot\left(\vec{B}\times\overbrace{\int\psi^{int^*}\vec{r}\psi^{int}dV}^{=\mathbf{0}}\right) = 0$$

$V = \vec{r}\cdot\vec{E}_V$

$$i\int\psi^{int^*}\vec{\mathbf{v}}\cdot\vec{\nabla}\psi^{ext}V\psi^{int}dV = i\vec{\mathbf{v}}\cdot\vec{\nabla}\psi^{ext}\int\psi^{int^*}\vec{r}\cdot\vec{E}_V\psi^{int}dV = i\vec{\mathbf{v}}\cdot\vec{\nabla}\psi^{ext}\vec{E}_V\cdot\overbrace{\int\psi^{int^*}\vec{r}\psi^{int}dV}^{=\mathbf{0}} = 0$$

Or in summary:

$$\left(-i\hbar\vec{\nabla}+\vec{\Sigma}\times\hbar\vec{\nabla}-e\vec{A}\right)^2\psi = \left(-i\hbar\vec{\nabla}-e\vec{A}\right)^2\psi - 2e\vec{A}\cdot\left(\vec{\Sigma}\times\hbar\vec{\nabla}\psi\right) - \hbar\left(\vec{\Sigma}\cdot\left(\vec{\nabla}\times\vec{A}\right)\right)\psi$$

$$\rightarrow \left(-i\hbar\vec{\nabla}-e\vec{A}\right)^2\psi + i2\hbar e\psi^{ext}B\cdot L + 2\hbar e\psi^{ext}B\cdot L_S = \left(-i\hbar\vec{\nabla}-e\vec{A}\right)^2\psi + i2\hbar e\psi^{ext}B\cdot\overbrace{(L+L_S)}^{\sigma}$$

But $(L + L_S)$ is the angular momentum total of the electron, and that can be considered the external, effective electron spin. This quantity should be identified with $\sigma = \frac{\hbar}{2}$h/2.

And this is the classical Pauli equation:

$$\underbrace{i\hbar\frac{\partial}{\partial t}|\varphi_\pm\rangle = \left[\frac{1}{2m}\left(\hat{\vec{p}}-q\vec{A}\right)^2+qV\right]\hat{1}|\varphi_\pm\rangle}_{Schrodinger\ equation} - \underbrace{\frac{q\hbar}{2m}\vec{s}\cdot\vec{B}|\varphi_\pm\rangle}_{Stern\ Gerlach\ term}$$

# 3. Chapter: Photon Soliton

In this chapter we present a possible scenario for the generation of an electromagnetic soliton, or 'photon', consisting in the generalization of the Maxwell equations to include vacuum polarization terms. In the previous chapter of this work we have considered the formation of solitons for the electron field starting from the Dirac-Poisson equations. The chapter 4 concerns an intuitive presentation of Field Quantization in general.

## 3.1. Maxwell equations:

### 3.1.1. Electromagnetic fields

In terms of the electromagnetic fields the equations are

$$\vec{\nabla} \cdot \vec{D} = \rho$$

$$\vec{\nabla} \cdot \vec{B} = 0$$

$$\vec{\nabla} \times \vec{E} = -\frac{\partial \vec{B}}{\partial t}$$

$$\vec{\nabla} \times \vec{H} = \frac{\partial \vec{D}}{\partial t} + \vec{j}$$

where $\rho \; and \; j$ are the charge and current densities respectively. In the absence of polarization charges and currents

$$\vec{B} = \mu_0 \vec{H}$$

$$\vec{D} = \epsilon_0 \vec{E}$$

The expressions can be re-expressed in free space in terms of a single field as:

$$\nabla^2 \vec{E} - \frac{1}{c^2}\frac{\partial^2 \vec{E}}{\partial t^2} = -\frac{1}{\epsilon_0}\left(-\vec{\nabla}\rho - \frac{1}{c^2}\frac{\partial \vec{j}}{\partial t}\right)$$

$$\nabla^2 \vec{B} - \frac{1}{c^2}\frac{\partial^2 \vec{B}}{\partial t^2} = -\mu_0 \vec{\nabla} \times \vec{j}$$

**Eq 110**

where the equation

$$-\nabla \times \nabla \times \vec{E} - \frac{1}{c^2}\frac{\partial^2 \vec{E}}{\partial t^2} = \frac{1}{\varepsilon_0}\frac{1}{c^2}\frac{\partial \vec{j}}{\partial t}$$

and the identity

$$\nabla \times \nabla \times \vec{E} = \vec{\nabla}\left(\vec{\nabla} \cdot \vec{E}\right) - \nabla^2 \vec{E}$$

together with the replacement

$$\nabla \cdot \vec{E} = \frac{\rho}{\varepsilon_0}$$

have been used. Because of the use of this last equation, the solutions of Eq 110 are fulfilling automatically Poisson equation.

### 3.1.2. Potential fields

The fields can be also specified in terms of the potentials. The electromagnetic potentials are defined by the equations:

$$\vec{E} = -\vec{\nabla}V - \dot{\vec{A}}$$
$$\vec{B} = \vec{\nabla} \times \vec{A}$$

The Maxwell's equations for the electromagnetic potentials in Lorentz gauge

$$\frac{1}{c}\frac{\partial V}{\partial t} + \vec{\nabla} \cdot \vec{A} = 0$$

read

$$\nabla^2 V - \frac{1}{c^2}\frac{\partial^2 V}{\partial t^2} = -\frac{\rho}{\varepsilon_0}$$

$$\nabla^2 \vec{A} - \frac{1}{c^2}\frac{\partial^2 \vec{A}}{\partial t^2} = -\mu_0 \vec{J}$$

**Eq 111**

or

$$\square \vec{A} = \frac{1}{\varepsilon_0 c^2}\vec{J}$$

$$\square V = \frac{1}{\varepsilon_0}\rho$$

where

$$\square = \left(\frac{1}{c^2}\frac{\partial^2}{\partial t^2} - \nabla^2\right)$$

### 3.1.3. Different notations, same equations, but really different interpretation?

Maxwell's equation for the potential fields in 4-vector relativistic notation is

$$\left(\frac{1}{c^2}\frac{\partial^2}{\partial t^2} - \nabla^2\right)A_\mu = \square A_\mu = \partial^\nu \partial_\nu A_\mu = \mu_0 j_\mu$$

**Eq 112**

where

$$\partial^\mu = \frac{\partial}{\partial x_\mu} = \left(\frac{1}{c}\frac{\partial}{\partial t}, -\vec{\nabla}\right), \qquad \partial^\nu \partial_\nu = \partial_\mu \partial^\mu = \square$$

and the Lorentz gauge equation is given by

$$\partial^\mu A_\mu = 0$$

where it is clear that the potential 4-vector $A_\mu$ is defined perpendicular to the $\partial^\mu$ differential operator.

In QED language one can use the following equivalences for the momentum operator $q$:

$$q = (q_0, -\vec{q}) \sim \left(\frac{1}{c}\frac{\partial}{\partial t}, -\vec{\nabla}\right) = \partial^\mu$$

$$q^2 = (q_0, -\vec{q}) \cdot (q_0, +\vec{q}) = q_0^2 - |\vec{q}|^2 = \square = \left(\frac{1}{c^2}\frac{\partial^2}{\partial t^2} - \nabla^2\right) = \partial^\nu \partial_\nu$$

This operator can be interpreted as adding a new notation for the electromagnetic field equations. In fact, on the mass shell $q^2 = 0$, therefore the 'on mass shell' fields are really satisfying Maxwell's equations in vacuum (Laplace equation) for the potential fields:

$$q^2 A_\mu(q) = 0$$

**Eq 113**

which is nothing else than

$$q^2 A_\mu(q) = 0 \quad \rightarrow \quad \square A_\mu = \left(\frac{1}{c^2}\frac{\partial^2}{\partial t^2} - \nabla^2\right) A_\mu = \partial^\nu \partial_\nu A_\mu = 0$$

while the Lorentz gauge condition reads

$$q^\mu A_\mu(q) = 0$$

In summary the following notations refer to the same differential operator, namely the d'Alembertian:

$$\boxed{\partial^\nu \partial_\nu \equiv \partial_\mu \partial^\mu \equiv \square \equiv q^2 \equiv \left(\frac{1}{c^2}\frac{\partial^2}{\partial t^2} - \nabla^2\right)}$$

### 3.1.4. Still another type of notation: Polarization sources

In the presence of polarization, the charge and current densities are given by

$$j = \frac{\partial P}{\partial t} + \nabla \times M + \sigma E \qquad \rho = -\nabla \cdot P$$

**Eq 114**

the sources can be expressed in terms of the polarization and magnetization density by

$$\vec{j}(\vec{r}) = \vec{j}_p(\vec{r}) + \vec{j}_m(\vec{r})$$

$$\vec{j}_p(\vec{r}) = \dot{\vec{P}}(\vec{r})$$

$$\vec{j}_m(\vec{r}) = \vec{\nabla} \times \vec{M}(\vec{r})$$

The relationship between the fields D, E, H and B goes over:

$$\vec{H} = \frac{1}{\mu_0}\vec{B} - \vec{M}$$

$$\vec{B} = \mu \vec{H}$$

$$\vec{D} = \epsilon_0 \vec{E} + \vec{P}$$

the equation for B reads:

$$\vec{\nabla} \times \vec{B} = \mu_0 \left[\vec{j} + \vec{\nabla} \times \vec{M}\right]$$

and for E:

$$\nabla^2 E - \frac{1}{c^2}\frac{\partial^2 E}{\partial t^2} = -\frac{1}{\varepsilon_0}\left(-\nabla\rho - \frac{1}{c^2}\frac{\partial J}{\partial t}\right) = -\frac{1}{\varepsilon_0}\left(\nabla(\nabla\cdot P) - \frac{1}{c^2}\frac{\partial^2 P}{\partial t^2} - \frac{1}{c^2}\frac{\partial}{\partial t}(\nabla\times M) - \frac{\sigma}{c^2}\frac{\partial E}{\partial t}\right)$$
$$= -\frac{1}{\varepsilon_0}\left(\nabla\times\nabla\times P + \nabla^2 P - \frac{1}{c^2}\frac{\partial^2 P}{\partial t^2} - \frac{1}{c^2}\frac{\partial}{\partial t}(\nabla\times M) - \frac{\sigma}{c^2}\frac{\partial E}{\partial t}\right)$$

Or

$$\nabla^2 D - \frac{1}{c^2}\frac{\partial^2 D}{\partial t^2} == -\frac{1}{\varepsilon_0}\left(\nabla\times\nabla\times P - \frac{1}{c^2}\frac{\partial}{\partial t}(\nabla\times M)\right)$$

**Eq 115**

The magnetic field follows the equation

$$\nabla^2 \vec{B} - \frac{1}{c^2}\frac{\partial^2 \vec{B}}{\partial t^2} = 4\pi\left\{\vec{\nabla}\times(\vec{\nabla}\times\vec{M}) - \frac{1}{c}\frac{\partial}{\partial t}(\vec{\nabla}\times\vec{P})\right\}$$

**Eq 116**

when $\sigma = 0$.

In a similar way, the equations for the potentials read:

$$\nabla^2 V - \frac{1}{c^2}\frac{\partial^2 V}{\partial t^2} = -\frac{\rho}{\varepsilon_0} = \frac{\nabla\cdot P}{\varepsilon_0}$$

$$\nabla^2 A - \frac{1}{c^2}\frac{\partial^2 A}{\partial t^2} = -\mu_0 j = -\mu_0\frac{\partial P}{\partial t} - \mu_0\nabla\times M - \mu_0\sigma E$$

Textbooks on Nonlinear Optics[41], typically simplify equation Eq 116 to:

$$\nabla^2 \vec{E} - \frac{1}{c^2}\frac{\partial^2 \vec{E}}{\partial t^2} = \frac{1}{\epsilon_0 c^2}\frac{\partial^2 \vec{P}}{\partial t^2}$$

$$\nabla^2 \vec{E} - \frac{1}{\epsilon_0 c^2}\frac{\partial^2}{\partial t^2}\vec{D} = 0$$

and therefore are not useful for a thorough analysis of nonlinear vacuum polarization effects.

## 3.2. Photon-matter interaction and effective equations:

For an energy eigenstate in the absence of magnetic fields the Schrödinger equation reduces to

$$\frac{\hbar^2}{2m}\nabla^2\psi = -(E - eV)\psi$$

**Eq 117**

The charge density is given by

$$\rho_e(r) = e\psi^*\psi(r)$$

**Eq 118**

Which can be inserted into the Poisson's equation to get the electrostatic field

$$\nabla^2 V(r) = -\frac{1}{4\pi\epsilon_0} e\psi^*\psi(r)$$

**Eq 119**

Taking a further Laplacian from Eq 119 we have

$$\nabla^4 V(r) = -\frac{1}{4\pi\epsilon_0} e\nabla^2\left(\psi^*\psi(r)\right) = -\frac{1}{4\pi\epsilon_0} e(\psi^*\nabla^2\psi + \psi\nabla^2\psi^* + 2\nabla\psi\cdot\nabla\psi^*) = -\frac{1}{4\pi\epsilon_0} e(\psi^*\nabla^2\psi + \cdots)$$

Replacing from Eq 85

$$\nabla^4 V(r) = \frac{1}{4\pi\epsilon_0} e\psi^* \frac{2m}{\hbar^2}(E - eV)\psi + \cdots = \frac{1}{4\pi\epsilon_0}\frac{2m}{\hbar^2} e\psi^*\psi(E - eV) + \cdots = \frac{2m}{\hbar^2} e\nabla^2 V(E - eV) + \cdots$$

From this last equation we can observe the appearance of:

a. non linear terms in $V$:

$$\frac{2m}{\hbar^2} e^2 V\nabla^2 V$$

b. higher derivatives:

$$\nabla^4 V$$

Then we can conclude that the existence of interaction with other fields, is equivalent in first approximation to an effective equation[42] including nonlinear terms and higher order derivatives[4].

## 3.3. Vacuum polarization

We are going to consider the effective modifications in the Maxwell equations generated by the presence of vacuum polarization. First to the one-loop one-photon problem leading to the so called Uehling potential, and then to the photon-photon scattering problem leading to the effective expressions known as Heisenberg-Euler Lagrangian.

### 3.3.1. One-loop Feynman diagram due to virtual electron-positron pair creation

The unperturbed photon propagator $D_{F\mu\nu}(q)$ is modified due to a one-loop Feynman diagram corresponding to virtual electron-positron pair creation, to order $e^2$, to[43]:

$$iD_{F\mu\nu}(q) = \frac{-4\pi i}{q^2 + i\varepsilon} g_{\mu\nu} \qquad \rightarrow \qquad iD'_{F\mu\nu}(q) = iD_{F\mu\nu}(q) + iD_{F\mu\lambda}(q)\frac{i\Pi^{\lambda\sigma}(q)}{4\pi} iD_{F\sigma\nu}(q) + \cdots$$

The unperturbed matrix element for scattering $M_{fi}$ due to the interaction of an electron with a field $A_\mu$ in the presence of the previous polarization term is modified to:

$$M_{fi} \sim e\bar{u}_f\gamma^\mu u_i A_\mu(q) \qquad \rightarrow \qquad M_{fi}^{VP} \sim M_{fi} + e\bar{u}_f\gamma^\mu u_i \frac{-4\pi i}{q^2 + i\varepsilon}\frac{i\Pi_{\mu\nu}(q)}{4\pi} A^\nu(q) = e\bar{u}_f\gamma^\mu u_i \left(\delta_{\mu\nu} + \frac{-4\pi i}{q^2 + i\varepsilon}\frac{i\Pi_{\mu\nu}(q)}{4\pi}\right) A^\nu(q)$$

**Eq 120**

From the textbook calculations by Greiner[43], it is found that

$$\Pi_{\mu\nu}(q) = -\frac{e^2}{3\pi}\ln\frac{\Lambda^2}{m^2} + \Pi^R(q^2)$$

also called the Uehling correction. Here

$$\Pi^R(q^2) = -\frac{e^2}{\pi}\frac{q^2}{m^2}\left(\frac{1}{15} + \frac{1}{140}\frac{q^2}{m^2} + \cdots\right)$$

where $e^2 \equiv \alpha$ stands for the fine structure constant. Making the replacements $e^2 \leftrightarrow \alpha, \quad m \leftrightarrow \frac{1}{\lambda_{Comp}}$ this last equation reads

$$\Pi^R\left(\frac{1}{c^2}\frac{\partial^2}{\partial t^2} - \nabla^2\right) = -\frac{\alpha}{\pi}\lambda_{Comp}^2\left(\frac{1}{c^2}\frac{\partial^2}{\partial t^2} - \nabla^2\right)\left(\frac{1}{15} + \frac{1}{140}\lambda_{Comp}^2\left(\frac{1}{c^2}\frac{\partial^2}{\partial t^2} - \nabla^2\right) + \cdots\right)$$

---

[4] Typical nonlinear equations derived from effective Lagrangian theories are the nonlinear Schrödinger equation and the nonlinear Klein Gordon equation, similar to the Born Infeld equation. We can also mention the similarity between the non-abelian monopole solution and the Uehling potential analyzed below.

The modified matrix element Eq 120 can be interpreted as generated by an effective or renormalized field $A_\mu^R$ given by

$$A_\mu^R = A_\mu^0 + iD_{F\mu\nu}(q)\frac{i\Pi^{\nu\sigma}(q)}{4\pi}A_\sigma^0 = A_\mu^0 + \frac{-4\pi i}{q^2}\frac{i}{4\pi}\overline{\Pi}(q^2)A_\mu^0 = \left(1 + \frac{\widetilde{\Pi}(q^2)}{q^2}\right)A_\mu^0$$

where $A_\mu^0$ is the value of the field in the absence of vacuum polarization. The equation followed by $A_\mu^R$ in the presence of charges is

$$\left(\frac{1}{c^2}\frac{\partial^2}{\partial t^2} - \nabla^2\right)A_\mu^R = j_\mu$$

or equivalently

$$\left(\frac{1}{c^2}\frac{\partial^2}{\partial t^2} - \nabla^2\right)A_\mu^0 = j_\mu + \frac{i\Pi_{\mu\nu}(q)}{4\pi}A^\nu$$

which allows the identification of a 'polarization current' given by[5]

$$j_\mu^{VP}(q) = \frac{i\Pi_{\mu\nu}(q)}{4\pi}A^\nu(q)$$

This current density is also called Uehling current density. This means that the original Maxwell equation Eq 112 in the absence of external sources should be modified as follows:

$$\left(\frac{1}{c^2}\frac{\partial^2}{\partial t^2} - \nabla^2\right)A_\mu = 0 \;\; goes\ over \;\; \left(\frac{1}{c^2}\frac{\partial^2}{\partial t^2} - \nabla^2\right)\left(1 - \frac{\alpha}{\pi}\left(\frac{1}{15} + \frac{\lambda_{Comp}^2}{140}\left(\frac{1}{c^2}\frac{\partial^2}{\partial t^2} - \nabla^2\right) + \cdots\right)\right)A_\mu = 0$$

The last equation can be rewritten as

$$\left(\frac{1}{c^2}\frac{\partial^2}{\partial t^2} - \nabla^2\right)\left(1 - \frac{1}{\pi}\left(\frac{\alpha}{15} + r_e^2\underbrace{\frac{1}{140\alpha}}_{0.979}\left(\frac{1}{c^2}\frac{\partial^2}{\partial t^2} - \nabla^2\right) + \cdots\right)\right)A_\mu = 0$$

---

[5] the current $j_\mu^{VP}(q)$ fulfills the condition of current conservation because the equation

$$q^\mu\Pi_{\mu\nu}(q) = 0$$

is valid due to gauge invariance.

or as

$$\left(\frac{1}{c^2}\frac{\partial^2}{\partial t^2}-\nabla^2\right)A_\mu \cong \underbrace{\frac{1}{\pi}\left(\frac{1}{c^2}\frac{\partial^2}{\partial t^2}-\nabla^2\right)\left(\frac{\alpha}{15}+r_e^2\left(\frac{1}{c^2}\frac{\partial^2}{\partial t^2}-\nabla^2\right)\right)A_\mu}_{j_\mu^{Ueh}}$$

**Eq 121**

This is an effective Maxwell equation in the presence of linear vacuum polarization effects. In this equation, one can recognize the Uehling vacuum polarization current density $j_\mu^{Ueh}$is given by

$$j_\mu^{Ueh}=\frac{1}{\pi}\left(\frac{1}{c^2}\frac{\partial^2}{\partial t^2}-\nabla^2\right)\left(\frac{\alpha}{15}+r_e^2\left(\frac{1}{c^2}\frac{\partial^2}{\partial t^2}-\nabla^2\right)\right)A_\mu$$

#### 3.3.1.1. Solutions, Uehling potential

For the Coulomb case the original equation in the presence of charges:

$$-\nabla^2 V=-4\pi\rho_{free}$$

is modified to

$$-\nabla^2 V+\frac{1}{\pi}\nabla^2\left(\frac{\alpha}{15}-r_e^2\nabla^2\right)V=-4\pi\rho_{free}$$

**Eq 122**

or

$$-\nabla^2 V=-4\pi\rho_{Uehling}-4\pi\rho_{free}$$

where the Uehling density is defined

$$\rho_{Uehling}=\nabla^2\frac{1}{\pi}\left(\frac{\alpha}{15}-r_e^2\nabla^2\right)V=\nabla\cdot\nabla\frac{1}{\pi}\left(\frac{\alpha}{15}-r_e^2\nabla^2\right)V$$

This last expression allows us to identify a polarization density $P_{Uehling}$ as

$$\rho_{Uehling}=\nabla\cdot P_{Uehling} \qquad \rightarrow \qquad P_{Uehling}=\nabla\frac{1}{\pi}\left(\frac{\alpha}{15}-r_e^2\nabla^2\right)V$$

We can see that the polarization term is linear in the fields. Now, Eq 122 can be rewritten as

$$-\nabla^2\left(1-\frac{1}{\pi}\left(\frac{\alpha}{15}-r_e^2\nabla^2\right)\right)V=-4\pi\rho_{free}$$

which allows to define a renormalized potential field $V^R$

$$V^R=\left[1-\frac{1}{\pi}\left(\frac{\alpha}{15}-r_e^2\nabla^2\right)\right]V$$

which follows the classical or original Coulomb equation

$$-\nabla^2 V^R=-4\pi\rho_{free}$$

**Eq 123**

and where the term

$$\frac{1}{\pi}\left(\frac{\alpha}{15} - r_e^2 \nabla^2\right)$$

can be interpreted as a renormalization factor.

*3.3.1.1.1. Case 1: charge density given at a point*

The well known solution to Eq 123 when

$$4\pi\rho_{free} = \delta(r)$$

is

$$V^R(r) = \frac{q}{r}$$

which implies

$$\frac{1}{\pi}\left(\frac{\alpha}{15} - r_e^2 \nabla^2\right) V = \frac{q}{r}$$

The solution of this last equation for V is the Uehling potential[44]. From Greiner the solution for V is:

$$V(r) \cong -\frac{Ze}{r}\left[1 + \frac{2\alpha}{3\pi}\left(\log\frac{\lambda_{Comp}}{r} - \frac{5}{6} - C\right)\right] \qquad mr \ll 1$$

$$V(r) \cong -\frac{Ze}{r}\left[1 + \frac{\alpha}{4\sqrt{\pi}}\frac{e^{-2r/\lambda_{Comp}}}{\left(r/\lambda_{Comp}\right)^{3/2}}\right] \qquad mr \gg 1$$

In the short distance limit the charge responsible for this potential is given by (Greiner)

$$\rho_{VP}^{-}(r) \cong Ze\frac{2\alpha}{3\pi}\frac{1}{\pi}\frac{1}{r^3}$$

It is to remark that the polarization charge density starts being appreciably different from a zero at a distance of the order of the Compton wavelength from the center of the electron. It is to note that both, the charge density and the potential fields diverge at the center of the electron. But show something important, they might be the source of a soliton-like structure residing in the interior of the electron. It is also important to remark that this problem has been treated historically on the basis of both, propagator techniques following Feynman in the 1950's and of partial differential equations in space and time on the original papers of Uehling and Weisskopf in the 1930's. The origin of the description in terms of a propagator can be found in the 1930's papers from Dirac and Heisenberg using the non-diagonal density R matrix techniques.

We see that the Uehling potential corresponds to a solution in the vicinity of a point charge, where the unmodified potential field would be $V = \frac{q}{r}$.

$$\frac{1}{\pi}\left(\frac{\alpha}{15} - r_e^2 \nabla^2\right) V = \frac{q}{r}$$

$$\frac{1}{\pi} r_e^2 \nabla^2 V = -\frac{q}{r} + \frac{\alpha}{15\pi} V$$

$$r_e^2 \nabla^2 V = -\frac{\pi}{r_e^2}\frac{q}{r} + \frac{\pi}{r_e^2}\frac{\alpha}{15\pi} V$$

The Uehling potential has the following shape:

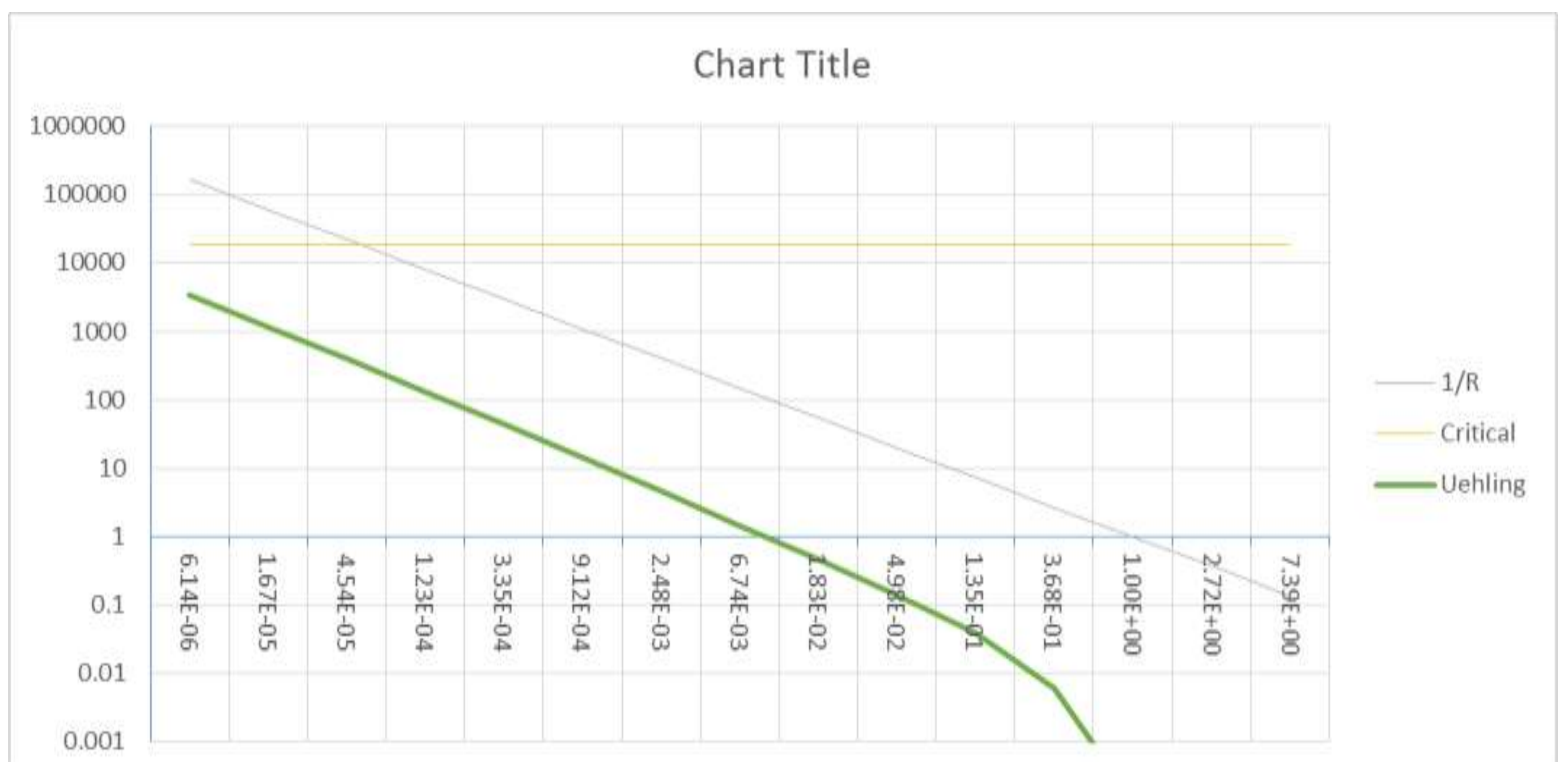

**Figure 13**

#### *3.3.1.1.2. Case 2: Extended source:*

The Uehling potential corresponds to the vacuum polarization generated by a point-like charge source. For an extended finite source, one can find the vacuum polarization field by convolution of the original Uehling potential with the source charge density[43]. For an electrostatic spherically symmetric source

$$j_0(r) \equiv \rho(r) = -Zeh(r)$$

with $\int d^3xh(r) = 4\pi \int_0^\infty drr^2h(r) = 1$

In the momentum or Fourier space, the convolution reduces to a multiplication:

$$U^{VP}(x) = \int \frac{d^3q}{(2\pi)^3} e^{i\vec{q}\cdot\vec{x}} \Pi^R(-q^2)\frac{4\pi}{q^2}\tilde{\rho}(\vec{q})$$

$$\tilde{\rho}(\vec{q}) = \int d^3xe^{-i\vec{q}\cdot\vec{x}}\,\rho(\vec{x})$$

after angular integration:

$$U^{VP}(x) = \frac{2}{\pi}\int d|q|\frac{\sin|q|r}{|q|r}\Pi^R(-q^2)\frac{4\pi}{q^2}\tilde{\rho}(|q|)$$

Greiner[43] calculates the Uehling potential for a source charge density described by the Fermi distribution function

$$\rho(r) = \frac{\rho_0}{1+e^{(r-R)/c}}$$

#### *3.3.1.1.3. Case 3: null source:*

It is interesting to study the solution to the case in pure vacuum, where $\rho_{Uehling} = 0$. In this case the equation for the potential reads:

$$\rho_{Uehling} = \nabla^2 \frac{1}{\pi} \underbrace{\left[\left(\frac{\alpha}{15} - r_e^2 \nabla^2\right) V\right]}_{V_0} = \frac{1}{\pi} \nabla^2 V_0 = 0$$

with a particular solution $V_0 = constant$. In particular, for $V_0 = 0$

$$r_e^2 \nabla^2 V = \frac{\alpha}{15} V$$

In spherical coordinates, assuming spherical symmetry the Laplacian is,

$$\nabla_r^2 = \frac{1}{r^2} \frac{\partial}{\partial r} \left( r^2 \frac{\partial V}{\partial r} \right)$$

and the equation reads

$$\frac{1}{r^2} \frac{\partial}{\partial r} \left( r^2 \frac{\partial V(r)}{\partial r} \right) = aV(r)$$

where

$$a = \frac{\alpha}{15 r_e^2}$$

Following Wolframalpha.com the solution is

$$V(r) = \frac{c_1 e^{-\sqrt{a} r}}{r} + \frac{c_2 e^{\sqrt{a} r}}{\sqrt{a} r}$$

In this case we see that the vacuum would support, a divergent soliton-like structure, based solely on first order vacuum polarization. The characteristic size of the soliton structure would be a few Compton wavelengths.

### 3.3.2. Nonlinear vacuum polarization, Euler Heisenberg Lagrangian:

We will consider the next order vacuum polarization effect, first analyzed by Heisenberg and Euler who expressed their result in terms of an effective Lagrangian expression in 1936[45]. The full Lagrangian for the electromagnetic field in the presence of vacuum polarization found by Euler and Heisenberg is given by[43] :

$$\mathcal{L} = \mathcal{L}_0 + \mathcal{L}'$$

where $\mathcal{L}_0$ is the free-field Lagrangian we have seen previously:

$$\mathcal{L}_0 = \frac{1}{8\pi} (E^2 - B^2)$$

and $\mathcal{L}'$ is given by:

$$\mathcal{L}'(\vec{B} \parallel \vec{E}) = \frac{m^4}{8\pi^2} \int_0^\infty d\eta \frac{e^{-\eta}}{\eta^3} \left[ -\tilde{E}\eta \cot \tilde{E}\eta \, \tilde{B}\eta \cot \tilde{B}\eta + 1 - \frac{1}{3} (\tilde{E}^2 - \tilde{B}^2) \eta^2 \right]$$

where

$$\tilde{B} = \frac{B}{B_{cr}} = \frac{|e|}{m^2} B \qquad B_{cr} = \frac{m^2}{|e|} = \frac{m^2 c^3}{|e| \hbar} = 4.4 \times 10^{13} \, Gauss = 4.4 \times 10^9 \, Tesla$$

$$\tilde{E} = \frac{E}{E_{cr}} = \frac{|e|}{m^2} E$$

For weak fields the first powers expansion in the fields is

$$\mathcal{L}'(\vec{B} \parallel \vec{E}) = \frac{1}{8\pi}\frac{e^4}{45\pi m^4}\left[(B^2 - E^2)^2 + 7\left(\vec{E}\cdot\vec{B}\right)^2\right]$$

As we can see this expression is provided as a function of the fields $\vec{E}$ and $\vec{B}$. This Lagrangian readily corresponds to the dipole representation where the Lagrangian is given by[46] the multipole form or the Power Zienau Wooley transformation:

$$L = \sum_{\alpha} \frac{m_\alpha \dot{\vec{r}}_\alpha^2}{2} - \sum_{\alpha<\beta} \frac{q_\alpha q_\beta}{4\pi\varepsilon_0\left|\vec{r}_\alpha - \vec{r}_\beta\right|} - \sum_{\alpha} \varepsilon_{Coul}^{\alpha} + \frac{\varepsilon_0}{2}\int d^3r\left[\vec{E}_\perp^2 - c^2\vec{B}^2\right] + \int d^3r\vec{M}\cdot\vec{B} + \int d^3r\vec{P}\cdot\vec{E}_\perp$$

where we can see that the Heisenberg-Euler Lagrangian can be interpreted as providing the last two terms in the previous expression. These terms are describing the interaction of the $\vec{E}$ and $\vec{B}$ fields with effective polarization and magnetization fields.

We recall here that the original purpose of the Lagrangian in classical physics

$$L = \int d^3r\, \mathcal{L}\left(A_j, \dot{A}_j, \partial_i A_j\right)$$

is to provide a basic expression from which the equations of motion, in this case the Maxwell equations, could be deduced by using the Lagrange equations

$$\frac{d}{dt}\frac{\partial\mathcal{L}}{\partial\dot{A}_j} = \frac{\partial\mathcal{L}}{\partial A_j} - \sum_{i=x,y,z} \partial_i \frac{\partial\mathcal{L}}{\partial\left(\partial_i A_j\right)}$$

This can be done when the Lagrangian is expressed, not as a function of the $\vec{E}$ and $\vec{B}$ fields, but as a function of the electromagnetic potentials $\vec{A}$ and $V$. Such a Lagrangian is called the standard Lagrangian which in terms of the potentials reads

$$\mathcal{L} = \left(\left(\dot{\vec{A}} + \vec{\nabla}V\right)^2 - c^2\left(\vec{\nabla}\times\vec{A}\right)^2\right)$$

From this Lagrangian one can derive the equations for the potentials in Coulomb gauge. To derive the equations in the Lorentz gauge a different version of the Lagrangian is used:

$$\mathcal{L}_R = \frac{\varepsilon_0}{2}\left[\dot{\vec{A}}^2 - \left(\frac{\dot{V}}{c^2}\right)^2 - c^2\sum_{ij}\left(\partial_i A_j\right)^2 + \left(\vec{\nabla}V\right)^2\right]$$

But in all cases one has to write the Lagrangian in terms of the potentials and its first derivatives in order to derive the correct equations of motion via the Lagrange equations. The reason for that is because the Lagrangian originally was defined as the difference between the kinetic and potential energies of the system. The kinetic energy was associated with the time and spatial derivatives of the fields, and the potential energy as a function of the field themselves. But in the Heisenberg-Euler Lagrangian, the potential energy is given as a function of the potential field derivatives, and therefore the sign of the terms obtained from the Lagrangian may not provide the right sign for the different terms in the wave equations. We can identify therefore two issues with the Euler-Heisenberg Lagrangian when trying to derive the wave equations or equations of motion for the fields: the Lagrangian is written in terms of the dipole formalism, and the sign of the different terms in the derived differential equation are not clear. The safest way to obtain the effective Maxwell equations is to identify the electric and magnetic dipoles from the Lagrangian expressions, and from there by using Eq 114 write the traditional Maxwell equations including the expressions for the polarization charges and currents. Following this approach, Jackson[47] identifies the first polarization terms as:

$$\vec{P}_{EH} = \frac{e_G^4\hbar}{45\pi m^4c^7}(E^2 - c^2B^2)\vec{E} + \cdots = -\frac{e_G^4\hbar}{45\pi m^4c^7}\left(\left(\dot{\vec{A}} + \vec{\nabla}V\right)^2 - c^2\left(\vec{\nabla}\times\vec{A}\right)^2\right)\left(\dot{\vec{A}} + \vec{\nabla}V\right) + \cdots$$

$$\vec{M}_{EH} = -\frac{e_G^4\hbar}{45\pi m^4c^7}(E^2 - c^2B^2)\vec{B} + \cdots = -\frac{e_G^4\hbar}{45\pi m^4c^7}\left(\left(\dot{\vec{A}} + \vec{\nabla}V\right)^2 - c^2\left(\vec{\nabla}\times\vec{A}\right)^2\right)\left(\vec{\nabla}\times\vec{A}\right) + \cdots$$

**Eq 124**

With the identifications

$$\alpha = \frac{e^2}{4\pi\epsilon_0\hbar c}, \qquad r_e = \frac{e^2}{4\pi\epsilon_0 mc^2}, \qquad E_{cl} = \frac{e}{4\pi\epsilon_0 r_e^2},$$

we can write

$$\frac{e_G^4\hbar}{45\pi m^4c^7}2E^2 = \underbrace{\frac{2}{45\pi}\frac{1}{\alpha}}_{1.938}\frac{E^2}{E_{cl}^2} \cong \frac{2}{E_{cl}^2}E^2$$

From these exprssions we can obtain the approximated charge and current densities as

$$\rho_{EH} = \vec{\nabla}\cdot\left[\frac{2}{E_{cl}^2}(E^2 - c^2B^2)\vec{E}\right] = \vec{\nabla}\cdot\left[\frac{2}{E_{cl}^2}\left(\left(\dot{\vec{A}} + \vec{\nabla}V\right)^2 - c^2\left(\vec{\nabla}\times\vec{A}\right)^2\right)\left(\dot{\vec{A}} + \vec{\nabla}V\right)\right]$$

$$\vec{j}_{\mu,EH} = -\frac{\partial\left[\frac{2}{E_{cl}^2}(E^2 - c^2B^2)\vec{E}\right]}{\partial t} - \vec{\nabla}\times\left[\frac{2}{E_{cl}^2}(E^2 - c^2B^2)\vec{B}\right]$$

$$= -\frac{\partial\left[\frac{2}{E_{cl}^2}\left(\left(\dot{\vec{A}} + \vec{\nabla}V\right)^2 - c^2\left(\vec{\nabla}\times\vec{A}\right)^2\right)\left(\dot{\vec{A}} + \vec{\nabla}V\right)\right]}{\partial t} - \vec{\nabla}\times\left[\frac{2}{E_{cl}^2}\left(\left(\dot{\vec{A}} + \vec{\nabla}V\right)^2 - c^2\left(\vec{\nabla}\times\vec{A}\right)^2\right)\left(\vec{\nabla}\times\vec{A}\right)\right]$$

In 'Appendix 3.1. Relative weight of the different terms.' we show that for fields moving with a speed close to the speed of light in vacuum, the right hand side terms of Eq 115 and Eq 116 are negligible in front of the left hand side. This doesn't mean that the vacuum polarization doesn't play any role in those equations, because they are still present in the definition of the fields $\vec{D}$ and $\vec{B}$.

As noted by Jackson[47], these polarization effects are apparent at distances of the order of the classical electron radius, which are in contrast with the Uehling polarization which was important at distances of the order of the Compton wavelength.

#### 3.3.2.1. Self-trapping solutions.

The system of equations Eq 115 and Eq 116 was considered by Marin Soljacic and Mordechai Segev[48], in the presence of the Euler-Heisenberg vacuum polarizations and were able to show that these equations admit soliton solutions in plane wave geometry. This was achieved by reducing them, under specific boundary conditions, to the nonlinear Schrödinger equation. For an alternative derivation of the field equations see also the work by Radozyki[49].

#### 3.3.2.2. Special solution for the divergence of D

We can see that the equation for the divergence of D in the absence of magnetic fileds reduced to

$$\vec{\nabla}\cdot\vec{D} = \vec{\nabla}\cdot\left(\vec{E} + 4\pi\vec{P}\right) = \vec{\nabla}\cdot\left(\vec{E} + 2\frac{2}{E_{cl}^2}E^2\vec{E}\right) = 0$$

This is the same as the differential equation followed by a dielectric in the presence of a nonlinear polarization. In fact, this equation corresponds to a dielectric where the tensor $\vec{D}$ and the electric field $\vec{E}$ are related through

$$\vec{D} = (1 + E^2)\vec{E}$$

In the absence of free charges the solution to the Maxwell equation is

$$\vec{\nabla} \cdot \vec{D} = 0 \qquad \rightarrow \vec{D} \approx \frac{1}{r^2}$$

In the limit for $1 \ll E^2$ , E scales as

$$\vec{E} \approx \frac{1}{r^{2/3}}$$

Which diverges, but its energy content remains finite. In fact, from Wolfram.com we get for the inversion of the dielectric constant $D(E) = E + E^3$ to the electric field $\vec{E}$:

$$E(D) = \frac{\sqrt[3]{\frac{2}{3}}}{\sqrt[3]{\sqrt{3}\sqrt{27D^2 + 4} - 9D}} - \frac{\sqrt[3]{\sqrt{3}\sqrt{27D^2 + 4} - 9D}}{\sqrt[3]{2}3^{2/3}}$$

**Eq 125**

In the limit for $r \rightarrow 0$ we can see that Eq 125 scales as

$$\vec{E} \approx \frac{1}{r^{2/3}}$$

as expected. It is assumed that the introduction of higher order nonlinear terms in the wave equation will prevent the divergences and provide a relationship between energy and mass.

## 3.4. Vacuum solitons

At this point we will consider the effects of having both vacuum polarization current terms together. We will see that by the simultaneous consideration of the Uehling and the Euler-Heisenberg terms, the effective Maxwell's equations for the electromagnetic fields leads to the so called non-linear Klein Gordon equation, which admits numerical soliton solutions in spherical geometry. The solutions we find in this section can also be applied to non-abelian fields, and to the Yang–Mills and Mass Gap Millennium Problem. In particular we find new types of soliton-particles moving at a speed virtually equal to the speed of light that can resolve the wave-particle paradox for the electromagnetic field.

The free-field Lagrangian for the original Maxwell equations is:

$$\mathcal{L}_0 = \frac{1}{8\pi}(E^2 - B^2) = -\frac{1}{4}F_{\mu\nu}^2$$

where

$$F_{\mu\nu} = \partial_\mu A_\nu - \partial_\nu A_\mu$$

The Uehling and Euler Heisenberg vacuum polarization effects have been expressed in terms of effective Lagrangians[50] :

$$\mathcal{L}_{eff} = -\frac{1}{4}F_{\mu\nu}^2 + \mathcal{L}_U + \mathcal{L}_{EH} + \cdots$$

where

$$\mathcal{L}_U = \frac{\alpha}{60\pi m^2} F_{\mu\nu} \Box F^{\mu\nu}; \qquad F_{\mu\nu} = \partial_\mu A_\nu - \partial_\nu A_\mu; \qquad \Box \equiv \partial_\mu \partial^\mu$$

with $\alpha = \frac{e^2}{4\pi}$ is the fine structure constant, and

$$\mathcal{L}_{EH} = \frac{2\alpha^2}{45m^4}\left[(E^2 - B^2)^2 + 7\left(\vec{E}\cdot\vec{B}\right)^2\right] = \frac{\alpha^2}{90m^4}\left[\left(F_{\mu\nu}F^{\mu\nu}\right)^2 + \frac{7}{4}\left(F_{\mu\nu}\tilde{F}^{\mu\nu}\right)^2\right]$$

It is traditionally argued that this type of equations and Lagrangian for space-time classical-type fields are deprived of any physical meaning other than an "effective" academic exercise, the real meaningful objects in the theory being 'field operators'. On the contrary we argue that using these equations and applying them to classical fields is as valid as solving the Schrödinger equations for the hydrogen atom in first quantization. The only missing parts in the properties of the atomic hydrogen when described by first quantization calculations are just very fine details (like the Lamb shift), which can be properly taken into account by including radiation reaction effects in the Schrödinger description. Additionally well-known interpretational problems present in the solution of the hydrogen atom in the frame of quantum electrodynamics are totally absent when using first quantization. Our purpose is to solve the system of equations derived from these generalized Lagrangians. It is however more clear to start with the expression for the generalized current densities.

In the absence of free sources, the potential field equations can be written as:

$$\left(\frac{1}{c^2}\frac{\partial^2}{\partial t^2} - \nabla^2\right) A_\mu = 4\pi j_{\mu,Uehling} + 4\pi j_{\mu,EH}$$

$$\left(\frac{1}{c^2}\frac{\partial^2}{\partial t^2} - \nabla^2\right) V = 4\pi \rho_{Uehling} + 4\pi \rho_{EH}$$

Taking the time derivative of the first equation, and the gradient of the second we find that the equation followed by the electric field $\vec{E}$ is :

$$\left(\frac{1}{c^2}\frac{\partial^2}{\partial t^2} - \nabla^2\right)\vec{E} = -\frac{1}{\pi}\left(\frac{1}{c^2}\frac{\partial^2}{\partial t^2} - \nabla^2\right)\left(\frac{\alpha}{15} + r_e^2\left(\frac{1}{c^2}\frac{\partial^2}{\partial t^2} - \nabla^2\right) - \frac{2\pi}{E_{cl}^2}(E^2 - c^2B^2)\right)\vec{E} + \frac{\partial \vec{\nabla}\times\left[\frac{2}{E_{cl}^2}(E^2 - c^2B^2)\vec{B}\right]}{\partial t} - \vec{\nabla}\times\vec{\nabla}\times\left[\frac{2}{E_{cl}^2}(E^2 - c^2B^2)\vec{E}\right]$$

**Eq 126**

which is the same, after discarding the Uehling terms, as the equations given in the paper by Mordechai Segev[48]. Noting that we have shown in 'Appendix 3.1. Relative weight of the different terms.' that when the solution is moving with a speed close to the speed of light the last two terms are negligible,

$$\frac{\partial \nabla\times\left[(E^2 - c^2B^2)\vec{B}\right]}{\partial t} - \nabla\times\nabla\times\left[(E^2 - c^2B^2)\vec{E}\right] \underset{v\to c}{\Longrightarrow} 0$$

we can write the equation in the absence of free charges as

$$\left(\frac{1}{c^2}\frac{\partial^2}{\partial t^2} - \nabla^2\right)\left[\left(1 + \frac{\alpha}{15\pi}\right) + \frac{r_e^2}{\pi}\left(\frac{1}{c^2}\frac{\partial^2}{\partial t^2} - \nabla^2\right) - \frac{2}{E_{cl}^2}(E^2 - c^2B^2)\right]\vec{E} = 0$$

**Eq 127**

with a particular solution

$$\left[\left(1+\frac{\alpha}{15\pi}\right)+\frac{r_e^2}{\pi}\left(\frac{1}{c^2}\frac{\partial^2}{\partial t^2}-\nabla^2\right)-\frac{2}{E_{cl}^2}(E^2-c^2B^2)\right]\vec{E}=0$$

**Eq 128**

The $\frac{\alpha}{15\pi}$ term can be understood as a renormalization of the electric field and the electric charge. In the Section 'Solitons motion and the speed of light' below we discuss the speed of motion achievable by soliton solutions to Eq 128. There we find that the soliton speed can be extremely close but never equal to the nominal speed of light in vacuum 'c'. Therefore, solutions to Eq 128 can be found initially in a frame at rest, and then Lorentz-transformed to the required moving frame. For a stationary solution in a frame at rest we have to solve

$$\left[\left(1+\frac{\alpha}{15\pi}\right)-\frac{r_e^2}{\pi}\nabla_{rest}^2-\frac{2}{E_{cl}^2}E_{rest}^2\right]\vec{E}_{rest}=0$$

**Eq 129**

Once we have solved Eq 129 in the rest frame we can obtain the solution in a moving frame by using the Lorentz transformations. Practically the transformations 'flatten' the solution profile in the direction of motion. One can observe that the energy content of the moving profile is larger than the energy content of the solution in the rest frame.

### 3.4.1. Plane geometry solution

Consider here the time independent case in the absence of magnetic fields. The Eq 129 reduces to

$$\frac{r_e^2}{\pi}\nabla^2 E=\left(1+\frac{\alpha}{15\pi}\right)E-\frac{2}{E_{cl}^2}E^3$$

**Eq 130**

with a constant solution in plane geometry $\nabla^2 E=0$ at

$$\left(1+\frac{\alpha}{15\pi}\right)E=\frac{2}{E_{cl}^2}E^3$$

which is fulfilled at $E=bE_{critical}$. We see that a primary effect of this term is that the nonlinear polarization imposes an upper limit at the critical value for the fields. This value can in principle be achieved at any distance from the center of the solution. We observe that this equation cannot be obtained by the Uehling or the Euler Heisenberg equations separated. One has to write both terms and solve for both terms simultaneously to find this solution.

### 3.4.2. Spherical symmetry, longitudinal solutions

We have considered time independent solutions in the absence of sources. This case corresponds to

$$\rho_{ext}=0; \qquad J=0$$

We have calculated two types of solitons using spherical symmetry $(r,\theta,\phi)$:

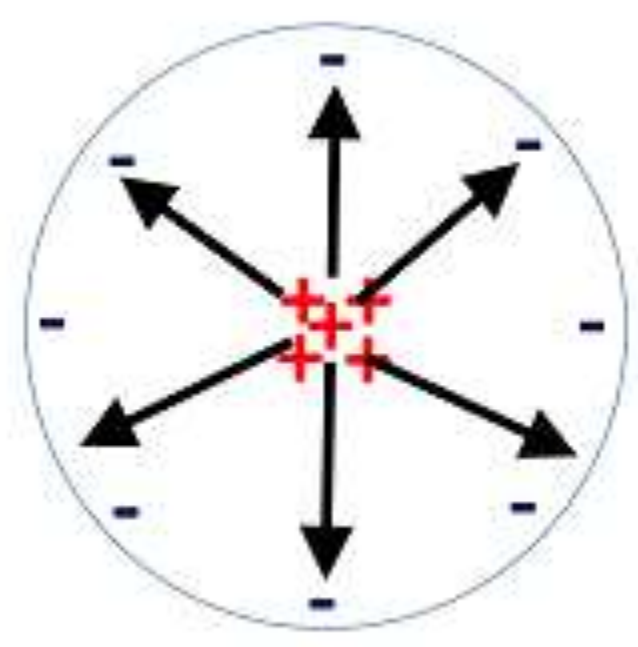

$$\vec{E}(\vec{r}) = \hat{r}\left|\vec{E}\right|(|\vec{r}|) = \hat{r}E(r)$$

We verify that the Laplacian reads

$$\nabla^2\vec{E} = \vec{\nabla}\cdot\left(\vec{\nabla}\vec{E}\right) = \hat{r}\cdot\frac{1}{r^2}\frac{\partial}{\partial r}\left(r^2\hat{r}\frac{\partial}{\partial r}\hat{r}E(r)\right) = \hat{r}\frac{1}{r^2}\frac{\partial}{\partial r}\left(r^2\frac{\partial}{\partial r}E(r)\right) = \hat{r}\nabla^2E(r)$$

This solution doesn't need of any external source. Typically this source will extend to a size of the classical radius and a maximum field of E critical.

We have solved numerically the equations for this case. This solution is possible only when the gradient of the electric field at the center of symmetry is null, and as a result the energy content of the solution is quantized. We find the profiles shown in Figure 14 and Figure 15.

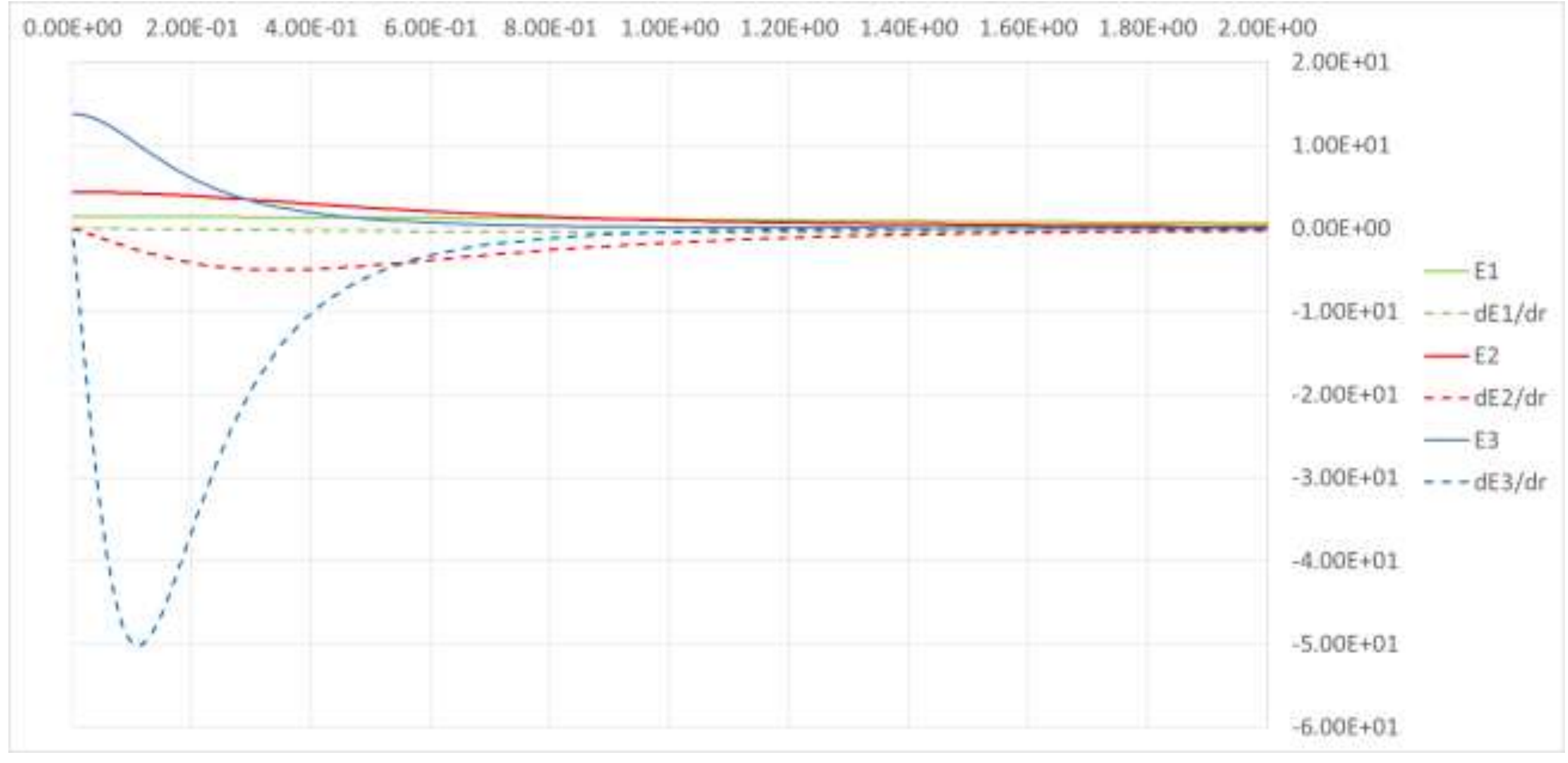

**Figure 14**

The mass parameter in each curve is given by

$E1{:}\,10^0$ $E2{:}\,10^1$ $E3{:}\,10^2$

in dimensionless units. The physical application of this solution is not clear because the electric field has a discontinuity at the origin; even when its magnitude is constant, its direction points radially outwards. The effective charge density for these profiles is given by

$$\rho_{effective} \cong \vec{\nabla}\cdot\vec{E} = \frac{1}{r^2}\frac{\partial}{\partial r}\left(r^2 E(r)\right) = \frac{1}{r^2}\left(2rE + r^2\frac{\partial E}{\partial r}\right) = \left(\frac{2}{r}E + \frac{\partial E}{\partial r}\right)$$

and from the numerical integration we find

$$\frac{\partial E}{\partial r} \underbrace{\rightarrow}_{r\to 0} 0 \qquad \frac{2}{r}E \approx \frac{2}{r} \underbrace{\rightarrow}_{r\to 0} \infty$$

Is there any relationship between the finite longitudinal field and the neutrino?

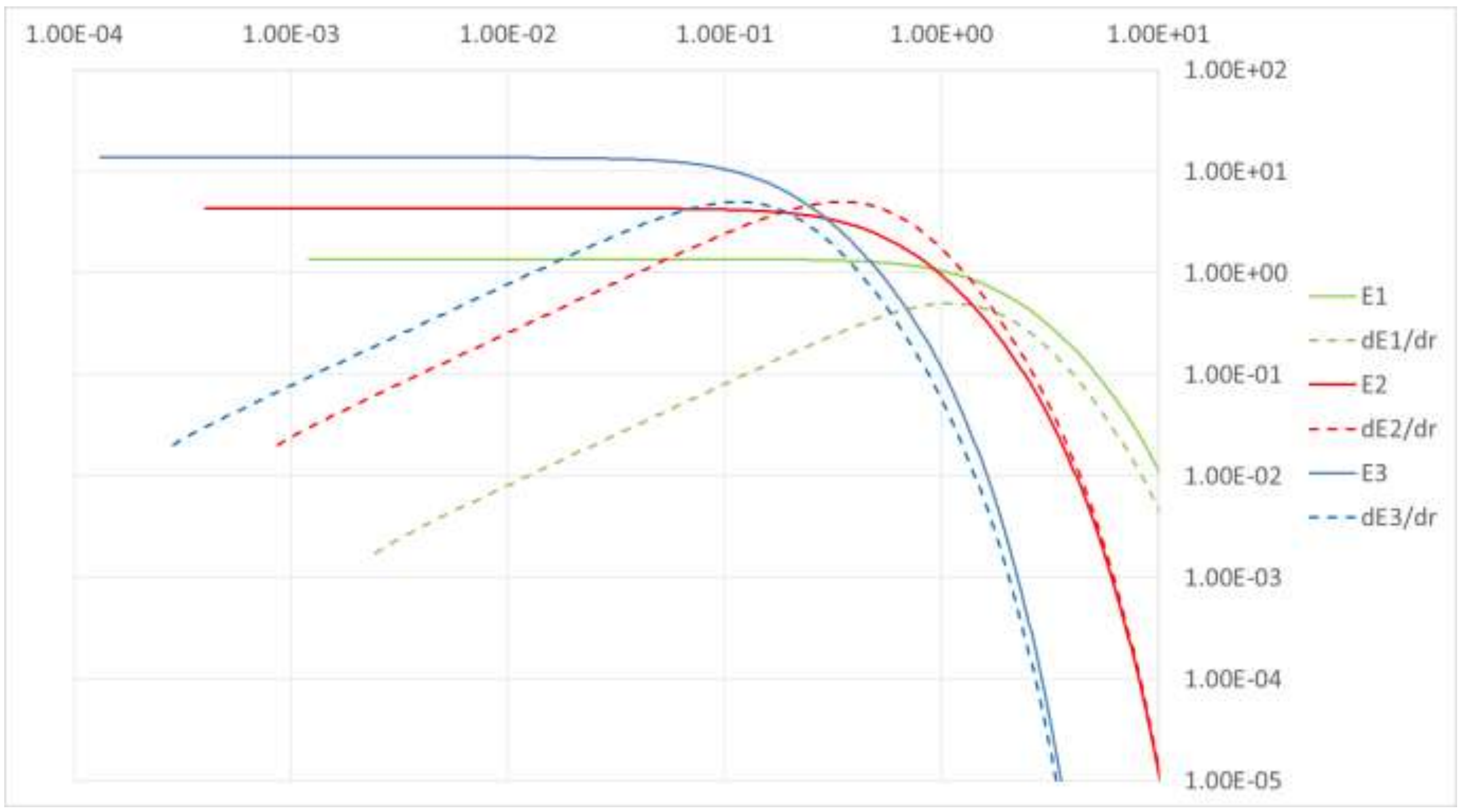

**Figure 15**

### 3.4.3. Spherical symmetry, transverse solutions

In this Section we consider transverse solutions to the Eq 127. We find solutions that correspond to a new type of soliton-particles moving at a speed virtually equal to the speed of light. These solutions are candidates to resolve the wave-particle paradox for the electromagnetic field. We decompose the electromagnetic fields into an 'internal' and an 'external' components following de Broglie's original 'double solution' model, and obtain a set of three equations, one for the external field, one for the internal field and one providing the 'guiding condition'. Nonlinearities can be confined to the internal equation. The division into the three equations is not unique, and can give rise to solitons following an external equation identical with the linear massless Maxwell equations, or to a massive equation similar to the linear Proca equation. Finally the internal solution remains non-linear and generate soliton-like structures.

Which solution is realized will be determined by the fulfillment of energy-momentum conservation for the entire system through entanglement. A variable mass can be interpreted as a relativistic scalar potential field. Entanglement can then generate a quantum potential acting on the photon.

#### 3.4.3.1. Null source. Soliton solutions-photons, de Broglie hypothesis

To simplify notation we write Eq 128 it as

$$d\nabla^2\vec{E} = d\frac{1}{c^2}\frac{\partial^2\vec{E}}{\partial t^2} + b\vec{E} - a(E^2 - c^2B^2)\vec{E}$$

**Eq 131**

At this point we are going to use de Broglie hypothesis that the field can be expressed as the product of an 'internal' field $E_i$ and an 'external' field $\vec{E}_e$ :

$$\vec{E} = \vec{E}_e E_i, \qquad \vec{B} = \vec{B}_e B_i$$

The internal functions are scalars, the external functions keep the vector character of the fields. Replacing this expression into Eq 131 and taking into account that

$$\nabla^2\vec{E} = \nabla^2 E_i\vec{E}_e + E_i\nabla^2\vec{E}_e + 2\nabla E_i\cdot\nabla\vec{E}_e$$

$$\frac{\partial^2\vec{E}}{\partial t^2} = \frac{\partial^2 E_i}{\partial t^2}\vec{E}_e + \frac{\partial^2\vec{E}_e}{\partial t^2}E_i + 2\frac{\partial E_i}{\partial t}\frac{\partial\vec{E}_e}{\partial t}$$

we obtain the following equation:

$$d\left(\nabla^2 E_i\vec{E}_e + E_i\nabla^2\vec{E}_e + 2\nabla E_i\cdot\nabla\vec{E}_e\right) = d\frac{1}{c^2}\left(\frac{\partial^2 E_i}{\partial t^2}\vec{E}_e + \frac{\partial^2\vec{E}_e}{\partial t^2}E_i + 2\frac{\partial E_i}{\partial t}\frac{\partial\vec{E}_e}{\partial t}\right) + b\vec{E}_e E_i - a(E^2 - c^2B^2)\vec{E}_e E_i$$

**Eq 132**

To simplify the analysis we assume that $E_i(r,t)$ is a real function, and $\vec{E}_e(r,t)$ is an energy eigenstate, this means $\vec{E}_e(r,t) = \vec{E}_{e,1}(r)e^{-i(\vec{k}(r)\cdot\vec{r}-\omega t)}$, with $\vec{E}_{e,1}(r)$ a real function, then

$$\frac{\partial\vec{E}_e}{\partial t} = i\omega\vec{E}_e(r,t)$$

$$\nabla E_e = -i\left(\vec{k}(r) + \nabla\vec{k}(r)\cdot\vec{r}\right)E_e(r,t) + \nabla E_{e,1}(r)e^{-i(\vec{k}(r)\cdot\vec{r}-\omega t)} == \left[-\underbrace{i\left(\vec{k}(r) + \nabla\vec{k}(r)\cdot\vec{r}\right)}_{imaginary} + \underbrace{\frac{\nabla E_{e,1}(r)}{E_{e,1}(r)}}_{real}\right]E_e(r,t)$$

$$\equiv i\nabla E_e^{Im} + \nabla E_e^{Re}$$

We also assume that $E_i(r,t)$ behaves as a soliton, namely that

$$E_i(r,t) = E_i^0(r_i,t)$$

$$r_i = r - r_1(t) \qquad \frac{dr_1}{dt} = \mathrm{v}$$

therefore

$$\nabla_r E_i = \nabla_{r_i} E_i^0$$

$$\frac{\partial E_i}{\partial t} = \nabla_{r_i} E_i^0 \cdot \mathrm{v} + \frac{\partial E_i^0}{\partial t}$$

We will assume that $\frac{\partial E_i^0}{\partial t}$ can be neglected in front of $\nabla_{r_i} E_i^0 \cdot \mathrm{v}$. Assuming $E_e(r,t)$ is a monochromatic field, that is $E_e(r,t) = E_{e,1}(r) e^{-i(\varphi(r)-\omega t)}$, with $E_{e,1}(r)$ a real function, then

$$\frac{\partial E_e}{\partial t} = i\omega E_e$$

and

$$\nabla E_e^{Im} = -\vec{\nabla}\varphi(r) E_e(r,t)$$

In the case of a plane wave $\varphi(r) = \vec{k} \cdot \vec{r}$ and $\vec{\nabla}\varphi = \vec{k}$. In the general case the guiding condition becomes

$$\vec{\mathrm{v}}_{soliton} = c^2 \frac{\nabla E_e^{Im}}{\frac{\partial E_e}{\partial t}}$$

or

$$\vec{\mathrm{v}}_{soliton} = \frac{c^2}{\omega} \vec{\nabla}\varphi = c \frac{\vec{\nabla}\varphi}{k}$$

**Eq 133**

We consider next different ways to solve Eq 132.

## 3.5. Massless Photons

The first case we consider is by writing Eq 131 in the form

$$d\left(E_i \nabla^2 E_e - \frac{1}{c^2}\frac{\partial^2 E_e}{\partial t^2} E_i\right) + d\left(2\nabla E_i \cdot [\nabla E_e^{Im} + \nabla E_e^{Re}] - 2\frac{1}{c^2}\frac{\partial E_i}{\partial t} \underbrace{\frac{\partial E_e}{\partial t}}_{imaginary}\right)$$
$$+ \left(d\left(\nabla^2 E_i E_e - \frac{1}{c^2}\frac{\partial^2 E_i}{\partial t^2} E_e\right) - bE_e E_i + a(E^2 - c^2 B^2) E_e E_i\right) = 0$$

or equivalently

$$dE_i\left(\nabla^2 E_e - \frac{1}{c^2}\frac{\partial^2 E_e}{\partial t^2}\right) + d\left(2\nabla E_i \cdot \nabla E_e^{Im} - 2\frac{1}{c^2}\frac{\partial E_i}{\partial t}\frac{\partial E_e}{\partial t}\right)$$
$$+ E_e\left(d\left(\nabla^2 E_i - \frac{1}{c^2}\frac{\partial^2 E_i}{\partial t^2}\right) - bE_i + a(E^2 - c^2 B^2) E_i + 2\nabla E_i \cdot \frac{\nabla E_e^{Re}}{E_e}\right) = 0$$

which shows that the equation can be split into three parts:

1. The usual linear Maxwell wave equation:

$$\nabla^2 E_e = \frac{1}{c^2}\frac{\partial^2 E_e}{\partial t^2}$$

that can be solved using the real boundary conditions for the field and choosing a normalization equal to 1. If the field is expressed in Cartesian coordinates, as we assume, this equation is fulfilled for each component independently.

2. The so called 'guiding condition' joining the motion of the inner solution or soliton to the gradient of the external solution, and therefore to the boundary conditions:

$$2\nabla E_i \cdot \nabla E_e^{Im} = 2\frac{1}{c^2}\frac{\partial E_i}{\partial t}\frac{\partial E_e}{\partial t}$$

$$\nabla E_i \cdot \left(\vec{k}(r) + \nabla\vec{k}(r)\cdot\vec{r}\right)E_e(r,t) = \frac{1}{c^2}\left(\nabla_{r_i}E_i^0 \cdot \mathrm{v} + \frac{\partial E_i^0}{\partial t}\right) i\omega E_e(r,t)$$

$$\nabla_{r_i}E_i^0 \cdot \left(\vec{k}(r) + \nabla\vec{k}(r)\cdot\vec{r}\right) = \frac{1}{c^2}\left(\nabla_{r_i}E_i^0 \cdot \mathrm{v} + \frac{\partial E_i^0}{\partial t}\right)\omega \rightarrow \begin{cases} \nabla_{r_i}E_i^0 \cdot \vec{k}(r) = \frac{1}{c^2}\nabla_{r_i}E_i^0 \cdot \mathrm{v}\omega \\ \nabla_{r_i}E_i^0 \cdot \nabla\vec{k}(r)\cdot\vec{r} = \frac{1}{c^2}\frac{\partial E_i^0}{\partial t}\omega \approx 0 \end{cases}$$

$$\rightarrow \quad \vec{k}(r) = \frac{1}{c^2}\vec{\mathrm{v}}\omega \qquad \vec{\mathrm{v}} = c\hat{k}(r)$$

which as shown by de Broglie, provides the motion of the soliton.
In most macroscopic cases $\nabla E_e^{Re} \ll \nabla E_e^{Im}$ and one can simply replace $\nabla E_e^{Im} \sim \nabla E_e$, $\nabla E_e^{Re} \sim 0.$ For a plane wave this expression is exact.

3. The inner equation,

$$d\nabla^2 E_i = d\frac{1}{c^2}\frac{\partial^2 E_i}{\partial t^2} + bE_i - a(E^2 - c^2B^2)E_i + \left(2\frac{\nabla E_i}{E_i}\cdot\frac{\nabla E_e^{Re}}{E_e}\right)E_i$$

From this equation we can see that $2\frac{\nabla E_i}{E_i}\cdot\frac{\nabla E_e^{Re}}{E_e}$ plays a roll analogous to an external potential energy that will modify the shape of the solution according to the variations of the external field.

The last equation is the nonlinear Klein Gordon equation which has the spherical soliton solution with any desired energy content less than the electron mass as we will see shortly. In the absence of external real sources, the solution for this equation will provide a photon soliton. The shape of these photon solitons is characterized by a typical length, provided by the classical electron radius, where the Heisenberg Euler effects become appreciable limiting the maximum strength of the electric and magnetic fields to its critical values. The second equation is always the same and determines the guidance condition. Different models for the trajectories of photon particles have been studied over time by many authors[51]. Finally the first equation define a massless dispersion relation.

### 3.5.1. Numerical Integration

If these external effective forces provided by $2\frac{\nabla E_i}{E_i}\cdot\frac{\nabla E_e^{Re}}{E_e}$ can be neglected, this equation can be solved numerically for the frame where the system is at rest using Eq 130

$$\frac{r_e^2}{\pi}\nabla^2 E = E - \frac{2}{E_{cl}^2}E^3$$

**Eq 134**

where we neglected the term $\frac{\alpha}{15\pi}$ in front of 1. Measuring the electric amplitude in units of

$$E_b = \frac{1}{\sqrt{2}} E_{cl} \cong 0.71 E_{cl}$$

and distances in units of

$$r_b = r_e \frac{1}{\sqrt{\pi}} \cong 0.56 r_e$$

we can bring $\frac{r_e^2}{\pi} \nabla^2 E = E - \frac{2}{E_{cl}^2} E^3$

Eq 134 into the dimensionless and normalized form:

$$\nabla^2 \tilde{E} = \tilde{E} - \tilde{E}^3$$

**Eq 135**

where $E = E_b \tilde{E}(\tilde{r}), \quad r = r_b \tilde{r}$ ,

We have solved numerically Eq 135 in spherical coordinates. With these units the energy can be written as

$$\varepsilon = \varepsilon_b \tilde{\varepsilon}$$

where

$$\varepsilon_b = {E_b}^2 {r_b}^3 = 0.089 mc^2$$

We have considered the specific case of a transverse wave where the external electric field $\vec{E}_e$ points in the z-direction, and where the internal field $E_i$has spherical symmetry as follows:

$$\vec{E}(\vec{r}) = \hat{k} \left| \vec{E} \right| (|\vec{r}|) = \hat{k} E(r)$$

as shown in the next figure:

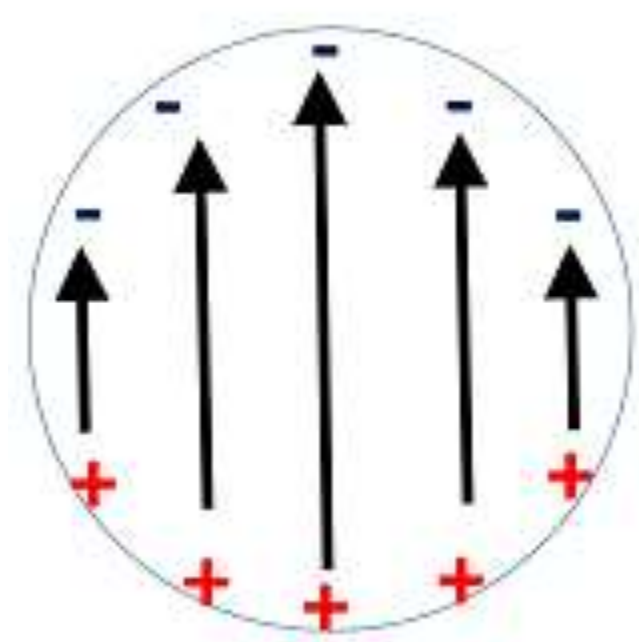

We verify that in this case the Laplacian reads

$$\nabla^2 \vec{E} = \hat{k} \frac{1}{r^2} \frac{\partial}{\partial r} \left( r^2 \frac{\partial E}{\partial r} \right) = \hat{k} \frac{1}{r^2} \left( 2r \frac{\partial E}{\partial r} + r^2 \frac{\partial^2 E}{\partial r^2} \right) = \hat{k} \left( \frac{2}{r} \frac{\partial E}{\partial r} + \frac{\partial^2 E}{\partial r^2} \right)$$

We solved numerically this case, the electric field at the center of symmetry can be different from null. The energy content of the soliton can be arbitrary. The profiles we find are shown in Figure 16 where we plot the value of $E$ and the magnitude of the derivative of the field $\left|\frac{\partial E}{\partial r}\right| = -\frac{\partial E}{\partial r}$ because the field gradient is negative.

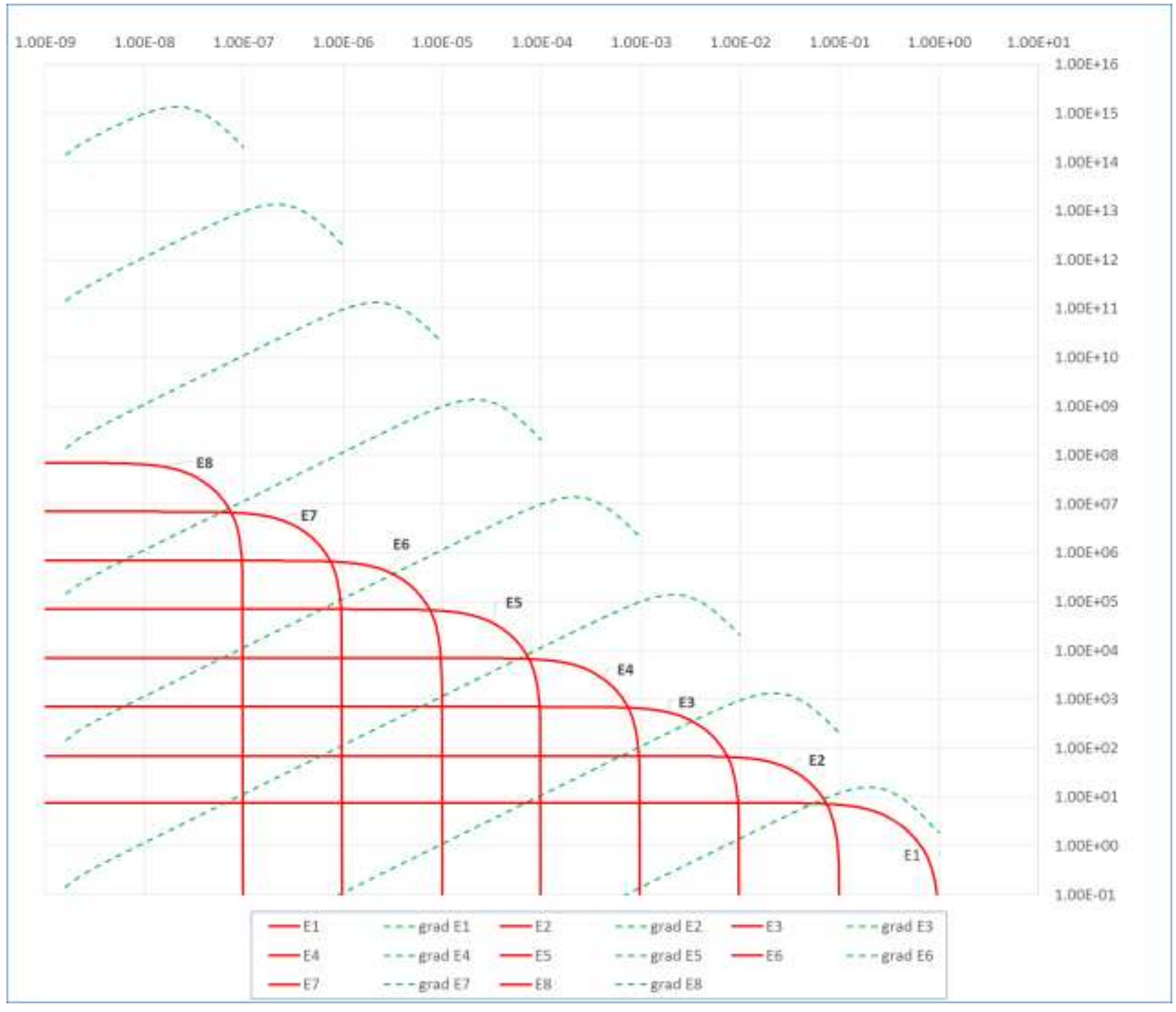

**Figure 16**

The effective charge density for these profiles is given by

$$\rho_{effective} \cong \vec{\nabla} \cdot \vec{E} = \frac{\partial E_z}{\partial z}$$

and from the numerical integration we find

$$\frac{\partial E_z}{\partial z} \underset{z \to 0}{\to} 0$$

changing sign with inversion of the z axis. When we calculate the values of the fields, energies and distances of the above solutions we find the values shown in Figure 17, where the units are electron mass for energy, Classical Electron Radius for distance, and critical electric field for the electrical field amplitude. Here we can see that the photon energy scales linearly with the radius of the soliton. For a soliton with energy equal to 1 electron mass we find a radius of approximately 1 classical electron radius and a central electric field close to 10 times the critical electric field value. A photon with an energy of about $E = 1 \times 10^{-5} mc^2$ corresponding to about 5 eV in the visible spectrum, has a radius of the order of $r \approx 1 \times 10^{-5}\, r_e$, where $r_e$ is the classical electron radius. Here we have used for the energy content of the soliton the classical expression

$$U = \int \frac{\varepsilon_0 E^2}{2} dr^3$$

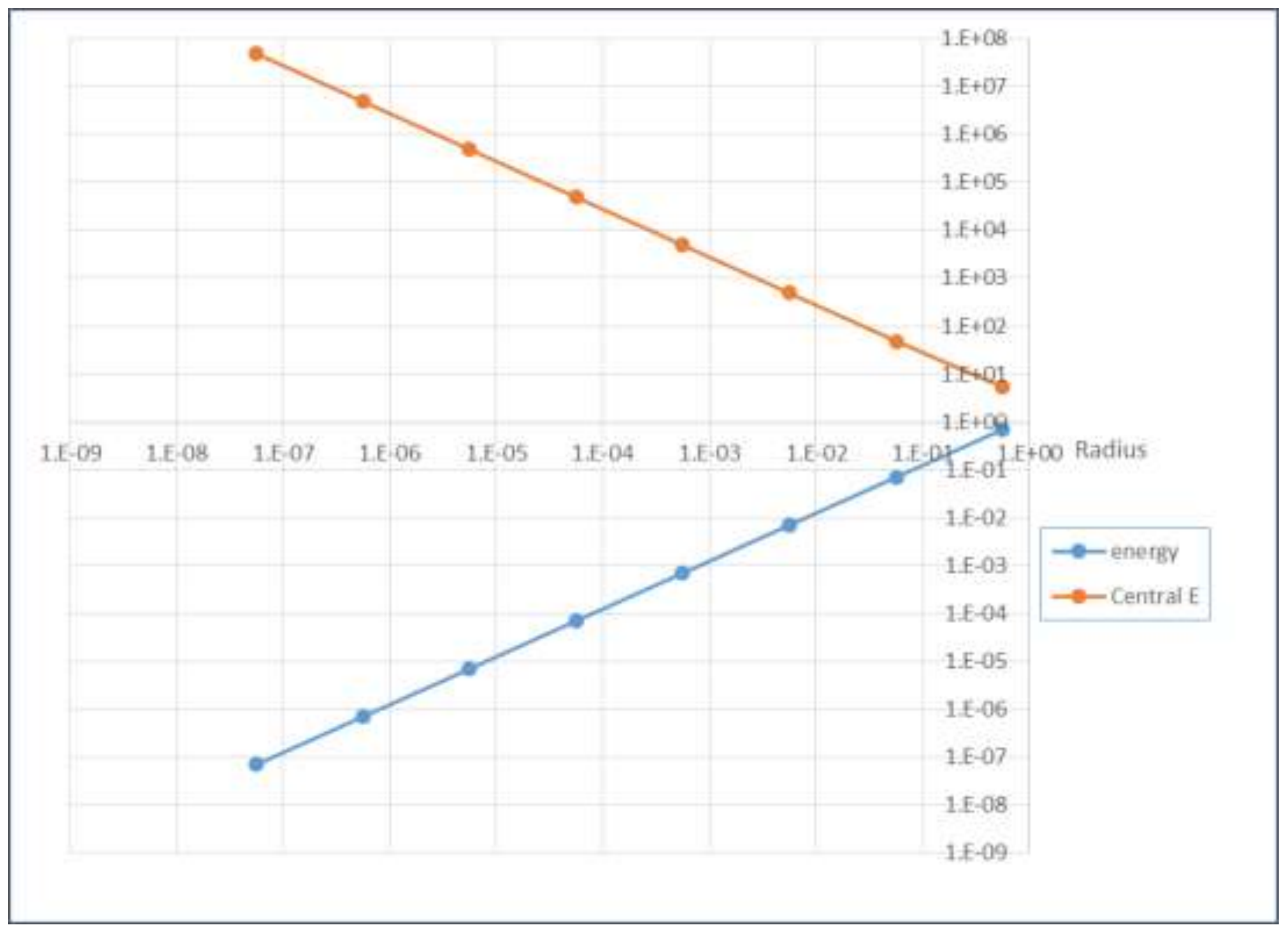

**Figure 17**

This assumption needs some future analysis.

We can see how the solutions to Eq 135 plotted in Figure 16 really approach a self-similar character as the amplitude of the electric field E grows higher that the electron critical field $E_{cl}$. Eq 135 is written taking the electric field in units of $E_{cl}$ and distances in units of $r_e$. When the electric field amplitude is much larger than $E_{cl}$ the linear term in Eq 135 can be neglected in front of $\tilde{E}^3$. In this case Eq 135 reduces to

$$\nabla^2 \tilde{E} = -\tilde{E}^3$$

The last equation in spherical coordinates reads

$$\frac{2}{r}\frac{\partial \tilde{E}}{\partial r} + \frac{\partial^2 \tilde{E}}{\partial r^2} + \tilde{E}^3 = 0$$

a trial solution of the form

$$\tilde{E} = a + br + cr^2 + dr^3 + er^4$$

for $r \ll 1$ provides the following values for the coefficients, taking $a$ an arbitrary value:

$$b = 0, c = -\frac{a^3}{6}, d = 0, e = \frac{a^5}{40},$$

and for the field and its derivative

$$\tilde{E} = a - \frac{a^3}{6}r^2 + \frac{a^5}{40}r^4$$

$$\frac{\partial \tilde{E}}{\partial r} = -\frac{a^3}{3}r + \frac{a^5}{10}r^3$$

in good agreement with the numerical simulations.

### 3.5.2. Self-similarity

Once a solution is found, the same 'profile' will be a solution when distances are scaled by a dimensionless factor $\alpha$ and simultaneously the electric field amplitude is scaled by a factor $1/\alpha$. This is a self-similar solution. The energy in that case will scale as $\varepsilon \propto E^2 r^3 \propto \alpha$ the same as the distance. Smaller solitons have smaller energy. In this case the role of the linear term will be merely to add a very low amplitude, exponentially decaying tail with dimensions of the classical electron radius. This tail will have an extremely small fraction of the total soliton energy content.

The transverse finite soliton can have any amplitude and energy. Transverse polarization solitons or photons can exist in vacuum; can we identify the soliton solution represented in Figure 16 as the photon-particle resolving the wave – particle duality for the electromagnetic fields?

We remark very strongly that we are not developing new physics here. We are only performing calculations based in the very same equations the full theory of QED has always utilized. The fact that nobody apparently has calculated this is simply because, when learning QED, this matters appear to the student as fully understood, and that nothing new can possibly be learned from here. It is our opinion that that feeling is wrong.

### 3.5.3. Guiding condition and Poynting vector

It can be seen that Eq 133 provides the value of the soliton velocity only in the direction of propagation of the external field. The guidance condition says nothing about the possibility of a component perpendicular to propagation. It was derived using the electric field $E$ only, without any reference to the magnetic field $B$. On the other hand, by using the radiation gauge, the vector potential field $A$ is enough to provide a full description of both, the electric and magnetic transverse radiation fields. This is an excellent approximation for multipole radiation and emission by systems with zero net charge. In the radiation gauge we have

$$\mu\vec{H} = \vec{\nabla} \times \vec{A} \qquad \vec{E} = -\frac{1}{c}\frac{\partial \vec{A}}{\partial t}$$

and the vector potential $\vec{A}$ fulfills the equations

$$\vec{\nabla} \cdot \vec{A} = 0$$

and

$$\nabla^2 A - \frac{1}{c^2}\frac{\partial^2 A}{\partial t^2} = 0$$

In this gauge, for a monochromatic field

$$\vec{E} = \frac{\partial \vec{A}}{\partial t} = i\omega\vec{A} \qquad \vec{A} = \frac{1}{i\omega}\,\vec{E} = \frac{1}{i\omega}E_{e,1}(r)e^{-i(\varphi(r)-\omega t)} = \hat{e}R(r)e^{i(\varphi(r)-\omega t)}$$

and therefore the fields $\vec{E}$ and $\vec{A}$ have the same gradient that can be written as

$$\vec{\nabla}\varphi = \frac{i}{2}\frac{\left[\left(\vec{\nabla}\vec{A}^*\right)\cdot\vec{A} - \left(\vec{\nabla}\vec{A}\right)\cdot\vec{A}^*\right]}{\vec{A}^*\cdot\vec{A}}$$

and then the soliton guidance condition can be written as

$$\vec{\mathrm{v}}_{soliton} = \frac{c^2}{\omega}\frac{i}{2}\frac{\left[\left(\vec{\nabla}\vec{A}^*\right)\cdot\vec{A} - \left(\vec{\nabla}\vec{A}\right)\cdot\vec{A}^*\right]}{\vec{A}^*\cdot\vec{A}}$$

where we mention that the last expression is real even with the presence of the imaginary unit factor. Schiller has shown that the Poynting vector

$$\langle\vec{S}\rangle = \frac{c}{16\pi}\left(\vec{E}^*\times\vec{H} + \vec{E}\times\vec{H}^*\right)$$

in terms of the vector potential A, in the radiation gauge is given by

$$\langle\vec{S}\rangle = \frac{i\omega}{16\pi\mu}\left(\left(\vec{\nabla}\vec{A}^*\right)\cdot\vec{A} - \left(\vec{\nabla}\vec{A}\right)\cdot\vec{A}^* + \vec{\nabla}\times\left(\vec{A}^*\times\vec{A}\right)\right)$$

**Eq 136**

and the velocity for propagation for the energy content of the field is given by

$$\vec{\mathrm{v}}_{energy} = \frac{\vec{S}}{U} = \frac{\vec{S}}{\vec{A}^*\cdot\vec{A}}$$

The last term in Eq 136 is different from zero, in circularly or elliptically polarized light with a non-uniform density distribution in the plane perpendicular to the direction of propagation[52]. It is zero however for plane polarized light. The first term corresponds to the translational motion, while the second to the 'spin' motion, the asymmetric part of the energy momentum stress tensor. Because the difference between $\vec{\mathrm{v}}_{energy}$, and $\vec{\mathrm{v}}_{soliton}$ is perpendicular to the direction of propagation, it is consistent to assume that the total soliton velocity is given by the Poynting formula, namely

$$\vec{\mathrm{v}}_{solitonR} = \vec{\mathrm{v}}_{energy}$$

Moreover, Schiller was also able to show that the phase and amplitude of the vector potential A fulfill the equations

$$\vec{\nabla}\cdot\left(R^2\vec{\nabla}\varphi\right) = 0$$

**Eq 137**

and

$$k^2\left(\epsilon\mu - \vec{\nabla}\varphi\cdot\vec{\nabla}\varphi\right)R + \nabla^2 R = 0 \qquad k = \frac{\omega}{c}$$

that can be rewritten as

$$(\nabla\varphi)^2 = \underbrace{\epsilon\mu}_{c^2} - \frac{\nabla^2 R}{k^2 R} \qquad (\nabla\varphi)^2 + \frac{\nabla^2 R}{k^2 R} = \underbrace{\epsilon\mu}_{c^2}$$

**Eq 138**

Here Eq 137 can be interpreted as the continuity equation for the fluid with density proportional to the energy density $R^2$, and velocity given by $\nabla\varphi$ as in the electron case. Eq 138 can be interpreted as an energy equation where we can recognize again the de Broglie-Bohm quantum potential[53,54]

$$U_Q = \frac{\nabla^2 R}{k^2 R}$$

### 3.5.4. Solitons motion and the speed of light

Let's analyze the value obtained for the speed of motion for the soliton given by the guiding condition when the solution to $\nabla^2 E_e - \frac{1}{c^2}\frac{\partial^2 E_e}{\partial t^2}$ is an energy eigenstate in vacuum. In that case the solution to the wave equation is a plane wave given by

$$E_e = ae^{i(\omega t - kr)}$$

and in this case the propagation velocity is given by the nominal speed of light in vacuum

$$\vec{\mathrm{v}} = c$$

Even in the most empty regions of the interstellar space, matter is present with a density as low as $10^{-4}$ ions per $cm^3$, see http://www.jb.man.ac.uk/pulsar/Education/Tutorial/tut/node14.html. Also it can be shown that also the big bang background radiation generates, via vacuum polarization precisely a background index of refraction larger than 1 as shown by Ravndal et al.[55]. As a result it can be argued that the index of refraction of the medium where light propagates can never be absolutely equal to 1, and any electromagnetic structure traveling in space propagates at a speed lower than the speed of light in vacuum c. The classical value for the speed of light in vacuum therefore can be seen as an academic mathematical limit never reached in practice by any physical object, including radiation fields themselves. As a corollary we can say that solitons travel at a speed that is smaller than c, and therefore they can be obtained as Lorentz transformations from the solutions calculated at rest. Additionally we can argue that the Coulomb attraction potential between the dipole charges composing the vacuum polarization field will provide the glue needed to keep the photon soliton from disintegrating in front of otherwise present diverging forces.

## 3.6. Non-abelian, nonlinear Photons

We can rearrange the Uehling terms in Eq 131 and obtain:

$$\left(d\left(E_i\nabla^2 E_e - \frac{1}{c^2}\frac{\partial^2 E_e}{\partial t^2}E_i\right) + \alpha\beta(E^2 - c^2B^2)E_eE_i\right) + 2d\left(\nabla E_i\cdot\nabla E_e - \frac{1}{c^2}\frac{\partial E_i}{\partial t}\frac{\partial E_e}{\partial t}\right)$$
$$+\left(d\left(\nabla^2 E_iE_e - \frac{1}{c^2}\frac{\partial^2 E_i}{\partial t^2}E_e\right) - bE_eE_i + a(1-\beta)(E^2 - c^2B^2)E_eE_i\right) = 0$$

where we have just added and subtracted the term $\alpha E_eE_i$; splitting the different terms:

$$\nabla^2 E_e - \frac{1}{c^2}\frac{\partial^2 E_e}{\partial t^2} + \beta(E^2 - c^2B^2)E_eE_i = 0$$

**Eq 139**

$$\nabla E_i\cdot\nabla E_e - \frac{1}{c^2}\frac{\partial E_i}{\partial t}\frac{\partial E_e}{\partial t} = 0$$

**Eq 140**

$$\nabla^2 E_i - \frac{1}{c^2}\frac{\partial^2 E_i}{\partial t^2} - bE_i + a(1-\beta)(E^2 - c^2B^2)E_i$$

**Eq 141**

And magic!! If we forget that these equations corresponds to two components of the same particle, but consider them as two different interacting particles, Eq 139 can be related to a massive gauge boson, Eq 141 to a Higgs boson equation , and Eq 140 represents the interaction between the particles. Have we identified an alternative interpretation for the origin of mass in the massive photons?

## 3.7. Massive Photons

If we comeback to Eq 131 and rearrange the Uehling terms we obtain:

$$\left(d\left(E_i\nabla^2 E_e - \frac{1}{c^2}\frac{\partial^2 E_e}{\partial t^2}E_i\right) - \beta E_e E_i\right) + 2d\left(\nabla E_i\cdot\nabla E_e - \frac{1}{c^2}\frac{\partial E_i}{\partial t}\frac{\partial E_e}{\partial t}\right) + \left(d\left(\nabla^2 E_i E_e - \frac{1}{c^2}\frac{\partial^2 E_i}{\partial t^2}E_e\right) - (b-\beta)E_e E_i + a(E^2 - c^2B^2)E_e E_i\right) = 0$$

where we have just added and subtracted the term $\beta E_e E_i$; splitting the different terms:

$$\nabla^2 E_e - \frac{1}{c^2}\frac{\partial^2 E_e}{\partial t^2} - \beta E_e = 0$$

**Eq 142**

$$\nabla E_i\cdot\nabla E_e - \frac{1}{c^2}\frac{\partial E_i}{\partial t}\frac{\partial E_e}{\partial t} = 0$$

**Eq 143**

$$\nabla^2 E_i - \frac{1}{c^2}\frac{\partial^2 E_i}{\partial t^2} - (b-\beta)E_i + a(E^2 - c^2B^2)E_i$$

We see from Eq 142 that we obtain the Proca equation. We note that the mass term, proportional to the field, for the external field is different to the one for the internal field.

## 3.8. Massive nonlinear Photons

If we comeback to Eq 131 and rearrange the Uehling terms we obtain:

$$\left(d\left(E_i\nabla^2 E_e - \frac{1}{c^2}\frac{\partial^2 E_e}{\partial t^2}E_i\right) - \beta E_e E_i + \gamma(E^2 - c^2B^2)E_e E_i\right) + 2d\left(\nabla E_i\cdot\nabla E_e - \frac{1}{c^2}\frac{\partial E_i}{\partial t}\frac{\partial E_e}{\partial t}\right) + \left(d\left(\nabla^2 E_i E_e - \frac{1}{c^2}\frac{\partial^2 E_i}{\partial t^2}E_e\right) - (b-\beta)E_e E_i + a(1-\gamma)(E^2 - c^2B^2)E_e E_i\right) = 0$$

where we have just added and subtracted the term $\beta E_e E_i$; $0$ and $\gamma(E^2 - c^2B^2)E_e E_i$splitting the different terms:

$$\nabla^2 E_e - \frac{1}{c^2}\frac{\partial^2 E_e}{\partial t^2} - \beta E_e + \gamma(E^2 - c^2B^2)E_e = 0$$

**Eq 144**

$$\nabla E_i\cdot\nabla E_e - \frac{1}{c^2}\frac{\partial E_i}{\partial t}\frac{\partial E_e}{\partial t} = 0$$

**Eq 145**

$$\nabla^2 E_i - \frac{1}{c^2}\frac{\partial^2 E_i}{\partial t^2} - (b-\beta)E_i + a(1-\gamma)(E^2 - c^2B^2)E_i = 0$$

We obtain the equations for a nonlinear Proca field. Eq 144 defines a massive, nonlinear dispersion relation characteristic of a nonlinear Proca equation, or a non-abelian massive photon.

**The last solutions considered in 3.7 and 3.8 could provide a model for the *Z* and *W* massive photons considered in Weak Interactions. Moreover, the interplay between the internal and external solutions may provide also a model** **for the Higgs boson as an effective particle description for the mass of the interaction bosons, and as a contribution to the electron mass as well.**

## 3.9. Conclusions

The most important contribution from this part of the work is the analysis and numerical integration of longitudinal and transverse fields with zero net charge densities. The transverse case in particular can provide a solution to the wave-particle duality for the electromagnetic field. In fact, the transverse solution, able to carry any amount of energy, is a perfect candidate to be the photon-soliton transferring energy from the source to the absorber[6]. The origin of these equations is the inclusion of the Uehling and the Euler-Heisenberg vacuum polarization terms to provide an effective generalization of the Maxwell's equations.

As a result of these calculations we find that massless as well as massive solutions are possible, the first one can be identified with the photon, while the second ones can be related with massive gauge bosons. The dispersion relation found in the different models is not unique and can be selected according to the conditions and requirements. The mass of these last particles is provided by the energy content of the vacuum polarization terms, and can be considered as a substitute for the Higgs-boson as a universal provider of mass. The attainment of massless and massive dispersion relations when applying the double solution method of de Broglie to these equations, allows to consider this model as relevant for the solution of the Millennium problem related to the Yang–Mills existence and mass gap problem.

In this model an effective position can be assigned to the photon solitons as long as the dimensions of the soliton are negligible in relation to the characteristic dimensions of the problem under consideration. Special mention can be given to the work of many authors who provided a theoretical basis to the concept of photon position operator.

## 3.10. Appendix 3.1. Relative weight of the different terms.

The property

$$\frac{\partial \nabla \times [a(E^2 - c^2B^2)B]}{\partial t} + \nabla \times \nabla \times [a(E^2 - c^2B^2)E] \underset{v \to c}{\Longrightarrow} 0$$

**Eq 146**

will be shown in the following. Let's start with the Lorentz transformations for the fields and coordinates:

$$t' = \gamma\left(t - \frac{vx}{c^2}\right); \qquad x' = \gamma(x - vt); \qquad y' = y; \qquad z' = z$$

$$\vec{E}'_{\parallel} = \vec{E}_{\parallel}\ ; \qquad \vec{B}'_{\parallel} = \vec{B}_{\parallel}\ ; \qquad \vec{E}'_{\perp} = \gamma\left(\vec{E}_{\perp} + \vec{v} \times \vec{B}\right); \qquad \vec{B}'_{\perp} = \gamma\left(\vec{B}_{\perp} - \frac{1}{c^2}\vec{v} \times \vec{E}\right)$$

where

$$\gamma \equiv \frac{1}{\sqrt{1 - v^2/c^2}}$$

---

[6] We can identify four levels for the electron photon interaction

- Compton/alpha = Bohr radius. Correspond to the hydrogen atom.
- Compton wavelength giving the electron radius from our calculations in the previous part of this work. Corresponds to the electron problem solved in our work. Because the field intensity is much lower than the critical value, nonlinear effects and the Uehling potential play the role of a small correction.
- Compton x alpha = classical electron radius, photon radius. We identify it with the photon soliton particle.
- Fermi energy: beta decay non abelian nonlinear fields. This level belongs to a different energy range, but the equations are the same as those found for the polarization effects in massless electromagnetic fields.

First let's find the transformation properties of the term

$$\left(\frac{1}{c^2}\frac{\partial^2}{\partial t^2}-\nabla^2\right)E$$

We start by noting that

$$\vec{\nabla}^2=\nabla_{\parallel}^2+2\nabla_{\perp}^2$$

$$\left(\vec{v}\cdot\vec{\nabla}\right)^2=|\vec{v}|^2\nabla_{\parallel}^2+0\times 2\nabla_{\perp}^2$$

$$\frac{\partial^2 E_i}{\partial t^2}=\frac{d\vec{v}}{dt}\cdot\vec{\nabla}E_i+\vec{v}\cdot\vec{\nabla}\frac{\partial E_i}{\partial t}=\frac{d\vec{v}}{dt}\cdot\vec{\nabla}E_i+\left(\vec{v}\cdot\vec{\nabla}\right)\left(\vec{v}\cdot\vec{\nabla}\right)E_i=\frac{d\vec{v}}{dt}\cdot\vec{\nabla}E_i+\left(\vec{v}\cdot\vec{\nabla}\right)^2E_i=\frac{d\vec{v}}{dt}\cdot\vec{\nabla}E_i+|\vec{v}|^2\cos\theta^2\,\vec{\nabla}^2E_i$$

$$=\frac{d\vec{v}}{dt}\cdot\vec{\nabla}E_i+|\vec{v}|^2\nabla_{\parallel}^2E_i$$

taking

$$\frac{d\vec{v}}{dt}\sim 0$$

we get

$$\frac{\partial^2 E_i}{\partial t^2}\cong|\vec{v}|^2\nabla_{\parallel}^2E_i$$

For a plane wave where the field depends on a moving coordinate system as

$$E=E(r_{rel}),\ \ being\ \ r_{rel}=r_{rest}-\mathrm{v}t$$

we obtain:

$$\left(\frac{1}{c^2}\frac{\partial^2}{\partial t^2}-\nabla^2\right)E=\frac{|\vec{v}|^2}{c^2}\nabla_{\parallel}^2E-\nabla^2E=\frac{|\vec{v}|^2}{c^2}\nabla_{\parallel}^2E-\nabla_{\parallel}^2E-2\nabla_{\perp}^2E=\frac{|\vec{v}|^2}{c^2}\nabla_{\parallel}^2E-\nabla_{\parallel}^2E-2\nabla_{\perp}^2E$$

$$=-\left(1-\frac{|\vec{v}|^2}{c^2}\right)\nabla_{\parallel}^2E-2\nabla_{\perp}^2E=-\frac{\left(1-\frac{|\vec{v}|^2}{c^2}\right)}{\left(1-\frac{|\vec{v}|^2}{c^2}\right)}\nabla_{\parallel\,rest}^2E-2\nabla_{\perp rest}^2E=-\nabla_{\parallel\,rest}^2E-2\nabla_{\perp rest}^2E=-\nabla_{rest}^2E$$

And we see that this holds independently from the symmetry of the function E. In principle one can take the limit of this equation for v->c but, the only divergence appearing in the magnitude of E, at the same time that the space and temporal width of the 'pulse' goes to zero keeping the energy content constant. However this is not required because absolute vacuum doesn't exist, according to Wikipedia.

For a field that at rest has no magnetic field:

$$B_{rest}=0\qquad B=\gamma\frac{\mathrm{v}}{c^2}E_{rest}\qquad\qquad E=\gamma E_{rest}$$

so that

$$\left(\frac{1}{c^2}\frac{\partial^2}{\partial t^2}-\nabla^2\right)E=-\gamma\nabla_{rest}^2E_{rest}$$

Let's go now to the term $(E^2-c^2B^2)$. Under the same conditions as before we find:

$$(E^2-c^2B^2)=\gamma^2E_{rest}^2\left(1-\frac{\mathrm{v}^2}{c^2}\right)=\frac{\left(1-\frac{\mathrm{v}^2}{c^2}\right)}{\left(1-\frac{\mathrm{v}^2}{c^2}\right)}E_{rest}^2=E_{rest}^2\underset{\mathrm{v}\to c}{\Longrightarrow}E_{rest}^2$$

The polarization terms transform as

$$\nabla \times \left(\nabla \times P - \frac{1}{c}\frac{\partial M}{\partial t}\right) = \nabla \times \nabla \times \frac{\xi}{4\pi} 2(E^2 - c^2B^2)E - \nabla \times \frac{\partial}{\partial t}\frac{\xi}{4\pi} 2(E^2 - c^2B^2)B$$

$$\rightarrow \frac{\nabla_{rest}}{\gamma} \times \frac{\nabla_{rest}}{\gamma} \times \frac{\xi}{4\pi} 2(E^2 - c^2B^2)E - \frac{\nabla_{rest}}{\gamma} \times \mathrm{v} \cdot \frac{\nabla_{rest}}{\gamma}\frac{\xi}{4\pi} 2(E^2 - c^2B^2)B =$$

$$= \frac{\nabla_{rest}}{\gamma} \times \frac{\nabla_{rest}}{\gamma} \times \frac{\xi}{4\pi} 2E_{rest}^2 \gamma E_{rest} - \frac{\nabla_{rest}}{\gamma} \times \mathrm{v} \cdot \frac{\nabla_{rest}}{\gamma}\frac{\xi}{4\pi} 2E_{rest}^2 \frac{\mathrm{v}}{c^2}\gamma E_{rest}$$

$$\cong \frac{\xi}{4\pi} 2\nabla_{rest} \times \frac{\nabla_{rest}}{\gamma} \times E_{rest}^3 \left(1 - \frac{\mathrm{v}^2}{c^2}\right) = \frac{\xi}{4\pi} 2\nabla_{rest} \times \nabla_{rest} \times E_{rest}^3 \left(1 - \frac{\mathrm{v}^2}{c^2}\right)^{\frac{3}{2}} \underbrace{\rightarrow}_{\mathrm{v}\rightarrow\mathrm{c}} 0$$

# 4. Chapter: Many Particles and Canonical Quantization Introduction

This paper is the fourth in a series of five papers entitled “Solitons and Quantum Behavior”. We discuss here a realistic interpretation of Multi-Particle First Quantization and of Second Quantization in relation to the double solution method pioneered by deBroglie in 1927.

Quantum mechanics achieved a huge advance in mathematical techniques, especially in the early 1950s during the development of QED (quantum electrodynamics), but with time it appeared that as higher the mathematical successes, the poorer the physical understanding of the processes described by the theory. In any experimental setup involving atomic/particle dimensions, there are sources of uncertainty that fall out of the experimentalist control: among them the initial position and velocity of the particles under study, the natural trembling to which apparently all subatomic particles are subject (related to quantum fluctuations, zero-point fluctuations, Zitterbewegung, etc.), and in the case of emission or creation of particles, the actual time of emission/creation is also uncertain experimentally. These ‘natural uncertainties’ forced the theoretical description of atomic processes to be of a statistical nature. In fact, the original, and since then traditional interpretation given to the solution of the Schrödinger equation was to provide the probability distribution to measure the position of the system in space. The statistical interpretation prevailed since the creation of quantum mechanics, and some theorists even postulated that this world is only of a probabilistic nature, deprived of a realistic objectivity. We will verify that quantum theory is able to provide a realistic description for individual events, and not only on a statistical basis.

Perhaps the best way to understand the realistic approach is by describing de Broglie’s original ideas. In traditional quantum mechanics a particle moving with a momentum *p* in the *z*-direction is represented by a “plane” wave-function,

$$\psi_r(x,y,z,t) = const.\, e^{\frac{2\pi i}{\hbar}(Wt-pz)}$$

which in first quantization will correspond to a solution of the Schrödinger, Klein Gordon, Pauli, Dirac or Maxwell equations, and in QED would be an incoming or outgoing “free state”. The amplitude of this wave is constant in space, and can therefore provide no indication about where the particle is located. The apparently best possibility was to attribute a statistical interpretation to this function, by providing a constant probability density to find the particle anywhere in space. However de Broglie noticed that the following function

$$\psi_r(x,y,z,t) = \frac{const.}{\sqrt{x^2+y^2+\frac{(z-\mathrm{v}t)^2}{1-\beta^2}}}\, e^{\frac{2\pi i}{\hbar}(Wt-pz)}$$

was also a solution of the Klein Gordon equation, but now it was describing what today would be called a “soliton” moving with the same speed as the previous solution. In contraposition with the first solution, this one provides an undisputed localization for the particle. This simple function has the potential to show the way to solve most of the paradoxes in the description and interpretation of Quantum Mechanics, including the wave-particle duality and particle interference among others. Despite all the advantages provided by this valid solution, it was totally neglected by the traditional interpretation and teachings of quantum mechanics.

The so called 'canonical or second quantization' is today’s version of quantum mechanics accepted as the best possible description of nature, and can be considered not only a generalization of first quantization, but also a master piece in mathematical notation and representation. In second quantization the fields and wave functions of the first quantization theory are replaced by ‘field operators’, following similar wave equations as their classical originators, but with a different interpretation: they are operators ‘acting’ on the state of the system under study. The mathematical properties of these operators provide an exceptional notational tool for deriving and writing down the wave equations in the presence of many types of particles. Our plan is to extend the ideas of the Double Solution of de Broglie, to the areas of many particles, QED and Quantum Optics. We will be using solitons that are found as solutions to the quantum equations when including self-interactions for electrons and photons. Let's us remark that the consideration of self-interactions is not something extraneous to canonical quantization, but it readily provides the basic elements of the theory of renormalization.

We will verify that the abstract formulation of Field Quantization in terms of Field Operators can be understood as a mathematical formalism able to generate the wave equations followed by 'first quantized' multi-particle fields in Fock space. We will also see that the inclusion of self-interactions and de Broglie's Pilot Wave Theory permits the interpretation of the solutions to those equations as a realistic (nonlocal) description of particle-field states and interaction potentials simultaneously.

We start Section 2 with a demonstration that Fock space is not a formalism that can be applied exclusively to Second Quantization but it can be used to provide an alternative description of classical states and interactions. In Section 3 we provide a "textbook" description of First Quantization to become familiar with the formalism of the multi-particle Schrödinger equation in Configuration Space. In Section 4 we discuss the double solution method in first quantization. In Section 5 we discuss Second Quantization from the point of view of Dirac-Heisenberg-Pauli associated to Field Operators on one hand, and in terms of Fock space and numerical functions on the other hand. The relationship linking both representations is given in Section 6, as a groundwork for understanding the real significance of Field Quantization given in Section 7. In Section 8 we discuss the achievement of ergodicity in single-particle systems. Finally in Section 9 we present the interpretation of Field Quantization in terms of the Double Solution model of de Broglie for electrons and photons.

## 4.1. Application of the Fock space in classical physics

It is a common belief that the mathematical tools used in quantum mechanics cannot describe an objective reality. To disproof this belief we start by showing how spinors, configuration space and Fock space can be used equally well as mathematical tools to describe classical objective physics.

It is easy to show that spinors can be used as a notational tool to rewrite second order differential equations in time as first order equations. In the classical Lagrangian formalism, or equivalently using Newton's law, the equation of motion for a harmonic oscillator is given by

$$\frac{d^2x}{dt^2} = \frac{-kx}{m}$$

Which is a second order differential equation in time. If we choose the Hamiltonian formalism, then the momentum and the position become two independent variables and their respective equations of motion are

$$\frac{dx}{dt} = \frac{p}{m}$$

$$\frac{dp}{dt} = -kx$$

These two equations can be combined into the single form

$$\frac{d}{dt}\binom{x}{p} = \begin{pmatrix} 0 & 1/m \\ -k & 0 \end{pmatrix}\binom{x}{p} = \binom{p/m}{-kx}$$

Where we have a first order differential equation acting on a two component spinor. Moreover, the equation takes the form

$$\frac{d}{dt}s = \hat{a}s$$

Where $\hat{a} = \begin{pmatrix} 0 & 1/m \\ -k & 0 \end{pmatrix}$ is now a matrix operator acting on the spinor $s = \begin{pmatrix} x \\ p \end{pmatrix}$. So, we see that the use of one or other formalism, associated with different notational tools modifies the order of the equation of motion, but it is clear that the physical interpretation behind both formalisms has to be the same.

As a second example, we will consider the description of a classical system in configuration space (space where each particle is assigned an independent set of coordinates describing them) and in Fock space (the open product of multiple configuration spaces with variable number of coordinates). Take a system composed by three frozen droplets composed of helium, xenon and radon localized inside a room. Each has a density that is a function of real space given by $\rho_{Argon}(r,t)$ , $\rho_{Xenon}(r,t)$, and $\rho_{Helium}(r,t)$

However, in order to be more explicit, we can rename the coordinates for the gases as

$$argon \rightarrow r_{Argon}$$

$$xenon \rightarrow r_{Xenon}$$

$$helium \rightarrow r_{Helium}$$

If we apply the logical 'AND' operation to these densities, we can define a generalized density for the entire system of three gases in configuration space by the product

$$\rho_{system}(r_{Argon}, r_{Xenon}, r_{Helium}, t) = \rho_{Argon}(r_{Argon}, t)\rho_{Xenon}(r_{Xenon}, t)\rho_{Helium}(r_{Helium}, t)$$

Which logically means that the density of argon is given by $\rho_{Argon}(r_{Argon}, t)$ and the density of xenon by $\rho_{Xenon}(r_{Xenon}, t)$ and the density of helium by $\rho_{Helium}(r_{Helium}, t)$. It is clear that here again we are in the presence of a simple notational tool to represent the state of the whole system composed of the three gases but now defined in configuration space.

Suppose some of the fields interacting with the particles are external, and the potentials seen by the particles are

$$V_{argon} = V_1(r_{argon})$$

$$V_{xenon} = V_2(r_{xenon})$$

$$V_{helium} = V_3(r_{helium})$$

Suppose additionally that the gases are ionized and they have a mutual interaction through the Coulomb potential:

$$V_{Xe,Ar}^{Cou} = V_4(r_{argon}, r_{xenon}) = \frac{q_{Xe}q_{Ar}}{4\pi\epsilon_0(r_{argon} - r_{xenon})}$$

$$V_{He,Ar}^{Cou} = V_5(r_{argon}, r_{helium}) = \frac{q_{He}q_{Ar}}{4\pi\epsilon_0(r_{argon} - r_{helium})}$$

$$V_{He,Xe}^{Cou} = V_6(r_{xenon}, r_{helium}) = \frac{q_{He}q_{Xe}}{4\pi\epsilon_0(r_{xenon} - r_{helium})}$$

Assume we have an extra potential energy that depends simultaneously on the positions of all three particles:

$$V_{extra} = V_7(r_{argon}, r_{xenon}, r_{helium})$$

whose explicit form we leave undetermined for simplicity. We can build the following 'generalized potential' energy as

$$V_{system}(r_{argon}, r_{xenon}, r_{helium}) = V_1(r_{argon}) + V_2(r_{xenon}) + V_3(r_{helium}) + V_4(r_{argon}, r_{xenon}) + V_5(r_{argon}, r_{helium}) + V_6(r_{xenon}, r_{helium}) + V_7(r_{argon}, r_{xenon}, r_{helium})$$

The actual force acting on each droplet can be found as the gradient of $V_{system}$ respect to the coordinate of that particle as follows:

$$F_i(r_i) = -\frac{\partial V_{system}(r_1, r_2, r_3)}{\partial r_i},$$

Moreover, we can see that $V_1(r_{argon}),\ V_2,(r_{xenon})\ and\ V_3(r_{helium})$ depend on a single variable, while $V_4(r_{argon}, r_{xenon}),\ V_5(r_{argon}, r_{helium})\ and\ V_6(r_{xenon}, r_{helium})$ depend on two variables, and finally $V_{extra}(r_{argon}, r_{xenon}, r_{helium})$ depends on three variables. If the potentials are not diverging functions and have integrable square magnitude, then we find that $V_{system}$ fulfills the definitions of a function in the Fock space for three particles!!!!!

It is clear again that the configuration and Fock space are just a notational convenience or mathematical convention. The particles and the forces are all present in real space. Additionally we can see that $V_4(r_{argon}, r_{xenon})$, $V_5(r_{argon}, r_{helium})$ and $V_6(r_{xenon}, r_{helium})$ are functions of the relative coordinate between the respective particles. As such cannot be reduced as a single product of a function of the coordinate of one gas times another function of the coordinate of the other gas. For reasons that will become clear later, we can call these potentials as "entangled potentials".

**So we see that configuration space can be used to describe a classical state, while Fock space can be used to describe classical interactions.**

As a very simple example, we can mention the classical Maxwell equation in the case where the source term is given by a classical charged field, like an ionized gas:

$$\nabla^2 \vec{A}(\vec{r}, t) - \frac{1}{c^2}\frac{\partial^2 \vec{A}(\vec{r}, t)}{\partial t^2} = -\mu_0 \vec{J}_{gas}(\vec{r}, t)$$

that can also be written as

$$\nabla^2 \vec{A}(\vec{r}_{photon}, t) - \frac{1}{c^2}\frac{\partial^2 \vec{A}(\vec{r}_{photon}, t)}{\partial t^2} = -\mu_0 \int \vec{J}_{gas}(\vec{r}_{gas}, t)\delta(\vec{r}_{photon} - \vec{r}_{gas})d\,\vec{r}_{gas}^{\,3}$$

here the left hand side of the equation can be associated with a state of a single electromagnetic field, while the right hand side can be associated with a single gas field.

In order to compare later with second quantization under the double solution model, we want to notice that we could also create a new 'universal' state function by the product of the function describing the system state times the function describing the mutual interactions between the particles. This description would be useful if we wanted to describe a state in a theory where the functional description of the particle mutual interactions would not be given by fixed functions as the Coulomb or Biot-Savart interactions, but would be given by expressions that themselves depend explicitly on time. Moreover, it is clear that we could provide an equally useful description of the universal state of the system if instead of writing the total potential energy function, we would write a function of it, as long as we know the rule to recover the original total potential function from that modified functional expression.

## 4.2. First Quantization

We provide here a summary of first quantization. Initially quantization was the name given to the theory dealing with the solution of the quantum wave equation (initially Schrödinger, later Klein Gordon, Pauli and Dirac) for individual particles (typically an electron) in real space and time. The generally accepted physical interpretation of the theory was that the square of the amplitude of the wave function provided the probability for finding the electron at a given position and at a given time. Multiparticle systems with a fixed number of particles, such as a group of electrons inside an atom, molecule, or a solid/liquid material, or of nucleons inside a nucleus, could also be described by a multiparticle wavefunction, defined in configuration space and whose evolution was described by a multiparticle Schrödinger equation. In this description electromagnetic fields had to be provided as external functions of space and time in a classical description allowing a continuous value for the electromagnetic energy content. This last property was in contradiction with the fact that the energy of the electromagnetic field should have values that are only multiples of $\hbar\omega$. This description sometimes is called semi classical description.

### 4.2.1. Single particle

The most basic equation is the single-particle Schrödinger equation

$$\frac{d\psi(r)}{dt} = \hat{H}\psi(r)$$

where $\hat{H}$ is an operator typically given by an expression like

$$\hat{H}\psi(r) = -\frac{\hbar^2}{2m}\nabla^2\psi(r) + V(r)\psi(r) = E\psi(r)$$

Where $V(r)$ is the scalar potential energy, as noticed before, given by external functions of position and time.

There are different equations describing the evolution of the electron wave function: the non-relativistic Schrödinger and Pauli equations, and the relativistic Klein Gordon and Dirac equations. The Schrödinger and Klein Gordon equations, describe the evolution of complex-scalar wave functions, the Pauli equation describes a two-component spinor, and the Dirac equation a four-component spinor. The Klein Gordon equation is a second order partial differential equation in time, while the other three are first order partial differential equation in time. The Dirac equation and the corresponding 4-spinors are considered to be the best representation of the electron. In single-particle first quantization it is generally assumed that the solutions $\psi(r)$ are found in real space-time, and the arguments of the solutions are the classical space coordinates.

### 4.2.2. Many particles. Lagrangian, Hamiltonian

There are many issues to face when considering the problem of many particles and the radiation field. The Schrödinger equation can be generalized without problems to many particles because the equation is first order in time and has no sources. If the multiparticle wavefunction is taken to be the product of *n* single wavefunctions, one per particle, then the equation can be split into *n* equations, each one of them reproducing a single particle Schrödinger equation. The Klein Gordon equation represents a difficulty because is a partial differential equation of second order in time that prevents the previous generalization. The Maxwell's equations for the potential fields are traditionally also of second order in time and show additionally a source term that prevent unitarity. For this reason electromagnetic fields were not included initially in the quantum formalism. To this difficulty one has to add the problem of negative frequencies and energies raised by second order differential equations in time characteristic of relativistic problems, and related to the existence of antiparticles.

The first generalization to a many particle equation was done in configuration space, where the coordinates for each particle receive a different nomination and are considered as independent orthogonal variables. In the absence of spin variables the multiparticle wavefunction has the form:

$$\psi(r_1, \cdots, r_i, \cdots r_n, t)$$

where every coordinate $r_i$ corresponds to a different particle. The number of particles remains constant in time. In principle this function can be any solution to the quantum equation compatible with the boundary conditions.

The most basic property of the solutions in configuration space is that the wavefunction for a given particle has not necessarily a single value at each point in space, but the value of the wavefunction depends simultaneously on the positions of all particles.

The Schrödinger equation for a system of many particles is:

$$i\hbar \frac{d\psi(r_1, \cdots, r_i, \cdots r_n, t)}{dt} = \hat{H}\left(r_1, \cdots, r_i, \cdots r_n, \nabla_{r_1}, \cdots, \nabla_{r_i}, \cdots \nabla_{r_n}, t\right)\psi(r_1, \cdots, r_i, \cdots r_n, t)$$

Where the Hamiltonian operator $\hat{H}$ in the case of an atomic system is given typically by

$$\hat{H}\psi = \sum_{i=1}^{n}\left(-\frac{\hbar^2}{2m}\nabla_{r_i}{}^2\psi - Ze^2\sum_R \frac{1}{\left|\vec{r}_i - \vec{R}\right|}\psi\right) + \frac{1}{2}\sum_{i\neq j}\frac{e^2}{\left|\vec{r}_i - \vec{r}_j\right|}\psi = E\psi$$

One way to solve this equation is to reduce the function $\psi$ to the product of functions for each particle variable:

$$\psi(r_1, \cdots, r_i, \cdots r_n, t) = \psi_1(r_1, t)\cdots\psi_i(r_i, t)\cdots\psi_n(r_n, t)$$

where each function is normalized to one. In the case of a differential equation of first order in time (as the Schrodinger equation) the general equation could be split in *n* equations, one per particle. The individual functions are typically eigenstates of a single particle Hamiltonian:

$$-\frac{\hbar^2}{2m}\nabla^2\psi_i(\vec{r}) + V^{ion}(\vec{r})\psi_i(\vec{r}) + \left[\sum_j d\vec{r}'\int\left|\psi_j(\vec{r}')\right|^2\frac{e^2}{|\vec{r}-\vec{r}'|}\right]\psi_i(\vec{r}) = \varepsilon_i\psi_i(\vec{r})$$

The system energy *E* will be found as a function of the individual particle eigenenergies $\varepsilon_i$.

In the case that some or all of the particles are identical, the corresponding part of the wave function has to be symmetrized or antisymmetrized, depending on the particles being bosons or fermions.

$$\psi(r_1, \cdots, r_i, \cdots r_n) = \begin{vmatrix} \psi_1(r_1) & \cdots & \psi_1(r_i) & \cdots & \psi_1(r_n) \\ \vdots & \vdots & \vdots & \vdots & \vdots \\ \psi_i(r_1) & \cdots & \psi_i(r_i) & \cdots & \psi_i(r_n) \\ \vdots & \vdots & \vdots & \vdots & \vdots \\ \psi_n(r_1) & \cdots & \psi_n(r_i) & \cdots & \psi_n(r_n) \end{vmatrix}_{\xi}$$

where $\xi$ indicates if the determinant or the permanent has to be taken. We note that this method will not provide all the solutions available to the original Hamiltonian, but only a stationary subset. However the solutions obtained in this way could be used as a basis to expand more general solutions.

The next step in generalization of the theory would be to use a relativistic wave equation. We note here that it is possible to write the single particle Klein Gordon (and also Maxwell) equations as a first order differential equation acting over a spinor as described in the book by Greiner[56]. In this form the multi-particle Klein Gordon equation can also be written as a first order differential equation in time acting on some multi-dimensional spinor representing the multi-particle Klein Gordon fields. This step was recognized however relatively recently. Historically the second step was taken in 1929 and consisted in writing the multi – particle Dirac equation, making use of a generalized configuration space adapted for 4-spinors. This was possible because the Dirac equation is also first order in time as the Schrödinger equation is. The difficulty here stayed in the fact that, as Breit[57] has shown, the state vector or description of the state of the system, should be given by a spinor with $4^n$ members, corresponding to the open product of the *n* 4-spinor single particle fields. Breit developed his model initially for the interaction between two electrons, but his model can be straightforwardly generalized to *n* particles. This would provide an excessive complexity for the solution of the equations involving higher number of particles. In this sense the development of second quantization provided a huge notational notation, allowing for a most manageable notation.

### 4.2.3. Entanglement

In order to introduce the concept of entanglement we are going to compare interference for a single particle in real space with interference for a system of particles in configuration space. For a single wavefunction in real space, one can have negative interference at a point $r_A$. Suppose that the wavefunction at the point $r_A$ has two contributions

$$\psi_A(r_A) = \psi_{A1}(r_A) + \psi_{A2}(r_A)$$

And at that point both contributions have the same magnitude but opposite sign,

$$\psi_{A1}(r_A) = -\psi_{A2}(r_A)$$

then

$$\psi_A(r_A) = |\psi_{A1}(r_A)| - |\psi_{A2}(r_A)| = 0$$

where the wavefunction at that point is null: $\psi_A(r_A) = 0$. The interpretation is that the probability to find the particle at $r_A$ is null: it will never be found (measured) at $r_A$.

In the presence of two particles *A* and *B*, in configuration space it is also possible to achieve negative interference at a point $(r_A, r_B)$, and find that the wavefunction is null at that point $(r_A, r_B)$. Assume we have two contributions again

$$\psi_A(r_A, r_B) = \psi_{A1}(r_A, r_B) + \psi_{A2}(r_A, r_B)$$

And at that point in configuration space both contributions have the same magnitude but opposite sign

$$\psi_{A1}(r_A, r_B) = -\psi_{A2}(r_A, r_B)$$

then

$$\psi_A(r_A, r_B) = |\psi_{A1}(r_A, r_B)| - |\psi_{A2}(r_A, r_B)| = 0$$

The interpretation of this is that the probability to find simultaneously particle *A* at $r_A$ and particle *B* at $r_B$ is null: never are we going to see that situation. This doesn't mean that particle *A* will never be at position $r_A$, or particle *B* will never be at $r_B$ but it means that we will never observe particle *A* at $r_A$ 'simultaneously' with particle *B* at $r_B$. And this happens for a pair of particles that otherwise would be considered independent and non-interacting. The interference phenomenon we just mentioned is an example of a more general property of multi-particle quantum mechanics known as 'entanglement'.

An example is provided by a system of two particles where the state of the system is given by

$$\psi(r_A, r_B) = e^{i(k_A \cdot r_A)} e^{-i(k_B \cdot r_B)} e^{-i\omega t} + e^{-i(k_A \cdot r_A)} e^{i(k_B \cdot r_B)} e^{-i\omega t} =$$

$$e^{i(k_A \cdot r_A - k_B \cdot r_B)} e^{-i\omega t} + e^{i(k_A \cdot r_A - k_B \cdot r_B)} e^{-i\omega t} = 2\cos(k_A \cdot r_A - k_B \cdot r_B)\, e^{-i\omega t}$$

It can be verified that this wavefunction is null every time the cos function is zero. The theorem known as Schmidt decomposition specifies that any state of two particles can be written as a sum of products of single particle functions. A special property of functions describing entangled states is that their Schmidt decomposition has always more than one term. It cannot be rewritten as a product of a function of $r_A$ times a function of $r_B$. We can observe that $\psi(r_A, r_B)$ is a function of the relative coordinate $k_A \cdot r_A - k_B \cdot r_B$. In this case the wavefunction will not depend on the position of particle *A* and particle *B* independently, but only as a function of its relative distance. This type of states appear to show a mutual interdependence between the particles, even when there is no apparent or classical interaction between them. In the theory of the quantum potential, this interaction can be made explicit by the existence of a mutual quantum potential energy between the particles that depends on their relative coordinate and is determined by the form of the wavefunction describing the state.

Let's consider a similar case, but now working with angular coordinates:

$$\psi(\alpha_A, \alpha_B) = \sin(k_A\alpha_A)\cos(k_B\alpha_B)\, e^{-i\omega t} - \cos(k_A\alpha_A)\sin(k_B\alpha_B)\, e^{-i\omega t} = \sin(k_A\alpha_A - k_B\alpha_B)\, e^{-i\omega t}$$

Here we see the same type of entanglement as before, but now related to an angular coordinate. In fact this example shows what is called the "singlet state" of two angular momenta. It shows that the orientation of the two angular momenta is not independent but between them exists a definite relationship. This example shows the basis over which is built most of the theory of composition of angular momenta to form compound particles. The sum or difference of the angular momenta of all particles conforming a compound particle may have a definite value, but not the values of the independent components.

A similar case of entanglement is also shown by identical particles following the symmetrization or anti-symmetrization postulate, providing the basis for the Pauli principle where two fermions cannot occupy the same position at the same time.

This effect could be intuitively understood however once we identify the wavefunction in configuration space as a quantum potential energy that can achieve a nonlocal character. In many cases the coordinate dependence of the wavefunction is on the relative coordinate $r_A - r_B$. In these cases the quantum potential would also depend on $r_A - r_B$, and conservation of momentum can be easily verified. Something similar holds with energy conservation. The conservation laws are not holding between the particles and the wavefunction, but between the different particles entangled through the wavefunction, in a nonlocal, 'instantaneous' way. Entanglement is a measurable property, and it is clear that in order to describe it, a general single equation is needed that will describe the evolution of all particles simultaneously. A multi-particle quantum theory must therefore provide such a general equation. We have seen above that entanglement can be described by multi-particle first quantization models for electrons.

### 4.2.4. Hydrogen atom

As an example of entanglement we can consider the hydrogen atom. The equation of motion for two particles as a function of the particle coordinates on one side and as a function of the center of mass and relative coordinates on the other side can be taken from the textbook by Cohen Tannoudji[58].

The Lagrangian as a function of the two particles is

$$\mathcal{L}(\vec{r}_1, \dot{\vec{r}}_1, \vec{r}_2, \dot{\vec{r}}_2) = T - V = \frac{1}{2} m_1 \dot{\vec{r}}_1^2 + \frac{1}{2} m_2 \dot{\vec{r}}_2^2 - V(\vec{r}_1 - \vec{r}_2)$$

And in center of mass and relative coordinates:

$$\mathcal{L}\left(\vec{r}_G, \dot{\vec{r}}_G, \vec{r}, \dot{\vec{r}}\right) = \frac{1}{2} M \dot{\vec{r}}_G^2 + \frac{1}{2} \mu \dot{\vec{r}}^2 - V(\vec{r})$$

where $\vec{r}_G$ is the center of mass coordinate and $\vec{r}$ is the relative coordinate between the nucleus and the electron. Similarly the Hamiltonian as a function of the two particles is

$$H = \frac{\vec{p}_1^2}{2m_1} + \frac{\vec{p}_2^2}{2m_2} + V(\vec{r}_1 - \vec{r}_2)$$

$$\frac{\partial}{\partial t} = \frac{1}{2} m_1 \nabla_1^2 + \frac{1}{2} m_2 \nabla_2^2 + V(r_2 - r_1)$$

And in center of mass and relative coordinates:

$$H = \frac{\vec{p}_G^2}{2M} + \frac{\vec{p}^2}{2\mu} + V(\vec{r})$$

$$\frac{\partial}{\partial t} = \frac{1}{2} M \nabla_{CM}^2 + \frac{1}{2} \mu \nabla_{rel}^2 + V(r_{rel})$$

If we choose the center of mass and relative coordinates as the independent variables, the CM is free. The state of the system can be written as a product of a function of the center of mass times a function of the relative coordinates:

$$|\varphi\rangle = |\chi_G\rangle \otimes |\omega_r\rangle$$

The CM solution is then the corresponding to a plane wave, for example

$$\chi_G(\vec{r}_G) = \frac{1}{(2\pi\hbar)^{3/2}} e^{\frac{i}{\hbar}\vec{p}_G \cdot \vec{r}_G}$$

With energy eigenvalues

$$E_G = \frac{\vec{p}_G^2}{2M}$$

The Hamiltonian dependent on the relative coordinate is

$$\mathcal{H}(\vec{r}, \vec{p}) = \frac{\vec{p}^2}{2\mu} - \frac{e^2}{r}$$

Which gets transformed into a Schrödinger equation

$$\left[-\frac{\hbar^2}{2\mu}\nabla^2 - \frac{e^2}{r}\right] \psi(\vec{r}) = E\psi(\vec{r})$$

With the well-known solutions

$$\varphi_{k,l,m}(\vec{r}) = R_{k,l}(r) Y_l^m(\theta, \varphi) = \frac{1}{r} u_{k,l}(r) Y_l^m(\theta, \varphi)$$

Now the point is that the coordinate $\vec{r}$ **IS NOT** a spatial coordinate, but it is a relative coordinate of the electron respect to the proton, and therefore this is an entangled solution as we have seen above. In classical physics the description of a mechanical problem in relative and CM variables is a mathematical technique to simplify the calculation. In quantum physics it implies a physical interaction leading to the generation of measurable entanglement between the particles.

## 4.3. Double Solution for First Quantization Multi-Particle Case

We have seen that in the case of a single particle the quantum potential (in the Klein Gordon description) is given by

$$Q=-\frac{\hbar^2}{8\pi^2 m}\left(\frac{\Box|\psi_{ext}(r)|}{|\psi_{ext}(r)|}\right)$$

where *r* is the coordinate of the particle, that could be associated in that case with real space.

We have seen that the multiple-particle Schrödinger equation in first quantization is a differential equation with a multi-particle solution function defined in configuration space. In the case of non-interacting particles, the general solution can be written as the product of *n* independent functions, each one solution of the single-particle Schrödinger equation. The double solution method that we have used for a single particle can be applied also to the multiple-particle case: the full function for *n* particles will be the product of *n* internal wave-functions, one per particle-coordinate, multiplying an external part. For example in the case of two independent fermion particles, the wavefunction would be

$$\begin{aligned}\psi(r_{e1},r_{e2},t)&=\psi_{int,1}(r_{e1},t)\psi_{int,2}(r_{e2},t)\times\psi_{ext}(r_{e1},r_{e2},t)\\&=\psi_{int,1}(r_{e1},t)\psi_{int,2}(r_{e2},t)\times\left[\psi_{ext,1}(r_{e1},t)\psi_{ext,2}(r_{e2},t)-\psi_{ext,2}(r_{e1},t)\psi_{ext,1}(r_{e2},t)\right]\end{aligned}$$

The difference with traditional quantum mechanics is that the symmetrization postulate applies only to the external wavefunction. For particles with a mutual interaction dependent on their relative separation as is the case of a bound state, the two-particle state can be written as

$$\begin{aligned}\psi(r_{e1},r_{e2},t)&=\psi_{int,1}(r_{e1},t)\times\psi_{int,2}(r_{e2},t)\times\psi_{ext,CM}\left(\frac{r_{e1}+r_{e2}}{2},t\right)\times\psi_{ext,REL}(r_{e1}-r_{e2},t)\\&\equiv\psi_{int,1}(r_{e1},t)\times\psi_{int,2}(r_{e2},t)\times\psi_{ext,CM}(R_{cm},t)\times\psi_{ext,REL}(r_{re1},t)\end{aligned}$$

This can be the case also for the usual hydrogen atom, with *e1* the nucleus and *e2* the electron. It can also be a pair of entangled electrons or of entangled photons. In this case each solitons have a real existence in space, but both share a common external wavefunction, and consequently a common quantum interaction, where energy and momentum is conserved, making it possible a realistic interpretation.

In the case when the spatial extension of the internal functions is negligible in front of the extension of the external wavefunction, the internal wave function can be approximated by a Dirac-delta-like function representing the position of the particle. The guidance condition from the double solution model shows that over these particles exists a Quantum potential energy, function of all coordinates at time *t* given by:

$$Q=-\frac{\hbar^2}{8\pi^2 m}\left(\frac{\Box|\psi_{ext}(r_{e1},r_{e2},t)|}{|\psi_{ext}(r_{e1},r_{e2},t)|}\right)$$

See papers from quantum potential and quantum statistics[59] for distinguishable particles.

***The internal solutions provide reality of presence for the particles while the external part represent interactions between the particles. As we have seen this function can be considered as the representation of a real system as shown in section 4.1 above, describing the state of the system and of the interaction energy between the particles. The difference with classical physics is that the interaction can be non-local. This constitute an important result of our work: the description in realistic terms of quantum mechanics.***

# 4.4. Second Quantization

A generalization was needed for First Quantization to consider many particles in a relativistic way, and the inclusion of the electromagnetic fields into the quantum formalism. In fact, it was known that the electromagnetic fields had many properties in common with electrons, such as behaving like particles with a definite amount of momentum and energy. As a result of using relativistic quantum equations, the appearance of solutions with negative energy, identified as antiparticles had to be treated in a unified way with the positive energy particles and providing consistency with experiments. An additional fact that needed to be accommodated in the theory was the existence of the so called 'vacuum fluctuations' or 'zero-point energy', already present in systems at zero temperature. Finally, the theory should be able to include and describe entanglement in its formalism. The present day theory that can handle in an appropriate way relativity, quantization of radiation, and many particle systems is called second or field or canonical quantization.

The inclusion of electromagnetic photons as valid 'particles' into the quantum theory required the creation of a new formalism, and of new differential equations able to handle a variable number of particles (photons), due to the emission and absorption processes. Later it was found that not only the number of photons, but also the number of the other particles present in the system, such as the electrons or protons could be variable as well. This implied that in the new theory, the number of particles had to be a space and time dependent variable following an equation of motion. These forced a further generalization of the theory known as Canonical or Second Quantization.

As the number of particles changes with time, the state of the system had to be described by a 'super' field function of n(t) variables, each variable representing a particle. As we will see, in some very special cases, the 'super' field function is just the product of individual wavefunctions, one for each particle. But this is not the general case. The equation of motion for this 'super' field is built on a new operator called the 'field operator'. As it was the case with the older differential operators associated with the dynamical properties of the particles, this field operator was also space and time dependent, and as mentioned, had its own equation of motion. This new required equation of motion was known as the Canonical or Second Quantization Schrodinger's equation. We will see more clearly this fact when writing that equation in the occupation number space.

In the following we will consider the formalism that was constructed to handle all these equations for the fields and the operators themselves. We can start analyzing the new theory by considering the operator Schrodinger's equation in Second Quantization. It is an equation where the operators themselves are the variables or functions of time. To understand what this means, let's consider first the classical Maxwell's equation with constant sources, namely constant current and charge densities. In this case the electromagnetic fields, solutions to the equation, are mostly waves evolving with time. In the equation of motion, the time and space derivatives as well as the source densities can be considered constant operator expressions, they don't change in time nor in space. The same picture happens when we consider the Schrodinger's equation for a single particle wavefunction (an electron or photon field, for example). In the following we are going to use indistinctly the word particle or field to refer to the same object.

The situation changes basically when we consider a variable number of particles (or fields) . In this case not only the sources become functions that will change with position and time, but the equation itself will change with space and time, becoming a time varying sum of powers and products of terms composed of partial derivatives of varying order, involving also a changing number of fields. Eventually the differential equation will have an infinite number of terms, and will become an infinite series in products and powers (order) of operators acting on an infinite number of fields.

Inside all this complexity there is however a well pre-established order. The evolution of the fields differential equations as well as the evolution of the operators composing each term in the equation follow a well-defined law. The equation for the individual operators appearing in the terms is called the Heisenberg operator equation of motion for that operator. The equation followed by the full equation of motion is the so called second quantization Schrodinger equation, where the 'variables' are operators and states.

This is the meaning of second quantization as we see it, it is the required theory to provide the additional equation of motion for the differential equations themselves, where the participating variables are not only numerical field functions, but also products of differential operator expressions of varying order and defined on a time and space dependent number of variables.

We borrow here the definition of second quantization as provided by Wikipedia:

*“**Second quantization** is a formalism used to describe and analyze quantum many-body systems. It is also known as canonical quantization in quantum field theory, in which the fields (typically as the wave functions of matters) are upgraded into field operators, following the similar idea that the physical quantities (position, momentum etc.) are upgraded into operators in first quantization. The key ideas of this method were introduced in 1927 by Dirac, and were developed, most notably, by Fock and Jordan later. In this approach, the quantum many-body states are represented in the Fock state basis, which are constructed by filling up each single-particle state with a certain number of identical particles. The second quantization formalism introduces the creation and annihilation operators to construct and handle the Fock states, providing useful tools to the study of the quantum many-body theory.”*

Canonical quantization in addition to provide a generalization to many particle quantum systems also handles self-interactions and self-mass. These last effects are however not provided through the natural description of classical physics but through the effects associated with the so called vacuum fluctuations. Many of the properties and effects analyzed in this way were attributed to second quantization, but they can be described in a first quantization theory susceptible of a realistic interpretation. In fact, it can be argue that the so called vacuum fluctuations effects are generated by the combination of self-interactions and the quantum potential.

In first quantization the Hamiltonian operator present in the Schrödinger equations is the classical particle Hamiltonian where variables are replaced by operators. The ‘novelty’ of Second Quantization over First Quantization consists of two steps. In first place the variables of interest are not more the position and momentum of the particle, but the field amplitude and momentum. In other words, the Hamiltonian operator is not derived from the classical particle Hamiltonian, but is now derived from the classical ‘field Hamiltonian’ where the fields are replaced by operators fulfilling certain commutation relations. And second the definition of the field operators in terms of creator and destructor operators multiplied by functions of space and time. Both steps together determine the structure of the differential equations followed by the wavefunctions in Fock space. This structure is different from the first quantization equivalents, allowing for example to the presence of sources and sinks in the differential equations.

### 4.4.1. Second quantization history

There were two periods of far reaching developments in the theory of second quantization: after its creation around 1930, and with the development of quantum electrodynamics or QED around 1950.

a. Around 1930.
   - a.1. In 1927 Dirac[60] quantized the electromagnetic field, he based his work on previous works by Born and Jordan[61] and quantized the transverse electromagnetic field, the Coulomb field was not included in this model. He used the fact that the Hamiltonian for the field has the same structure as the Hamiltonian for the amplitude of a harmonic oscillator. Therefore, applied the quantum solution for a harmonic oscillator to the electromagnetic field. The primary interpretation of the model was in terms of the discretization of energy in the electromagnetic field, equivalent to the discretization of energy in the harmonic oscillator. However, he went farther and re-interpreted these discrete energy levels as representing the existence of radiation particles. In this sense it could be said to be a multiparticle description of the electromagnetic field in terms of “photons”. He defined commutation relations for the electromagnetic field amplitude in parallel with the commutation relations for the amplitude of the harmonic oscillator. This description was given a statistical nature, where distribution functions could be defined for the value of the electromagnetic field phase and amplitude and for the number of radiation or light particles.

a.2. Later Heisenberg, Pauli et al.[62] generalized this technique to other systems, such as bosons in general and fermions in 1929-1930, which was called Canonical or Second Quantization. Here the wave amplitude for bosons was quantized by defining commutation relations similar to those for the electromagnetic field and harmonic oscillators, while for the fermion fields anticommutation rules were defined. These last rules were equivalent with the prescription that a multi-particle fermion wave function has to be antisymmetrized.

a.3. In 1930 Landau and Peierls[63] developed a quantum electrodynamics in configuration space. In his theory they developed the operators involved in the process of radiation emission and absorption that played a similar role of Dirac's operators in 1927. Their theory was more intuitive than the previous models but was discontinued after their initial paper.

a.4. Fock[64] in 1932 showed that Heisenberg's canonical quantization was mathematically equivalent to a multi-particle theory defined in a generalized configuration space with a variable number of particles, and where the creation and destruction field operators could be understood in terms of operators coupling functions possessing different number of particles. This theory was in some aspects similar to 1930 Landau and Peierls' quantum electrodynamics.

a.5. Weisskopf in 1934 applied perturbation theory to the problem of electron self-mass and self-energy using both the electrostatic classical interaction for the scalar potential $V$ and the quantized fields for the electromagnetic potential $A$. He utilized non relativistic theory of perturbations and obtained the same results as those obtained later, using full relativistic techniques in the 1950s for a particle at rest.

a.6. Heisenberg, Euler, Uehling and others calculated around 1935 vacuum polarization effects, which are nonlinear contributions to the Maxwell equations, referred to as self-energy of the photons, and that are produced because at higher electromagnetic energy densities, the creation of electron-positron pairs is possible.

What was missing from the 1930s theory was a relativistic perturbation theory. To make the perturbation theory relativistic, one needs to consider expressions that possess symmetry between space and time, they must treat space and time on the same footage. Lagrangian/ Hamiltonian theories leading to a wave equation do not possess such symmetry because the wave equation selects a particular frame of reference and calculates the evolution of the wave for all space at a constant time.

b. Around 1950.

b.1. More than 20 years after the original work of Dirac, independently Feynman, Schwinger and Tomonaga developed a fully relativistic version known as QED taking fully account for antiparticle (positron) physics[65]. Two different approaches were taken to develop the relativistic perturbation theory. Schwinger and Tomonaga generalized the formalism to allow for a '4-dimensional' wave equation, where the 'time slice' could be arbitrary. Feynman provided a new definition for many particle quantum mechanics, including radiation, but starting with the Huygens principle in real space and time. He was able to show how the traditional Schrödinger equation could be derived from the expression of the classical action function. This derivation could be seen in some sense as a redefinition of quantum mechanics where the basic principle was assigned to the Huygens principle instead to the Schrodinger equation. The relativistic Feynman method was simpler, based in the propagator function which is a classical 4 dimensional concept symmetric in space and time.

b.2. QED renormalization and self-interactions in 1950.

The theories of Schwinger, Tomonaga and Feynman allowed for the development of a perturbation theory following all relativistic prescriptions. This however didn't mean that the traditional perturbation methods used around 1930 were wrong. In fact the newly developed relativistic techniques corroborated the previous works by Weisskopf and others. What the new method brought was a more powerful and elegant mathematical treatment. However in many cases the original old method was more intuitive, transparent and with more physical sense.

c. 1967: A model of Leptons[66] were developed by Weinberg providing unification for weak and electromagnetic interactions. In this model the zero mass of the electromagnetic field appeared as a result of a proper intensity and phase relationship at the time of emission of the *B* and *W* gauge boson fields generating a massless field.

In the remaining of Section 4.4 'Second Quantization' we follow a 'textbook' description of second quantization, allowing for a comparison of the Heisenberg-Dirac and Fock formalisms. In Sections 4.5 'Dirac-Fock relationship' and 4.6 'Field Quantization Significance' it will become evident that the formulation of Field Quantization in terms of field operators can be understood as a symbolic formalism able to provide sets of differential equations linking multi-particle fields and wavefunctions with different number of particle-coordinates. An intuitive analogy would be a field description of various multi-component charged fluids acting as sources for multi-component electromagnetic fields, each component of the electromagnetic field having a well-defined energy content. Finally in Section 4.8 'Double solution in Canonical Quantization' we show how a realistic-intuitive interpretation of this formalism can be devised.

### 4.4.2. Areas of application

We are going to consider only two main areas of research in canonical quantization: QED and Quantum Optics. QED is a relativistic theory working typically with the scattering of a small number of electrons and photons, moving in space under the influence of their mutual interactions. Typical examples are:

- Coulomb Scattering of Electrons
- Scattering of an Electron off a Free Proton, the Effect of Recoil
- Scattering of Identical Fermions
- Electron-Positron Scattering, Bhabha Scattering and Muon Pair Creation
- Scattering of Polarized Dirac Particles
- Bremsstrahlung
- Compton Scattering, the Klein-Nishina Formula
- Annihilation of Particle and Antiparticle

QED also considers radiation reaction, self-interactions and self-mass problems however not through the natural description provided in classical physics but through the effects associated with its interaction with vacuum or the so called vacuum fluctuations:

- Electron-Positron Scattering in Fourth Order
- Vacuum Polarization
- Self-Energy of the Electron
- The Vertex Correction

The customary way to solve scattering problems in QED is by giving the initial and final states of the system. In that way one is considering only those particular components of the fields that start and end respectively in the given initial and final states, restricting the calculation to the paths connecting those states exclusively. That may provide some interesting insights, like the virtual fields involved in the mutual electron scattering appear to be massive, but in general they tend to provide a restricted view of the general physical process.

Quantum Optics works typically with non-relativistic atomic systems, where the electrons are not free, but confined inside a volume, that could be an atom, a molecule, a crystal, a gas, etc. The number of electrons and photons can be very large. Over time it was concerned with the following problems[67,68] :

- 1960: Powerful and coherent sources of light: laser. Light Amplifiers, what is the ultimate limit in noise to signal ratio? Quantum fluctuations. Statistical properties of light, photon statistics. Incoherent nonlinear optics: frequency mixing and conversion.
- 1970: source field and radiation reaction, shown to be equivalent to vacuum fluctuations and commutation relations. Fields following sources. Resonance fluorescence. Use of Density matrix, Bloch equations and rate equations.
- 1980-90: Applications of nonlinear optics: squeeze states. Transmission of information at lower signal levels. Entanglement, initially generated by parametric spontaneous down conversion crystals. No signaling theorem, no cloning theorem. Coherent nonlinearities: nonlinear susceptibility.
- 1990-today: Quantum information applications.

### 4.4.3. Dirac-Heisenberg-Pauli representation

We provide here the second quantization formalism as developed initially by Dirac in 1927. Later the electromagnetic quantization model was used for other types of bosons, and also for fermions. Dirac described the state of the electromagnetic field in terms of its Eigenmodes. He used the fact that the Hamiltonian for the Eigenmodes had the same structure as the Hamiltonian for the amplitude of a harmonic oscillator and applied the quantum solution for a harmonic oscillator to the electromagnetic field. In analogy with the quantum harmonic oscillator, he showed how excitations can be created or destroyed by the creation and destruction operators, how a given state can be generated by the multiple action of the creation operator on the 'ground state', defined as the state with no excitations. Taking the harmonic oscillator as a model, Dirac defined creation and destruction operators following the commutation relations

$$[a, b^+] = ab^+ - b^+a = \delta_{ab}$$

Later analogous anticommutation rules for fermions were defined in the form:

$$\{a, b^+\} = ab^+ + b^+a = \delta_{ab} \qquad \rightarrow \qquad ab^+ + b^+a - 2b^+a = ab^+ - b^+a = [a, b^+] = \delta_{ab} - 2b^+a$$

Using these operators, the state of the field with n excitations in the single Eigen mode of type i is defined by:

$$\left| n_{i\,k_i} \right\rangle = \frac{\hat{a}_i^{\dagger\,n}}{\sqrt{n_i!}} \left| vac \right\rangle$$

**Eq 147**

A great advantage of second quantization is that it allows to concentrate on those parts of the wavefunction that are non-null, by specifying the state of the system by Eq 147. When we compare the extension of this expression with the $4^n$ terms appearing in Breit expression we can see that this is a huge notational advantage. A more complex state where we have $n_{ai}$ excitations of type $a_i$ and $n_{bi}$ excitations of type $b_i$ is given by:

$$|state\rangle = |\cdots n_{ai} n_{bi} \cdots\rangle = \sum_i \frac{a_i^{\dagger n_{ai}}}{\sqrt{n_{ai}!}} \frac{b_i^{\dagger n_{bi}}}{\sqrt{n_{bi}!}} \cdots |vac\rangle$$

Where a sum over the index *i* is assumed in order to allow for a possible entanglement. The most fundamental evolution equation of his theory is a Schrödinger-like equation:

$$\frac{d|state\rangle}{dt} = \hat{H}|state\rangle$$

**Eq 148**

Where $\widehat{H}$ is the Hamiltonian operator for second quantization, obtained from the classical field Hamiltonian, by replacing the field variables by 'field operators'. In this way in an energy eigenstate, the frequency of 'oscillation' of the quantum state is given by the **energy content of the field**. This can be considered the QUANTIZATION CONDITION of the theory and can be thought to be the deepest meaning of Eq 148.

In a Hamiltonian eigenstate it is the energy content of the fields which is conserved and the fields acquire the primary role of the theory. Because the Hamiltonian operator acting on the state is not defined in terms of differential or algebraic operators, like gradient, multiplication or alike, the Schrödinger equation should not be considered a 'real' differential equation, but a template or matrix from which to obtain explicit differential equations in each particular case. In fact, Bjorken and Drell in the Preface to their book 'Relativistic Quantum Mechanics' consider that the quantum "field theory formalism ... may be viewed more as a superstructure than as a foundation" [69] This is very similar to the utilization or use of templates in modern software development techniques and languages. In that case the template is a piece of code that can be copied and modified to fit a specific situation. In fact, once the operator equations are sandwiched between initial and final states, the equations for the wave functions or fields in Fock space become evident. The same applies to Hamiltonian or Lagrangian expressions. In this way explicit differential equations can be achieved by encapsulation of the operator expression between states of the system, which are providing at the same time the boundary conditions required by the solution of the differential equations.

The second quantization Hamiltonian operator appearing in Eq 148 for the case of the Schrödinger equation is given by

$$\widehat{H} = \int \left( \frac{\hbar^2}{2m} \nabla\widehat{\psi}^* \cdot \nabla\widehat{\psi} + V(r,t)\widehat{\psi}^*\widehat{\psi} \right) d^3r$$

It has the same shape as the classical field Hamiltonian, but with the classical fields replaced by 'field operators' as shown, for example in the book by Greiner[70] or Abrikosov[71].

**The field Hamiltonian is not a local expression, but involves the integral of the different fields composing the system over the entire space. The enforcement of energy-momentum conservation by this nonlocal Hamiltonian over the whole space, in cooperation with collective and relative coordinates are responsible for entanglement and nonlocal interactions, even between disjoint portions of the system.**

In the nonrelativistic theory, the field operators are defined as a function of the creation and destruction operators by

$$\widehat{\psi}(\vec{r},t) = \sum_i \hat{a}_i(t) u_i(\vec{r})$$

$$\widehat{\psi}^\dagger(\vec{r},t) = \sum_i \hat{a}_i^\dagger(t) u_i^*(\vec{r})$$

**Eq 149**

where $u_i$ are the single- particle eigenfunctions corresponding to the Eigen modes. By making use of these relations, it can be found that the field Hamiltonian takes the form

$$\widehat{H} = \int d^3r \widehat{\psi}^\dagger(\vec{r},t) \left( -\frac{\hbar^2}{2m} \nabla^2 + V(r) \right) \widehat{\psi}(\vec{r},t) = \sum_i \hat{a}_i^\dagger \hat{a}_i \, \varepsilon_i$$

**Eq 150**

in terms of the harmonic-oscillator-like creation and destruction operators. The quantity $\varepsilon_i$ is the energy associated with the eigenstate *i*.

For use in future sections we present here the definition of the field operator for the second quantization of the Klein Gordon equation.

$$\hat{\psi}(\vec{r},t) = \int d^3p \left(\hat{a}_{\vec{p}} u_{\vec{p}}(\vec{r},t) + \hat{b}^{\dagger}_{\vec{p}} u^*_{\vec{p}}(\vec{r},t)\right)$$

$$\hat{\psi}^{\dagger}(\vec{r},t) = \int d^3p \left(a^{\dagger}_{\vec{p}} u^*_{\vec{p}}(\vec{r},t) + \hat{b}_{\vec{p}} u_{\vec{p}}(\vec{r},t)\right)$$

The main difference is the appearance of a second term $\hat{b}_{\vec{p}}, \hat{b}^{\dagger}_{\vec{p}}$ that corresponds to the creation and destruction of antiparticles, represented by solutions of the equations with negative energy. They correspond to the second degree of freedom due to the fact that the Klein Gordon equation is a second order differential equation in time, in opposition to the Schrödinger equation which is first order in time. As we will see later, they give rise to a source term associated with positrons in the wave equation for the electrons and vice versa.

The field operator for the second quantization of the electromagnetic field in the relativistic version is given by

$$\hat{A}^{\mu}(\vec{r},t) = \int \frac{d^3k}{\sqrt{2\omega_k(2\pi)^3}} \sum_{\lambda=0}^{3} \left(\hat{a}_{\vec{k}\lambda} \epsilon^{\mu}(\vec{k},\lambda) u_{\vec{k}\lambda}(\vec{r},t) + \hat{a}^{\dagger}_{\vec{k}\lambda} \epsilon^{\mu}(\vec{k},\lambda) u^*_{\vec{k}\lambda}(\vec{r},t)\right)$$

**Eq 151**

where $\epsilon^{\mu}(\vec{k},\lambda)$ are the polarization unit vectors and $u_{\vec{k}\lambda}(\vec{r},t)$ the eigenstates solutions as defined by the boundary conditions; in plane geometry $\vec{k}$ labels the different plane waves, and $u_{\vec{k}\lambda}(\vec{r},t)$ means $u_{\vec{k}\lambda}(\vec{r},t) = e^{-i(\vec{k}\cdot\vec{r}-\omega t)}$. The interesting point here is the presence of the second term containing $\hat{a}^{\dagger}_{\vec{k}\lambda}$ . This term apperas because in the classical Maxwell's equations the fields $A^{\mu}(x)$ are real.

Based on the creation and destruction commutation relations, in the case of Bosons the following commutation relations are found between the field operators:

$$\left[\hat{\psi}(\vec{r},t), \hat{\psi}^{\dagger}(\vec{r}',t)\right] = \delta(\vec{r}-\vec{r}')$$

$$\left[\hat{\psi}(\vec{r},t), \hat{\psi}(\vec{r}',t)\right] = \left[\hat{\psi}^{\dagger}(\vec{r},t), \hat{\psi}^{\dagger}(\vec{r}',t)\right] = 0$$

In the case of the Dirac equation the following definition is given for the field operator:

$$\hat{\psi}(\vec{r},t) = \sum_{s=1}^{2} \int d^3p \left(\hat{b}(\vec{p},s)\chi(\vec{p},s) u_{\vec{p}}(\vec{r},t) + \hat{d}^{\dagger}(\vec{p},s)\chi'(\vec{p},s) u^*_{\vec{p}}(\vec{r},t)\right)$$

where $\hat{b}^{\dagger}$ is the creator of an electron, $\hat{b}$ the destructor of an electron, $\hat{d}^{\dagger}$ the creator of a positron, and $\hat{d}$ the destructor of a positron. The index s can have the values 1 (spin up) and 2 (spin down). The 4-spinor $\chi(\vec{p},s)$ corresponds to an electron with spin s, and $\chi'(\vec{p},s)$ is a 4-spinor associated with a positron of spin s. We see that it adds the spin degree of freedom to the Klein Gordon field.

The electron fields, being fermions fulfil the following anticommutation relations:

$$\left\{\hat{\psi}(\vec{r},t), \hat{\psi}^{\dagger}(\vec{r}',t)\right\} = \delta(\vec{r}-\vec{r}')$$

$$\left\{\hat{\psi}(\vec{r},t), \hat{\psi}(\vec{r}',t)\right\} = \left\{\hat{\psi}^{\dagger}(\vec{r},t), \hat{\psi}^{\dagger}(\vec{r}',t)\right\} = 0$$

Returning for simplicity to the nonrelativistic case, the wave equations for the field operator $\hat{\psi}$ can be derived from the Hamiltonian Eq 150 by the Heisenberg equation of motion involving the commutators:

$$\frac{d}{dt}\hat{\psi} = [\hat{H}, \hat{\psi}] + \frac{\partial}{\partial t}\hat{\psi}$$

**Eq 152**

The wave field appearing in the Hamiltonian is proportional to the creation operators:

$$\hat{\psi}^* \sim a^\dagger$$

Therefore the second quantization Schrödinger equation has the form

$$\frac{d|vac\rangle}{dt} = \hat{H}|vac\rangle \sim a^\dagger|vac\rangle$$

With a formal solution of the type

$$|state(t)\rangle \sim e^{a^\dagger t}|vac\rangle = a^\dagger t|vac\rangle + \frac{1}{2}{a^\dagger}^2 t^2|vac\rangle + \frac{1}{3!}{a^\dagger}^3 t^3|vac\rangle + \cdots$$

$$\sim = a(t)|1part\rangle + b(t)|2part\rangle + b(t)|3part\rangle + \cdots$$

Which shows that in this formalism the number of particles tends to grow continuously, limited only by energy conservation. The coefficients *a, b, c*, … change with time due to interactions with other terms containing a higher or lower number of particles. Each term in general will be modified until the entire system may reach an equilibrium state.

There is a double standard in the evolution equations of second quantization. On one side, the differential equations governing the evolution of the fields are given by the Heisenberg equations of motion Eq 152, by the commutators of the field operators with the Hamiltonian, and can be of any order in the fields. They can be linear in the fields or non-linear. On the other hand the Schrödinger Eq 148 governing the 'state' of the system is unitary. This is required because the eigenvalue of that equation is just the field energy of the entire system, and that has to be a constant of the system. The energy of the state in turn is proportional to the norm of the state, implying that the state cannot change its norm and therefore the evolution has to be unitary. This requirement then is a consequence of the conservation of energy. Conservation of energy is not required for each individual field of the system, but only for the system in its totality, implying the different Heisenberg equation of motion for partial fields don't need to be unitary or linear.

A more general state is given by

$$|state(t)\rangle = \sum_{n,\tau} c_{a,\tau}(t)|n_1 \cdots n_i \cdots n_a\rangle = a(t)|n_1 \cdots n_i \cdots n_a\rangle + b(t)|n_1 \cdots n_i \cdots n_b\rangle + c(t)|n_1 \cdots n_i \cdots n_c\rangle + \cdots$$

**Eq 153**

Where $n_a$ refers to the number of excitations in the state $a$ and so on. The coefficients $c_{a,\tau}(t)$ are time dependent and are interpreted as providing the probability and phase of the state $|n_1 \cdots n_i \cdots n_a\rangle$ respect to the rest of the system.

The density matrix is defined by the products of the coefficients: the diagonal elements are the norm square of them:

$$|a|^2 \quad |b|^2 \quad |c|^2$$

they represent the probability to have this configuration. The non-diagonal members are the following

$$ab^* \; a^*b \;\; ac^* \; a^*c \ldots \ldots$$

They are related to the intensity and coherence properties of the transition or emission process. It is just a mathematical construct that can reproduce interference and coherence properties. They form the so called 'density matrix' which is very used in Quantum Optics, of great utility when solving the problem of the atomic two-level system interacting with the radiation field, using the Bloch optical equations.

Up to this point we are in the presence of a very abstract formulation from which is very difficult to find a clear interpretation in terms of real particles or fields. What we can say in parallelism to the workings of operators in first quantization is that once a state is given, the operators can be used to find statistical properties of the observables quantities associated with that operator. In this case the field amplitude and field phase. So the average value of these quantities, their standard deviation, or the statistical distribution for them could be calculated. Also, as we have already seen, the time evolution of the system will be provided by the action of a Hamiltonian constructed from the above operators. But the question about how the concept of particles fits in this formalism, or what is the relationship between this theory and the first quantization formalism still remains unclear.

### 4.4.4. Fock representation

The previous questions were partially answered by Fock in 1932. He upgraded the configuration space into what is known today as "Fock space". In first place he considered the states of the system: instead of restricting the state to be described by wavefunctions with a prefixed number of particle coordinates, he allowed to specify the state as the sum of wavefunctions dependent on a different number of particle coordinates. For example, a valid state could be given by the sum of a wavefunction of *n* photons and *i* electrons plus a wavefunction of *n-2* photons and *i+1* electrons. This multiparticle wavefunctions followed the same construction rules in terms of single particle wavefunctions as in first quantization in terms of determinants and permanents. As a second achievement Fock was able to write the creation and destruction operators as 'matrix' operators acting on the state of the system in terms of single-particle wavefunctions. He was able to show that his formalism was fully equivalent to the Heisenberg-Dirac's formulation. We will see later that the creation and destruction operators are related to the existence of particle sources and sinks in analogy with classical electromagnetism.

A state that in the Heisenberg-Dirac representation is defined by an expression like

$$|\psi(t)\rangle = a(t)|n_1 \cdots n_i \cdots n_a\rangle + b(t)|n_1 \cdots n_i \cdots n_b\rangle + c(t)|n_1 \cdots n_i \cdots n_c\rangle + \cdots$$

in Fock space will be given by

$$\psi(r_i, \cdots, r_n, t) = a(t)\psi_a\left(r_1, \cdots, r_{n_a}, t\right) + b(t)\psi_b\left(r_1, \cdots, r_{n_b}, t\right) + c(t)\psi_c\left(r_1, \cdots, r_{n_c}, t\right) + \cdots$$

**Eq 154**

where each wave function $\psi_i\left(r_1, \cdots, r_{n_i}, t\right)$ is defined in the same way as in the first quantization case by the determinant or permanent:

$$\psi(r_1, \cdots, r_i, \cdots r_n) = \begin{vmatrix} \psi_1(r_1) & \cdots & \psi_1(r_i) & \cdots & \psi_1(r_n) \\ \vdots & \vdots & \vdots & \vdots & \vdots \\ \psi_i(r_1) & \cdots & \psi_i(r_i) & \cdots & \psi_i(r_n) \\ \vdots & \vdots & \vdots & \vdots & \vdots \\ \psi_n(r_1) & \cdots & \psi_n(r_i) & \cdots & \psi_n(r_n) \end{vmatrix}_{\xi}$$

where each individual wavefunction could represent an eigenstate of the energy for the particular particle Hamiltonian.

The next step was to find the expression for the creation and destruction operators appearing in the Hamiltonian producing the transition from $n$ to $n \pm 1$ particles states. The solution from Fock was the following, using the notation

$$|\psi_1, \cdots, \psi_n\rangle = \psi(r_i, \cdots, r_n, t)$$

Given a one particle state $|\phi\rangle$, the creation operator $\hat{a}^\dagger(\phi)$ is defined by its action on an arbitrary state in Fock space as follows

$$\hat{a}^\dagger(\phi)|\psi_1, \cdots, \psi_n\rangle = |\phi, \psi_1, \cdots, \psi_n\rangle$$

The destruction operator $\hat{a}(\phi)$ is then defined as the adjoint of $\hat{a}^\dagger(\phi)$

$$\begin{aligned}
\langle \chi_1, \cdots, \chi_{n-1}|\hat{a}(\phi)|\psi_1, \cdots, \psi_n\rangle &= \langle \psi_1, \cdots, \psi_n|\hat{a}^\dagger(\phi)|\chi_1, \cdots, \chi_{n-1}\rangle^* = \langle \psi_1, \cdots, \psi_n|\phi, \chi_1, \cdots, \chi_{n-1}\rangle^* \\
&= \begin{vmatrix} \langle\psi_1|\phi\rangle & \langle\psi_1|\chi_1\rangle & \cdots & \langle\psi_1|\chi_{n-1}\rangle \\ \vdots & \vdots & \cdots & \vdots \\ \langle\psi_n|\phi\rangle & \langle\psi_n|\chi_1\rangle & \cdots & \langle\psi_n|\chi_{n-1}\rangle \end{vmatrix}_\xi^* = \sum_{k=1}^{n} \xi^{k-1}\langle\psi_k|\phi\rangle \begin{vmatrix} \langle\psi_1|\chi_1\rangle & \cdots & \langle\psi_1|\chi_{n-1}\rangle \\ \vdots & (no\ \psi_k) & \vdots \\ \langle\psi_n|\chi_1\rangle & \cdots & \langle\psi_n|\chi_{n-1}\rangle \end{vmatrix}_\xi^* \\
&= \sum_{k=1}^{n} \xi^{k-1}\langle\psi_k|\phi\rangle\langle\chi_1, \cdots, \chi_{n-1}|\psi_1, \cdots \tilde{\psi}_k \cdots, \psi_n\rangle
\end{aligned}$$

where $\tilde{\psi}_k$ means that $\psi_k$ is absent. From here the expression for the destruction operator is evidently:

$$\hat{a}(\phi)|\psi_1, \cdots, \psi_n\rangle = \sum_{k=1}^{n} \xi^{k-1}\langle\phi|\psi_k\rangle|\psi_1, \cdots \hat{\psi}_k \cdots, \psi_n\rangle$$

and found expressions for these operators in terms of determinants and permanents. He also showed that these operators fulfill the following commutation relations

$$[\hat{a}_\alpha, \hat{a}_\beta] = [\hat{a}_\alpha^\dagger, \hat{a}_\beta^\dagger] = 0 \qquad [\hat{a}_\alpha, \hat{a}_\beta^\dagger] = \delta_{\alpha\beta},$$

for bosons, and the anticommutation relations

$$\{\hat{a}_\alpha, \hat{a}_\beta\} = \{\hat{a}_\alpha^\dagger, \hat{a}_\beta^\dagger\} = 0 \qquad \{\hat{a}_\alpha, \hat{a}_\beta^\dagger\} = \delta_{\alpha\beta},$$

**Eq 155**

for fermions, which are the same relations fulfilled by the creation and destruction operators in Heisenberg's theory, showing that both models where mathematically equivalent and could be applied to the same physical problem. Also it can be shown that these commutation rules are what is needed for boson wavefunctions to be symmetric against permutation of particle indexes, and fermion wavefunctions to be antisymmetric against permutation of particle indexes. In this way he was able to find an interpretation for the existence of anticommutation relations. All this was done by Fock without any reference to the harmonic oscillator!

In the Dirac representation a system with a fixed number of particles was defined as an eigenstate of the Hamiltonian, now in the Fock representation it can be given a much more intuitive interpretation: as long as the number of coordinates $r_i$ remains constant and equal to '*n*', we are in the presence of an eigenstate of '*n*', or a state with constant number of particles. This representation allows for the description of systems were particles can be created or absorbed, as in electron-positron pair creation, or with photons. In this model the state becomes independent of the Eigenmodes of the field as was initially in the Dirac theory. It can be seen that a state with a particle can be defined in any wavefunction associated with it, not just in energy Eigenmodes. This fact can be accommodated in the Dirac representation by defining superposition states associated with the creation and destruction operators. One of the most important creation operators are for creation of a particle in the momentum state or in the position state. They are related by

$$\hat{a}^\dagger(\vec{p}) = \int d^d r \hat{a}^\dagger(\vec{x}) e^{i\vec{p}\cdot\vec{x}}$$

$$\hat{a}^\dagger(\vec{x}) = \int \frac{d^d p}{(2\pi)^d} \hat{a}^\dagger(\vec{p}) e^{-i\vec{p}\cdot\vec{x}}$$

We have seen that in order to obtain entanglement, it is clear that a single equation involving the whole of all the particles in the system is required. Second quantization Schrödinger Eq 148 for the state of the whole system fulfills this requirement. Deriving wave equations from the field Hamiltonian provides equations for the individual fields.

## 4.5. Dirac-Fock relationship

### 4.5.1. States in Fock space

In order to find the formal relationship between the Dirac and Fock representations, let's start considering a state consisting of $n_a$ particles in state *a* . This state is given by

$$|n_a\rangle = \frac{\hat{a}^{\dagger\, n_a}}{\sqrt{n_a!}} |vac\rangle$$

In order to regain the field as function of space and time one can use powers of the field operator

$$\hat{\psi}(x,t) = \int d^3p \left( \hat{a}_p(t) u_p(x,t) + \hat{a}_p^\dagger(t) u_p^*(x,t) \right) = \sum_i \left( \hat{a}_i(t) u_i(x,t) + \hat{a}_i^\dagger(t) u_i^*(x,t) \right)$$

encapsulated between the $|n_a\rangle$ state and the vacuum state. This is equivalent to the Fock projection operator $P_{Fock}$

$$P_{Fock} = \langle vac| \sum_{n=1}^{n_{max}} \frac{1}{\sqrt{n!}} \prod_{i=1}^{n} \hat{\psi}(r_i)$$

For another representation of scalar QED in Configuration space see[72]. Let's see some examples:

*One particle in state a*

$$|1_a\rangle = \hat{a}^\dagger |vac\rangle$$

$$\left[ \langle vac| \sum_{n=1}^{n_{max}} \prod_{i=1}^{n} \hat{\psi}_a(r_i) \right] |1_a\rangle \rightarrow \langle vac|\hat{\psi}_a(r_1)|1_a\rangle = \langle vac|\hat{a}\psi_a(r_1)\hat{a}^\dagger|vac\rangle = \langle vac|\psi_a(r_1)\left(1 + \hat{a}^\dagger\hat{a}\right)|vac\rangle$$

$$= \langle vac|\psi_a(r_1)|vac\rangle = \psi_a(r_1)$$

*Two bosons in state a*

$$|2_a\rangle = \frac{\hat{a}^{\dagger\,2}}{\sqrt{2!}}|vac\rangle$$

$$\left[\langle vac| \sum_{n=1}^{n_{max}} \frac{1}{\sqrt{n!}} \prod_{i=1}^{n} \hat{\psi}_a(r_i)\right] |2_a\rangle \to \frac{1}{\sqrt{2}}\langle vac|\hat{\psi}_a(r_2)\hat{\psi}_a(r_1)|2_a\rangle = \frac{1}{\sqrt{2}}\langle vac|\hat{a}^2\psi_a(r_2)\psi_a(r_1)|2_a\rangle$$

$$= \frac{1}{2}\langle vac|\psi_a(r_2)\psi_a(r_1)\hat{a}(1+\hat{a}^\dagger\hat{a})\hat{a}^\dagger|vac\rangle = \frac{1}{2}\langle vac|\psi_a(r_2)\psi_a(r_1)(\hat{a}\hat{a}^\dagger + \hat{a}\hat{a}^\dagger\hat{a}\hat{a}^\dagger)|vac\rangle$$

$$= \frac{1}{2}\left\langle vac\middle|\psi_a(r_2)\psi_a(r_1)\left((1+\hat{a}^\dagger\hat{a}) + (1+\hat{a}^\dagger\hat{a})(1+\hat{a}^\dagger\hat{a})\right)\middle|vac\right\rangle = \frac{2}{2}\langle vac|\psi_a(r_2)\psi_a(r_1)|vac\rangle$$

$$= \psi_a(r_2)\psi_a(r_1)$$

*One particle in state a and one particle in state b*

$$|1_a, 1_b\rangle = \hat{a}^\dagger\hat{b}^\dagger|vac\rangle$$

where the field operators are given by

$$\left[\langle vac| \sum_{n=1}^{n_{max}} \frac{1}{\sqrt{n!}} \prod_{i=1}^{n} \hat{\psi}_a(r_i)\right] |1_a, 1_b\rangle \to \langle vac|\hat{\psi}(r_2)\hat{\psi}(r_1)|1_a, 1_b\rangle = \left\langle vac\middle|\hat{\psi}_a(r_2)\frac{\left(\hat{a}\psi_a(r_1) + \hat{b}\psi_b(r_1)\right)}{\sqrt{2}}\middle|1_a, 1_b\right\rangle$$

$$= \frac{1}{\sqrt{2}}\langle vac|\left(\hat{a}\psi_a(r_2) + \hat{b}\psi_b(r_2)\right)(\psi_a(r_1)|1_b\rangle + \psi_b(r_1)|1_a\rangle)$$

$$= \frac{1}{\sqrt{2}}\langle vac|(\psi_a(r_2)\psi_b(r_1)|vac\rangle + \psi_b(r_2)\psi_a(r_1)|vac\rangle) = \frac{1}{\sqrt{2}}\left(\psi_a(r_2)\psi_b(r_1) + \psi_b(r_2)\psi_a(r_1)\right)$$

*Two photons in state a and one photon in state b is*

$$|2a, 1b\rangle = \frac{\hat{a}_a^{\dagger\,2}}{\sqrt{2}}\hat{a}_b^\dagger|vac\rangle$$

$$\left[\langle vac| \sum_{n=1}^{n_{max}} \frac{1}{\sqrt{n!}} \prod_{i=1}^{n} \hat{\psi}_a(r_i)\right] |2_a, 1_b\rangle$$

$$\to \sqrt{2}\left(u_a(r_1,t)u_a(r_2,t)u_b(r_3,t) + u_a(r_1,t)u_a(r_3,t)u_b(r_2,t) + u_a(r_2,t)u_a(r_3,t)u_b(r_1,t)\right)$$

*One electron in a and another in b:*

$$|1a, 1b\rangle = \hat{a}_a^\dagger\hat{a}_b^\dagger|vac\rangle$$

$$\left[\langle vac| \sum_{n=1}^{n_{max}} \frac{1}{\sqrt{n!}} \prod_{i=1}^{n} \hat{\psi}_a(r_i)\right] |1_a, 1_b\rangle \to \frac{1}{\sqrt{2}}\left(u_a(r_1,t)u_b(r_2,t) - u_b(r_1,t)u_a(r_2,t)\right)$$

The same technique can be used when having different particles.

*One electron in a and one photon in b:*

$$|1ea, 1phb\rangle = \hat{a}_{e,a}^\dagger\hat{a}_{ph,b}^\dagger|vac\rangle$$

$$\left[\langle vac| \sum_{n,m=1}^{n_{max}} \frac{1}{\sqrt{n!}} \frac{1}{\sqrt{m!}} \prod_{i,j=1}^{n} \hat{\psi}_a(r_i) \hat{V}(r_j, t)\right] |1ea, 1phb\rangle \underbrace{\rightarrow}_{configuration\ space} u_a(r_e, t) V_b(r_{ph}, t)$$

where we call $V_b(r_{ph}, t)$ the photon field or wavefunction.

### 4.5.2. Hamiltonian in Fock space

The Fock projection operator $P_{Fock}$ can be used not only to find the representation of the Dirac state into the Fock wavefunctions, but also one can use this technique to find the expression for the Second Quantization Schrodinger equation. It is also possible to find the Fock representation of the Lagrangian and Hamiltonian starting from its representation in terms of field operators. Here we present some examples for the Hamiltonian expression in Fock space.

#### *4.5.2.1. Non-relativistic electron and radiation field*

The classical theory of Lagrangian, energy density, wave equations for single particle quantum mechanics and the interaction with the electromagnetic field can be found in the book by Barut[73], for a Schrödinger electron it is:

$$H = \int \left[\frac{1}{2}\left(\left(-\frac{1}{c}\dot{\vec{A}} - \vec{\nabla}V\right)^2 + \left(\vec{\nabla} \times \vec{A}\right)^2\right) - e\psi^*\psi V + \frac{ie\hbar}{2m}\left(\psi^*\vec{\nabla}\psi - \psi\vec{\nabla}\psi^*\right) \cdot \vec{A}\right] d^3\vec{r} + \int \left(\frac{\hbar^2}{2m}\vec{\nabla}\psi^* \cdot \vec{\nabla}\psi\right) d^3\vec{r}$$

**Eq 156**

The case for many electrons and photons will be

$$H = \sum_j \int \left[\frac{1}{2}\left(\left(-\frac{1}{c}\dot{\vec{A}}(\vec{r}_{ph,j}) - \vec{\nabla}V(\vec{r}_{ph,j})\right)^2 + \left(\vec{\nabla} \times \vec{A}(\vec{r}_{ph,j})\right)^2\right)\right] d^3\vec{r}_{ph,j}$$
$$+ \sum_{i,j} \int_{r_{e,i}} \int_{r_{ph,j}} \left[-e\psi^*(\vec{r}_{e,i})\psi(\vec{r}_{e,i})V(\vec{r}_{ph,j}) + \frac{ie\hbar}{2m}\left(\psi^*(\vec{r}_{e,i})\vec{\nabla}\psi(\vec{r}_{e,i}) - \psi(\vec{r}_{e,i})\vec{\nabla}\psi^*(\vec{r}_{e,i})\right) \cdot \vec{A}(\vec{r}_{ph,j})\right] \delta(\vec{r}_{ph,j}$$
$$- \vec{r}_{e,i}) d^3\vec{r}_{ph,j}\, d^3\vec{r}_{e,i} + \sum_i \int \left(\frac{\hbar^2}{2m}\vec{\nabla}\psi^*(\vec{r}_{e,i}) \cdot \vec{\nabla}\psi(\vec{r}_{e,i})\right) d^3\vec{r}_{e,i}$$

**Eq 157**

#### *4.5.2.2. Relative coordinates*

Two electrons

If we write the term of the Lagrangian corresponding to the mutual interaction between two electrons in the Coulomb approximation we get,

$$\iint \psi_1^*\psi_1 V(\vec{r}_2 - \vec{r}_1)\psi_2^*\psi_2 d^3\vec{r}_1 d^3\vec{r}_2$$

Where the interaction between the 2 electrons is

$$V(\vec{r}_2 - \vec{r}_1)$$

This interaction is dependent only on the relative coordinate. And we have the same case as for the hydrogen atom treated previously.

### One electron and one photon

The Lagrangian for an electron and a field can be written as

$$\iint \vec{A}(\vec{r}_{ph}) \cdot \psi^* \vec{\nabla}\psi \quad \delta(\vec{r}_{ph} - \vec{r}_e) d^3\vec{r}_e d^3\vec{r}_{ph} + \iint V(\vec{r}_{ph}) \delta(\vec{r}_{ph} - \vec{r}_e) \psi^* \ \psi \quad d^3\vec{r}_e d^3\vec{r}_{ph}$$

where we see that the interaction between the electron and the field is given by $\delta(\vec{r}_{ph} - \vec{r}_e)$, also dependent only on the relative coordinate. However is not a simple task to write the kinetic energy of the system in terms of the relative and center of mass coordinates between the electron and the photon. The interaction in terms of the relative coordinate reads:

$$\iint V(\vec{r}_{rel}) \psi_{rel}^* \psi_{rel} \psi_{CM}^* \psi_{CM} d^3\vec{r}_{rel} d^3\vec{r}_{CM} = \int \psi_{CM}^* \psi_{CM} d^3\vec{r}_{CM} \int V(\vec{r}_{rel}) \psi_{rel}^* \psi_{rel} d^3\vec{r}_{rel} = \int V(\vec{r}_{rel}) \psi_{rel}^* \psi_{rel} d^3\vec{r}_{rel}$$

The function for the photon and electron can be expressed through the Schmidt decomposition as described in the paper by Eberly et al[74] .

$$\iint V(\vec{r}_{ph}) \delta(\vec{r}_{ph} - \vec{r}_e) \psi^* \ \psi \quad d^3\vec{r}_e d^3\vec{r}_{ph} = \iint \sum_{i=1}^{n} V_i(\vec{r}_{ph}) \delta(\vec{r}_{ph} - \vec{r}_e) \psi_i^* \psi_i d^3\vec{r}_e d^3\vec{r}_{ph}$$

That is an entangled state.

### Two photons

A case of two-photon entanglement generation could be:

$$H = \int_{r_e} \int_{r_{ph,1}} \int_{r_{ph,2}} \left[ \frac{ie\hbar}{2m} \left( \psi^*(\vec{r}_e) \vec{\nabla} \psi(\vec{r}_e) - \psi(\vec{r}_e) \vec{\nabla} \psi^*(\vec{r}_e) \right) \cdot \left( \vec{A}_a(\vec{r}_{ph,1}) \vec{A}_b(\vec{r}_{ph,2}) - \vec{A}_b(\vec{r}_{ph,1}) \vec{A}_a(\vec{r}_{ph,2}) \right) \right] \delta(\vec{r}_{ph,1} - \vec{r}_e) \delta(\vec{r}_{ph,2} - \vec{r}_e) d^3\vec{r}_{ph,1} d^3\vec{r}_{ph,2} \, d^3\vec{r}_e$$

Or better yet:

$$H = \int_{r_e} \int_{r_{ph,1}} \int_{r_{ph,2}} \left[ \frac{ie\hbar}{2m} \left( \psi^*(\vec{r}_e) \vec{\nabla} \psi(\vec{r}_e) - \psi(\vec{r}_e) \vec{\nabla} \psi^*(\vec{r}_e) \right) \cdot \vec{A}_{ext}(\vec{r}_{ph,1} + \vec{r}_{ph,2}) \right] \delta\left( \frac{\vec{r}_{ph,1} + \vec{r}_{ph,2}}{2} - \vec{r}_e \right) \delta(\vec{r}_{ph,1} - \vec{r}_{ph,2}) d^3\vec{r}_{ph,1} d^3\vec{r}_{ph,2} \, d^3\vec{r}_e$$

Where the photon coordinates selected are the center of mass and relative position, which is similar also to the electron-positron entangled pair generation. See section on electron-positron in next chapter.

When one is not working with individual particle coordinates, but with any other set of coordinates, one has to make the transformation from the individual coordinates to the new collective coordinates like relative and center of mass, and rewrite the Lagrangian and Hamiltonian in terms of this new coordinates like

$$\vec{\nabla}_1, \vec{\nabla}_2 \quad \rightarrow \quad \vec{\nabla}_{CM}, \vec{\nabla}_{rel}$$

As done in classical physics (See Goldstein, Ruth, etc.)

# 4.6. Field Quantization Significance

## 4.6.1. Schrödinger equation in Fock space

The purpose of Second Quantization is to find solutions to Eq 148 that we rewrite here

$$\frac{d|state\rangle}{dt} = \widehat{H}_{SQ}|state\rangle$$

There are two typical cases, the first one is to find stationary, steady solutions, also called energy eigenstates. The other one is given an initial state that is not an eigenstate of the full Hamiltonian, find the state at a later time. A special situation of this last case is when the Hamiltonian operator can be decomposed in two parts, the first of them has known steady solutions, and the second part is small compared with the first one, and is known as a 'perturbation' that can be time dependent or time independent. In that case, if the initial state is one of the energy eigenstates, the problem can be solved by finding the components of the final state in terms of all energy eigenstates. The amplitude of each one of these components is related to the so called 'transition probability' between the stationary states due to the perturbation. The most typical problems in QED (scattering) belong to this last case. Most of the scattering QED problems are readily time dependent, but they are handled as stationary problems for simplicity.

There are many methods to find these solutions. Here we are going to mention only three of them.

The first one consists in working with the general commutation properties of the field operators. This technique has its roots very early in quantum mechanics, in fact Schrödinger and Pauli found solutions to the harmonic oscillator[75] and hydrogen atom[76] problems respectively without solving the Schrodinger differential equation, but by working exclusively with the operator properties as early as 1926. The harmonic oscillator problem is very used in text books on quantum optics to solve the generalized harmonic oscillator Hamiltonian adapted to the radiation field. This is done in order to provide a solution without the need to use differential equations, which in the case of field operators have a dubious meaning. The drawback of using this technique is the higher degree of abstraction, making it very difficult to understand really what it is being calculated.

The second method is to use the projector we have defined previously and enclose the operator Schrödinger equation between the actual state and vacuum, and obtain a differential equation in time acting on wavefunctions in Fock space.

A third method was found by Feynman, who showed an alternative way to find the contribution to the final state at a given point in space and time, due to an initial state at other point in space and time. That contribution is called the propagator, and is equivalent to solving the Green function for the differential equation. It can be understood as a consequence of the classical Huygens principle for the propagation of waves. Feynman showed that the final state contribution, could be found as an integral over all possible paths of the classical action for the motion of a classical particle between both points. This formulation has two very deep consequences: first he showed that one doesn't need at all the field operator formalism of canonical quantization, but any calculation could be performed making use of only classical concepts associated with differential and integral operators. From there one can derive the second consequence that is what gives rise to the particular names given by Feynman to his papers: "Space-time approach to non-relativistic quantum mechanics" and "Space-time approach to quantum electrodynamics": and states that second quantization can be fully formulated working with objective functions in real space-time. This last consequence for us seems to be one of the most important contributions from Feynman's work, however it seems to have been overlooked in present-day text books.

The same technique used to reduce the abstract 'quantum state' $|n\rangle$ into a multi-coordinates space time function as in 4.5.1 'States in Fock space' can be used to obtain expressions for the Hamiltonian and for the wave equation in terms of individual particle wavefunctions. If the wave function is given in Fock space by

$$\psi(r_i,\cdots,r_n,t) = a(t)\psi_a\left(r_1,\cdots,r_{n_a},t\right) + b(t)\psi_b\left(r_1,\cdots,r_{n_b},t\right) + c(t)\psi_c\left(r_1,\cdots,r_{n_c},t\right) + \cdots$$

then the generalized Schrödinger equation will provide the equation of motion for the coefficients $a(t), b(t), c(t)\cdots,$. Solving for the $a$ coefficients is equivalent to solving for the density matrix. Many approximations have been developed to simplify the problem and find solutions to the laser equations for example. By using the Fock projection operator on the Second Quantization Schrodinger equation we obtain

$$\left[\langle vac|\sum_{n=1}^{n_{max}}\frac{1}{\sqrt{n!}}\prod_{i=1}^{n}\hat{\psi}_a(r_i)\right]\frac{d}{dt}\widehat{\Psi}^{n_e}(r,t)\widehat{\mathrm{A}}^{n_{ph}}(r,t)|s\rangle =$$

$$\left[\langle vac|\sum_{n=1}^{n_{max}}\frac{1}{\sqrt{n!}}\prod_{i=1}^{n}\hat{\psi}_a(r_i)\right]\widehat{H}_{sq}\widehat{\Psi}^{n_e}(r,t)\widehat{\mathrm{A}}^{n_{ph}}(r,t)|s\rangle$$

Which should give

$$\frac{d}{dt}\sum_i \psi^{n_{ai}}(r,t)\cdots A^{n_{bi}}(r,t)\cdots = \sum_i \widehat{H}_{sq,i}\psi^{n_{ai}}(r,t)\cdots A^{n_{bi}}(r,t)\cdots$$

The equations for the fields present source terms in the relativistic case as we will see later, as in the case of the electromagnetic fields. Source terms were absent from the multi particle equations of first quantization.

***In summary, the difference between the first quantization multi-particle description and canonical quantization is that in the first case the system consists only of electron wavefunctions with a fix number of particle coordinates, while the electromagnetic fields are given as external fields. For second quantization, the system is composed of electrons, positrons and photons described by relativistic equations whose solutions consists of functions with variable number of particle coordinates, and no external fields are required. The greatest significance of the field operators in Field Quantization can be found in the fact that they allow for a highly condensed and efficient way to provide a set of differential equations relating a large number of fields and coordinates.***

This property can be appreciated in quantum optics, where a complete set of equations for the different possible states with a variable number of photons and different atomic populations can be derived from Eq 148. On the other hand the traditional problems considered in QED involve a single electron and photon. In most of the QED cases the equation derived from the operator formalism is not much different from the corresponding equation in first quantization as long as self-field interactions are not considered. In many cases that similitude creates the temptation to believe that the field operators are the real objects of the theory and not just a formal template from which to obtain the actual working equations and wavefunctions.

### 4.6.2. Heisenberg's equation in Fock space

In a similar way as done with the Schrodinger equation above, the field operators' based Heisenberg equations of motion can be translated into an equation in Fock space for the actual functions $\psi_a\left(r_1,\cdots,r_{n_a},t\right)$, and also for the single particle functions $\psi^{n_{ai}}(r,t)\cdots A^{n_{bi}}(r,t)$ that can be its constituents.

The corresponding Heisenberg equation of motion would be

$$\left[\langle vac| \sum_{n=1}^{n_{max}} \frac{1}{\sqrt{n!}} \prod_{i=1}^{n} \hat{\psi}_a(r_i)\right] \frac{d\hat{\psi}(r,t)}{dt} |\cdots n_i \cdots, t\rangle = \left[\langle vac| \sum_{n=1}^{n_{max}} \frac{1}{\sqrt{n!}} \prod_{i=1}^{n} \hat{\psi}_a(r_i)\right] [\hat{\psi}(r,t), \hat{H}_{sq}] |\cdots n_i \cdots, t\rangle$$

## 4.7. Self-Interactions, and Nonlinear Optics

Before continuing we need to give again a precise meaning of what is understood under Canonical Quantization. We repeat the definition given in Wikipedia that reads:

"When the canonical quantization procedure is applied to a field, such as the electromagnetic field, the classical field variables become *quantum operators*.

The classical equations of motion of a field are typically identical in form to the (quantum) equations for the wave-function of *one of its quanta*.

Quantum mechanically, the variables of a field (such as the field's amplitude at a given point) are represented by operators on a Hilbert space. In general, all observables are constructed as operators on the Hilbert space, and the time-evolution of the operators is governed by the Hamiltonian, which must be a positive operator.

The classical Hamiltonian is found to be

$$H(\phi, \pi) = \int dx \left[\frac{1}{2}\pi^2 + \frac{1}{2}(\partial_x \phi)^2 + \frac{1}{2}m^2\phi^2 + V(\phi)\right].$$

Canonical quantization treats the variables $\phi(x)$ and $\pi(x)$ as operators with canonical commutation relations at time $t = 0$, given by

$$[\phi(x), \phi(y)] = 0, \quad [\pi(x), \pi(y)] = 0, \quad [\phi(x), \pi(y)] = i\hbar\delta(x - y).$$ "

In summary to quantize a system, take the classical field Hamiltonian and interpret the fields not as space-time functions, but as operators in Hilbert space fulfilling the prescribed commutation relations of the particular case in hand. Then obtain the equation of motion for the field operators by the commutator between the field operator and the Hamiltonian. This equations of motion are called Heisenberg equations and for a single particle are almost always identical in form with the classical wave equations (Schrödinger, Klein Gordon, Dirac, Maxwell). We repeat the definition of this procedure here, because in many textbooks on nonlinear effects on second quantization, this procedure is simplified and many of the terms that should appear in the equations are neglected.

### Self-interactions

We have seen that the introduction of self-interactions induce the introduction of nonlinear terms in the wave equations. This is not the only place where nonlinearities appear in quantum mechanics. The field equation for the electromagnetic fields can be non-linear, as is the case for the Kerr effect. Also the solution to the Schrodinger equation of canonical quantization is a series in multiple products of the different fields composing the quantum system. The Schrodinger equation is

$$\frac{d|\psi(t)\rangle}{dt} = \hat{H}|\psi(t)\rangle$$

With a formal solution in the Schrödinger picture of the type

$$|\psi(t)\rangle \sim e^{\hat{H}\times(t-t_0)}|\psi(t_0)\rangle = \hat{H}(t-t_0)|\psi(t_0)\rangle + \frac{1}{2}\hat{H}^2(t-t_0)^2|\psi(t_0)\rangle + \frac{1}{3!}\hat{H}^3(t-t_0)^3|\psi(t_0)\rangle + \cdots$$

**Eq 158**

if the initial time is taken to be $t_0$. Here the state $|\psi(t)\rangle$ is an abstract state defined in the occupation space or Hilbert space of the states of the system. Eq 158 represents the Taylor expansion of the solution as a function of time. This expansion presents all possible products of creation and destruction operators, each one multiplied by a field function. From the classical Hamiltonian for the interaction of an electron and the radiation field is given by Eq 156, the Hamiltonian operator for Canonical Quantization in the Schrodinger picture is given by

$$\hat{H} = \int \left[\frac{1}{2}\left(\left(-\frac{1}{c}\hat{\vec{\Pi}} - \vec{\nabla}\hat{V}\right)^2 + \left(\vec{\nabla}\times\hat{A}\right)^2\right) - e\hat{\psi}^*\hat{\psi}\hat{V} + \frac{ie\hbar}{2m}\left(\hat{\psi}^\dagger\vec{\nabla}\hat{\psi} - \hat{\psi}\vec{\nabla}\hat{\psi}^\dagger\right)\cdot\hat{\vec{A}}\right]d^3\vec{r} + \int\left(\frac{\hbar^2}{2m}\vec{\nabla}\hat{\psi}^\dagger\cdot\vec{\nabla}\hat{\psi}\right)d^3\vec{r}$$

where we remember that for the Schrödinger case

$$\hat{\psi}(\vec{r},t) = \sum_i \hat{b}_i u_i(\vec{r},t)$$

and

$$\hat{A}^\mu(\vec{r},t) = \int \frac{d^3k}{\sqrt{2\omega_k(2\pi)^3}}\sum_{\lambda=0}^{3}\left(\hat{a}_{\vec{k}\lambda}\epsilon^\mu(\vec{k},\lambda)u_{\vec{k}\lambda}(\vec{r},t) + \hat{a}^\dagger_{\vec{k}\lambda}\epsilon^\mu(\vec{k},\lambda)u^*_{\vec{k}\lambda}(\vec{r},t)\right)$$

in a finite volume the integral goes over a finite sum over the transverse fields:

$$\hat{\vec{A}}_\perp(\vec{r},t) = \sum_k\sqrt{\frac{\hbar}{2\omega_k\varepsilon_0 V}}\left(\hat{a}_k\vec{A}_k(\vec{r},t) + \hat{a}_k^\dagger\vec{A}_k^*(\vec{r},t)\right)$$

$$\hat{\vec{E}}_\perp(\vec{r},t) = \frac{d\hat{\vec{A}}(\vec{r},t)}{dt} = i\sum_k\sqrt{\frac{\hbar\omega_k}{2\varepsilon_0 V}}\left(\hat{a}_k\epsilon_k\vec{A}_k(\vec{r},t) - \hat{a}_k^\dagger\epsilon_k\vec{A}_k^*(\vec{r},t)\right)$$

$$\hat{\vec{\Pi}}(\vec{r},t) = c\hat{\vec{E}}_\perp(\vec{r},t)$$

where $\vec{A}(\vec{r},t)$ and $\vec{\Pi}(\vec{r},t)$ are classical functions defined in real space-time, describing the state of the field, typically eigenstates defined by the boundary conditions. Similarly for the Schrödinger field operator. In the Schrödinger picture where the creation and destruction operators $\hat{a}$ and $\hat{a}^\dagger$ are time independent, they fulfil the commutation relations

$$\left[\hat{a}_\alpha,\hat{a}_\beta\right] = \left[\hat{a}_\alpha^\dagger,\hat{a}_\beta^\dagger\right] = 0 \qquad \left[\hat{a}_\alpha,\hat{a}_\beta^\dagger\right] = \delta_{\alpha\beta},$$

We don't need to consider the Hamiltonian terms for free fields, because they just define the eigenstates and eigenenergies of the system in the absence of interactions. The most interesting part, providing for the evolution of the system and the exchange of particles is given by the interaction term, which in the absence of static fields becomes

$$\hat{H}_i = \frac{ie\hbar}{2m}\int\left[\hat{\psi}^\dagger(\vec{r},t)\left(\vec{\nabla}\hat{\psi}(\vec{r},t)\right) - \left(\vec{\nabla}\hat{\psi}^\dagger(\vec{r},t)\right)\hat{\psi}(\vec{r},t)\right]\cdot\hat{\vec{A}}(\vec{r},t)d^3\vec{r} = \int\hat{\vec{j}}(\vec{r},t)\cdot\hat{\vec{A}}(\vec{r},t)d^3\vec{r}$$

Where all what we have done is to replace the classical fields by its corresponding operators and have defined the current density operator

$$\hat{j}(\vec{r},t) = -\frac{ie\hbar}{2m}\left[\hat{\psi}^\dagger\left(\vec{\nabla}\hat{\psi}\right) - \left(\vec{\nabla}\hat{\psi}^\dagger\right)\hat{\psi}\right]\sum_{i,j}\hat{b}_j^\dagger\hat{b}_i\vec{j}_{ji}(\vec{r},t)$$

In Appendix 4.1. Nonlinear Hamiltonian in second quantization we show that the square of the Hamiltonian operator can be written as

$$\widehat{H}_i^{\,2} = -\left(\frac{e\hbar}{2m}\right)^2 \frac{\hbar}{2\varepsilon_0 V} \sum_{i,j,i',j'} \hat{b}_{j'}^{\dagger}\hat{b}_{i'}\hat{b}_i^{\dagger}\hat{b}_j \left[\sum_k \frac{1}{\omega_k} \iint \left(\vec{j}_{j'i'}(\vec{r},t)\cdot\vec{A}_k(\vec{r},t)\vec{j}_{ij}(\vec{r}',t)\cdot\vec{A}_k^*(\vec{r}',t)\right) d^3\vec{r}d^3\vec{r}'\right]$$

where the classical current density is

$$\vec{j}_{ji}(\vec{r},t) = u_j^*(\vec{r},t)\vec{\nabla}u_i(\vec{r},t) - \left(\vec{\nabla}u_j^*(\vec{r},t)\right)u_i(\vec{r},t)$$

When the nonlinear Hamiltonian is sandwiched between 1 electron state $\left\langle 1el\middle|\widehat{H}_i^{\,2}\middle|1el\right\rangle$ it takes the form $\left\langle 1el\middle|\hat{b}_j^{\dagger}\hat{b}_i\hat{b}_i^{\dagger}\hat{b}_j\middle|1el\right\rangle$. This Hamiltonian has stationary terms in a state with no photons and a single electron only in two cases:

- terms of the form $\hat{b}_j^{\dagger}\hat{b}_j\hat{b}_i^{\dagger}\hat{b}_i$, non-exchange case, where gives the classical Hamiltonian term corresponding to the magnetostatic Biot-Savart self-interaction[77]:

  $$\left\langle a\middle|\widehat{H}_i^{\,2}\middle|a\right\rangle = \frac{1}{8\pi}\int d^3\vec{r}\int d^3\vec{r}' \sum_j \frac{j_{a\mu}(\vec{r},t)j_j^{\mu}(\vec{r}',t)}{|\vec{r}-\vec{r}'|}$$

  A similar consideration applies to the term with the scalar potential which provides the equation including the electrostatic self-interaction. In particular the term where $\hat{b}_i^{\dagger}\hat{b}_i\hat{b}_i^{\dagger}\hat{b}_i = 1$ is thrown away in traditional quantum mechanics, because is diverging and 'uninteresting'. **However it is clear that it corresponds to a self-interaction term, the construction and destruction operators can be seen as a purely technical manipulation in order to derive the equations of motion for the fields.**

- terms of the form $\hat{b}_j^{\dagger}\hat{b}_i\hat{b}_i^{\dagger}\hat{b}_j$with $i \neq j$, exchange case, the intermediate state is not the same as the initial one, this means the electron density suffers a sort of trembling that is responsible, as in the Lamb shift effect for the achievement of ergodicity in the probability distribution for the position of the electron in quantum mechanics. Because the action of the operators implies going through the vacuum state, this type of interaction energy is said to be due to vacuum fluctuations.

Once we have sandwiched the Hamiltonian between states, we are left with an analytical expression from which we can obtain the differential equation. It should be equivalent to find the Heisenberg's equation of motion and then sandwich it to obtain the analytical form.

It is well known that $\widehat{H}_i^{\,2}$ corresponds in first approximation to the Coulomb electrostatic and Biot-Savart magnetostatic self-energies of the electron[77]. Let's analyzed this fact in more detail. The traditional power series expansion used in most textbooks on QED for the electron self-mass calculations, can provide valid results only under the condition that the 'correction' to the set of states used as a basis for the expansion is very small, or equivalently that the full solution is approximately known from beforehand. This technique works well in the hydrogen atom case because the solution to the problem are very closely known as provided by the original Schrödinger equation, and effectively the Lamb shift is a very small perturbation. I ask here: what would be the result of the calculation if plane waves were used instead of the well-known atomic wavefunctions as basis for the perturbation expansion? Exactly the same situation appears in this case, however plane waves are typically used as solutions for the electron wavefunctions. Let's remember again how canonical quantization is applied to the hydrogen atom: first the allowed quantized states are found by solving the first quantization Schrödinger equation for the electron in the presence of the nuclear field, and later the second quantization operator formalism is applied to these states. There is nothing wrong with that, it is the right way to proceed. We proposed to apply the same procedure here. First solve the self-interaction problem for the Dirac and Maxwell equations in first quantization, and later apply the second quantization formalism to the solutions found in this way. This is what we have done in the Chapter 2 of this series, where we have solved this problem at all orders, not in a first order approximation, and found the electron soliton solutions. The same applies exactly to the problem of the photon self-interaction provided by the Uehling and Euler-Heisenberg vacuum polarization terms. In this last case we have found the photon soliton solutions that can solve the wave-particle duality for the electromagnetic fields.

In a similar way it can be shown that any term having the same number of creation and destruction operators implies a (nonlinear) self-interaction of the system.

So we see that self-interactions are indeed inside the formalism of Canonical Quantization. This is the term that appears in some of the problems on nonlinear contributions to the index of refraction in quantum optics. It is this term that we consider in the case of the soliton solutions for the electron and photon.

## Kerr effect

A similar analysis can be done for Maxwell's equation in the presence of a nonlinear polarization $\vec{P}^{NL}$. A typical coherent nonlinear effect in crystals is produced by the Kerr effect. This case is provided by the effective 3rd order susceptibility $\tilde{\chi}^{(3)}_{ijkl}$ where the polarization is given by:

$$\vec{P}^{NL} \ \rightarrow \ P_i \ = \tilde{\chi}^{(3)}_{ijkl} E_j E_k E_l$$

In this case the effective interaction Hamiltonian has an additional contribution of the form:

$$H_{NL} = a \int \vec{P}^{NL}(\vec{r},t) \cdot \vec{E}(\vec{r},t) d^3\vec{r} = a \int \tilde{\chi}^{(3)}_{ijkl} E_i(\vec{r},t) E_j(\vec{r},t) E_k(\vec{r},t) E_l(\vec{r},t) d^3\vec{r}$$

and the classical Maxwell equation, as well as the second quantization Heisenberg equation for the electric field operator read:

$$\nabla^2 \vec{E}(\vec{r},t) - \frac{1}{c^2} \frac{\partial^2 \vec{D}(\vec{r},t)}{\partial t^2} = \frac{4\pi}{c^2} \frac{\partial^2 \vec{P}^{NL}(\vec{r},t)}{\partial t^2}$$

where $\vec{D}(\vec{r},t)$ is the linear dielectric field in the presence of a linear polarization. In the simplest isotropic case, we have $\vec{P}^{NL} = \frac{3}{4}\tilde{\chi}^{(3)} \left|\vec{E}\right|^2 \vec{E}$ , where the factor $\frac{3}{4}$ is related to the symmetry properties of the Kerr crystal. In this case the canonical Hamiltonian is proportional to the fourth power of the operator $\hat{\vec{\mathrm{E}}}(\vec{r},t)$, and the Heisenberg equation is

$$\nabla^2 \hat{\vec{E}}(\vec{r},t) - \frac{n^2}{c^2} \frac{\partial^2 \hat{\vec{E}}(\vec{r},t)}{\partial t^2} = \frac{4\pi}{c^2} \frac{3}{4} \tilde{\chi}^{(3)} \frac{\partial^2 \left( \left( \hat{\vec{E}} \cdot \hat{\vec{E}} \right) \hat{\vec{E}}(\vec{r},t) \right)}{\partial t^2}$$

In Appendix 4.2. Kerr effect. we show that by sandwiching between 1 photon state and vacuum the equation goes over:

$$\nabla^2 \vec{\mathrm{E}}(\vec{r},t) - \frac{n^2}{c^2} \frac{\partial^2 \vec{\mathrm{E}}(\vec{r},t)}{\partial t^2} = \frac{4\pi}{c^2} \frac{3}{4} \tilde{\chi}^{(3)} \frac{\partial^2 \left( \vec{\mathrm{E}}^*(\vec{r},t) \cdot \vec{\mathrm{E}}(\vec{r},t) \right) \vec{\mathrm{E}}(\vec{r},t)}{\partial t^2}$$

which is the same as the classical Kerr equation responsible for the self-focusing effect among others. The solution of the problem can be done in the Schrödinger or the Heisenberg representation. Both are equivalent. In the first one the Schrödinger equation in Hilbert state has to be solved, in the second case one can solve the Heisenberg equations of motion for the different fields present in the system.

In traditional textbooks on quantum optics this coherence-preserving term is mostly not considered, leaving the student with the impression that 'classical' nonlinear effects in field propagation have no place in Second Quantization, that products of field operators in the Hamiltonian or quantum equations are related exclusively to the creation and destruction of particles. This is merely an oversimplification of the problem and as we have presently seen it is wrong. Second Quantization maintains non-linear terms at the single photon level in a similar way as classical nonlinear optics.

As one of the 'lessons learned' from nonlinear quantum optics, we can mention that the quantization of nonlinear modes should not be based in solutions of linear equations[78].

## 4.8. Double solution in Canonical Quantization

In traditional Canonical Quantization, the field $E$ for photons and $\psi$ for electrons fulfills a linear differential equation in partial derivatives, identical with the usual Schrödinger-type or Maxwell-type equations. Being a linear equation, it allows for the solutions to be undefined to the order of a multiplicative factor, or normalization factor. In the case of the electron the solutions are normalized to unity to make them compatible with a probabilistic interpretation. In the case of the Maxwell equations the fields are normalized to possess an energy content of $E = h\nu$. The normalization against $h\nu$ for photons is also a key ingredient needed for the explanation of spontaneous emission in quantum optics and QED[79]. In both cases this normalization is in some sense arbitrary, and defined in order to make compatible the solution with the experimental results and the particular interpretation given to the theory. In the present work, the external part of the field solutions fulfill the same linear equations as the traditional case, reason why de Broglie called his model the 'double solution' model. But in the double solution model these external parts are not the full solution to the problem. The complete solution is given by $\vec{E} = E_{int}\vec{E}_{ext}\ and\ \psi = \psi_{int}\psi_{ext}$, and are the solution to a nonlinear equation. The amplitude of the total solution $E\ and\ \psi$ are not undefined but are given as a result of the complete solution to the equation. In this way by assigning a particular value for the normalization of the external fields $\vec{E}_{ext}\ and\ \psi_{ext}$, this normalization will enter into the expression for the remaining part of the equation, this time the nonlinear equation fulfilled by the internal field $E_{int}, \psi_{int}$ in such a way that the total function $\vec{E} = E_{int}\vec{E}_{ext}, \psi = \psi_{int}\psi_{ext}$ remains unaffected.

Once we start considering self-interactions the probability-type solutions where the solitons are absent, are not solutions any more. The only existing solutions include a soliton. The probability type solutions to linear equations exist only under the assumption that self-interactions doesn't exist at all. The fact that some of the solutions to soliton equations may not be stable is not a reason for the soliton idea to be wrong. It only shows that more terms need to be included in the equation to provide stability. For example in the photon case one should add the attractive interaction energy between the different charge densities, and that would prevent the soliton from dismemberment.

In the case of second quantization, we have seen that the state of the system can be fully represented by a general function in Fock space, where the number of particles is not fixed, and where electrons, positrons and photons are all admitted as particles. In the presence of electrons and electromagnetic fields the general relationship between the standard and the Fock representation of second quantization is given by

$$\left[\langle vac| \sum_{n,m=1}^{n_{max}} \frac{1}{\sqrt{n!}}\frac{1}{\sqrt{m!}} \prod_{i,j=1}^{n} \hat{\psi}_a(r_i)\hat{V}\left(r_j,t\right)\right]|s\rangle \underbrace{\rightarrow}_{\substack{configuration \\ space}}$$

$$F\left(r_1^e \cdots r_{n_e}^e, r_1^{ph} \cdots r_{n_{ph}}^{ph}, t\right) = \sum_{i,j} \psi^{n_{ai}}(r_i^e, t) \cdots A^{n_{bi}}\left(r_j^{ph}, t\right) \cdots$$

For the case of three photons the traditional solution in Fock space would be of the form

$F\left(r_1^{ph}, r_2^{ph}, r_3^{ph}, t\right) = aG_a\left(r_1^{ph}, t\right) + bG_b\left(r_1^{ph}, r_2^{ph}, t\right) + cG_c\left(r_1^{ph}, r_2^{ph}, r_3^{ph}, t\right) + \cdots$

Generalizing the Double Solution model for a single particle to the case of many particles, we consider the function of the type

$G\left(r_1^{ph}, r_2^{ph}, r_3^{ph}, t\right) = f_a^{int}\left(r_1^{ph}, t\right) f_b^{int}\left(r_2^{ph}, t\right) f_c^{int}\left(r_3^{ph}, t\right) F^{ext}\left(r_1^{ph}, r_2^{ph}, r_3^{ph}, t\right)$

Where $f_a^{int}\left(r_i^{ph}, t\right)$ are the internal functions corresponding to each particle present in the system, and

$F^{ext}\left(r_1^{ph}, r_2^{ph}, r_3^{ph}, t\right) = F\left(r_1^{ph}, r_2^{ph}, r_3^{ph}, t\right)$

is the external function equal to the traditional function described above. This function is a solution of the second quantization equations when self-interactions are included in the formulation. $F^{ext}$ is the external wavefunction generating the de Broglie's Guiding Condition which may be nonlocal. Electrons and photons have their own guiding condition.

***We can see that the field operator is a Placeholder for the actual wavefunctions and fields defined in Fock space when finding Hamiltonian and Lagrangian expressions or the differential equations.***

## 4.9. Conclusion

We have verified that the existence of solitons can make possible a realistic interpretation of Multi-particle First Quantization and of Second Quantization. We showed that the formalisms of configuration space and Fock space are compatible with a realistic description of the quantum system. The only non-classical property of this description is the existence of non-local interactions between the particles that are components of entangled states. Entanglement however is a very common property of quantum states and has its roots in the conservation laws when the solution to the multi-particle wave function is given in terms of collective and/or relative coordinates. Canonical quantization allows the incorporation of the electromagnetic field on the same footing as all other types of particles. The Canonical Formalism can be fully understood as a template model from which the actual differential equations for the multi-particle fields in Fock space can be derived. Contrary to common belief, we have seen that the Canonical Quantization includes nonlinear field equations on the same footing as the commonly referred multi-particle interactions described by the Feynman diagrams. These nonlinear equations represent self-interactions between the fields and particles. The entire formalism of Canonical Quantization can be fully understood in the frame of real space-time.

In the Fock space or configuration space, the wavefunction depends on the actual position of all particles, and therefore the quantum potential implies nonlocal interactions among them. The reason why the quantum force can be independent of the amplitude of the wavefunction and still comply with energy and momentum conservation is because the force is not between the particles and the wavefunction, but between the particles among themselves or with the physical boundary condition. For example, it is the first screen where the slits are drilled who receives the energy and recoil from the particles in a two slit experiment, when they suffer alterations in their trajectory before reaching the final screen. The wavefunction serves only as an intermediary.

In the double solution model, no particle interpretation is needed, but it is automatically provided. See efforts of other people[72].

The different quantum states in quantum optics, coherent, thermal, n-photon number, all define different inter-photon potentials that are responsible for the quantum optics statistics. The state-functions in Fock space define the external multi photon functions, and with them, the potentials.

As a side note let's notice that multiple dimensions in string models can be related to the internal degrees of freedom of the particles.

## 4.10. Appendix 4.1. Nonlinear Hamiltonian in second quantization

Start with the Hamiltonian given by

$$\hat{H}_i = \frac{ie\hbar}{2m}\int \left[\hat{\psi}^\dagger(\vec{r},t)\left(\vec{\nabla}\hat{\psi}(\vec{r},t)\right) - \left(\vec{\nabla}\hat{\psi}^\dagger(\vec{r},t)\right)\hat{\psi}(\vec{r},t)\right]\cdot\hat{\vec{A}}(\vec{r},t)d^3\vec{r} = \int \hat{\vec{j}}(\vec{r},t)\cdot\hat{\vec{A}}(\vec{r},t)d^3\vec{r}$$

Where all what we have done is to replace the classical fields by its corresponding operators and have defined the current density operator

$$\begin{aligned}\hat{\vec{j}}(\vec{r},t) &= -\frac{ie\hbar}{2m}\left[\hat{\psi}^\dagger\left(\vec{\nabla}\hat{\psi}\right) - \left(\vec{\nabla}\hat{\psi}^\dagger\right)\hat{\psi}\right] = -\frac{ie\hbar}{2m}\left[\sum_j \hat{b}_j^\dagger u_j^*(\vec{r},t)\left(\vec{\nabla}\sum_i \hat{b}_i u_i(\vec{r},t)\right) - \left(\vec{\nabla}\sum_j \hat{b}_j^\dagger u_j^*(\vec{r},t)\right)\sum_i \hat{b}_i u_i(\vec{r},t)\right] = \\ &= -\frac{ie\hbar}{2m}\left[\sum_{i,j}\hat{b}_j^\dagger\hat{b}_i u_j^*(\vec{r},t)\vec{\nabla}u_i(\vec{r},t) - \sum_{i,j}\hat{b}_j^\dagger\hat{b}_i\left(\vec{\nabla}u_j^*(\vec{r},t)\right)u_i(\vec{r},t)\right] \\ &= -\frac{ie\hbar}{2m}\sum_{i,j}\hat{b}_j^\dagger\hat{b}_i\left(u_j^*(\vec{r},t)\vec{\nabla}u_i(\vec{r},t) - \left(\vec{\nabla}u_j^*(\vec{r},t)\right)u_i(\vec{r},t)\right) = -\frac{ie\hbar}{2m}\sum_{i,j}\hat{b}_j^\dagger\hat{b}_i\vec{j}_{ji}(\vec{r},t)\end{aligned}$$

and the classical current density

$$\vec{j}_{ji}(\vec{r},t) = u_j^*(\vec{r},t)\vec{\nabla}u_i(\vec{r},t) - \left(\vec{\nabla}u_j^*(\vec{r},t)\right)u_i(\vec{r},t)$$

the Hamiltonian squared is

$$\begin{aligned}\hat{H}_i^{\,2} &= -\left(\frac{e\hbar}{2m}\right)^2\int \left[\hat{\psi}^\dagger(\vec{r},t)\left(\vec{\nabla}\hat{\psi}(\vec{r},t)\right) - \left(\vec{\nabla}\hat{\psi}^\dagger(\vec{r},t)\right)\hat{\psi}(\vec{r},t)\right]\cdot\hat{\vec{A}}(\vec{r},t)d^3\vec{r} \\ &\quad\times\int \left[\hat{\psi}^\dagger(\vec{r}',t)\left(\vec{\nabla}\hat{\psi}(\vec{r}',t)\right) - \left(\vec{\nabla}\hat{\psi}^\dagger(\vec{r}',t)\right)\hat{\psi}(\vec{r}',t)\right]\cdot\hat{\vec{A}}(\vec{r}',t)d^3\vec{r}' \\ &= \int \hat{\vec{j}}(\vec{r},t)\cdot\hat{\vec{A}}(\vec{r},t)d^3\vec{r}\int \hat{\vec{j}}(\vec{r}',t)\cdot\hat{\vec{A}}(\vec{r}',t)d^3\vec{r}' = \iint \hat{\vec{j}}(\vec{r},t)\cdot\hat{\vec{A}}(\vec{r},t)\hat{\vec{j}}(\vec{r}',t)\cdot\hat{\vec{A}}(\vec{r}',t)d^3\vec{r}d^3\vec{r}'\end{aligned}$$

The product of the current density and field operator is given by

$$\begin{aligned}\hat{\vec{j}}(\vec{r},t)\cdot\hat{\vec{A}}(\vec{r},t) &= -\frac{ie\hbar}{2m}\sum_{i,j}\hat{b}_j^\dagger\hat{b}_i\vec{j}_{ji}(\vec{r},t)\sum_k\sqrt{\frac{\hbar}{2\omega_k\varepsilon_0 V}}\left(\hat{a}_k\vec{A}_k(\vec{r},t) + \hat{a}_k^\dagger\vec{A}_k^*(\vec{r},t)\right) \\ &= -\frac{ie\hbar}{2m}\sqrt{\frac{\hbar}{2\varepsilon_0 V}}\sum_{i,j,k}\sqrt{\frac{1}{\omega_k}}\hat{b}_j^\dagger\hat{b}_i\vec{j}_{ji}(\vec{r},t)\cdot\left(\hat{a}_k\vec{A}_k(\vec{r},t) + \hat{a}_k^\dagger\vec{A}_k^*(\vec{r},t)\right)\end{aligned}$$

$$\begin{aligned}&\hat{\vec{j}}(\vec{r},t)\cdot\hat{\vec{A}}(\vec{r},t)\hat{\vec{j}}(\vec{r}',t)\cdot\hat{\vec{A}}(\vec{r}',t) \\ &\qquad = -\left(\frac{e\hbar}{2m}\right)^2\frac{\hbar}{2\varepsilon_0 V}\sum_{i,j,k,i',j',k'}\sqrt{\frac{1}{\omega_k}}\sqrt{\frac{1}{\omega_{k'}}}\hat{b}_j^\dagger\hat{b}_i\hat{b}_{j'}^\dagger\hat{b}_{i'}\vec{j}_{ji}(\vec{r},t)\cdot\left(\hat{a}_k\vec{A}_k(\vec{r},t) + \hat{a}_k^\dagger\vec{A}_k^*(\vec{r},t)\right)\vec{j}_{j'i'}(\vec{r}',t) \\ &\qquad\cdot\left(\hat{a}_{k'}\vec{A}_{k'}(\vec{r}',t) + \hat{a}_{k'}^\dagger\vec{A}_{k'}^*(\vec{r}',t)\right)\end{aligned}$$

$$= -\left(\frac{e\hbar}{2m}\right)^2 \frac{\hbar}{2\varepsilon_0 V} \sum_{i,j,k,i',j',k'} \sqrt{\frac{1}{\omega_k}} \sqrt{\frac{1}{\omega_{k'}}} \hat{b}_j^\dagger \hat{b}_i \hat{b}_{j'}^\dagger \hat{b}_{i'} \left(\hat{a}_k \vec{j}_{ji}(\vec{r},t)\cdot\vec{A}_k(\vec{r},t) + \hat{a}_k^\dagger \vec{j}_{ji}(\vec{r},t)\cdot\vec{A}_k^*(\vec{r},t)\right)\left(\hat{a}_{k'}\vec{j}_{j'i'}(\vec{r}',t)\cdot\vec{A}_{k'}(\vec{r}',t) + \hat{a}_{k'}^\dagger \vec{j}_{j'i'}(\vec{r}',t)\cdot\vec{A}_{k'}^*(\vec{r}',t)\right)$$

$$= -\left(\frac{e\hbar}{2m}\right)^2 \frac{\hbar}{2\varepsilon_0 V} \sum_{i,j,k,i',j',k'} \sqrt{\frac{1}{\omega_k}} \sqrt{\frac{1}{\omega_{k'}}} \hat{b}_j^\dagger \hat{b}_i \hat{b}_{j'}^\dagger \hat{b}_{i'} \Big(\hat{a}_k \vec{j}_{ji}(\vec{r},t)\cdot\vec{A}_k(\vec{r},t)\hat{a}_{k'}\vec{j}_{j'i'}(\vec{r}',t)\cdot\vec{A}_{k'}(\vec{r}',t) + \hat{a}_k^\dagger \vec{j}_{ji}(\vec{r},t)\cdot\vec{A}_k^*(\vec{r},t)\hat{a}_{k'}\vec{j}_{j'i'}(\vec{r}',t)\cdot\vec{A}_{k'}(\vec{r}',t) + \hat{a}_k \vec{j}_{ji}(\vec{r},t)\cdot\vec{A}_k(\vec{r},t)\hat{a}_{k'}^\dagger\vec{j}_{j'i'}(\vec{r}',t)\cdot\vec{A}_{k'}^*(\vec{r}',t) + \hat{a}_k^\dagger \vec{j}_{ji}(\vec{r},t)\cdot\vec{A}_k^*(\vec{r},t)\hat{a}_{k'}^\dagger\vec{j}_{j'i'}(\vec{r}',t)\cdot\vec{A}_{k'}^*(\vec{r}',t)\Big)$$

If we keep to simplify the number of photons constant in $\left\langle ph\middle|\hat{H}_i^{\,2}\middle|ph\right\rangle$ only the third term survives.

$$= -\left(\frac{e\hbar}{2m}\right)^2 \frac{\hbar}{2\varepsilon_0 V} \sum_{i,j,k,i',j',k'} \sqrt{\frac{1}{\omega_k}} \sqrt{\frac{1}{\omega_{k'}}} \hat{b}_j^\dagger \hat{b}_i \hat{b}_{j'}^\dagger \hat{b}_{i'} \Big(\hat{a}_k^\dagger \vec{j}_{ji}(\vec{r},t)\cdot\vec{A}_k^*(\vec{r},t)\hat{a}_{k'}\vec{j}_{j'i'}(\vec{r}',t)\cdot\vec{A}_{k'}(\vec{r}',t) + \hat{a}_k \vec{j}_{ji}(\vec{r},t)\cdot\vec{A}_k(\vec{r},t)\hat{a}_{k'}^\dagger\vec{j}_{j'i'}(\vec{r}',t)\cdot\vec{A}_{k'}^*(\vec{r}',t)\Big)$$

for same initial and final states I'=j and k=k', i=j' the case j'=I', i=j later

$$= -\left(\frac{e\hbar}{2m}\right)^2 \frac{\hbar}{2\varepsilon_0 V} \sum_{i,j,k} \frac{1}{\omega_k} \hat{b}_j^\dagger \hat{b}_i \hat{b}_i^\dagger \hat{b}_j \Big(\hat{a}_k^\dagger \hat{a}_k \left(\vec{j}_{ji}(\vec{r},t)\cdot\vec{A}_k^*(\vec{r},t)\vec{j}_{ij}(\vec{r}',t)\cdot\vec{A}_k(\vec{r}',t)\right) + \hat{a}_k \hat{a}_k^\dagger \left(\vec{j}_{ji}(\vec{r},t)\cdot\vec{A}_k(\vec{r},t)\vec{j}_{ij}(\vec{r}',t)\cdot\vec{A}_k^*(\vec{r}',t)\right)\Big)$$

If there are no photons initially, $\hat{a}_k\hat{a}_k^\dagger = 1, \hat{a}_k^\dagger\hat{a}_k = 0$ and the term simplifies to

$$= -\left(\frac{e\hbar}{2m}\right)^2 \frac{\hbar}{2\varepsilon_0 V} \sum_{i,j,k} \frac{1}{\omega_k} \hat{b}_j^\dagger \hat{b}_i \hat{b}_i^\dagger \hat{b}_j \left(\vec{j}_{ji}(\vec{r},t)\cdot\vec{A}_k(\vec{r},t)\vec{j}_{ij}(\vec{r}',t)\cdot\vec{A}_k^*(\vec{r}',t)\right)$$

and

$$\hat{H}_i^{\,2} = -\left(\frac{e\hbar}{2m}\right)^2 \frac{\hbar}{2\varepsilon_0 V} \sum_{i,j} \hat{b}_j^\dagger \hat{b}_i \hat{b}_i^\dagger \hat{b}_j \left[\sum_k \frac{1}{\omega_k} \iint \left(\vec{j}_{ji}(\vec{r},t)\cdot\vec{A}_k(\vec{r},t)\vec{j}_{ij}(\vec{r}',t)\cdot\vec{A}_k^*(\vec{r}',t)\right) d^3\vec{r}\,d^3\vec{r}'\right]$$

## 4.11. Appendix 4.2. Kerr effect.

The Heisenberg equation for the Kerr effect is

$$\nabla^2 \hat{\vec{E}}(\vec{r},t) - \frac{n^2}{c^2}\frac{\partial^2 \hat{\vec{E}}(\vec{r},t)}{\partial t^2} = \frac{4\pi}{c^2}\frac{3}{4}\tilde{\chi}^{(3)}\frac{\partial^2\left(\left(\hat{\vec{E}}\cdot\hat{\vec{E}}\right)\hat{\vec{E}}(\vec{r},t)\right)}{\partial t^2}$$

or sandwiching between 1 photon state and vacuum:

$$\left\langle vac \middle| \left( \nabla^2 \left(\hat{a}\vec{\mathrm{E}}(\vec{r},t) + \hat{a}^{\dagger}\vec{\mathrm{E}}(\vec{r},t)\right) - \frac{n^2}{c^2}\frac{\partial^2\left(\hat{a}\vec{\mathrm{E}}(\vec{r},t) + \hat{a}^{\dagger}\vec{\mathrm{E}}^*(\vec{r},t)\right)}{\partial t^2} \right) \middle| 1ph \right\rangle$$

$$= \left\langle vac \middle| \frac{4\pi}{c^2}\frac{3}{4}\tilde{\chi}^{(3)}\frac{\partial^2\left(\left[\left(\hat{a}\vec{\mathrm{E}}(\vec{r},t) + \hat{a}^{\dagger}\vec{\mathrm{E}}^*(\vec{r},t)\right)\cdot\left(\hat{a}\vec{\mathrm{E}}(\vec{r},t) + \hat{a}^{\dagger}\vec{\mathrm{E}}^*(\vec{r},t)\right)\right]\left(\hat{a}\vec{\mathrm{E}}(\vec{r},t) + \hat{a}^{\dagger}\vec{\mathrm{E}}^*(\vec{r},t)\right)\right)}{\partial t^2} \middle| 1ph \right\rangle$$

$$\nabla^2\vec{\mathrm{E}}(\vec{r},t) - \frac{n^2}{c^2}\frac{\partial^2\vec{\mathrm{E}}(\vec{r},t)}{\partial t^2}$$

$$= \left\langle vac \middle| \frac{4\pi}{c^2}\frac{3}{4}\tilde{\chi}^{(3)}\frac{\partial^2\left(\left[\left(\hat{a}\vec{\mathrm{E}}(\vec{r},t) + \hat{a}^{\dagger}\vec{\mathrm{E}}^*(\vec{r},t)\right)\cdot\left(\hat{a}\vec{\mathrm{E}}(\vec{r},t) + \hat{a}^{\dagger}\vec{\mathrm{E}}^*(\vec{r},t)\right)\right]\left(\hat{a}\vec{\mathrm{E}}(\vec{r},t) + \hat{a}^{\dagger}\vec{\mathrm{E}}^*(\vec{r},t)\right)\right)}{\partial t^2} \middle| 1ph \right\rangle$$

The interesting part is again given by $\left(\hat{a}\vec{\mathrm{E}}(\vec{r},t) + \hat{a}^{\dagger}\vec{\mathrm{E}}^*(\vec{r},t)\right)\cdot\left(\hat{a}\vec{\mathrm{E}}(\vec{r},t) + \hat{a}^{\dagger}\vec{\mathrm{E}}^*(\vec{r},t)\right)$ that is of the form

$$\left(\hat{a}\vec{\mathrm{E}}(\vec{r},t) + \hat{a}^{\dagger}\vec{\mathrm{E}}^*(\vec{r},t)\right)\cdot\left(\hat{a}\vec{\mathrm{E}}(\vec{r},t) + \hat{a}^{\dagger}\vec{\mathrm{E}}^*(\vec{r},t)\right)$$

$$= \vec{\mathrm{E}}(\vec{r},t)\cdot\vec{\mathrm{E}}(\vec{r},t)\hat{a}^2 + \vec{\mathrm{E}}^*(\vec{r},t)\cdot\vec{\mathrm{E}}^*(\vec{r},t){\hat{a}^{\dagger}}^2 + \vec{\mathrm{E}}^*(\vec{r},t)\cdot\vec{\mathrm{E}}(\vec{r},t)(2n+1)$$

Where here the first term destroys two photons and oscillates in the complex plane with a frequency of $-2\omega$, the second term creates two photons and oscillates in the complex plane with a frequency of $2\omega$ and the last term is the product of the magnitude squared of the fields, leaving the number of original photons constant.

$$\left\langle vac \middle| \left[\left(\hat{a}\vec{\mathrm{E}}(\vec{r},t) + \hat{a}^{\dagger}\vec{\mathrm{E}}^*(\vec{r},t)\right)\cdot\left(\hat{a}\vec{\mathrm{E}}(\vec{r},t) + \hat{a}^{\dagger}\vec{\mathrm{E}}^*(\vec{r},t)\right)\right]\left(\hat{a}\vec{\mathrm{E}}(\vec{r},t) + \hat{a}^{\dagger}\vec{\mathrm{E}}^*(\vec{r},t)\right) \middle| 1ph \right\rangle =$$

$$\langle vac|\left[\vec{\mathrm{E}}^*(\vec{r},t)\cdot\vec{\mathrm{E}}(\vec{r},t)(2n+1)\right]\vec{\mathrm{E}}(\vec{r},t)\hat{a}|1ph\rangle + \langle vac|\left[\vec{\mathrm{E}}(\vec{r},t)\cdot\vec{\mathrm{E}}(\vec{r},t)\right]\vec{\mathrm{E}}^*(\vec{r},t)\hat{a}^2\hat{a}^{\dagger}|1ph\rangle$$

In the first term $\hat{a}|1ph\rangle = |vac\rangle$, leaving *n = 0*, in the second term $\hat{a}^2\hat{a}^{\dagger}|1ph\rangle = 2|vac\rangle$. Neglecting the second term by energy considerations the Maxwell equation reads:

$$\nabla^2\vec{\mathrm{E}}(\vec{r},t) - \frac{n^2}{c^2}\frac{\partial^2\vec{\mathrm{E}}(\vec{r},t)}{\partial t^2} = \frac{4\pi}{c^2}\frac{3}{4}\tilde{\chi}^{(3)}\frac{\partial^2\left(\vec{\mathrm{E}}^*(\vec{r},t)\cdot\vec{\mathrm{E}}(\vec{r},t)\right)\vec{\mathrm{E}}(\vec{r},t)}{\partial t^2}$$

# 5. Chapter: Electrons and Photons

## 5.1. Electrons

This paper is the fifth in a series of five papers entitled “Solitons and Quantum Behavior”.  In this chapter we review some problems of the interaction between electrons and photons from the point of view of the double solution and solitons.

### 5.1.1. Feynman boundary conditions

In traditional QED textbooks, to obtain the electron positron pair creation cross section, one solves the Compton problem and then applies a space-time rotation, to obtain the desired cross section based on relativistic symmetry properties. This leaves unclear the physical process, what's going on. The historical origin of the traditional calculation is very interesting: Wheeler came up with the idea that the electrons and positrons in the universe where in reality a single electron going back and forth from the temporal beginning of the universe to the temporal end of the universe, each time covering the path of each one of the electrons in the universe.  When coming back from the future to the past the single electron was seen as a positron.

Feynman sympathized with that idea (Wheeler had been his thesis advisor) and tried to formulate an interpretation to the Dirac electron model that were totally symmetric respect to time inversion.  In this way he designed his diagrams really reversible in time, interpreting negative energy electrons as positrons going back in time.  There is however a drawback to develop such an interpretation: in the real world waves naturally start in a localized region of space and if they are not confined, they would expand forever towards infinity.  The opposite process, namely a wave starting at infinity and converging towards a point never occurs naturally.  This fact creates an essential asymmetry in the boundary conditions for real waves. The initial condition is always a localized wave expanding towards infinity, never the opposite case.  This condition can clearly be seen when working in spherical geometry.  Real solutions are always of the form

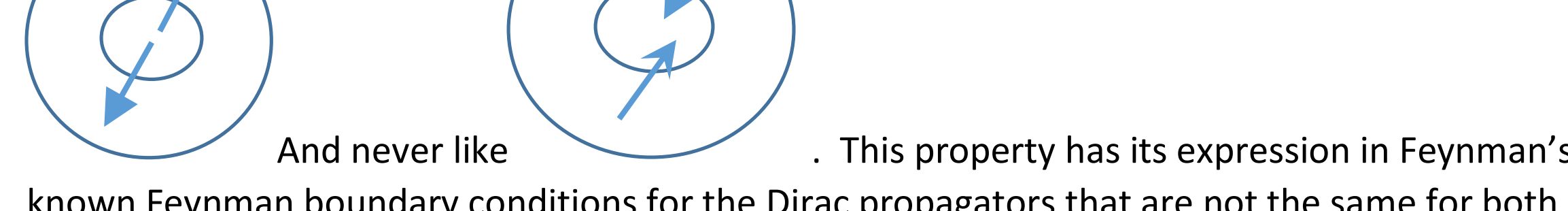

And never like . This property has its expression in Feynman’s theory in the well-known Feynman boundary conditions for the Dirac propagators that are not the same for both signs of the energy as shown in the next figure[80]

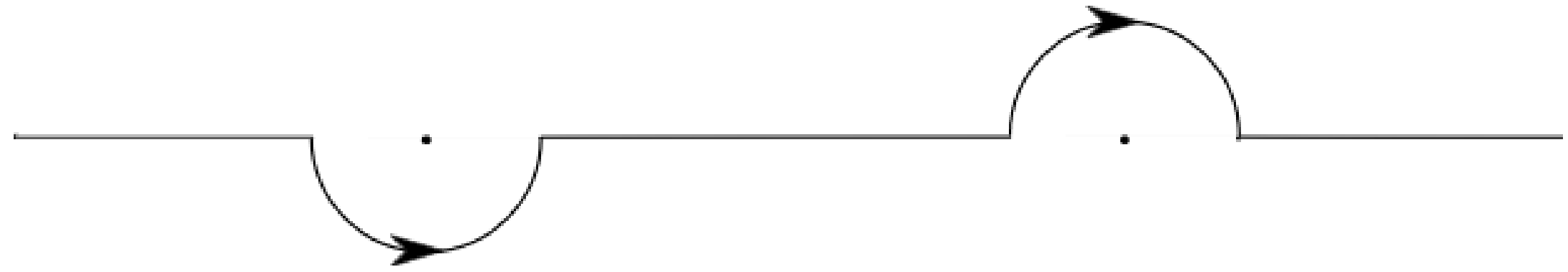

Apparently to dissimulate this behavior, Feynman developed his theory exclusively in plane geometry.  When expressed in terms of plane waves the boundary conditions for the diagrams appear to be symmetric, but when using spherical coordinates, positive energy solutions are diverging waves, while negative energy solutions are converging waves.

One can describe the electron-positron theory in the following two ways:

a) One can use Second Quantization considering two solutions: one with negative charge corresponding to an electron, and one with positive charge corresponding to a positron, where as usual in Second Quantization both particles have positive energies, or
b) Use a single First Quantization Dirac equation, accept positive as well as negative energy solutions, identify positive energy solutions with electrons and negative energy solutions with positrons, but using Feynman's boundary conditions, namely running positron solutions on a mirror and backwards in time, changing the sign of the spin,… as shown in text books on QED.

It can be argued that Feynman special boundary conditions for the solutions of the Dirac equation is just a mathematical tool based on the general symmetry properties of the Dirac equation. We can see then why the Feynman boundary condition is required: running the solutions in the time direction implied by the Feynman diagrams, spherical electron and positron wavefunctions will both start its motion in a localized region of space and both will expand with time. As we noticed before this behavior is not self-evident when working with plane waves.

One of the most important contributions from Feynman to QED is related to the title of his papers[81], where he express the firmly belief that all quantum systems could be described in real physical space by using classical expressions. In fact, he developed quantum mechanics on the basis that the propagator or Green function for the wave amplitude could be obtained as an integral of the classical action function of a classical particle following all possible paths between the initial and final points. In fact in his work no operator is defined for the particle fields, all states and all interactions are described by well-defined numerical functions explicitly defined in space and time.

He also stated that when two paths were possible between the initial and final points, the amplitude contribution should be given by the sum of both contributions, therefore allowing for the possibility of interference. Also the amplitude for a process involving independent paths, related to two independent particles for example, should be given by the product of the amplitudes of the individual processes. These two last rules are equivalent to the OR and the AND logical operations for probabilities, but applied to amplitudes instead of probabilities themselves.

The propagator technique can be used in the 'space or 'momentum' representations. The momentum representation is the most commonly used because the equations simplify enormously, which is nothing else than a consequence of the Fourier decomposition in plane waves. Plane waves are the eigenstates of the energy operator in the absence of sources or interactions in the entire space, namely "free particles". Many attempts have been developed to provide free particles states with special meaning. From our point of view, these attempts only generate confusion, because they seem to overlook the fact that plane waves are only a mathematical tool, without any further meaning that just representing the Fourier components of a single wavefunction with any complicated geometry. Readily, it is this full wavefunction and not its individual parts, which provides the information about the state of the electron. Feynman work showed that all the results obtained using the canonical quantization description in terms of creation and destruction operators can also be described with events and fields evolving in real space-time.

### 5.1.2. Sources and antiparticles in the effective equations

The First Quantization Klein Gordon and Dirac equations have 'obviously' no sources. However the effective equations derived from second quantization do have sources. We are going to verify this assertion in the following. To show our remarks, we consider for simplicity the Hamiltonian operator for the Klein Gordon equation in the presence of an electron and positron as provided by second quantization:

$$H_{field} = \int \Bigg[ -e\left(\hat{a}_{el}^{\dagger} u_{el}^{*} + \hat{b}_{pos} u_{pos}\right)\left(\hat{a}_{el} u_{el} + \hat{b}_{pos}^{\dagger} u_{pos}^{*}\right)\left(V(r) + \frac{A^2(r)}{2m}\right) + \frac{ie\hbar}{2m}\left(\left(\hat{a}_{el}^{\dagger} u_{el}^{*} + \hat{b}_{pos} u_{pos}\right)\nabla\left(\hat{a}_{el} u_{el} + \hat{b}_{pos}^{\dagger} u_{pos}^{*}\right) - \left(\hat{a}_{el} u_{el} + \hat{b}_{pos}^{\dagger} u_{pos}^{*}\right)\nabla\left(\hat{a}_{el}^{\dagger} u_{el}^{*} + \hat{b}_{pos} u_{pos}\right)\right)\cdot A(r)\Bigg] d^3r$$

We encapsulate the Hamiltonian between two states to get the equation for the electron.

$$(\langle vac| + \langle 1ph| + \langle 1e| + \langle 1p| + \langle 1e, 1p|)H_{field}(|vac\rangle + |1ph\rangle + |1e\rangle + |1p\rangle + |1e, 1p\rangle)$$

$$\begin{aligned}\rightarrow\ \ &(\langle vac| + \langle 1ph| + \langle 1e| + \langle 1p| \\ &+ \langle 1e, 1p|) \int \Bigg[ -e\left(\hat{a}_{el}^{\dagger}u_{el}^{*} + \hat{b}_{pos}u_{pos}\right)\left(\hat{a}_{el}u_{el} + \hat{b}_{pos}^{\dagger}u_{pos}^{*}\right)\left(V(r) + \frac{A^2(r)}{2m}\right) \\ &+ \frac{ie\hbar}{2m}\left(\left(\hat{a}_{el}^{\dagger}u_{el}^{*} + \hat{b}_{pos}u_{pos}\right)\nabla\left(\hat{a}_{el}u_{el} + \hat{b}_{pos}^{\dagger}u_{pos}^{*}\right) - \left(\hat{a}_{el}u_{el} + \hat{b}_{pos}^{\dagger}u_{pos}^{*}\right)\nabla\left(\hat{a}_{el}^{\dagger}u_{el}^{*} + \hat{b}_{pos}u_{pos}\right)\right) \\ &\cdot A(r)\Bigg] d^3r\, (|vac\rangle + |1ph\rangle + |1e\rangle + |1p\rangle + |1e, 1p\rangle)\end{aligned}$$

We get

$$\begin{aligned}H_{field} = \int \Bigg[ &-e\left(u_{el}^{*} + u_{pos}\right)\left(u_{el} + u_{pos}^{*}\right)\left(V(r) + \frac{A^2(r)}{2m}\right) + \frac{ie\hbar}{2m}\left(\left(u_{el}^{*} + u_{pos}\right)\nabla\left(u_{el} + u_{pos}^{*}\right) - \left(u_{el} + u_{pos}^{*}\right)\nabla\left(u_{el}^{*} + u_{pos}\right)\right) \\ &\cdot A(r)\Bigg] d^3r = \int j^{\mu}A_{\mu}d^3r\end{aligned}$$

**Eq 159**

where in the last equation we are using the relativistic notation for the current density.

The equation generated for the electron field by this Hamiltonian is

$$\frac{\partial u_{el}}{\partial t} = -e\left(V(r) + \frac{A^2(r)}{2m}\right)\left(u_{el} + u_{pos}^{*}\right) + \frac{ie\hbar}{2m}\nabla\left(u_{el} + u_{pos}^{*}\right)\cdot A(r)$$

or

$$\frac{\partial u_{el}}{\partial t} = \underbrace{\left[-e\left(V(r) + \frac{A^2(r)}{2m}\right) + \frac{ie\hbar}{2m}A(r)\cdot\nabla\right]}_{Hamiltonian\ effective\ operator}u_{el} - \underbrace{e\left(V(r) + \frac{A^2(r)}{2m}\right)u_{pos}^{*} + \frac{ie\hbar}{2m}A(r)\cdot\nabla u_{pos}^{*}}_{source\ term}$$

In the last equation we can clearly identify a source term for the electron field given by the overlapping of an electromagnetic field and a positron field:

$$\rho_e = -e\left(V(r) + \frac{A^2(r)}{2m}\right)u_{pos}^{*} + \frac{ie\hbar}{2m}\nabla u_{pos}^{*}\cdot A(r)$$

And a similar equation for the positron. If we consider the case of photon emission from an electro-positron pair, we see from Eq 159 that the interaction term $j^{\mu}A_{\mu}$ has a term proportional to $u_{el}^{\mu}u_{pos}^{\mu}A_{\mu}^{\dagger}$ that destroys an electron and a positron and creates a photon with an energy equal to the sum of the energies of the two original particles. Here we can find the reason for the strange definition of the relativistic field operators (in both Dirac and Klein Gordon equations)

$$\hat{\psi}(\vec{r}, t) = \int d^3p\left(\hat{a}_{\vec{p}}u_{\vec{p}}(\vec{r}, t) + \hat{b}_{\vec{p}}^{\dagger}u_{\vec{p}}^{*}(\vec{r}, t)\right)$$

$$\hat{\psi}^{\dagger}(\vec{r}, t) = \int d^3p\left(a_{\vec{p}}^{\dagger}u_{\vec{p}}^{*}(\vec{r}, t) + \hat{b}_{\vec{p}}u_{\vec{p}}(\vec{r}, t)\right)$$

where $\hat{\psi}$ hast two terms, one proportional to a positron creation operator $\hat{b}_{\vec{p}}^{\dagger}$ and the electron destruction operator $\hat{a}_{\vec{p}}$, while in the non-relativistic field operator the antiparticle term is missing. Correspondingly for $\hat{\psi}^{\dagger}$. The purpose of the additional positron operator in the electron field operator is to make possible a description of the electron-positron annihilation and pair creation processes as shown in the previous paragraph. On the other hand, the relativistic symmetry properties allow the Feynman's formalism where the charge density is given by

$$e(u_{el}^* + u_{pos})(u_{el} + u_{pos}^*) = e\left(u_{el}^* u_{el} + u_{pos} u_{pos}^* + \underbrace{u_{el}^* u_{pos}^* + u_{pos} u_{el}}_{2\,\mathrm{Re}(u_{pos}u_{el}) \sim cos(2\omega t)}\right)$$

and the charge current by

$$\frac{ie\hbar}{2m}\left((u_{el}^* + u_{pos})\nabla(u_{el} + u_{pos}^*) - (u_{el} + u_{pos}^*)\nabla(u_{el}^* + u_{pos})\right)$$

Assuming an oscillation of the positive energy solution as $ae^{-i\omega t}$ and of the negative energy solution of the form $be^{i\omega t}$, we obtain the time evolution of the charge and current densities as

$$\left(ae^{-i\omega t} + be^{i\omega t}\right)\left(a^* e^{i\omega t} - b^* e^{-i\omega t}\right) = |a|^2 - |b|^2 - ab^* e^{-2i\omega t} + a^* b e^{2i\omega t}$$

For *a, b* real: $|a|^2 - |b|^2 + ab \cdot cos(2\omega t)$

Oscillating at a frequency $f = 2\omega$ which is different from what one would obtain by using a simple non-relativistic canonical field definition

$$\left(ae^{-i\omega t} + be^{-i\omega t}\right)\left(a^* e^{i\omega t} - b^* e^{i\omega t}\right) = |a|^2 - |b|^2 - ab^* + a^* b$$

this last expression, for *a, b* real: $|a|^2 - |b|^2$ is time independent, and not oscillating at the frequency found before.

It can be mentioned that in the case of pair creation, the resulting particles are not independent, but entangled due to the conservation of energy and momentum in the creation process.

In 1966 Julian Schwinger proposed the description of electron-positron generation by the use of sources[82], but his interpretation of QED was totally rejected by his colleagues at that time, in a similar way as how de Broglie's double solution was rejected almost 40 years before.

### 5.1.3. Vacuum fluctuations. Ergodicity

The historical development of the concept of vacuum fluctuations is interesting in itself. One of the basic results of quantum mechanics is the uncertainty relations between the statistical distributions of complementary variables, derived from their commutation properties and interpreted originally by Heisenberg. In its original form, this principle states that any measurements on complementary variables will be irremediably linked with a statistical distribution in its results given by the uncertainty inequalities. The uncertainty in momentum (or velocity) appears to be inversely proportional to the spatial confinement dimensions of the system. Initially the first interpretation provided by Heisenberg and Born among others was that the dispersion in measurement results was somehow, directly or indirectly produced by the measuring apparatus. This could be conceptualized as an external arrangement interacting with the measured system that in a random and uncontrollable way modifies the environmental conditions seeing by the quantum system and thereby modifies the measurement outcome. However later it was further analyzed the fact that the uncertainty relations are a consequence of the commutation properties and therefore should be present even in the absence of external interactions. They are really an intrinsic property of the quantum system. One possible 'physical interpretation' for this fact would be that quantum systems of atomic dimensions suffer from what could be called 'Quantum fluctuations', some sort of random walk, always present, independently of any measurement attempt[83]. This random walk would be also responsible for the achievement of ergodicity in the probability distribution derived from the wavefunction in the traditional interpretation of quantum mechanics. In fact in the traditional interpretation of quantum mechanics, the solution to the quantum wave equations is seeing as providing a probability density distribution for the results of measurements for the position of a single quantum particle. In fact these 'quantum fluctuations' have physical reality and have been experimentally measured.

Many different models have been proposed as source for this quantum fluctuations present not only in the motion of particles, but also in the evolution of fields, like electromagnetic waves. The traditional interpretation in second quantization and QED assigns the source of this fluctuations to the commutation properties of the fields in the vacuum state. However this interpretation is just shifting an answer into a new question: why commutation properties are related to vacuum fluctuations? Aren't commutation relationships just formal expressions used to derive the equations of motion? A much more 'physical' insight was obtained by relatively modern works, where it has been shown that the second quantization vacuum fluctuations provide exactly the same result as the introduction of self-interactions and radiation reaction in the first quantization differential equations[84]. This provides a straightforward interpretation: the self-interaction of a charged particle with its own electromagnetic fields provides some sort of resonant, stationary state characterized by a random walk, as described by Welton's model[85]. In a hidden-variable theory the interpretation would be that the action of the radiation reaction fields plus the quantum potential (forces) cancel partially each other. In so doing they imply an energy addendum responsible for a contribution to the self-energy, vertex Feynman diagrams, Lamb shift, and random walk associated with the quantum fluctuations. It has to be marked that the random walk in the quantum 'particle' is also directly associated with fluctuations in the amplitude and phase of the wave-function or quantum field associated with that particle. Also, in the case of charged particles, the field for which they are source will suffer also of these fluctuations.

The correct inclusion of self-interaction and radiation reaction in the first quantization theory incorporates spontaneous emission, removes unwanted electron runaway solutions and explain the existence of the kinetic energy present even in the ground state of physical systems, intimately linked to the conservation of the commutation relations in matter-radiation interactions. As known, in the nonrelativistic limit the vector potential $A$ of radiation reaction has

a) a term proportional to the velocity, which generates a mass renormalization, and according to Cohen Tannoudji, at least part of this energy appears as an additional oscillation on the particle trajectory; and
b) A term proportional to the acceleration. In classical physics this term, depending on the relative phase between the acceleration and the velocity can give rise to radiation damping or to the unwanted runaway solutions.

The quantum radiation reaction problem differs from the classical calculation in the existence of the quantum potential. The part proportional to the acceleration is the responsible for spontaneous emission and bremsstrahlung, while the part proportional to the velocity increases the energy of the particle by an additional kinetic energy associated with a random walk, responsible for the electron ergodicity. It also prevents run away behavior.

We have seen in the Chapter2 related to the electron model and Dirac equation, that the equation relating the second 2-spinor to the first one in the 4-spinor representing the electron in the Dirac equation is given by

$$\chi = \frac{c\vec{\boldsymbol{\sigma}} \cdot \vec{\pi}\varphi - i\hbar \dfrac{\partial \chi}{\partial t}}{\left( i\hbar \dfrac{\partial \psi}{\psi \partial t} - e\mathrm{V} + mc^2 \right)}$$

We can see here that even in the presence of an energy eigenstate, the spinor $\chi$ is proportional to its own time derivative $\frac{\partial \chi}{\partial t}$. This relationship is similar to the appearance of terms proportional to the velocity and acceleration in the equation of motion for a charged particle in the presence of radiation reaction. This term generates a spin zitterbewegung as shown by Barut and Hestenes. As a result the spin is continuously trembling and is the responsible for the well-known algebra of spin, with non-zero components in all directions at all times. The similarity between the spin and the two-level model, makes it clear that the solutions of the two level system also explain many aspects of the spin property.

### 5.1.4. Lamb shift

We are going to consider the Lamb shift, attributed traditionally to the vacuum fluctuations, but can also be analyzed in terms of radiation reaction[84]. Using the quantum self-field approach to QED, it is shown that for a two level atom model, the equations of motion for the coefficients of the upper $C_1$ and lower $C_2$ energy eigenstates in front of the radiation reaction forces are [86]

$$i\dot{C}_1 = \frac{\alpha\Lambda}{\pi}C_1 - \frac{2\alpha}{3\pi}\omega_0^3|\mathbf{x}_{21}|^2\left[-ln\left(\frac{\Lambda}{\omega_0}\right) + i\pi\omega_0\right]|C_2|^2C_1$$

$$i\dot{C}_2 = \frac{\alpha\Lambda}{\pi}C_2 - \frac{2\alpha}{3\pi}\omega_0^3|\mathbf{x}_{21}|^2\left[-ln\left(\frac{\Lambda}{\omega_0}\right) + i\pi\omega_0\right]|C_1|^2C_2$$

$$\Lambda = \mathrm{m}$$

These equations are nonlinear differential equations, because we are considering the self-field produced by radiation reaction. In simplified form the equations are:

$$i\frac{dC_1}{dt} = C_1|C_2|^2Ve^{-i\omega_0 t}$$

$$i\frac{dC_2}{dt} = C_2|C_1|^2V^*e^{i\omega_0 t}$$

equivalent to

$$\frac{d^2C_2}{dt^2} = \left(-i\frac{dC_2}{dt} + \omega_0C_2\right)|C_1|^2V^*e^{i\omega_0 t}$$

From Wolfram.com we get the analytical solution to this equation, using as input:

$$y''(t) - \left(-iy'(t) + \omega_0 y(t)\right)Ve^{i\omega_0 t} = 0$$

we obtain the solution:

$$y(t) = c_2e^{-\frac{Ve^{i\omega_0 t}}{\omega_0}} - i\frac{c_1e^{-\frac{Ve^{i\omega_0 t}}{\omega_0}}E_i\left(\frac{Ve^{i\omega_0 t}}{\omega_0}\right)}{\omega_0}$$

where $E_i(x)$ is the exponential integral function. So the nonlinear equation admits a solution with constant amplitude due to the combined effect of radiation reaction and the quantum potential. It can be argue that in a more realistic multi-level model, the constant $c_2$ really is composed of a sum of a great number of terms with different frequencies. The addition of all those contributions with random phase distribution generates a random walk as described in Welton's model and the book by Mandel and Wolf. This quantum walk maintains the ergodicity of the electron cloud continuously.

### 5.1.5. Self-energy and mass renormalization

Weisskopf[87] in 1934 studied the effect of vacuum fluctuations and Coulomb self-interaction for the electron from the point of view of Second Quantization. Weisskopf was using the traditional, non-relativistic theory of perturbations, where all calculations are done at a constant time, and no time integration is performed. His result was obviously independent of time, and if one performs a time integration, its interpretation should not be modified. He obtained the same result as that given in the 1950s by the Feynman diagrams for an electron at rest. However the significance of Weisskopf calculations for their physical interpretation is large.

Because Weisskopf performed his work before the formal development of present day QED, he still was open to intuitive images for the electron. He was able to find a 'density profile' decaying exponentially with the Compton wavelength for $\frac{r}{r_{Compton}} \gg 1$ and like $\frac{1}{r^{5/2}}$ for $\frac{r}{r_{Compton}} \ll 1$.

Because Weisskopf calculation were done at a "constant time" condition, one can arguably interpret that the 'density profile' just obtained will apply to at least two measurements performed at the same time. However, this is just some way of declaring that the electron really occupies a volume in space that classically would be called the 'electron size'. The idea that the electron remains as a point particle doesn't sound a logical argumentation in this case. From Weisskopf model we see that there is no theoretical reason why the electron should be considered a particle with zero size in space. We also find that there is no experimental foundation that would prevent the interpretation of the electron as an elastic object with a finite size given by the Compton wavelength as was shown by Compton himself[88] before discovering his famous energy-momentum conservation law in electron-photon scattering.

We have performed a numerical calculation of the Dirac electron with self-fields and the exponential decay found by Weisskopf is the same we find in our solutions. The central part of our solution corresponds to the full solution of the problem, not just an approximation at first order. Experiments where an electron at low velocities can traverse a pinhole with a dimension smaller than the Compton size are unknown. Our calculations confirm that the electron has a finite radius of the order of the Compton size and a finite mass and charge that can be identified with the renormalization values. Therefore our interpretation is that there is no valid theoretical or experimental reason that would imply the electron occupies no volume in space.

### 5.1.6. Coulomb Scattering

From Bjorken and Drell the S scattering matrix for Coulomb scattering is given by[89]

$$S_{fi} = \frac{-ie^2}{V^2}(2\pi)^4\delta^4(P_f - P_i + p_f - p_i)\sqrt{\frac{m^2}{E_f E_i}}\sqrt{\frac{M}{E_f E_i}}\left[\bar{u}(p_f, s_f)\gamma_\mu u(p_i, s_i)\right]\frac{1}{(p_f - p_i)^2 + i\epsilon}\left[\bar{u}(P_f, S_f)\gamma^\mu u(P_i, S_i)\right]$$

This is equivalent to say the final state is given by

$$\begin{aligned} F(\vec{r}_1, \vec{r}_2, t) &= e^{-iEt/\hbar}\int g(\vec{p})e^{-i\vec{p}\cdot\vec{r}_1}e^{-i(\vec{P}_{cm}-\vec{p})\vec{r}_2}d\vec{p} \quad \approx \quad \sum f(\vec{r}_1)g(\vec{r}_2) \\ &= e^{-iEt/\hbar}F(\vec{R}_{cm})G(\vec{r}_{rel}) \qquad \text{entangled state} \end{aligned}$$

Which is a function of the center of mass times a function of the relative coordinate. This function can in general be decomposed as 'sums' of products of functions of the independent variables as shown by the Schmidt decomposition. This is an entangled state. It is also the same principle we can see behind the Clebsh-Gordan coefficients for angular momentum. They provide the relationship between the group and the individual.

In this example we verify that conservation laws create entanglement between the particles after they leave the interaction region.

The result shown above is a first order approximation. The trajectories predicted by the different orders in approximation can be seen in the next picture. When the WKB approximation doesn't hold, the first approximation is like the dipole approximation, where the source for scattered waves is a point, in higher order approximations the source would take a more 'realistic' shape. We can see that asymptotic solutions far from the scattering nucleus are a good approximation. Differences appear only close to the scattering center as seen in the following picture

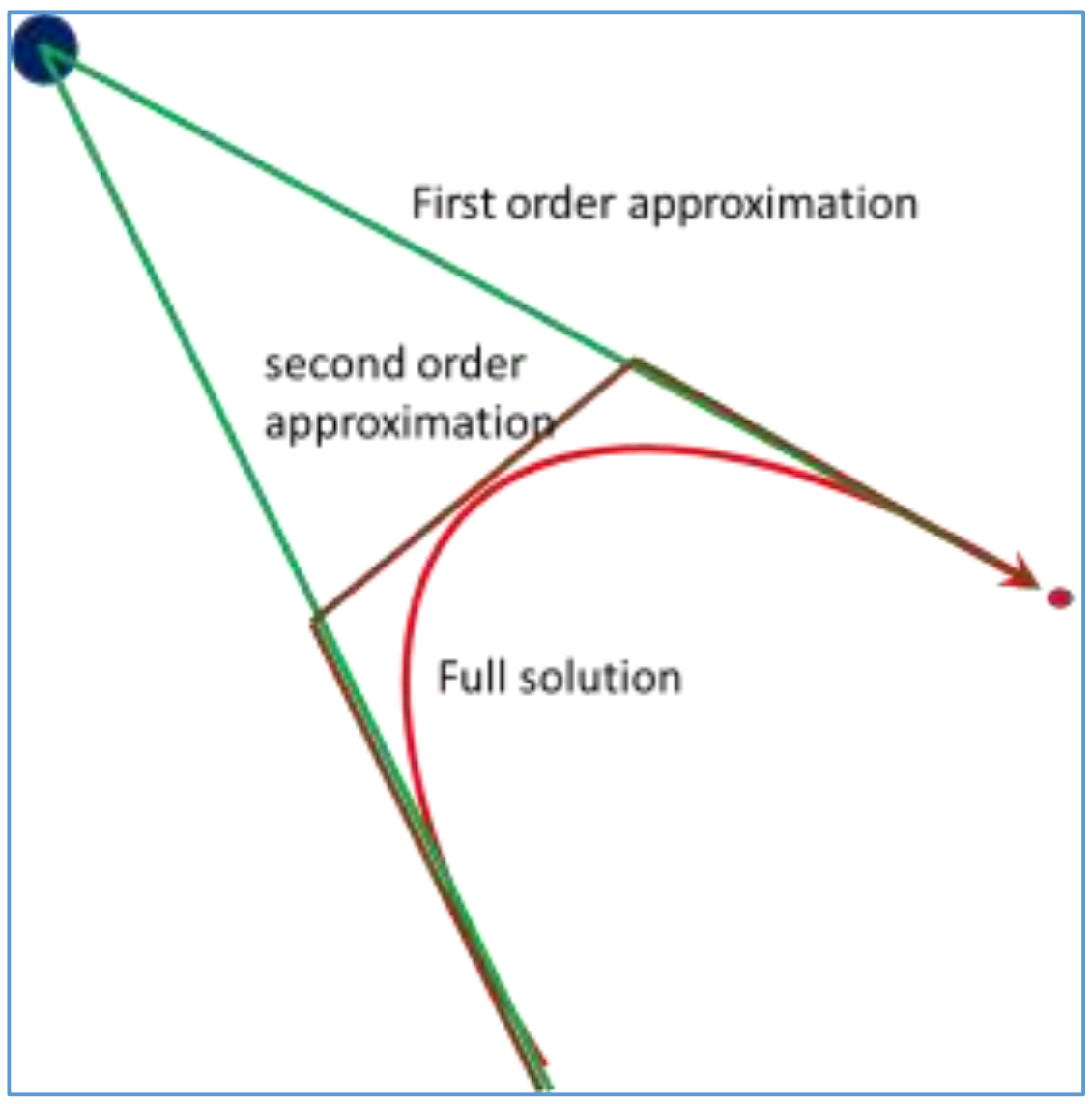

For a similar analysis of the Compton effect, identifying the appearance of a finite electron radius in the propagator formalism, applicable to all QED calculations of scattering cross sections, see Section 5.4.4 on the Compton Effect.

## 5.2. Photons

Classical Maxwell's equations have a very distinguishing property: they include sources and sinks, implying that the evolution of a possible wave packet would not be unitary: the integral under the field will not be conserved over time. When trying to unify electromagnetic fields and the electron wavefunctions, it was needed to reconcile this property with the case of electrons, where the norm of the wavefunction was conserved over time. The solution already implied in 1927 Dirac's model was that the normalization of the quantum photon states is constant, the energy of a monochromatic field at all times is $E = \hbar\omega$, what changes with time is the probability for finding the photon, and it is this probability what is represented by the time dependent norm of the solutions of the classical Maxwell's equations. This probability will show up in a statistical description, in the presence of many 'photons'.

### 5.2.1. Quantum Optics

It can be verified that the equations of motion for the electromagnetic field operators in the case of a single photon state, as provided by the Heisenberg equations of motion are practically identical with the classical Maxwell equations. Their interpretation however is different. The solutions to the Heisenberg equations of motion can be considered as the average value of the "total" or "collective" field as seen by all other particles. It is important to clarify that this collective field in general has not the same shape as the individual fields associated with each individual photon $r_{ph}^{1}, \cdots r_{ph}^{i}, \cdots r_{ph}^{n}$ composing the system in Fock space. The total field can be considered as the sum of all the individual fields associated with each individual photon or group of photons in Fock space. The square of the amplitude of the average of the field operator being proportional to the probability to "measuring" (or finding) the photons at that position.

The mathematical formalism of second quantization and quantum optics allows to find the average value for the macro field, and also the statistical distribution of multi photon measurements (coincidence and correlations measurements). The collective wave function is provided by the boundary conditions imposed to the evolution of the system. The existence of these functions, is the basis of entanglement, and generate non local interactions (of the quantum force) between the photons. They are responsible for the existence of the quantum statistics of photons (bosons).

### 5.2.2. Solitons

Already in 1924 Slater, Bohr and Kramers[90], following an original proposal by Slater[91] considered again Einstein's model where light was readily composed of corpuscles provided of energy and momentum. This time the photon particles were guided by virtual fields that were filling space even when atoms were in stationary states. Soon afterwards de Broglie[92] proposed a model, where he showed how a "guidance condition" for the particles could be derived from the wave equations.

In the double solution model, the energy content of the electromagnetic field is proportional to the square of the total field, which is the product of the external and the internal fields. The internal fields have a limited extension and decay exponentially outside of the range of the photon soliton. Therefore outside of the extension of the internal fields, the energy content of the electromagnetic field is practically zero, even when the external fields extend to infinity.

### 5.2.3. Single-Photon Interference

In the following pictures we are to compare the effects of first order interference and second order interference through entanglement. In these pictures we see a simulation on the trajectory of a photon according to the double solution model in the presence of first order (classical) interference:

$$\Psi(\vec{r}_1,t)=\left(1/\sqrt{2}\right)\left(\psi_a(\vec{r}_1,t)+\psi_b(\vec{r}_1,t)\right)$$

In the following geometry

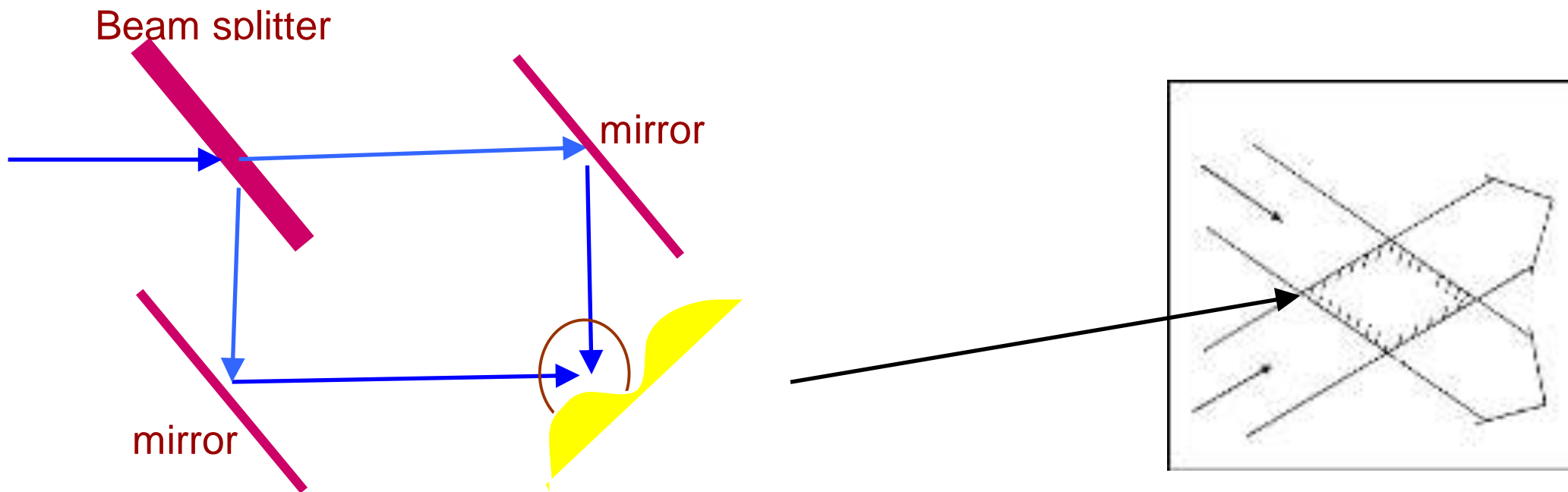

The simulation shows:

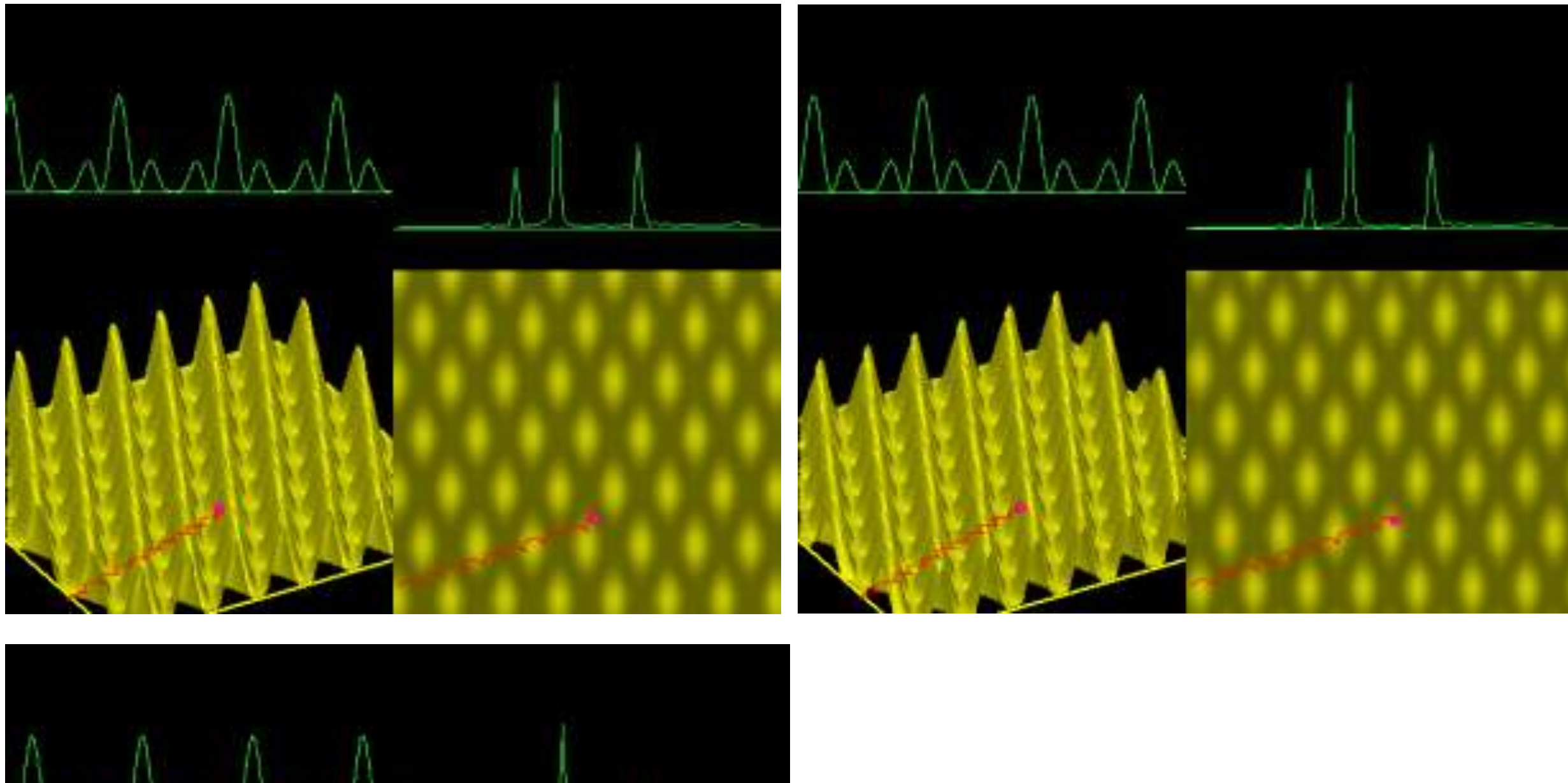

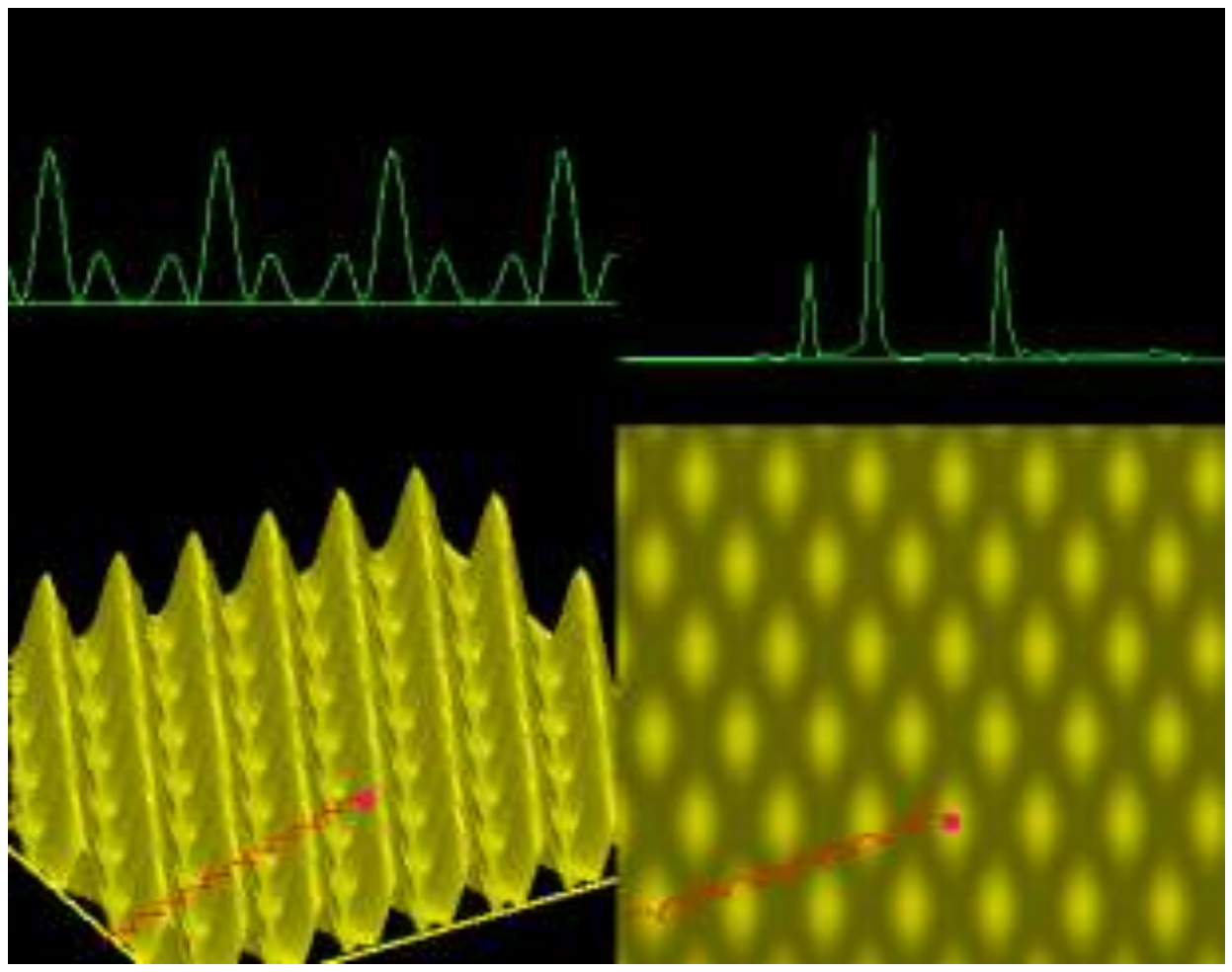

We can observe that the photon follows on the average to motion of the interference pattern from the lower left corner to the upper right corner. Moreover the pattern of the interference fringes remains constant in time.

The upper left corner shows the profile of the interference pattern following a line. On the upper right corner it is plotted the Fourier composition of the interference pattern following a line. We see that it remains unmodified over in time.

### 5.2.4. Bi-photon

The 2-photon wavefunction (diphoton) can be defined as[93]

$$\psi(\vec{r}_1, t_1;\ \vec{r}_2, t_2) \equiv \langle 0|E_2^+(\vec{r}_2, t_2)E_1^+(\vec{r}_1, t_1)|\Psi\rangle$$

This function doesn't need to be of the product form:

$$\Psi = f(r_1)\cdot g(r_2)$$

But it can be a function of the center of mass times a function of the relative coordinate like

$$\Psi(\tau_1, \tau_2) = \mathrm{v}(\tau_1 + \tau_2)\otimes u(\tau_1 - \tau_2) \qquad \tau_j = t_j - \frac{r_j}{c} \qquad j = 1,2$$

Which is an entangled state observed experimentally[93], exactly equal to what we have seen before for electron or atomic particles.

### 5.2.5. Entangled photons

In the next picture we display a simulation following the trajectories, as predicted by the double solution model for a pair of entangled photons according to the double solution model in the presence of second order interference:

$$\Psi(\vec{r}_1,\vec{r}_2,t)=\left(1/\sqrt{2}\right)\left[\psi_a(\vec{r}_1,t)\varphi_a(\vec{r}_2,t)+\psi_b(\vec{r}_1,t)\varphi_b(\vec{r}_2,t)\right]$$

in the following geometry:

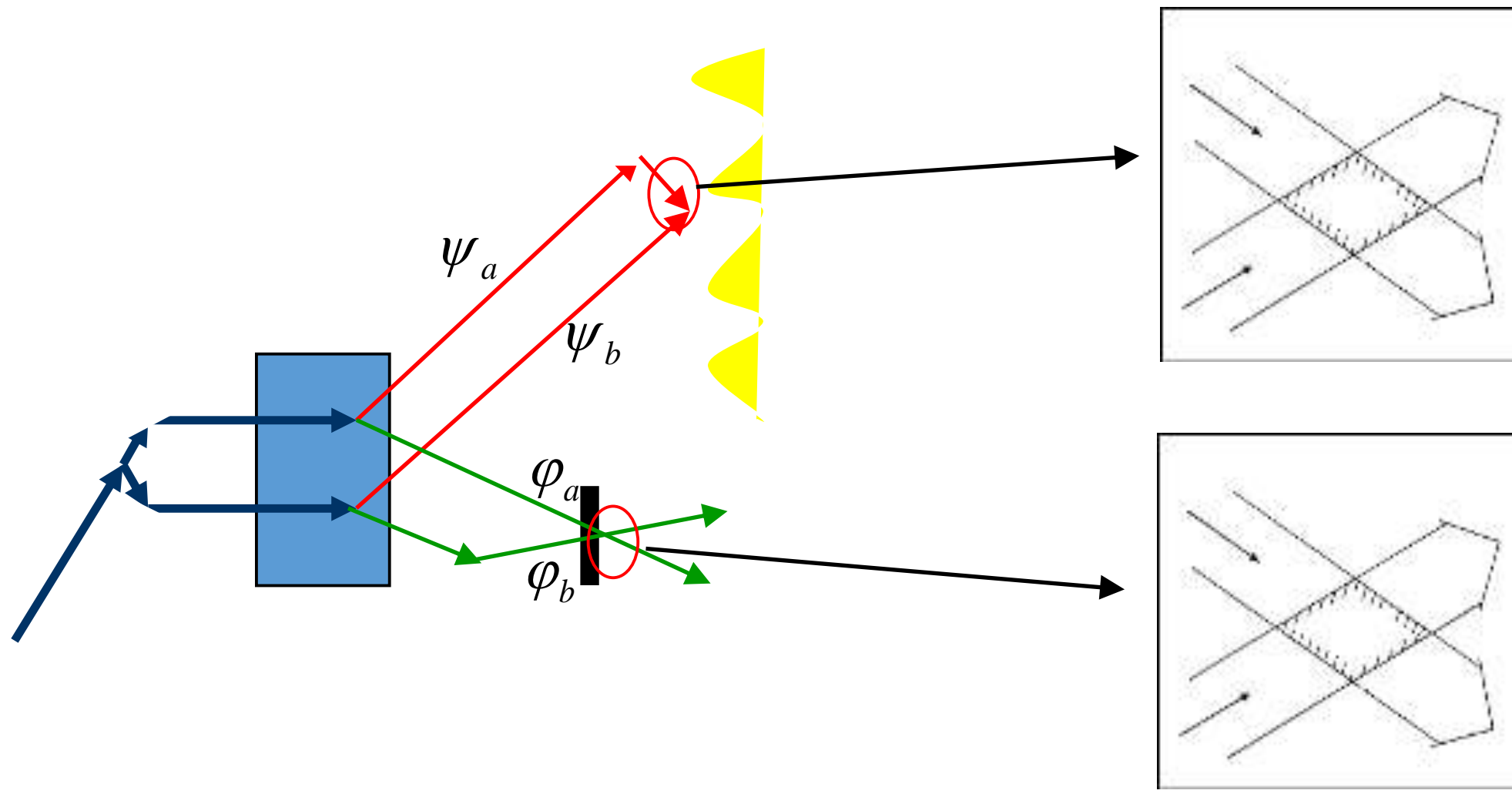

In the following pictures we show on the left the fields and trajectories from one photon, and on the right the fields and trajectories from the other photon:

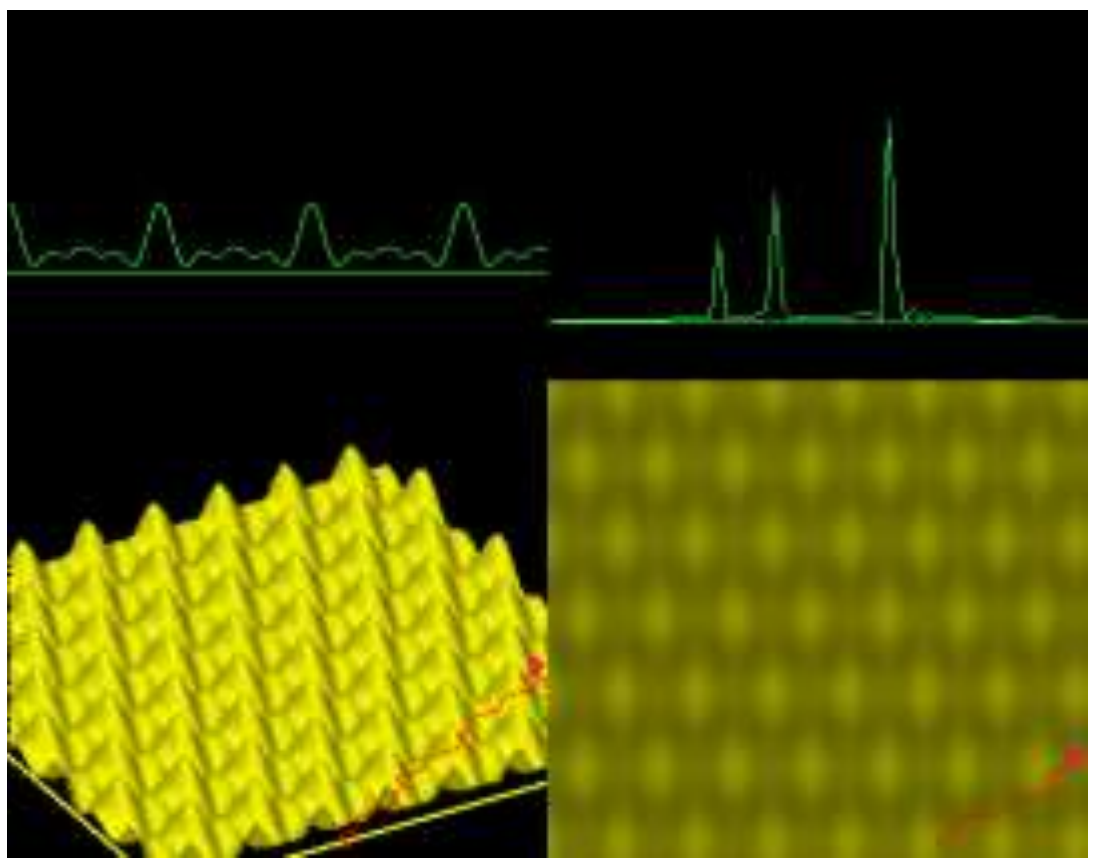

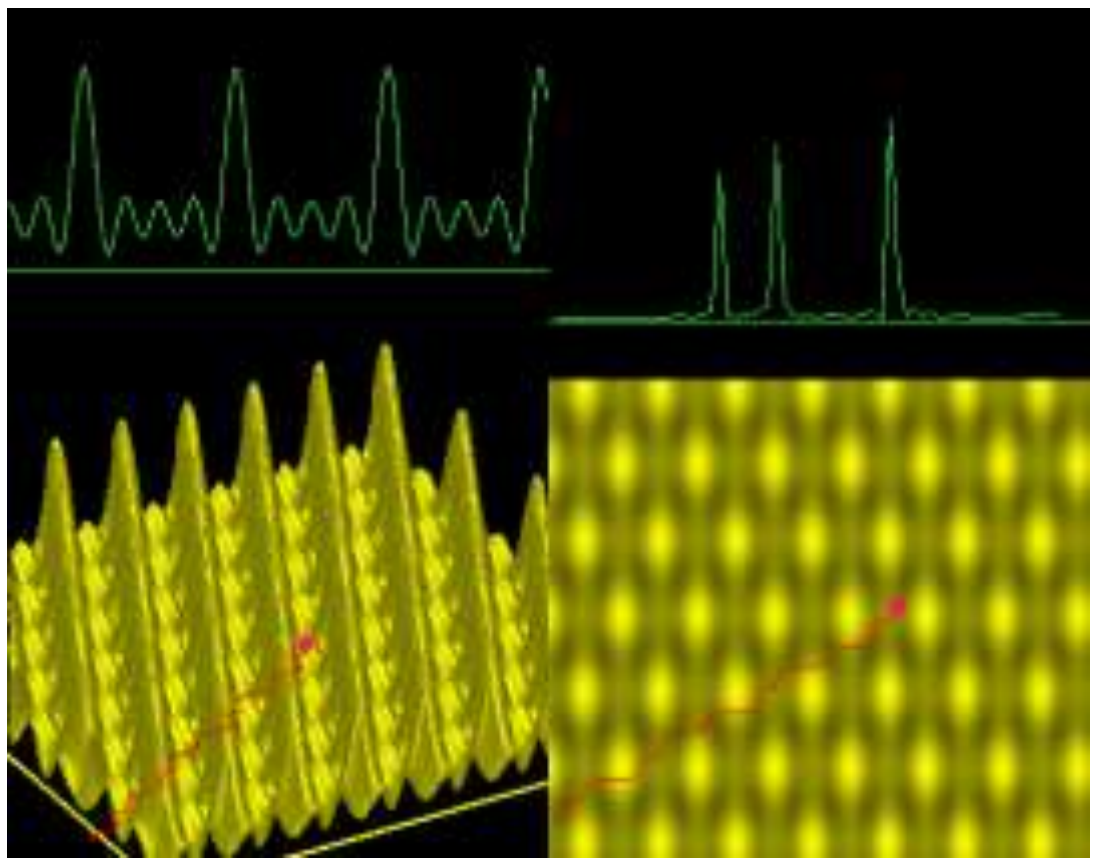

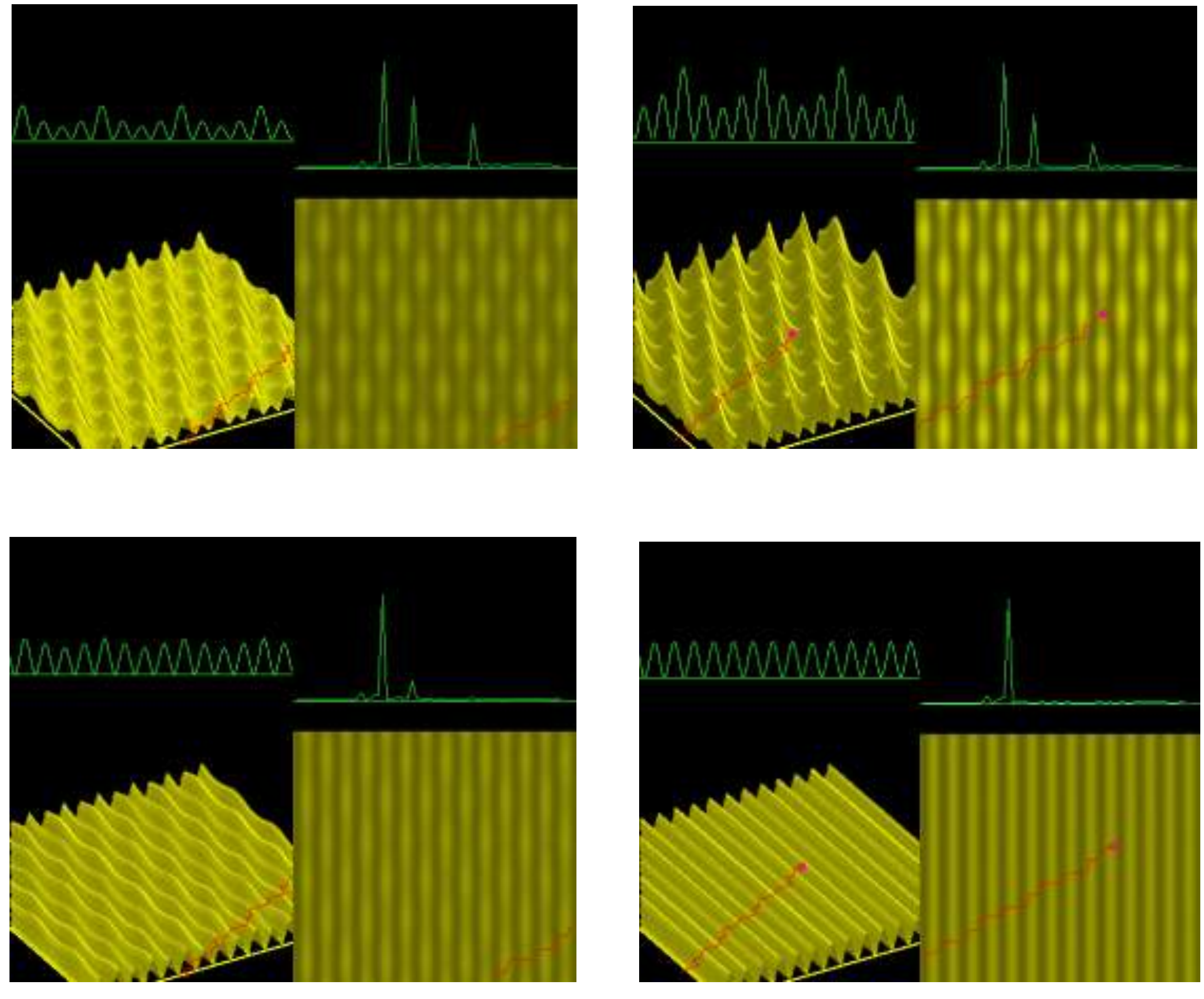

We can observe here, that entanglement modifies not only the trajectory of the photon, but also the full electromagnetic field interference pattern.

On the upper right corner it is plotted the Fourier composition of the interference pattern following a line. We see that in this case the patterns changes over time, however they remain synchronous for both photons, the slices shown on the upper left corner are different for both photons.

## 5.3. Spontaneous emission. Double Solution Model

The double solution model solves the question: what is the source of the electromagnetic fields, the wave or the particle? The equations of quantum electrodynamics take the electron wavefunction as the source for the electromagnetic fields. In unbounded states this wavefunction can reach interstellar dimensions, but all experiments with charged particles show that the source of the electric fields remains concentrated, as in the Geiger counter. The answer provided by our model shows that both answers are equivalent: as we have seen the electron field forms solitons with properties similar to a localized particle.

### 5.3.1. Near and far fields, virtual and real photons

In classical optics and electrodynamics, as well as in QED and Quantum Optics, the electromagnetic fields are typically classified in two groups as

| Free Fields | Attached Fields | Theoretical description |
|---|---|---|
| Far Fields | Near fields | **Classical Optics** |
| Real, on shell, massless Photons | Virtual, massive Photons | **QED** |
| Radiation, divergence free | Electrostatic, curl free | **Quantum Optics, Coulomb/Radiation Gauge** |
| Radiation | Electrostatic + Magnetostatic | **Quantum Optics, Pauli-Fierz-Kramers Gauge** |

It is customary to define a division between 'attached' and 'free' fields. In the Coulomb or radiation gauge the attached fields are the electrostatic fields derived from an electrostatic scalar potential while the free fields are derived from the transverse vector potential. In the Pauli-Fierz-Kramers gauge the attached fields include the magnetostatic Biot Savart fields. It is clear that these divisions are mathematically convenient, but non relativistic. In relativistic quantum descriptions the division is done in terms of 'virtual' and 'real' fields. The Einstein definition for the free fields is that they are 'on shell' fields. In other words, they are plane waves, and in this form they are used to calculate photon emission or scattering by Weisskopf emission theory. On the other hand, the classical expression for the fields in terms of charge and current densities, as shown by Jefimenko[94], is given by:

$$\vec{E} = \frac{1}{4\pi\varepsilon_0}\int\left[\left(\frac{[\rho]}{r^3} + \frac{1}{r^2 c}\left[\frac{\partial\rho}{\partial t}\right]\right)\vec{r} - \frac{1}{rc^2}\left[\frac{\partial\vec{j}}{\partial t}\right]\right] dV_{source}$$

$$\vec{B} = \frac{\mu_0}{4\pi}\int\left(\frac{[\vec{j}]}{r^2} + \frac{1}{rc}\left[\frac{\partial\vec{j}}{\partial t}\right]\right)\times\hat{r}\, dV_{source}$$

where squared brackets mean source retarded values at a time

$$t_{source} = t_{observer} - \frac{r}{c}$$

and

$$\vec{r} = \vec{r}_{observer} - \vec{r}_{source}$$

By making use of the continuity equation for the charge density

$$\frac{\partial\rho}{\partial t} + \vec{\nabla}\cdot\vec{j} = 0$$

the fields can be rewritten as Jefimenko described in page 21, eq. (2-2.12)

$$\vec{E} = \frac{1}{4\pi\varepsilon_0}\int\left(\underbrace{\frac{[\rho]\hat{r} - (1 - 2\hat{r}\hat{r}\,\cdot)\,[\vec{j}]/c}{r^2}}_{\substack{velocity\ and\ density\\ fields}} - \underbrace{\frac{(1 - \hat{r}\hat{r}\,\cdot)}{rc^2}\left[\frac{\partial\vec{j}}{\partial t}\right]}_{\substack{acceleration\\ fields}}\right) dV'$$

$$\vec{B} = \frac{\mu_0}{4\pi}\int\left(\underbrace{\frac{[\vec{j}]}{r^2}}_{\substack{velocity\\ fields}} + \underbrace{\frac{1}{rc}\left[\frac{\partial\vec{j}}{\partial t}\right]}_{\substack{acceleration\\ fields}}\right)\times\hat{r}\, dV'$$

where one can clearly identify two fields, one dependent on the charge and current densities and the other dependent on the source acceleration. The first term decays as $1/_{r^2}$ while the acceleration fields decay as $1/_{r}$. When explicitly calculated in the case of a point particle they reduce to the fields derived from the Liénard–Wiechert potentials:

$$\vec{E} = \frac{q}{4\pi\varepsilon_0}\left( \underbrace{\frac{\left(\hat{r}-\vec{\beta}\right)}{\gamma^2\left(1-\hat{r}\cdot\vec{\beta}\right)^3 r^2}}_{\substack{velocity\ and\ density\\ fields}} + \underbrace{\frac{\hat{r}\times\left(\hat{r}-\vec{\beta}\right)\times\dot{\vec{\beta}}}{c\left(1-\hat{r}\cdot\vec{\beta}\right)^3 r}}_{\substack{acceleration\\ fields}} \right)$$

$$\vec{B} = \frac{q\mu_0}{4\pi}\int\left( \underbrace{\frac{c\left(\vec{\beta}\times\hat{r}\right)}{\gamma^2\left(1-\hat{r}\cdot\vec{\beta}\right)^3 r^2}}_{\substack{velocity\\ fields}} + \underbrace{\frac{\hat{r}\times\hat{r}\times\left(\hat{r}-\vec{\beta}\right)\times\dot{\vec{\beta}}}{c\left(1-\hat{r}\cdot\vec{\beta}\right)^3 r}}_{\substack{acceleration\\ fields}} \right)\times\hat{r}\,dV' = \frac{\hat{r}}{c}\times\vec{E}$$

where all source quantities are at retarded time and

$$\vec{\beta} = \frac{\vec{\mathrm{v}}}{c} \qquad\qquad \gamma = \frac{1}{\sqrt{1-\beta^2}}$$

As Jackson[95] remarks, the fields are "naturally" separated in static or velocity fields and radiation or acceleration fields. In other words, he identifies the attached fields with the velocity, and the free fields with the radiation fields. The same fields can be expressed in full relativistic notation as done by Teitelboim[96], who analyzed the momentum-energy contents of the two types of fields. The total electromagnetic field tensor is given by

$$F^{\mu\nu} = F^{\mu\nu}_{velocity} + F^{\mu\nu}_{accel} = \underbrace{\frac{q}{\rho^2}\frac{[\mathrm{v}^\mu]r^\nu - [\mathrm{v}^\nu]r^\mu}{\rho}}_{\substack{velocity\ and\ density\\ fields}} + \underbrace{\frac{q}{\rho}\left(\frac{[a_\mu]r^\mu}{\rho}\frac{\left([\mathrm{v}^\mu]r^\nu - [\mathrm{v}^\nu]r^\mu\right)}{\rho} + \frac{[a^\mu]r^\nu - [a^\nu]r^\mu}{\rho}\right)}_{\substack{acceleration\\ fields}}$$

where

$$r^\mu = x^\mu_{obs} - \left[x^\mu_{source}\right] \qquad r^\mu r_\mu = 0 \qquad \rho = -[\mathrm{v}_\mu]r^\mu \qquad F^{01} = +E_z$$

This tensor follows the equations

$$\partial_\mu F^{\mu\nu} = -4\pi j^\nu \qquad \partial_\lambda F_{\mu\nu} + \partial_\mu F_{\nu\lambda} + \partial_\nu F_{\lambda\mu} = 0 \qquad j^\mu(x) = q\int_{-\infty}^{+\infty} d\tau\,\delta\big(x - z(\tau)\big)\mathrm{v}^\mu(\tau)$$

Teitelboim remarks that the separation is relativistically invariant because the two fields are tensors. They follow the equations:

$$\partial_\mu F^{\mu\nu}_{velocity}(x) = -\partial_\mu F^{\mu\nu}_{accel}(x) = \frac{2q}{\rho^3}\frac{[a_\mu]r^\mu}{\rho}r^\nu$$

$$\partial^\lambda F^{\mu\nu}_{vel,acc} + \partial^\mu F^{\nu\lambda}_{vel,acc} + \partial^\nu F^{\lambda\mu}_{vel,acc} = 0$$

Due to motion a field component perpendicular to the propagation vector is generated proportional to the charge and position of the charge that decreases as $\frac{1}{r^3}$. This component is more clearly seen in the dipole approximation. The fields generated by a charge following an oscillatory motion in the dipole approximation are:

$$\vec{E} = \left\{ \underbrace{\frac{[p](3(\hat{n}\cdot\hat{r})\hat{r}-\hat{n})}{r^3} + \frac{[q]\hat{r} + [\dot{p}](3(\hat{n}\cdot\hat{r})\hat{r}-\hat{n})}{cr^2}}_{\substack{velocity\ and\ density\\ fields}} + \underbrace{\frac{[\ddot{p}]\big((\hat{n}\cdot\hat{r})\hat{r}-\hat{n}\big)}{c^2 r}}_{\substack{acceleration\\ fields}} \right\}$$

$$\vec{B} = \left\{ \underbrace{\frac{[\dot{p}]}{cr^2}}_{\substack{velocity\\ fields}} + \underbrace{\frac{[\ddot{p}]}{c^2 r}}_{\substack{acceleration\\ fields}} \right\} (\hat{n}\times\hat{r})$$

**Eq 160**

We observe that the photon speed as given by the ratio between the Poynting vector and the energy density

$$\vec{v}_{energy} = \frac{\vec{S}}{U} = \frac{\vec{S}}{\vec{A}^{*}\cdot\vec{A}}$$

**Eq 161**

obtained from Eq 160 for the fields, approaches zero in the proximity of the source and reaches asymptotically the nominal speed of light in the radiation fields[97]. However we have seen in Chapter 3 on the Photon Solitons, that a residual index of refraction larger than 1 is present in the entire universe, due to the very thin but always finite background plasma density and eventually due also to the big bang background radiation field.

### 5.3.2. Massless and Massive Photons

We have seen in **Chapter 3: Massless Photons** that the double solution model decompose the electromagnetic field as the product of an internal field $E_i$ and an external field $E_e$: $E = E_i E_e$.

The external field fulfills the equation:

$$\nabla^2 E_e - \frac{1}{c^2}\frac{\partial^2 E_e}{\partial t^2} = 0$$

This equation defines a massless dispersion relation. The internal part fulfills a nonlinear equation due to the presence of vacuum polarization effects. The solution to this equation is of the soliton type, decaying exponentially with distance from the center of the photon soliton. This solution can be applied to free photons, in the far fields.

In particle scattering processes, where two charges approach each other and overlap their near fields, there is a definitive massive momentum exchange. This fact can be handled by the massive photons that we have mentioned in **Chapter 3 Massive Photons** where the external field, defining the particle dispersion relation fulfills the equation:

$$\nabla^2 E_e - \frac{1}{c^2}\frac{\partial^2 E_e}{\partial t^2} - \beta E_e = 0$$

Here the parameter $\beta$ can be understood as an effective mass term fully equivalent to a scalar potential energy. Depending on the boundary and entanglement conditions the external equation will describe particles with different masses and momenta. Which one is realized will be determined by the conservation of energy-momentum for the entire system through entanglement.

This is related to the description of scattering in terms of massive virtual photons, as given in QED textbooks. In the case of an electron scattered by a nucleus, the nucleus in first approximation will remain at rest, in the absence of bremsstrahlung, a massive photon will be exchanged between the nucleus and the electron, the electron and the photon will follow entangled trajectories. In the case of two electrons scattering, again neglecting bremsstrahlung, both electrons will emit a massive photon each, and each photon will be eventually absorbed by the other electron. All the two electrons and the two photons will follow entangled trajectories. This explains the counter-intuitive concept presented in QED textbooks that the photon is in a superposition state that includes components emitted by each one of the electrons taking part in the interaction, and possibly moving back in time. Additionally the mass of each photon can be variable to maintain the entire energy momentum conservation.

It is customary to solve the problem by providing the initial and final states. This is totally equivalent to providing additional parameters in the initial state. For example in scattering specifying the final scattering direction is equivalent to provide the initial impact parameter, they are related one to one. In traditional quantum mechanics the initial impact parameter is not available because the internal soliton is not defined. Therefore it is replaced by the final scattering angle, with no other physical interpretation other than a technical prescription.

### 5.3.3. Dipole decay revisited:

#### 5.3.3.1. Classical Model

Assuming a point-dipole approximation for an atomic system oscillating and radiating energy in the form of electromagnetic radiation, the solution of the problem shows the dipole starts oscillating at a given amplitude and decays exponentially with time, as its energy is depleted. The dipole energy is transferred continuously to the field, and through radiation reaction, the dipole suffers a force producing a gradual slowing down of the oscillation amplitude. The dipole amplitude goes as

$$d(t) = e^{-t/\Gamma} \cos \omega t$$

**Eq 162**

with $\Gamma = \tau\omega^2$ is the decay time constant and

$$\tau = \frac{2}{3}\frac{e^2}{mc^3}$$

is the time required for light to travel a distance equal to the classical electron radius. In first approximation the far field goes as

$$\varphi(r,t) \cong e^{-\left(t-\frac{r}{c}\right)/\Gamma} \frac{\cos \omega \left(t - \frac{r}{c}\right)}{kr} \sin\theta = \frac{d\left(t - \frac{kr}{\omega}\right)}{kr} \sin\theta$$

where

$$kr = \frac{\omega r}{c} \qquad \rightarrow \qquad \frac{r}{c} = \frac{kr}{\omega}$$

and $\theta$ is the angle between the dipole and the radius vector of the observation point. However we emphasize here that this is just a first order approximation, and in order to get a more complete understanding of the emission process we need to go one step further. H. A Lorentz[98] in 1909 developed a model where the electron occupied a finite volume in space. There he was able to show that the decay in the energy content of the electron oscillation while radiating the electromagnetic field could be explained as the mutual retarded action of the different parts of the electron distribution. In the dipole approximation the electron has no volume, possess no size in space, and this 'self-force' cannot be calculated in a straightforward way[99]. The result of the calculation however can be described as a delay of the electromagnetic respect to the oscillating dipole. This delay can be found by energy conservation arguments also under the point dipole approximation as follows:

From classical electromagnetic energy, the power transferred to infinity by an accelerated charged particle is given by Larmor power formula[100]

$$P_{Rad} = \frac{2}{3}\frac{e^2}{c^3}\dot{\mathrm{v}}^2$$

where $\dot{\mathrm{v}}$ is the particle acceleration. By energy conservation this power has to be equal to the long-term work per unit time done by the oscillating charge against the field, given by

$$P_{Work} = \mathrm{v}F = -e\mathrm{v}\mathrm{E} = \frac{2}{3}\frac{e^2}{c^3}\dot{\mathrm{v}}^2$$

where $\mathrm{E}$ is the electric field amplitude.

Let's take

$$x = x_0 \cos \omega_0 t \qquad \dot{x} = \mathrm{v} = -\omega_0 x_0 \sin \omega_0 t \qquad \ddot{x} = \dot{\mathrm{v}} = -\omega_0^2 x_0 \cos \omega_0 t$$

If we take in first approximation

$$-eE = F = m\dot{\mathrm{v}}(\omega_0 t + \alpha)$$

where $\alpha$ is the phase difference between the field and the dipole oscillation.

$$\dot{\mathrm{v}}(\omega_0 t + \alpha) = -\omega_0^2 x_0 \cos(\omega_0 t + \alpha) = -\omega_0^2 x_0 \{\cos(\omega_0 t)\cos(\alpha) - \sin(\omega_0 t)\sin(\alpha)\}$$

We find

$$\int_0^T \sin \omega_0 t \cos \omega_0 t \, dt = 0$$

$$\int_0^T \sin \omega_0 t \sin \omega_0 t \, dt = \frac{1}{2}T$$

The average power lost per period is then given by

$$P_{work} = \frac{m}{T}\int_0^T \dot{x}\dot{\mathrm{v}}dt = \frac{m}{T}\int_0^T -\omega_0 x_0 \sin \omega_0 t \, \omega_0^2 x_0 \sin(\omega_0 t)\sin(\alpha)\, dt = -\frac{m}{T}\omega_0^3 x_0^2 \sin(\alpha)\int_0^T \sin \omega_0 t \sin(\omega_0 t)\, dt$$

$$= -m\omega_0^3 x_0^2 \sin(\alpha)\frac{T}{2T} = -\frac{m\omega_0^3 x_0^2 \sin(\alpha)}{2}$$

the radiated power to infinity is given by

$$P_{rad} = -\frac{2}{3}\frac{e^2}{c^3}\dot{\mathrm{v}}^2 \cong -\frac{1}{3}\frac{e^2}{c^3}\omega_0^4 x_0^2$$

by equating both powers we get

$$P_{work} = P_{rad} \qquad \rightarrow \qquad \frac{m\omega_0^3 x_0^2 \sin\alpha}{2} = \frac{1}{3}\frac{e^2}{c^3}\omega_0^4 x_0^2$$

allowing for the determination of the angle $\alpha$:

$$\sin\alpha = \frac{2}{3}\frac{e^2}{mc^3}\omega_0 = \frac{r_e}{2\pi\lambda} = \frac{\tau}{2\pi T}$$

it can be verified that $\sin\alpha \cong \tan\alpha \cong \alpha \ll 1$ for optical frequencies, with $\tau$ given as before.

The fractional average energy emitted per unit time $\frac{dE/dt}{E}$is $\sin\alpha/_T = \tau\omega_0^2 = \Gamma$ as required by Eq 162. Also a frequency shift can be found given by

$$\Delta\omega \approx \frac{1}{2}\tau^2\omega_0^3$$

Therefore the next order approximation is given by

$$d(t) = e^{-t/_\Gamma}\cos(\omega + \Delta\omega)t$$

and

$$\varphi(r,t) \cong \frac{d(t - cr + \alpha)}{(k + \Delta k)r}\sin\theta$$

We remark here that the decay time constant $\Gamma$ as well as the dephasing between the field and dipole depend exclusively on the frequency and not on the amplitude of the oscillation. The radiated power on the other hand depends quadratically on the amplitude of the oscillation.

From the previous physical description it is clear that the line width is just the Fourier decomposition of the temporal evolution of the field amplitude, and consequently the coupling with the continuum of modes is an effect, and not a cause for radiation emission. This last point can be made clearer from the following considerations. Suppose an external force provides a compensation for the radiation reaction forces, and the decay doesn't occur at all. In this case the motion of the particle and the electromagnetic fields are strictly monochromatic, the line width is null, but the radiation force still exists and energy is transferred to the electromagnetic field and propagates toward infinity. The coupling with the continuum plays no role in this case. As long as the electric field amplitude decays with distance as $E \sim \frac{1}{r}$ energy will be delivered to infinite. However if the spatial decay occurs faster with distance, the energy will remain confined inside a given volume. When energy propagates to infinity, it is clear that its source is the external force used to keep the amplitude of the dipole oscillation constant.

#### 5.3.3.2. Second Quantization Model:

The equivalent problem in Quantum Mechanics corresponds to spontaneous emission, and was first solved by Weisskopf in 1930[101], where the decay is induced by the coupling with the continuum of states. The final result from Weisskopf model is the expression for the spontaneous emission decay time given by

$$A_{21} \equiv 2\gamma \equiv \frac{1}{\tau_R} = \frac{e^2\omega_0^3 D_{12}^2}{3\pi\epsilon_0\hbar c^3}$$

for a transition between an upper state 2 and a lower state 1, with a dipole matrix element $D_{12}$, which for an hydrogen transition is of the order of Bohr's radius. If we find the power emitted considering that the total energy to be emitted is $\hbar\omega_0$, we obtain

$$P_{21} = \hbar\omega_0 A_{21} = \frac{e^2 \omega_0^4 D_{12}^2}{3\pi\epsilon_0 c^3}$$

If we compare this value with the classical expression for the emitted power we find

$$P_{21} = \frac{e^2 \omega_0^4 D_{12}^2}{3\pi\epsilon_0 c^3} = \frac{1}{3\pi} \frac{e^2}{\epsilon_0 c^3} \dot{v}^2$$

which is completely compatible with the classical result as long as the value for the oscillation amplitude is given by

$$X \sim \sqrt{\frac{1}{2\pi}} D_{12} \cong a_0$$

with $a_0$ the Bohr's radius[102]. We see here that this quantum solution is not modifying the differential equations or the solutions provided by classical electrodynamics, they only provide the frequencies at which atomic transitions takes place, and what energy is available at those frequencies. We can understand why quantization cannot modify this value, because it is readily a consequence of the amount of energy delivered to infinity by the radiation field. From this fact one can deduce that second quantization does not provide a different model to classical electromagnetism for emission as noticed by Mandel et al[103]. It is possible to argue that, at least in respect to the spontaneous emission case the classical and fully quantum mechanical models are just two distinct mathematical models describing the same set of solutions. Something similar to the discussion that took place in the XVIII century respect to the solution of mechanical problems in terms of vectorial forces or in terms of energy fluid densities[104].

This full analogy between the classical and quantum theories can be better appreciated when solving the problem in the Heisenberg representation, particularly in the case of the Source-Field approach[105]. Here it can be fully verified that in first approximation the equations fulfilled by the dipole and electromagnetic fields, as well as their respective solutions are the same as in the classical case. The interpretation is however different. It is assumed now that the continuous functions are not providing a description of physical events taking place in real space-time, but just a statistical distributions for the probabilities of the events to occur. In this sense the spontaneous emission decay time $A_{21}$ is interpreted as the probability per unit time for the atom to suffer a transition from an upper to a lower state, or for a photon to be measured after being emitted in that transition.

#### 5.3.3.3. Dipole with constant amplitude

We have mentioned previously the case where the amplitude of the dipole oscillation is maintained constant by the presence of an external force that counteracts the effects of radiation reaction. In this case the energy content of the dipole is kept constant by some external source. The dipole amplitude is

$$d(t) = D_0 \cos \omega t$$

where the amplitude $D_0$ is a constant independent of time. Another case where the amplitude of oscillation can remain constant on the average is when the radiated electromagnetic fields cannot deliver its energy to infinity. This can be achieved when the fields are confined inside a finite volume of space or decay to infinity faster than $E \propto \frac{1}{r}$. Because the amplitude of the oscillation remains constant, the oscillation is monochromatic, and no coupling to the continuum of modes of the electromagnetic field takes place. The line width is null. In this case the total energy of the system can be split in a steady state between the dipole oscillation and the field. This problem can be solved in a classical or in a fully quantum formalism. In this last case, steady solutions to the dipole equations can be obtained just by diagonalization of the matrix elements[106], or by setting to zero the time dependence of the coefficients of the density matrix[107]. These solutions have been found by Jaynes and Cummings in 1963[108]; they consider a radiation field enclosed by a cylindrical reservoir of perfect reflecting walls. Similarly Cohen Tannoudji et al considered the radiation field in a multimode coherent state representing a ring laser where the beam follows a closed path with the help of mirrors[109]. This type of models gave rise to the concept of 'dressed atoms' or 'dressed states' [110] consisting of a steady-state dipole oscillating at a given frequency, immersed in the same field it produces, which decays with distance faster than $E \propto \frac{1}{r}$ and therefore keeps the energy confined[111]. This method allows to find the eigenenergies and eigenstates of the system, other methods involve the use of canonical transformations to remove the interaction terms from the Hamiltonian[112]. In all cases the solutions depend on the geometry of the device, involving the amplitude of the electromagnetic wave and the volume of the enclosure. In general in the presence of single photons, the eigenstates are only slightly different from the original eigenstates in the absence of radiation. An upper state with no photon and a lower state with a single photon present are possible. A small oscillation is present in both cases, but with an amplitude of oscillation of the order of the atomic radius multiplied by powers of the fine atomic constant. The energy is not depleted from the dipole system because no net physical work is performed by the dipole on the field. Anyway the possible radiation emission that could result from this oscillation is negligible during the time of spontaneous emission.

We want to mention here that a third possibility is presented when considering nonlinear equations for the electromagnetic field and the energy remains confined inside a finite volume by the presence of solitons.

### 5.3.4. Photon soliton model

From single-events spontaneous emission processes observed experimentally[113], it is clear that the transition is not a gradual process, but it is discontinuous. The atom remains in the upper state for a while, and then suffers an abrupt transition to a lower state emitting a photon, in a process reminiscent of Bohr 'quantum jumps' model. The statistical distribution of the decay processes corresponds to a constant decay probability per unit time, reproducing the continuous exponential decay characteristic of classical electrodynamics radiation-reaction damping. The exponential decay can be observed also in first order temporal coherence measurements, where the exponential decay can be explained by a decay in amplitude of the wave, or by a random change with constant probability per unit time in the phase of the wave.

Practically no theoretical work has been pursued trying to reproduce the single-atom behavior, although modeling works dedicated to the possibility to observe jumps and simulations have been performed[114]. It is clear from the experimental behavior that a model trying to explain the jump, must rely in the existence of quasi-stationary states, following the original idea of Niels Bohr, where the atomic system should spend most of the time before and after the physical transition. We can argue that the physical transition may occur due to nonlinear effects present in the wave evolution; this is the path we are going to follow.

As we have seen in Eq 161 using the fields from Eq 160 we obtain after a brief calculation that the soliton propagation velocity will reach asymptotically the nominal speed of light as the distance between the soliton and the emitting dipole is much larger than the field wavelength.

Our model can describe the emission process as follows. We start with the dipole in the upper steady state described by some of the above models, with the presence of a minuscule oscillation. We note that once we have included the self-interactions in the description of the electromagnetic field, linear solutions to the Maxwell equations are not more allowed. It will always be possible the appearance of a small volume in space where the amplitude of the wave is large enough, for the self-interaction to be larger than any external interaction with the sources, specially closed to the real dipole where the local fields are dominant. The value of the electric field at the source is of the order of 1/100 times the critical field value. In this region a small soliton will be formed that eventually will agglomerate the energy content of the field after been accelerated to a speed close to the nominal speed of light. We remember that this soliton is a product of the classical linear solution to the external sources and boundary conditions, multiplied by the amplitude of the nonlinear part, solution to the self-interaction equation. Being close to the dipole source and following the velocity field provided by the Poynting vector, it will initially move slowly compared with the speed of light in vacuum. Depending on the original position of the soliton, it will take a time of the order of the spontaneous emission constant to start a trajectory that eventually will exit the local fields and reach the radiation zone. In doing so, it will increase his speed approaching the speed of light. From the Lorentz transformations for the fields,

$$E'_{\parallel} = E_{\parallel}$$
$$B'_{\parallel} = B_{\parallel}$$
$$E'_{\perp} = \gamma\left(E_{\perp} + \frac{v}{c}B_{\perp}\right)$$
$$B'_{\perp} = \gamma\left(B_{\perp} - \frac{v}{c}E_{\perp}\right)$$

where $\gamma = \frac{1}{\sqrt{1-\frac{v^2}{c^2}}}$.

The energy $\varepsilon$ of a soliton-type solution moving in the x-direction, with the field pointing in the z-direction will scale as

$$\varepsilon' \propto \left({E'_z}^2 + {B'_z}^2\right)\Delta x'\Delta y'\Delta z'$$

where $\Delta x'\Delta y'$ remains constant at the same value they had at rest, while

$$\Delta z' = \frac{\Delta z}{\gamma}$$

$$\varepsilon'\left(\frac{\mathrm{v}}{c}\right) \propto \gamma^2\left(E_z^2 + \left(\frac{\mathrm{v}}{c}\right)^2 E_z^2\right)\Delta x\Delta y\frac{\Delta z}{\gamma} = \gamma E_z^2\left(1 + \left(\frac{\mathrm{v}}{c}\right)^2\right)\Delta x\Delta y\Delta z \approx \gamma\varepsilon = \frac{\varepsilon}{\sqrt{1-\frac{\mathrm{v}^2}{c^2}}} \underbrace{\rightarrow}_{\mathrm{v}\rightarrow c} \infty$$

where $\varepsilon$ is the energy at rest. We recognize that the energy contents in the soliton increases with speed even diverging when reaching the speed of light.

Related to the center of mass coordinate, at the moment of emission, the atom-photon momentum and energy must be conserved inside the limits of the uncertainty principle. This is translated into the following energy equation

$$|a|^2\hbar\omega + \varepsilon'\left(\frac{\mathrm{v}}{c}\right) = \hbar\omega$$

where *a* is the amplitude for the upper atomic level and $\omega$ the transition frequency. It is assumed that initially the atom is in the upper state and the photon soliton is at rest in the neighborhood of the emitting electron. This equation is enforced through entanglement between the photon and the atomic system. As the photon soliton approaches the boundary between the near and far fields, it approaches gradually the speed of light *c.* In order to conserve energy, the amplitude *a* will decrease from its initial value 1 to a final value 0. At the same time the soliton will accelerate and increase its energy until reaching the maximum available value of $\hbar\omega$. Once the soliton is in the far fields, the atomic system will reach the lower transition level, will stop oscillating, and the photon soliton will reach its final speed, very close to the nominal speed of light[115]. Once the soliton reaches a distance of the order of one wavelength from the atomic system, it will reaches its final state as a real free photon. The transition from slow moving-low energy soliton to full developed photon traveling at the speed of light may take place in just one or two oscillation periods. After this, the emission by spontaneous emission has occurred, a real photon has been emitted.

The absorption process is the inverse of the emission process. Through entanglement the energy is transferred from the photon to the atom as the soliton slows down in the near fields.

## 5.4. Radiation Scattering and Resonance fluorescence; two level model

As another example of the quantum potential action in the interaction of electrons and electromagnetic fields, we are going to consider a classic problem in quantum optics: a two level atom irradiated by an electromagnetic field[116].

The two level model of quantum optics is based on the density matrix formalism, where the non-diagonal elements represent the atomic current operators, source of the electromagnetic fields.

### 5.4.1. Harmonic oscillator subject to an electromagnetic field

The classical equation of motion is

$$m\ddot{x}_a + m\omega_0^2 x_a = eE(\boldsymbol{r_a}t)$$

On the other hand the equations obtained from the Bloch equations in the quantum optics model can be written as[117]:

$$m\ddot{x}_a + m\omega_0^2 x_a = eE\alpha\left(\frac{1}{2} - \frac{\mathfrak{En}}{\hbar\omega_0}\right) = eE - \underbrace{\alpha\left(eE\frac{1}{2} + \frac{\mathfrak{En}}{\hbar\omega_0}\right)}_{\substack{Quantum\\ Force}}$$

$$\frac{d\mathfrak{En}}{dt} = eE\dot{x}_a$$

In this case however, the oscillator force represented by $\omega_0^2$ is the quantum force, and is not a static spring like force, but a time dependent oscillatory force generated by the superposition of the two oscillating waves at the lower and upper transition levels. We identify an additional force term of quantum origin. Here $\mathfrak{En}$ is the energy of the system and $\hbar\omega_0$ the energy difference between the two levels. Where we see explicitly the action of the quantum potential.

Of course the interpretation given in quantum optics to the different symbols (letters) in the last equations is in terms of populations and non-diagonal matrix elements and field operators, but at the end of the day, the equation obtained is very similar to the classical one, and we wish to make a comparison in terms of classical objects with an additional contribution from the quantum force.

The solutions to the Optical Bloch equations are represented typically in the so called Bloch sphere. Typical solutions follow an 8-shaped orbit on the surface of the sphere[117].

### 5.4.2. Damped oscillator

The classical equation of motion for the damped harmonic oscillator is[117]

$$\ddot{x} - \frac{2}{\tau_0}\dot{x} + \omega_a^2 x = \frac{e}{m}E$$

With

$$E = 2\varepsilon\cos\omega t$$

Assume

$$x = x_0\left[u\cos\omega t - \mathrm{v}\sin\omega t\right]$$

The equation reduces to

$$\underbrace{\dot{\mathrm{v}}+\frac{u\omega}{2}}_{acceleration}=\underbrace{\frac{u}{2\omega}\omega_a^2}_{restoring\ force}\underbrace{-\frac{\mathrm{v}}{\tau_0}+\frac{\dot{u}}{\omega\tau_0}}_{dissipation}-\underbrace{\frac{e}{m\omega x_0}}_{driving\ force}\varepsilon$$

This forces can be represented by a vector diagram:

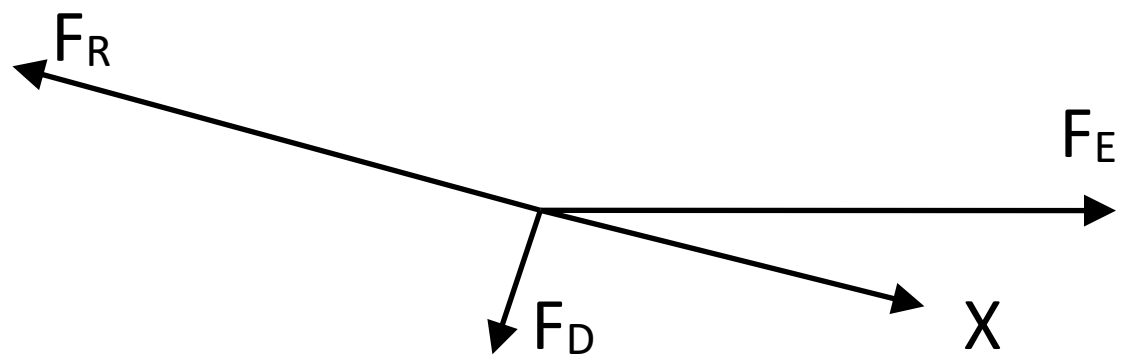

X = resultant force = mass x acceleration

$F_R$ = restoring force

$F_D$ = dissipation

$F_E$ = driving force

In equilibrium all forces cancel. The atom absorbs energy from the electromagnetic field which is reradiated.

The Bloch equations in the quantum optics model can be reduced now to the following equation:

$$\underbrace{\dot{\mathrm{v}}+\frac{u\omega}{2}}_{acceleration}=\underbrace{\frac{u}{2\omega}\omega_a^2}_{restoring\ force}\underbrace{-\frac{\mathrm{v}}{\tau_0}+\frac{\dot{u}}{\omega\tau_0}}_{dissipation}-\underbrace{\frac{e}{m\omega x_0}}_{driving\ force}\varepsilon+\underbrace{\frac{2we}{m\omega x_0}}_{Quantum\ force}\varepsilon=\underbrace{\frac{u}{2\omega}\omega_a^2}_{restoring\ force}\underbrace{-\frac{\mathrm{v}}{\tau_0}+\frac{\dot{u}}{\omega\tau_0}}_{dissipation}-\underbrace{(1-2w)}_{dephasing}\underbrace{\frac{e}{m\omega x_0}}_{driving\ force}\varepsilon$$

Which is similar to the classical equation, with the presence of an additional force of quantum origin. Here *w* is the population of the upper level. We see that the quantum force opposes the driving force and is proportional to twice the upper population (where $w=1/2$). This quantum force act as an additional dephasing between the electron oscillation and the external field. When the upper population is zero (the atom in the ground state) the quantum force is null, and then start to grow up proportional to the population of the excited level. If the atom is fully in the excited state, the total force has the same magnitude as the driving force but opposite direction. While $w<1/2$ the forces diagram is similar to the classical one, the total force is smaller than the classical force. It can be represented by:

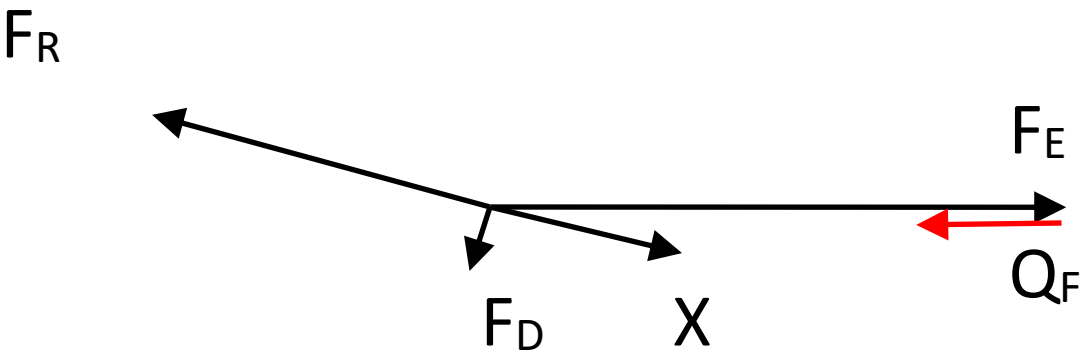

The symbols are the same as before plus

$Q_F$ = quantum force

So, the diagram simply rescales proportional to the new total force.

When the atom is in the upper state we have and the position, restoring force and dissipation force change direction. The atom gives energy to the electromagnetic field, is the laser effect.

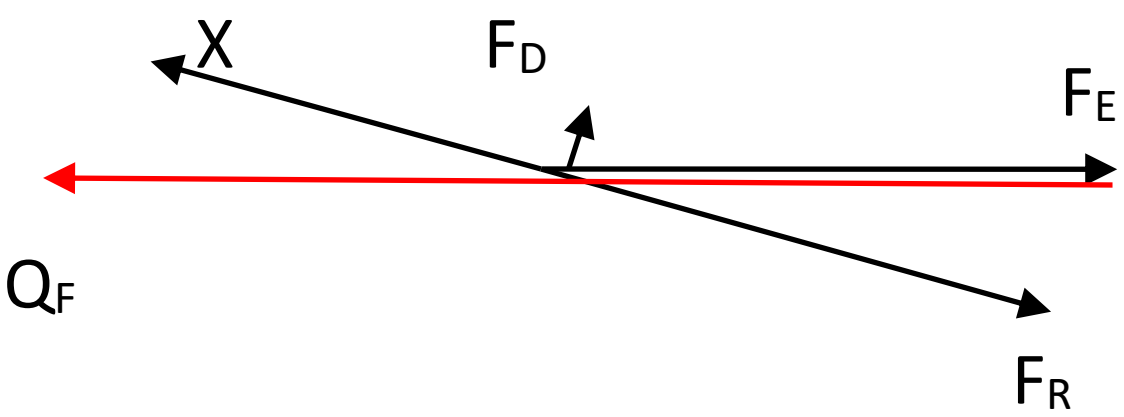

### 5.4.3. Photon dynamics during scattering

In the previous Sections we have analyzed the motion of the atomic dipole moment during the scattering process. The next question is how behaves the electromagnetic fields and the photons present in it. We will analyzed that from the point of view of the double solution model.

The critical point in order to apply the double solution model is to understand and clearly define the state basis in which the solutions to the quantum equations are sought.

A typical system for a two-level atomic system in the presence of a monochromatic incident beam is traditionally described by

$$|\psi(t)\rangle = a_{init}(t)|gr, n, 0k\rangle + a_{int}(t)|exc, n-1{,}0k\rangle + a_{final}(t)|gr, n-1{,}1k\rangle$$

Here

$$|gr, n, 0k\rangle = |gr\rangle|n\,\rangle|0k\rangle$$

is the state where the atomic system is in the ground state, the initial incident bean has *n* photons and there is no scattered photons.

$$|exc, n-1{,}0k\rangle = |exc\rangle|n-1\rangle|0k\rangle$$

is the state where the atomic system is in the excited state, the initial incident bean has *n-1* photons and there is no scattered photons.

$|gr, n-1, 1k\rangle = |gr\rangle|n-1\rangle|1k\rangle$

is the state where the atomic system is in the ground state, the initial incident bean has *n-1* photons and there is one scattered photon in the mode $k$. It this model the atomic states $|gr > \; and \; |exc >$ are generally taken in first approximation as the atomic states in the absence of radiation, the incident beam state is taken to be a photon-number or coherent plane wave corresponding to *n* photons, and the photon scattered *k* state as a plane wave with wave vector *k*. The first characteristic one can notice with this model is the apparent absence of any dynamical behavior. In the classical solution to light scattering one can observe an oscillating dipole, and a coherent superposition of an incident and scattered waves, all oscillating at the common frequency $\omega$. It is obvious that the classical solution conserves energy, the energy content of the atomic polarization energy and the scattered energy are taken from the shadow left by the atom on the incident beam. A similar behavior can be observed in the semiclassical description obtained by using a first quantization-two-level atomic system and classical radiation fields. On the other hand, the second quantization description is composed of three apparently 'static', not coherent states. One can argue that this last description does not correspond to experimental measurements. If we consider those frequencies where the dynamics of the scattering system can be observed, it can be recognized that all three elements, the incident beam, the scattered radiation, and the scattering center, are all oscillating coherently, as described by the classical solution. Additionally, the prescription of the scattered photon states as plane waves with wave vector *k* seems extremely arbitrary, when we consider that at optical frequencies, the scattered wave is in first approximation a spherical wave, oscillating at the same frequency as the incident one. Plane waves would play merely the role of Fourier components of the final field.

Attempts have been made to provide a Quantum Optics model closer to the classical solutions by working with different gauges and performing different canonical transformations, like the Pauli-Kramers-Fierz or the Power-Zienau-Wooley transformations, as in the case of spontaneous emission[118,119,120].

Traditionally radiation field states are not associated with a Fock space and particle coordinates as massive bosons are. Experimentally Geiger counters are an irrefutable proof that gamma rays are confined objects. Also our numerical calculations with the de Broglie's double solution model with self-interactions shows the formation of soliton structures that are a good approximation to photon particles. Therefore we see no principle objections to associate particle coordinates in Fock space to the state description of the electromagnetic field in second quantization. Once this is done, the definition of a radiation state with a given number of photons can be described by a set of functions depending on a fix number of coordinates, as done for massive bosons in first quantization. This can be applied to a whole scattering event, including both incident and scattered fields as both can be assigned to the same photon coordinate. Moreover, if the atom-photon system has a constant amount of energy content, they can be considered an eigenenergy state and be used as the basis states to consider in the scattering problem.

At this point we can describe this problem in terms of the double solution model. First we note that we believe that coherent scattering of a photon by a confined electron should not be very different from the figure shown in Compton Effect. We can also argue that if the field operator equations are the same as classical, at least some solutions should be the same.

For coherent scattering, the Floquet states are a candidate to represent the entire process, with a constant number of particles[116].

For the scattering process where a single photon is scattered incoherently by the atomic system we propose to take the states as follows:

1. For the initial state:
    a. For the photons:

We argue that the number of photons is given by the number of fully developed and free photon solitons present in the double solution alternative wave functions. This model makes the definition of the 'photon state' totally independent of the mode of the field being excited, and allows for a photon to share many modes simultaneously.

Second, we take as the external fields the full solution provided by the classical model oscillating at the same frequency and coherent with the incident beam. This includes both, the incident beam and the spherical dipole-like spherical scattered emission.

b. For the atomic system:

The solution to the semiclassical Bloch equations for the given incident field. Photons that leave the initial beam and move into regions where the expanding fields are dominant, still belong to the initial state and their fields are coherent with the initial beam. The initial state will be given by the atomic part multiplying the internal photon solitons $A_{\omega}^{int,i}$ times the external photon wave function $A_{inc,\omega}^{ext} + A_{scatt,\omega}^{ext}$ with an incident and a scattered part:

$$\Psi_{init}(t) = \underbrace{\left[a_1(t)\psi_{gr}(r_e,t) + a_2(t)\psi_{exc}(r_e,t)\right]}_{effective\ ground\ state} \times \underbrace{\cdots A_{\omega}^{int,i}\left(r_{ph,i},t\right)\cdots}_{n\ times} \times \underbrace{\cdots\left[A_{inc,\omega}^{ext}\left(r_{ph,i},t\right) + A_{scatt,\omega}^{ext}\left(r_{ph,i},t\right)\right]\cdots}_{n\ times}$$

2. For the excited intermediate state:

We assume that one of the incident solitons has been 'absorbed' by the irradiated atomic system, The transition from the initial state to the intermediate state happens when a photon trajectory takes the photon soliton into the atomic system, where the soliton is absorbed by the attached atomic field, and the atomic system is promoted to the excited state. The transition from the intermediate state to the final state takes place when a new photon soliton is generated from the atomic system and leaves the atom as a fluorescent photon. The transition from the intermediate state to the final state takes place in analogy with the description given for spontaneous emission: entanglement due to energy conservation assures that as the emitted photon reaches the speed of light in vacuum, the atomic system returns to the effective ground state.

a. For the photons:

We take the same state as before, but with one photonic soliton less, and the same external fields as before in the case of many incident photons, or no photon field if the incident photon was a single one.

b. For the atomic system:

The atomic system in its excited state as provided by the Bloch equations in the presence of external fields.

$$\Psi_{int}(t) = \underbrace{\left[a_3(t)\psi_{gr}(r_e,t) + a_4(t)\psi_{exc}(r_e,t)\right]}_{effective\ excited\ state} \times \underbrace{\cdots A_{\omega}^{int,i}\left(r_{ph,i},t\right)\cdots}_{n-1\ times} \times \underbrace{\cdots\left[A_{inc,\omega}^{ext}\left(r_{ph,i},t\right) + A_{scatt,\omega}^{ext}\left(r_{ph,i},t\right)\right]\cdots}_{n-1\ times}$$

3. For the final state:

The same as the initial state, but with a photonic soliton that has moved from a region where the incident fields are dominant to a region where the scattered fields are dominant, outside of the extension of the incident beam. Some of the scattered photonic photons may belong to fields oscillating at a different frequency than the original beam frequency.

$$\Psi_{init}(t) = \underbrace{\left[a_1(t)\psi_{gr}(r_e,t) + a_2(t)\psi_{exc}(r_e,t)\right]}_{effective\ ground\ state} \times \underbrace{\cdots A_{\omega}^{int,i}(r_{ph,i},t)\cdots}_{n-1\ times} A_{\omega_{scatt}}^{int,n}(r_{ph,n},t)$$
$$\times \underbrace{\cdots\left[A_{inc,\omega}^{ext}(r_{ph,i},t) + A_{scatt,\omega}^{ext}(r_{ph,i},t)\right]\cdots}_{n-1\ times} A_{scatt,\omega_{scatt}}^{ext}(r_{ph,n},t)$$

In our model we consider states with different number of particles only those states where the number of solitons really changes. Absorption and emission change the number of photons but coherent scattering no[121].

Our model allows us to visualize the use of photon creation and destruction operators in quantum optics such as the product $\hat{b}^{\dagger}\hat{a}$ as a formal description for the modification of the state of a particle from state *a* to state *b*, and in some cases, just as a zero order approximation for the gradual transition of the field from the initial state $|a\rangle$ to the final state $|b\rangle$. In such cases it has nothing to do with the actual destruction and recreation of a particle. That would be a very crude approximation leading to a misinterpretation and perhaps contributing to the generation of confusion. In coherent scattering the photon is not absorbed, just changes direction.

### 5.4.4. Compton Effect

In the scattering problems calculated in Quantum Electrodynamics, propagators are used to find the scattering cross section for an incident particle or wave by the field produced by a second particle or scattering center. In most actual experiments the scattering center is at rest. The initial state is represented by a plane wave incident from a given direction and with a given energy. The differential cross section is calculated for the outgoing particle at a given final direction and energy. This final state is typically described also by an outgoing plane wave. It is noted that energy-momentum is conserved during the scattering processes, implying that the scattering center and the scattered particle are found in an entangled final state, with a correlated scattered-recoiling motion. The expression for the different particle propagators are expressed as Fourier series in eigenfunctions of the free particle Hamiltonian, typically also plane waves[122].

Let's compare this experiment with a typical optical diffraction experiment involving a directed beam of light as it is scattered by ultrasound waves in some medium. Here scattering takes place mainly in two very definite directions: one due to reflection, and the other is known as first order diffraction peak. Higher orders are possible, but are of decreasing intensity[123]. This is a result of the explicit one-dimensional geometry.

On the other hand, in QED problems the scattering cross-section is typically a continuous function of direction, implying that it is not a one-directional or plane problem. In classical physics a continuous differential cross section is achieved by considering three-dimensional geometries, like expanding the outgoing fields by spherical waves and describing the scattering center either as a point or as some sizeable object with a given volume. It is precisely the coherent interference of the wavelets generated at different regions of the scattering center what produce the different angular distributions. It is essential that the scattering center be of a finite size in order to produce interference. It cannot be interpreted as a 'probability distribution' for a point scattering center, which would produce incoherent addition of wavelets instead of a coherent final result. A probability distribution is however the traditional interpretation provided by quantum mechanics to the scattering center wavefunction.

Moreover, at lower incident energies or larger wavelengths the QED solution approaches the classical result. This means that somehow in the QED formalism, the coherent addition of waves produced by a finite size scattering center is implied in the mathematics of the theory, despite the traditional probability interpretation. We will attempt to find this hidden geometry in the formalism.

Experimentally in Compton scattering, there are three effects that cannot be explained classically assuming a point-like fixed scattering center:

a. The differential cross section angular distribution presents a dependence that cannot be explained by a point-like scattering center.
b. As a function of incoming wavelength, the total scattering cross section does not follow the classical Thomson law, but scattering increases with increasing frequencies until reaching a maximum near the so called Compton wavelength ~2.42x10$^{-12}$ meters. It decreases at higher frequencies.
c. The frequency of the scattered radiation is not equal to the frequency of the incident beam and has an angular dependence that cannot be explained in a non-quantum model including finite energy-momentum conservation.

The first one attempting to provide a theoretical explanation for these effects was Arthur Compton who in 1919 was able to provide an approximate solution for effects a. and b., by assuming that the electron was really a classical sizeable object with a size of the order of the Compton radius or wavelength[124]. In 1923 he was able to explain the puzzling directional dependence of the scattered radiation frequency by assuming that the incoming light and the electron scatterer were both particles with a given energy and momentum related to frequency and wavelength through the de Broglie relations[125]. This last model was independent of the size of the particles and could be also applied to point particles. This last property was followed in the future by all quantum descriptions of Compton scattering. However, the Compton model could also be applied to particles with a finite size.

The classical intensity scattered at an angle $\theta$ with the incident beam, assuming a point-like electron, is given by classical electromagnetic theory as:

$$I_\theta = I\frac{e^4(1+\cos^2\theta)}{2\ell^2 m^2 c^4}$$

**Eq 163**

where $\ell$ is the distance at which the scattered beam is collected. The integrated fraction of incident energy removed from the incident beam by the electron is

$$\frac{E_s}{I} = \frac{8\pi e^4}{3m^2c^4}\frac{\sin^4\left(\frac{2\pi a}{\lambda}\right)}{\left(\frac{2\pi a}{\lambda}\right)^4} \quad \underset{a\ll\lambda}{\underbrace{\rightarrow}} \quad \frac{8\pi e^4}{3m^2c^4} \;\; \textit{Thomson scattering}$$

**Eq 164**

This model works well at longer wavelengths described by Thomson scattering, but does not correspond to the experimentally measured values when the incident wavelength approaches the Compton wavelength. In his 1919 paper, Compton was able to show that the corresponding distribution for an electron with a shape of a flexible spherical shell of radius $a$ is given by

$$I_\theta = I\frac{e^4(1+\cos^2\theta)}{2\ell^2 m^2 c^4}\left\{\sin^2\left(\frac{4\pi a}{\lambda}\sin\frac{\theta}{2}\right)^2 \Big/ \left(\frac{4\pi a}{\lambda}\sin\frac{\theta}{2}\right)^2\right\}$$

and for a ring-like electron

$$I_\theta = I\frac{e^4(1+\cos^2\theta)}{2\ell^2 m^2 c^4}\left\{1-\alpha\left(\frac{4\pi a}{\lambda}\sin\frac{\theta}{2}\right)^2+\beta\left(\frac{4\pi a}{\lambda}\sin\frac{\theta}{2}\right)^4-\cdots\right\}$$

The integrated absorption coefficient in both cases can be expressed by an expression of the form

$$\frac{E_s}{I} = \frac{8\pi e^4}{3m^2c^4}\left\{\alpha-\beta\left(\frac{4\pi a}{\lambda}\right)^2+\gamma\left(\frac{4\pi a}{\lambda}\right)^2-\cdots\right\} \quad \underset{a\ll\lambda}{\underbrace{\rightarrow}} \quad \frac{8\pi e^4}{3m^2c^4} \;\; \textit{Thomson scattering}$$

Contrary to the distribution described by Eq 163 and Eq 164, the last expressions provided a very close description to the available experimental results. The best fit to the experimental results was achieved when the electron radius was assumed to be close to $2x10^{-12}$ meters known as Compton radius.

Even when they do not provided a perfect description of the experimental results, they showed a good approximation, and what is more important, they provided a completely clear and intuitive description of the phenomenon. One can observe that the key signature for the existence of a characteristic length '*a*' where something significantly important takes place, in this case the sharp edge of the electron object, is that both, the differential cross section as well as the total cross section depend on the incident wavelength measured on units of the characteristic length '*a*'. This property can be seen to be a direct consequence of one of the fundamental theorems of Dimensional Analysis, namely the π-Theorem[126]. Compton identified this characteristic length with the wavelength having his name.

In his 1923 paper Compton[125] merged electromagnetic scattering theory and quantum mechanics by postulating that the scattering could be described by the "collision" of two mechanical particles like billiard balls, one corresponding to the electron and the other corresponding to the 'photon' envisaged by Einstein in 1905, each particle carrying mass, energy and momentum. For the photon these mechanical properties were related to the wave properties by the de Broglie relations. In this way Compton was able to derive a direction-dependent frequency shift for the photon particle that was caused by the change in momentum and energy given by

$$\Delta\lambda(\theta) = \lambda'(\theta) - \lambda = a(1 - \cos\theta) \qquad a = \frac{h}{mc} \qquad Compton\,radius$$

Here again we can observe that the wavelengths are specified in units of the characteristic length, now identified as the Compton radius or wavelength. This was probably the first step towards the future development of quantum electrodynamics. In the traditional QED representation, the starting point for the description is given by the scattering matrix formalism[122]

$$S_{fi}^{Compton} = \frac{e^2}{V^2}\sqrt{\frac{m^2}{E_f E_i}}\frac{1}{\sqrt{2\vec{k}\cdot 2\vec{k}'}}(2\pi)^4\delta^4\left(p_f + k' - p_i - k\right)\times$$

$$\bar{u}\left(p_f, s_f\right)\left[(-i\not{\epsilon}')\frac{1}{\not{p}_i + \not{k} - m}(-i\not{\epsilon}) + (-i\not{\epsilon})\frac{1}{\not{p}_i - \not{k}' - m}(-i\not{\epsilon}')\right]u(p_i, s_i)$$

**Eq 165**

From the scattering matrix expression, the differential scattering cross section is derived:

$$d\sigma = \frac{e^4 m}{(2\pi)^2 2kE_i|\vec{\mathrm{v}}|}\times$$

$$\int\left|\bar{u}\left(p_f, s_f\right)\left[\epsilon'\frac{1}{\not{p}_i + k - m}\not{\epsilon} + \epsilon\frac{1}{\not{p}_i - k' - m}\not{\epsilon}'\right]u(p_i, s_i)\right|^2\delta^4\left(p_f + k' - p_i - k\right)\frac{m\,d^3p_f}{E_f}\frac{d^3k'}{2k'}$$

**Eq 166**

where we can see the explicit integration over intermediate virtual electron and photon states. The final result, for the total cross section as a function of frequency is given by the Klein Nishina cross section:

$$\sigma = 2\pi\frac{\alpha^2}{m^2}\left[\frac{1+\gamma}{\gamma^3}\left(\frac{2\gamma(1+\gamma)}{1+2\gamma} - \ln(1+2\gamma)\right) + \frac{1}{2\gamma}\ln(1+2\gamma) - \frac{1+3\gamma}{(1+2\gamma)^2}\right]$$

where

$$\gamma = \left(\frac{\lambda}{r_C}\right)^{-1}$$

where $r_C$ is the Compton radius or Compton wavelength. It is clear that the Compton radius is the characteristic length of this problem, but it is very difficult to identify its role in the problem as a sizeable property of the electron. In Eq 165 we can observe that the main factor in the determination of the matrix element is the appearance of the propagator operator $S_F$[127]. The free particle propagator $S_F(x',x)$ is defined by the expression

$$(i\nabla\!\!\!/' - m_0)S_F(x',x) = \delta^4(x'-x)$$

The Fourier transform being given by

$$S_F(x',x) = S_F(x'-x) = \int \frac{d^4p}{(2\pi)^4} e^{-ip\cdot(x'-x)} S_F(p)$$

$$S_F(p) = \frac{p\!\!\!/ + m_0}{p^2 - m_0^2}$$

We recognize that the denominator operator $p^2 - m_0^2$ stands for $\partial^\mu\partial_\mu + m_0^2$ or $\frac{1}{c^2}\frac{\partial^2}{\partial t^2} - \nabla^2 + \frac{m_0^2c^2}{\hbar^2}$. When the application of this operator on a given function gives a null result, we are in the presence of the Klein Gordon equation:

$$\frac{1}{c^2}\frac{\partial^2}{\partial t^2}\psi - \nabla^2\psi + \frac{m_0^2c^2}{\hbar^2}\psi = 0$$

**Eq 167**

AS implied by Eq 166, the final state is found by integration of all possible intermediate virtual states weighted by an operator proportional to $S_F(p)$. We see that the overwhelming contribution to the construction of the cross section comes from those wavefunctions that fulfill Eq 167, where the application of the propagator operator produce a diverging effect. Typically it is said that the solutions of Eq 167 for free particles are given by plane waves

$$\psi(\vec{r},t) = Ae^{i(\vec{k}\cdot\vec{r}-\omega t)}$$

with $\omega$ and $\vec{k}$ real quantities. But plane waves don't cover all possibilities. Any function $\psi(\vec{r},t)$ fulfilling the conditions

$$\frac{\partial^2\psi}{\partial t^2} = z_\omega^2\psi$$

$$\nabla^2\psi = z_k^2\psi$$

$$\frac{z_\omega^2}{c^2} - z_k^2 = \frac{m^2c^2}{\hbar^2}$$

will work equally well, where $z_\omega$ and $z_k$ are complex numbers, and this includes spherical and cylindrical coordinates as well. Explicitly we can write

$$z_\omega = \pm\frac{1}{\tau} - i\omega \qquad \vec{z}_k = -\frac{\hat{r}}{r_C} \pm i\vec{k}$$

as complex quantities with a real and imaginary part. The imaginary parts $\omega$ and $\vec{k}$ take care of momentum and energy conservation during the scattering process. In a stationary solution, like the problem solved in QED, one can set $\frac{1}{\tau} = 0$, so the real part of $z_\omega$ vanishes. A typical, well known time independent solution of the Klein Gordon equation at rest different from a space constant is given in spherical coordinates by the Yukawa potential:

$$\psi_{Yukawa}(\vec{r}) = A\frac{e^{-\frac{r}{r_C}}}{r}$$

fulfilling the equation in spherical coordinates when no time dependence is present. This type of solutions provide the role of the characteristic length we are looking for. If the scattering center is at rest and the problem is stationary we allow for a finite frequency $\omega \neq 0$ but take $\frac{1}{\tau} = 0$. We recognize that a spherical coordinate system is the most appropriate for the description of the system and we see that the wavefunction that best describes the Compton scattering process in QED is given by

$$\psi(\vec{r}, t) = A\frac{e^{-\frac{r}{r_C}+ikr}}{r}e^{-i\omega t} = A\frac{e^{-\left(\frac{1}{r_C}+ik\right)r-i\omega t}}{r}$$

This function has a space dependence provided by the spherically decaying exponential, compatible with a confining boundary condition. This provides the solution to the puzzle in QED: in the propagator formalism, exponentially decaying functions are not explicitly identified but are however implicitly present as solutions providing an acceptable geometrical representation of the system. In this case $r_C$ assumes the role of the characteristic length responsible for the angular and total cross section properties of the scattering distribution. And that happens in all QED scattering problems involving a mass-dependent propagator. In this case the scattering center (and also the incident particle if it is massive) is implicitly described by an exponentially decaying wavefunction. The characteristic decay length is given by the corresponding particle Compton radius. This length represents the smooth edge of the scattering center in complete harmony with our electron model. It is to note that this exponentially decaying edge for the electron shape had been found similarly in the 1930 work of Weisskopf[128] on electron self-interactions[129]. By using this type of solutions, the entire scattering problem starts making sense in the picture of QED. And the same model applies for the treatment of other types of problems, like bremsstrahlung, electron-electron and electron-positron scattering, and pair creation.

We note that in a similar way as the case of particle scattering described in ‘Section 5.1.6 Coulomb Scattering’, from Eq 165 we can identify the final state as given by an entangled state

$$\begin{aligned} F\left(\vec{r}_e, \vec{r}_{ph}, t\right) &= e^{-i Et/\hbar}\int g\left(\vec{p}\right) e^{-i\vec{p}\cdot\vec{r}_e} e^{-i\left(\vec{P}_{cm}-\vec{p}\right)\cdot\vec{r}_{ph}}\, d\vec{p} \quad \approx \quad \sum f\left(\vec{r}_e\right) g\left(\vec{r}_{ph}\right) \\ &= e^{-i Et/\hbar} F\left(\vec{R}_{cm}\right) G\left(\vec{r}_{rel}\right) \quad \text{entangled state} \end{aligned}$$

A graphical representation of the electron-photon trajectories given by the Quantum force due to photon-electron entanglement in Compton Effect are shown in the following picture

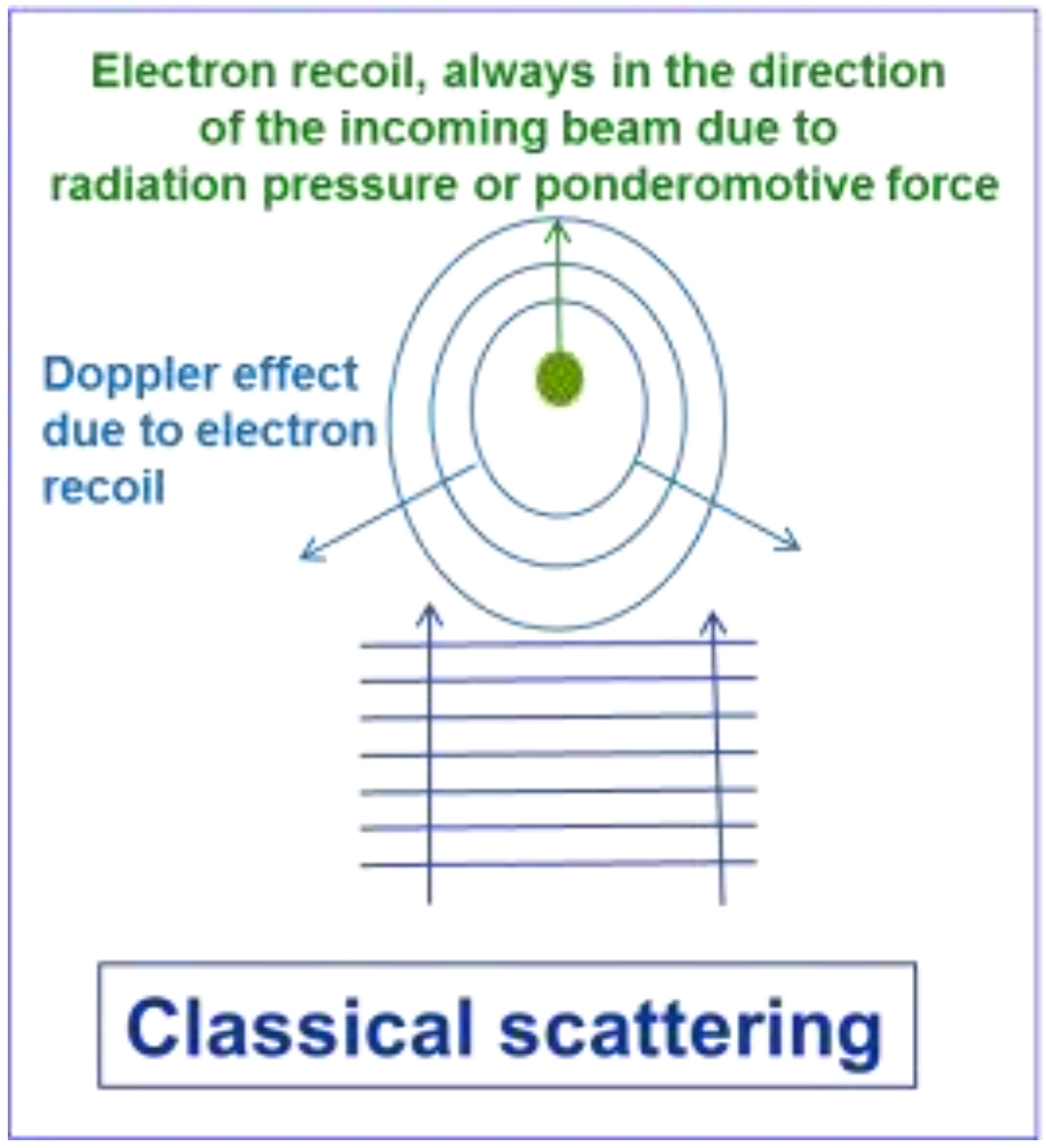

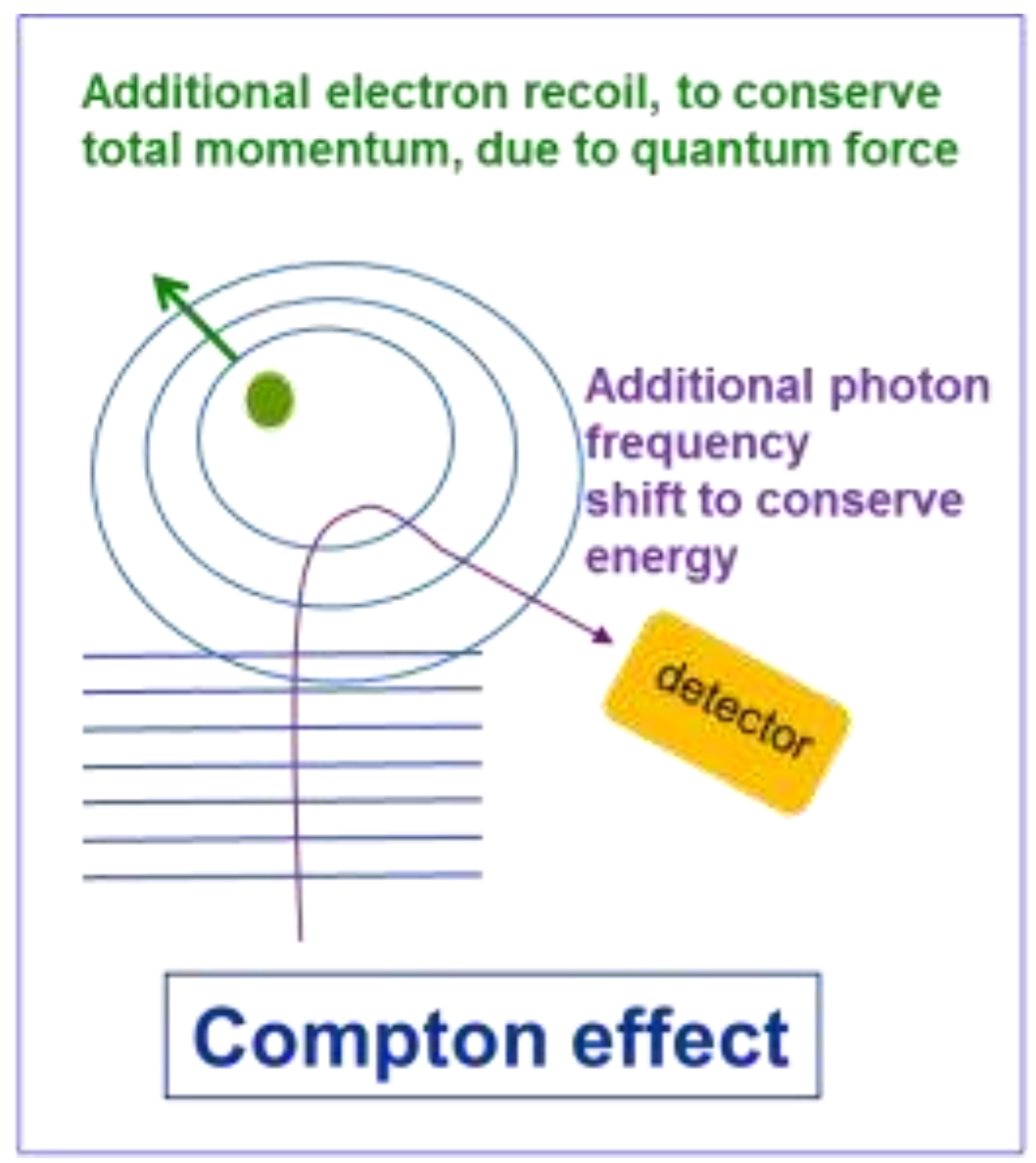

In the Compton Effect, entanglement generates an effective force between the electron and the photon. Not only the quantum potential acting on the particles depends on the actual position of all particles, but also the wave fields themselves depend on the actual position of all particles. In this way the asymmetries present in the wavelength and frequency of the scattered photon can be correlated with the correlated scattered electron. These asymmetries are absent from a classical description.

### 1.1.1. Radiation scattering simulations

Once the photon-soliton particle has been developed in the electromagnetic field, their trajectories in space can be calculated. In fact, when the "Bloch vector" moves upwards, the atomic polarization has a phase respect to the incoming beam so that energy is flowing towards the atom. The photons are absorbed by the atomic system. After absorption the atom will decay in an emission process, where the newly created photon has not exactly the same energy as the incoming photon. The scattering is mostly incoherent, like the case of Compton scattering. In the other case, when the "Bloch vector" moves downwards, the atomic polarization has a phase respect to the incoming beam so that energy is flowing out of the atom, photon trajectories would be scattered away by the atomic system. The scattered photons are part of the coherent component of scattering. The following pictures show simulations of scattering and absorption.

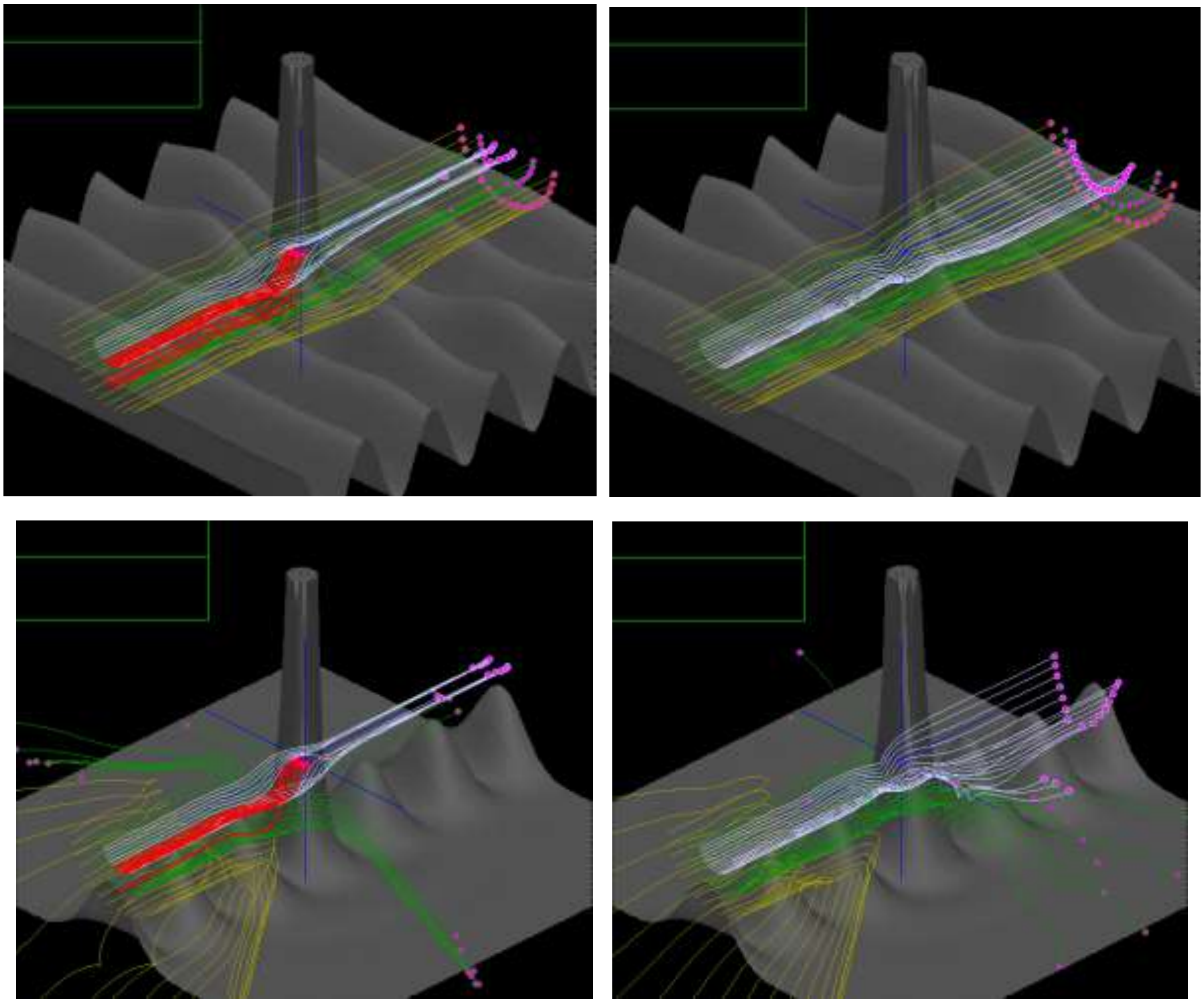

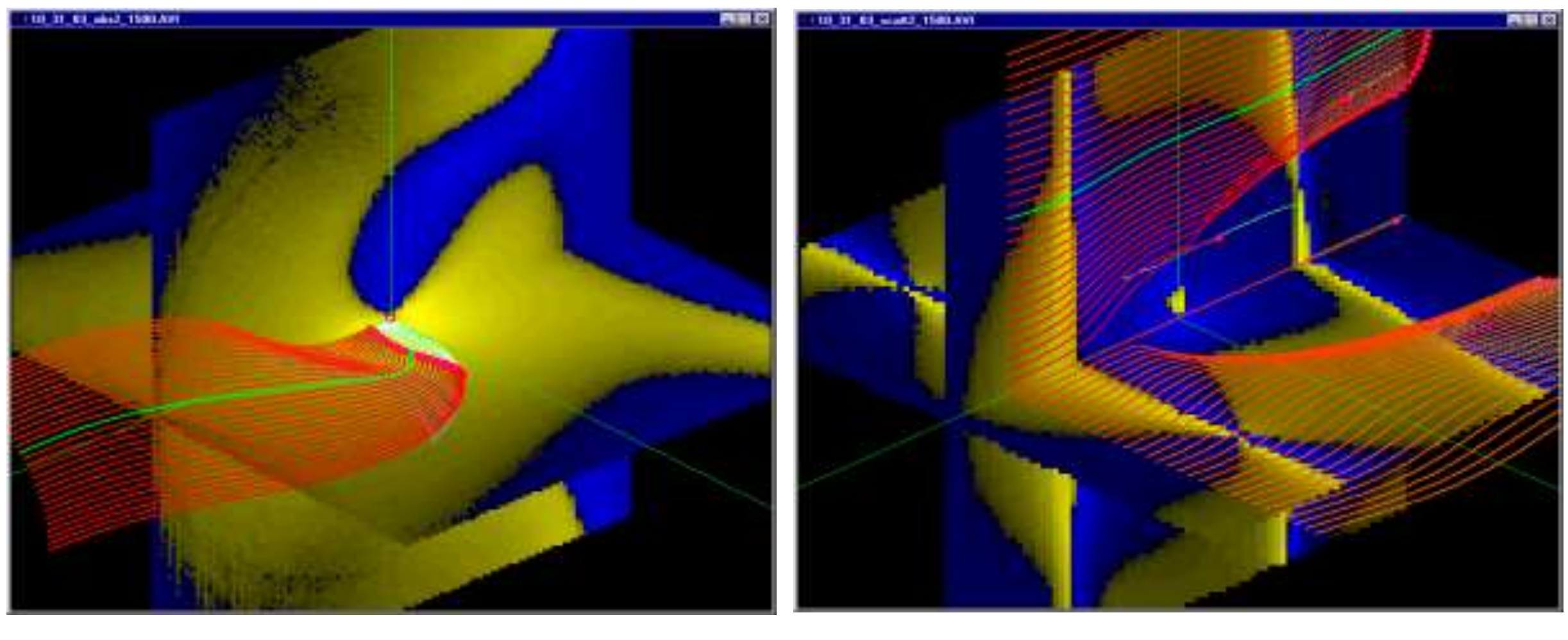

Absorption Scattering

## 1.2. Conclusion

In this chapter we have provided a realistic interpretation for some examples of the interaction between electrons and photons. This interpretation is based on the existence of solitons as solutions to the electron and photon equations in the presence of self-interactions. We saw that the Dirac equation in Canonical Quantization naturally includes a source term provided by the product of a positron field and the electromagnetic field. We have presented simulations of photon trajectories in the vicinity of an oscillating dipole. We analyzed the case of spontaneous emission and found that it can be described in terms of a quantum jump-like process, where a newly born soliton is accelerated to the speed of light and acquires the full energy of the transition in a period of time of the order of an oscillation period, leaving the proximity of the atomic system and becoming free. The quantum potential is the responsible for the dephasing that appears between the dipole oscillation and electromagnetic field, generating the different periods of absorption and emission in the different steps of resonance scattering.

The difference with many particle first quantization is a highly optimized notation, the possibility for the presence of sources in the equations of motion and the introduction of self-interactions directly into the equations. Note that source terms cannot be present in the First Quantization equation because they are not directly proportional to the wavefunction.

---

[1] H. Nikolic, *Superluminal velocities and nonlocality in relativistic mechanics with scalar potential*, arXiv:1006.1986 (2010). H. Nikolic, *Relativistic Quantum Mechanics and Quantum Field Theory*, Chapter 8. of the book "Applied Bohmian Mechanics: From Nanoscale Systems to Cosmology", edited by X. Oriols and J. Mompart (Pan Stanford Publishing, 2012); ISBN: 978-981-4316-39-2, Copyright 2012 by Pan Stanford Publishing Pte. Ltd, arXiv:1205.1992 (2012). H. Nikolic, *Time and probability: From classical mechanics to relativistic Bohmian mechanics*, arXiv:1309.0400 (2013).

[2] Margaret Hawton and T. Melde, *Photon number density operator iE*A*, Phys. Rev. A 51, 4186 (1995). Margaret Hawton, *Photon position operator with commuting components*, Phys. Rev. A 59, 0954 (1999). Margaret Hawton, *Photon wave functions in a localized coordinate space basis*, Phys. Rev. A 59, 3223 (1999). Margaret Hawton and William E. Baylis, *Photon position operators and localized bases*, Phys. Rev. A 64, 012101 (2001). Margaret Hawton, *Photon wave mechanics and position eigenvectors*, Phys. Rev. A 75, 062107 (2007). Margaret Hawton, *Photon position eigenvectors lead to complete photon wave mechanics*, SPIE 6664 (2007).

[3] Andrey Akhmeteli, *One real function instead of the Dirac spinor function,* J. Math. Phys. 52, 082303 (2011).

[4] W. Pauli, *On the hydrogen spectrum from the standpoint of the new quantum mechanics*, Zeitschrift für Physik 36, 336 (1926).

[5] P.A.M. Dirac, *Quantum mechanics and a preliminary investigation of the hydrogen atom,* Proc. Roy. Soc. London A110, 561 (1926). P.A.M. Dirac, *The elimination of the nodes in quantum mechanics,* Proc. Roy. Soc. London A111, 281 (1926).

[6] Max Born and P. Jordan, *Zur Quantenmechanik*, Zeitschrift für Physik 34, 858 (1925).

[7] Iwo Bialynicki-Birula and Zofia Bialynicka-Birula, Quantum Electrodynamics, Pergamon Press, Oxford 1975.

[8] Edwin Schrödinger, *Quantizierung als Eigenwert Problem*, Ann. d. Physik 79, 361 (1926). Edwin Schrödinger, *Quantizierung als Eigenwert Problem. (Zweite Mitteilung*), Ann. d. Physik 79, 489 (1926). Edwin Schrödinger, *Über das Verhältnis der Heisenberg-Born-Jordanschen Quantenmechanik zu der meinen*, Ann. d. Physik 79, 734 (1926). Edwin Schrödinger, *Quantisierung als Eigenwertproblem (Dritte Mitteilung: Störungstheorie, mit Anwendung auf den Starkeffekt der Balmerlinien*), Ann. d. Physik 80, 437 (1926). Edwin Schrödinger, *Quantisierung als Eigenwertproblem (Vierte Mitteilung*), Ann. d. Physik 81, 109 (1926).

[9] E. Madelung, *Quantentheorie in Hydrodynamischer Form*, Zeitschrift für Physik 40, 332 (1926).

[10] C. G. Darwin, *The Zeeman Effect and Spherical Harmonics*, Proc. Roy. Soc. London A 115, 001 (1927).

[11] W. Gerlach and O. Stern, *Der Experimentelle Nachweis des magnetischen Moments des Silberatoms*, Zeitschrift für Physik 8, 110 (1921). W. Gerlach and O. Stern, *Der experimentelle Nachweis der Richtungsquantelung in Magnetfeld*, Zeitschrift für Physik 09, 349 (1922).

[12] W. Pauli Jr., *Zur Quantenmechanik des magnetischen Elektrons*, Zeitschrift für Physik 43, 601(1927).

[13] W. Greiner*, Relativistic Quantum Mechanics Wave Equations*, Third edition, Springer, Berlin, 1987.

[14] For traditional descriptions of the electron see:
A. O. Barut, *A theory of particles of spin one-half*, Ann. Phys. (N.Y.) 5, 095 (1958). Ralph F. Guertin and Eugene Guth, *Zitterbewegung in Relativistic Spin-0 and -1/2 Hamiltonian Theories*, Phys. Rev. D 7, 1057 (1973). J. A. Lock, *The Zitterbewegung of* *a free localized Dirac particle*, Am. J. Phys. 47, 797 (1979). A. O. Barut and A. J. Bracken, *Zitterbewegung and the internal geometry of the electron*, Phys. Rev. D 23, 2454 (1981). A. O. Barut and A. J. Bracken, *Magnetic-moment operator of the relativistic electron*, Phys. Rev. D 24, 3333 (1981). A. O. Barut, *The Zitterbewegung and the Einstein A Coefficient of Spontaneous Emission*, in *Old and New Questions in Physics, Cosmology, Philosophy, and Theoretical Biology*, A. van der Merwe, ed., Plenum, New York, 1983. James A. Lock, *Relativistic invariance and Zitterbewegung*, Am. J. Phys. 52, 223 (1984). A. O. Barut and Nino Zanghi, *Classical Model of the Dirac Electron*, Phys. Rev. Lett. 52, 2009 (1984). A. O. Barut and W. Thacker, *Zitterbewegung of the electron in external fields*, Phys. Rev. D 31, 2076 (1985). A O Barut and M Pavsic, *Quantisation of the classical relativistic zitterbewegung in the Schrödinger picture*, Class. Quantum Grav. 4, L131 (1987). A. Barut and M. Pavsic, *Classical Model of the Dirac Electron in curved space*, Class. Quantum Grav. 4, L41 (1987). A. O. Barut and I. H. Duru, *Exact solutions of the Dirac equation in spatially flat Robertson-Walker space-times*, Phys. Rev. D 36, 3705 (1987). A. O. Barut, *Lorentz-Dirac equations and energy conservation for radiating electrons*, Phys. Lett. A 131, 011 (1988). G. C. Sherry, *Zitterbewegung revisited*, Eur. J. Phys. 09, 301 (1988). A. O. Barut and N. Unal, *Generalization of the Lorentz-Dirac equation to include spin*, Phys. Rev. A 40, 5404 (1989). A. O. Barut, *Excited states of zitterbewegung*, Phys. Lett. B 237, 436 (1990). V. J. Menon, *Spin conservation in one-particle Dirac theory*, Am. J. Phys. 60, 928 (1992). M Mattes and M Sorg, *Kinematics of Dirac's spinor field*, J. Phys. A: Math. Gen. Phys. 26, 3013 (1993). A O Barut and M G Cruz, *On the Zitterbewegung of the relativistic electron*, Eur. J. Phys. 15, 119 (1994). M. Czachor and A. Posiewnik, *What happens to spin during the SO(3)->SE(2) contraction?*, arXiv:quant-ph/9501017 (1995). C. A. Uzes and A. O. Barut, *Spinor Field as Elementary Excitations of a System of Scalar Fields*, Found. Phys. 28, 741 (1998). Jens Bolte and Rainer Glaser, *Zitterbewegung and semiclassical observables for the Dirac equation*, J. Phys. A: Math. Gen. Phys. 37, 06359 (2004). P. Krekora, Q. Su, and R. Grobe, *Relativistic Electron Localization and the Lack of Zitterbewegung*, Phys. Rev. Lett. 93, 043004 (2004). B. Altschul and Don Colladay, *Velocity in Lorentz-violating fermion theories*, Phys. Rev. D 71, 125015 (2005). Shun-Qing Shen, *Spin Transverse Force on Spin Current in an Electric Field*, Phys. Rev. Lett. 95, 187203 (2005). P. Brovetto, V. Maxia, M. Salis, *Some remarks about mass and Zitterbewegung substructure of electron*, arXiv:quant-ph/0512047 (2005).

[15] Edwin Schrödinger, *Über die Kraftfreie Bewegung in der relativistischen Quantenmechanik*, Berliner Berichte 24, 418 (1930).

[16] Works on classical and general descriptions of the electron can be found in:

C. Lanczos, *Die tensoranalytischen Beziehungen der Diracschen Gleichung*, Zeitschrift für Physik 57, 447 (1929). C. Lanczos, *On the covariant formulation of Dirac's equation*, Zeitschrift für Physik, 57, 474 (1929). C. Lanczos, *The conservation laws in the field theoretical representation of Dirac's theory*, Zeitschrift für Physik 57, 484 (1929). F. J. Belinfante, *On the current and the density of the electric charge, the energy, the linear momentum and the angular momentum of arbitrary fields*, Physica 7, 0449 (1940). S. J. Czyzak, *An Interpretation of the Variable Mass in Wessel's Theory of the Electron*, Am. J. Phys. 22, 335 (1954). I, Tadao NAKANO*, A Relativistic Field Theory of an Extended Particle*, Prog. Theor. Phys. 15, 333 (1956). M. Francis Halbwachs, *Sur le mouvement de la goutte relativiste de Bohm et Vigier*, C. R. Acad. Sci. Paris, 245, 1298 (1957). M. Francis Halbwachs, *Étude de la goutte de Bohm et Vigier en relation avec le formalisme hydrodynamique de Dirac-Takabayasi*, C. R. Acad. Sci. Paris 245, 1392 (1957). M. Francis Halbwachs, *Sur un cas particulier du mouvement de la goutte de Bohm et Vigier*, C. R. Acad. Sci. Paris 245, 1513 (1957). MM. Burhan, Cahit Unal et Jean-Pierre Vigier, I*ntroduction des paramètres relativistes d'Einstein-Kramers et de Cayley-Klein dans la théorie relativiste des fluides dotés de moment cinétique interne (spin*), C. R. Acad. Sci. Paris 245, 1785 (1957). MM. Burhan Cahit Unal et Jean-Pierre Vigier, *Introduction des paramètres relativistes d'Einstein-Kramers et de Cayley-Klein dans l'hydrodynamique relativiste du fluide à spin de Weyssenhoff*, C. R. Acad. Sci. Paris 245, 1890 (1957). D. Bohm and J.P. Vigier, *Relativistic hydrodynamics of rotating fluid masses*, Phys. Rev. 109, 1882 (1958). V. Bargmann, Louis Michel, V. L. Telegdi, *Precession of the Polarization of Particles Moving in a Homogeneous Electromagnetic Field*, Phys. Rev. Lett. 2, 435 (1959). David BOHM, Pierre HILLION, Takehiko TAKABAYASI and Jean-Pierre VIGIER, *Relativistic Rotators and Bilocal Theory*, Prog. Theor. Phys. 23, 496 (1960). H. C. Corben, *Spin in Classical and Quantum Theory*, Phys. Rev. 121, 1833 (1961). Zvi Grossmann and Asher Peres, *Classical Theory of the Dirac Electron*, Phys. Rev. 132, 2346 (1963). P. F. Browne, *Electron spin and radiative reaction*, Ann. Phys. (N.Y.) 59, 254 (1970). Herbert Jehle, *Relationship of Flux Quantization to Charge Quantization and the Electromagnetic Coupling Constant*, Phys. Rev. D 3, 0306 (1971). Herbert Jehle, *Flux Quantization and Particle Physic*s, Phys. Rev. D 6, 0441 (1972). L. P. Staunton, Sean Browne, *Classical limit of relativistic positive-energy theories with intrinsic spin*, Phys. Rev. D 12, 1026 (1975). D. Gutkowski, M. Moles, and J. P. Vigier, *Hidden-Parameter theory of the extended Dirac electron. I. Classical theory*, Nuovo Cimento B 39, 193 (1977). Herbert C. Corben, *Quantized relativistic rotator*, Phys. Rev. D 30, 2683 (1984). M Pavsic, *On the consistent derivation of rigid particles from strings*, Class. Quantum Grav. 7, L187 (1990). J. Rawnsley, *Exact symplectic structures and a classical model for the Dirac electron*, Lett. Math. Phys. 24, 331 (1992). Herbert C. Corben, *Factors of 2 in magnetic moments, spin–orbit coupling, and Thomas precession*, Am. J. Phys. 61, 0551 (1993). I. Bialynicki-Birula, *Simple relativistic model of a finite-size particle*, Phys. Lett. A 182, 346 (1993). Herbert C. Corben, *Structure of a spinning point particle at rest*, Int. J. Theor. Phys. 34, 019 (1995). Heiko J. Herrmann, G. Rueckner, W. Muschik, H.-H. v. Borzeszkowski, *Spin Axioms in Relativistic Continuum Physics*, Found. Phys. 34, 1005 (2004). Yuri A. Rylov*, Is the Dirac particle composite*?, arXiv:physics/0410045 (2004). Yuri A. Rylov, *Is the Dirac particle completely relativistic?*, arXiv:physics/0412032 (2004).

[17] Works on the tensorial description of the Dirac equation can be found in:

*solution theory*, Annales de la Fondation Louis de Broglie 12, 001 (1987). J.P. Vigier, *Explicit Mathematical Construction of Relativistic Nonlinear de Broglie Waves Described by Three-Dimensional (Wave and Electromagnetic) Solitons “Piloted” (Controlled) by Corresponding Solutions of Associated Linear Klein-Gordon and Schrödinger Equations*, Found. Phys. 21, 0125 (1991). G Horton, C Dewdney, *de Broglie's pilot wave theory for the Klein-Gordon equation*, arXiv:quant-ph/0109059 (2001). Ali Shojai, Fatimah Shojai*, About Some Problems Raised by the Relativistic Form of De-Broglie–Bohm Theory of Pilot Wave*, Phys. Scr. 64, 413 (2001). C.R. Courtney, L.J. Frasinski, *Coulomb imaging of H"2^+ nuclear wavefunctions using the pilot-wave theory of quantum mechanics*, Phys. Lett. A 318, 0030 (2003). W. Struyve, H. Westman, *A minimalist pilot-wave model for quantum electrodynamics*, Proc. Roy. Soc. London A 463, 3115 (2007).

[24] A.Granik, *On an Elementary Derivation of the Hamilton-Jacobi Equation from the Second Law of Newton*, arXiv:physics/0309059 (2003).

[25] D. Bohm, *A suggested interpretation of the quantum theory in terms of "hidden variables": Part I*, Phys. Rev. 85, 166 (1952). D. Bohm, A suggested interpretation of the quantum theory in terms of "hidden variables": Part II, Phys. Rev. 85, 180 (1952).

[26] Peter Holland, *Quantum Theory of Motion: An Account of the De Broglie-Bohm Causal Interpretation of Quantum Mechanics*, Cambridge University Press, 1995.

[27] Louis de Broglie, *The theory of measurement in wave mechanics*, Gauthier-Villars, Paris 1957.

[28] For descriptions of quantum mechanics and the quantum equations in terms of the quantum potential see:
R. Schiller, *Quasi-Classical Theory of the Spinning Electron*, Phys. Rev. 125, 1116 (1962). R. Schiller, *Quasi-Classical Theory of a Relativistic Spinning Electron*, Phys. Rev., 128, 1402 (1962). Ralph Schiller and Melvin Schwartz, *Structure-Independent Electrodynamics in the Electric-Dipole Approximation*, Phys. Rev. 126, 1582 (1962). Kenneth Rafanelli and Ralph Schiller, *Classical Motions of Spin-1/2 Particles*, Phys. Rev. 135, B279 (1964). Ralph Schiller, *Motion of a Dipole-Quadrupole System*, Phys. Rev. 169, 1282 (1968). F. Halbwachs, J. M. Souriau, and J. P. Vigier, *Le groupe d'invariance associe aux rotateurs relativistes et la theorie bilocale*, J. Phys. et Radium 22, 393 (1981). P. R. Holland, *Relativistic algebraic spinors and quantum motions in phase space*, Found. Phys. 16, 701 (1986). C. Dewdney, P. R. Holland and A. Kyprianidis, *What happens in a spin measurement?*, Phys. Lett. A 119, 259 (1986). C. Dewdney, P. R. Holland and A. Kyprianidis, *A causal account of non-local Einstein-Podolsky-Rosen spin correlations*, J. Phys. A: Math. Gen. Phys. 20, 4717 (1987). C. Dewdney, P. R. Holland and A. Kyprianidis, *A quantum potential approach to spin superposition in neutron interferometry*, Phys. Lett. A 121, 105 (1987). Dewdney, P. R. Holland and A. Kyprianidis and J. P. Vigier, *Spin and non-locality in quantum mechanics*, Nature 336, 536 (1988). P.R. Holland, *Causal interpretation of Fermi fields*, Phys. Lett. A 128, 009 (1988). Matej Paviz*, The quantization of a point particle with extrinsic curvature leads to the Dirac equation*, Phys. Lett. B 221. 264 (1989). P.R. Holland, *The Dirac equation in the de Broglie-Bohm theory of motion*, Found. Phys. 22, 1287 (1992). Anthony Challinor, Anthony Lasenby, Stephen Gull and Chris Doran, *A relativistic, causal account of a spin measurement*, Phys. Lett. A 218, 128 (1996). Anthony Challinor, Anthony Lasenby, Shyamal Somaroo, Chris Doran and Stephen Gull, *Tunnelling times of electrons*, Phys. Lett. A 227, 143 (1997). P. R. Holland, *New trajectory interpretation of quantum mechanics*, Found. Phys. 28, 881 (1998). C.R. Leavens and R. Sala Mayato, *Superluminal systematic particle velocity in relativistic stochastic Bohmian mechanics*, Phys. Lett. A 263, 001 (1999). P. R. Holland, *Uniqueness of paths in Quantum Mechanics*, Phys. Rev. A 60, 4326 (1999). Holland, Peter and Brown, Harvey R., *The non-relativistic limits of the Maxwell and Dirac equations: the role of Galilean and gauge invariance*, philsci-archive.pitt.edu/archive 00000999 (2002). C. Colijn, E.R. Vrscay, *Spin-dependent Bohm trajectories for hydrogen eigenstates*, Phys. Lett. A 300, 0334 (2002). Yuri A. Rylov, *Classical description of pair production*, arXiv:physics/0301020 (2003). Partha Ghose, Manoj K. Samal, Animesh Datta, *Bohmian picture of Rydberg atoms*, Phys. Lett. A 322, 0277 (2003). Harvey R. Brown, Peter Holland, *Simple applications of Noether's first theorem in quantum mechanics and electromagnetism*, arXiv:quant-ph/0302062 (2003). C. Colijn and E. R. Vrsc, Spin-Dependent Bohm Trajectories for Pauli and Dirac Eigenstates of Hydrogen, Found. Phys. 16, 303 (2003). Peter Holland and Chris Philippidis, *Implications of Lorentz covariance for the guidance equation in two-slit quantum interference*, Phys. Rev. A 67, 062105 (2003). C Colijn and E R Vrscay , *Spin-dependent Bohm trajectories associated with an electronic transition in hydrogen*, J. Phys. A: Math. Gen. Phys. 36, 4689 (2003). Caroline Colijn, *The de Broglie-Bohm Causal Interpretation of Quantum Mechanics and its Application to some Simple Systems*, Thesis, University of Waterloo 2003. C. Colijn and E. R. Vrscay, *Quantum relaxation in hydrogen eigenstates and two-state transitions*, Phys. Lett. A 327, 113 (2004). Peter Holland, *Computing the wavefunction from trajectories: particle and wave pictures in quantum mechanics and their relation*, Ann. Phys. (N.Y.) 315, 503 (2005). Peter Holland, *Hidden variables as computational tools: the construction of a relativistic spinor field*, Found. Phys. 36, 369 (2005).

[29] For works on the ‘hidden variables’ interpretation of the Klein Gordon equation:
Ya. P. Terletski and J. P. Vigier, *On the physical meaning of negative probabilities*, Sov. Phys. JETP 13, 356 (1961). J.P. Vigier, *Interpretation geometrique et physique de la formule du guidage en relativite generale*, C. R. Acad. Sci. Paris A 266, 598 (1968). Michael G. Fuda and Edward Furlani, *Zitterbewegung and the Klein paradox for spin-zero particles*, Am. J. Phys. 50, 0545 (1982). N. Cufaro-Petroni, C. Dewdney, P. Holland, T. Kiprianidids and J. P. Vigier, *Elimination of negative probabilities within the causal stochastic interpretation of quantum mechanics*, Phys. Lett. A 106, 368 (1984). N. Cufaro Petroni, C. Dewdney, P. Holland, A. Kyprianidis and J. P. Vigier, *Causal space-time paths of individual distinguishable particle motions in N -body quantum systems: Elimination of negative probabilities*, Lett. Nuovo Cimento 42, 285 (1985). P. R. Holland and J.P. Vigier, *Positive probabilities and* *the principle of equivalence for spin-zero particles in the causal stochastic interpretation of quantum mechanics*, Nuovo Cimento 88 B, 20 (1985). A. Kyprianidis, *Particle trajectories in relativistic Quantum Mechanics*, Phys. Lett. A 111, 111 (1985). P. R. Holland, *Geometry of dislocated de Broglie waves*, Found. Phys. 17, 345 (1987). P. R. Holland, A. Kyprianidis, J. P. Vigier*, Trajectories and causal phase-space approach to relativistic quantum mechanics*, Found. Phys. 17, 531 (1987). A. Kyprianidis, *Scalar time parametrization of relativistic quantum mechanics: The covariant Schrődinger equation*, Phys. Reports 155, 001 (1987). P.R. Holland, A. Kyprianidis, *Quantum potential, uncertainty and the classical limit*, Ann. Inst. H. Poincare 49, 325 (1988). Edward R. Floyd, *Progress in a trajectory interpretation of the Klein-Gordon equation*, Int. J. Theor. Phys. 27, 273 (1988). J. P. Vigier, *Real Physical Paths in the Quantum Mechanical Equivalence of the Einstein-de Broglie and Feynman Points of View in Quantum Mechanics*, 3rd. Int. Symp. Quant. Mech., Tokyo 1989. p. 140. P. N. Kaloyerou and J. P. Vigier, *Evolution time Klein-Gordon equation and derivation of its non-linear counterpart*, J. Phys. A: Math. Gen. Phys. 22, 0663 (1989). J.P. Vigier, *Explicit Mathematical Construction of Relativistic Nonlinear de Broglie Waves Described by Three-Dimensional (Wave and Electromagnetic) Solitons "Piloted" (Controlled) by Corresponding Solutions of Associated Linear Klein-Gordon and Schrõdinger Equations*, Found. Phys. 21, 0125 (1991). P.R. Holland, *The de Broglie-Bohm theory of motion and quantum field theory*, Phys. Reports 224, 95 (1993). F. Selleri, *Incompatibility between local realism and quantum mechanics for pairs of neutral kaons*, Phys. Rev. A 56, 3493 (1997). George Horton, Chris Dewdney and Alexei Nesteruk, *Time-like flows of energy momentum and particle trajectories for the Klein-Gordon equation*, J. Phys. A: Math. Gen. Phys. 33, 7337 (2000). G Horton, C Dewdney, *de Broglie's pilot wave theory for the Klein-Gordon equation*, arXiv:quant-ph/0109059 (2001). G Horton, C Dewdney, *A non-local, Lorentz-invariant, hidden-variable interpretation of relativistic quantum mechanics based on particle trajectories*, arXiv:quant-ph/0110007 (2001). Ali Shojai, Fatimah Shojai, *About Some Problems Raised by the Relativistic Form of De-Broglie–Bohm Theory of Pilot Wave*, Phys. Scr. 64, 413 (2001). Chris Dewdney, George Horton, *Relativistically invariant extension of the de Broglie-Bohm theory of quantum mechanics*, J. Phys. A: Math. Gen. Phys. 35, 10117 (2002). R. Tumulka, *Comment on ``Time-like flows of energy-momentum and particle trajectories for the Klein-Gordon equation''*, J. Phys. A: Math. Gen. Phys. 35, 7961 (2002). Philippe Bechouche, Norbert J. Mauser, Sigmund Selberg, *Nonrelativistic limit of Klein-Gordon-Maxwell to Schrödinger-Poisson*, arXiv:math-ph/0202201 (2002). Ali Mostafazadeh, *A Physical Realization of the Generalized PT-, C-, and CPT-Symmetries and the Position Operator for Klein-Gordon Fields*, arXiv:quant-ph/0307059 (2003). James Baker-Jarvis and Pavel Kabos*, Modified de Broglie approach applied to the Schrödinger and Klein-Gordon equations*, Phys. Rev. A 68, 042110 (2003). George Horton, Chris Dewdney, *A relativistically covariant version of Bohm's quantum field theory for the scalar field,* J. Phys. A: Math. Gen. Phys. 37, 11935 (2004). George Horton, Chris Dewdney, *Energy momentum flows for the massive vector field*, arXiv:quant-ph/0609198 (2006).

[30] R. Schiller, *Quasi-Classical Theory of the Spinning Electron*, Phys. Rev. 125, 1116 (1962).

[31] Additional work on Clifford algebras can be found in:

V. L. Figueiredo, E. Capelas de Oliveira, W. A. Rodrigues*, Covariant, algebraic, and operator spinors*, Int. J. Theor. Phys. 29, 0371 (1990). W. A. Rodrigues Jr., and E.C. Oliveira, *Dirac and Maxwell equations in the Clifford and Spin-Clifford bundles*, Int. J. Theor. Phys. 29, 0397 (1990). W. A. Rodrigues Jr., and V. I. Figuereido, *Real spin-Clifford bundle and the spinor structure of space-time*, Int. J. Theor. Phys. 29, 0413 (1990). Adolfo Maia, Jr. , Erasmo Recami , Waldyr A. Rodrigues, Jr. and Marcio A. F. Rosa, *Magnetic monopoles without string in the Kähler–Clifford algebra bundle: A geometrical interpretation*, J. Math. Phys. 31, 0502 (1990). J. R. R. Zeni and W. A. Rodrigues Jr., *A thoughtful study of Lorentz transformations by Clifford algebras*, Int. J. Modern Phys. A 7, 1793 (1992). J. Vaz Jr. and W. A. Rodrigues Jr, *On the equivalence of Dirac and Maxwell equations, and Quantum Mechanics*, Int. J. Theor. Phys. 32, 945 (1993). Matej Pavi, Erasmo Recami and Waldyr A. Rodrigues, Jr G. Daniele Maccarrone, Fabio Raciti and Giovanni Salesi, *Spin and electron structure*, Phys. Lett. B 318, 481 (1993). Waldyr A. Rodrigues, Jr. and Jayme Vaz, Jr., *About zitterbewegung and electron structure*, Phys. Lett. B 318, 623 (1993). Jayme Vaz, Jr. and Waldyr A. Rodrigues, Jr., *Zitterbewegung and the electromagnetic field of the electron*, Phys. Lett. B 319, 203 (1993). Giovanni Salesi and Erasmo Recami, *Field theory of the spinning electron and internal motions*, Phys. Lett. A 190, 137 (1994). Yuri A. Rylov, *Dirac equation in terms of hydrodynamic variables*, Advances in Applied Clifford Algebras 05, 001 (1995). Erasmo Recami, Giovanni Salesi, *Field Theory of the Electron Spin and Zitterbewegung*, arXiv:hep-th/9508168 (1995). J. Vaz, Jr., W. A. Rodrigues, Jr., *Maxwell and Dirac theories as an already unified theory*, arXiv:hep-th/9511181 (1995). W. A. Rodrigues, Jr., J. Vaz, Jr., *Subluminal and superluminal solutions in vacuum of the Maxwell equations and the massless Dirac equation*, arXiv:hep-th/9511182 (1995). M. Pavsic, Erasmo Recami, W.A. Rodrigues Jr, *Electron structure, Zitterbewegung, and the new non-linear Dirac-like equation*, Hadronic J. 18, 098 (1995). Jayme Vaz Jr, *The Barut and Zanghi model, and some generalizations*, Phys. Lett. B 344, 149 (1995). Jayme Vaz Jr, *A spinning particle model including radiation reaction*, Phys. Lett. B 345, 448 (1995). Erasmo Recami, Giovanni Salesi, *Kinematics and hydrodynamics of spinning particles*, arXiv:quant-ph/9607025 (1996). G. Salesi, *Spin and Madelung fluid*, Modern Physics Letters A 11, 1815 (1996). Giovanni Salesi, Erasmo Recami, *The spinning electron: Hidrodynamical formulation, and quantum limit, of the Barut-Zanghi theory*, Found., Phys. Lett. 10, 533 (1996). 1996 Giovanni Salesi, Erasmo Recami, *Field theory of the spinning electron: Internal motions*, arXiv:hep-th/9607211 (1996). Giovanni Salesi, Erasmo Recami, *About the kinematics of spinning particles*, arXiv:hep-th/9607213 (1996). Giovanni Salesi, Erasmo Recami, *Velocity field and operator in (non relativistic) quantum mechanics*, Found. Phys. 28, 763 (1998). Rodrigues, W.A. and Vaz, J., *From electromagnetism to relativistic quantum mechanics*, Found. Phys. 28, 789 (1998). G.Salesi, E.Recami, H.Hernandez F., L.C.Kretly, *Hydrodynamics of Spinning Particles*, arXiv:hep-th/9802106 (1998). Erasmo Recami,

Giovanni Salesi, *Kinematics and hydrodynamics of spinning particles*, Phys. Rev. A 57, 0098 (1998). Giovanni Salesi, Erasmo Recami, *Spin effects on the cyclotron frequency for a Dirac electron*, Phys. Lett. A 267, 219 (2000). Giovanni Salesi, *Non-Newtonian Mechanics*, Int. J. Modern Phys. A 17, 347 (2002). Waldyr a. Rodrigues jr., *The relation between Maxwell, Dirac, and the Seiberg-Witten equations*, International Journal of Mathematics and mathematical Sciences 43, 2707 (2003). G. Salesi, *Non-relativistic classical mechanics for spinning particles*, Int. J. Modern Phys. A 20, 2027 (2005).

[32] Peter Holland and Chris Philippidis, *Implications of Lorentz covariance for the guidance equation in two-slit quantum interference*, Phys. Rev. A 67, 062105 (2003).

[33] R. Schiller, *Quasi-Classical Theory of the Spinning Electron*, Phys. Rev. 125, 1116 (1962). R. Schiller, *Quasi-Classical Theory of a Relativistic Spinning Electron,* Phys. Rev., 128, 1402 (1962).

[34] Works on self-field solitons can be found in:

Nathan Rosen, *A Field Theory of Elementary Particles*, Phys. Rev. 55, 94 (1939). Finkelstein, *On the Quantization of a Unitary Field Theory*, PR 75, 1079 (1949). Maria J. Esteban, Vladimir Georgiev and Eric Séré, *Stationary solutions of the Maxwell-Dirac and the Klein-Gordon-Dirac equations*, Calculus of Variations and Partial Differential Equations 4, 265 (1996). Philippe Bechouche, Norbert J. Mauser, Sigmund Selberg, *Nonrelativistic limit of Klein-Gordon-Maxwell to Schrödinger-Poisson*, arXiv:math-ph/0202201 (2002). F. A. Kaempffer, *Elementary Particles as Self-Maintained Excitations*, Phys. Rev. 099, 1614 (1955). B. H. Armstrong, *Anomalous Solutions to the Dirac and Schrödinger Equations for the Coulomb Potential*, Phys. Rev. 130, 2506 (1963). T. Wakano, *Intensely Localized Solutions of the Classical Dirac-Maxwell Field Equations*, Prog. Theor. Phys. 35, 1117 (1966). J. R. Carver, *The Quantum Electrostatic Electron*, Am. J. Phys. 41, 552 (1973). Naoki Itoh, *Radiation reaction due to magnetic dipole radiation*, Phys. Rev. A 43, 1002 (1991). A. Garrett Lisi, *A Solution of the Maxwell-Dirac Equations in 3+1 Dimensions*, J. Phys. A: Math. Gen. Phys. 28, 5384 (1995). Moshe Flato, Jacques C.H. Simon, Erik Taflin, *Asymptotic completeness, global existence and the infrared problem for the Maxwell-Dirac equations*, arXiv:hep-th/9502061 (1995). Maria J. Esteban, Vladimir Georgiev and Eric Séré, *Stationary solutions of the Maxwell-Dirac and the Klein-Gordon-Dirac equations*, Calculus of Variations and Partial Differential Equations 4, 265 (1996). Michael D. Crisp, *Relativistic neoclassical radiation theory*, Phys. Rev. A 54, 0087 (1996). Abenda, *Solitary waves for Maxwell-Dirac and Coulomb-Dirac models,* Ann. Inst. H. Poincare A68, 229 (1998). C. Sean Bohun and F. I. Cooperstock, *Dirac–Maxwell Solitons*, Phys. Rev. A 60, 4291 (1999). Felix Finster, Joel Smoller and Shing-Tung Yau, *Particle-like solutions of the Einstein-Dirac-Maxwell equations*, Phys. Lett. A 259, 431 (1999). Coclite, *A Multiplicity Result for the Linear Schrödinger-Maxwell Equations with Negative Potential*, arXiv:math/0109017 (2001). Chris Radford, *The Stationary Maxwell-Dirac Equations*, arXiv:math-ph/0112037 (2001), D. Ranganathan, *A self consistent solution to the Einstein Maxwell Dirac Equations* , arXiv:gr-qc/0306090 (2003). Philippe Bechouche, Norbert J. Mauser, Sigmund Selberg*, On the asymptotic analysis of the Dirac-Maxwell system in the nonrelativistic limit*, arXiv:math/0303079 (2003). I.D. Feranchuk, *Nonperturbative Description Of The Mass And Charge Renormalization In Quantum Electrodynamics*, arXiv:hep-th/0309072 (2003). I.D.Feranchuk, S.I.Feranchuk, *Soliton-like excitation in large-alpha QED*, arXiv:math-ph/0605028 (2006). Ilya D. FERANCHUK and Sergey I. FERANCHUK, *Self-Localized Quasi-Particle Excitation in Quantum Electrodynamics and Its Physical Interpretation*, Symmetry, Integrability and Geometry 3, 117 (2007).

[35] Nonlinear theories:

R. Finkelstein, R. Lelevier, and M Ruderman, *Nonlinear Spinor Fields*, Phys. Rev. 83, 326 (1951). R. Finkelstein, C. Fronsdal, and P. Kaus, *Nonlinear Spinor Field*, Phys. Rev. 103, 1571 (1956). T. W. B. Kibble, *Lorentz Invariance and the Gravitational Field*, J. Math. Phys. 2, 212 (1961). U. Enz, *Discrete Mass, Elementary Length, and a Topological Invariant as a Consequence of a Relativistic Invariant Variational Principle*, Phys. Rev. 131. 1392 (1963). G. Derrick, *Comments on Nonlinear Wave Equations as Models for Elementary Particles*, J. Math. Phys. 5, 1252 (1964). Gerald Rosen, *Particlelike Solutions to Nonlinear Scalar Wave Theories*, J. Math. Phys. 6, 1269 (1965). G. Rosen, *Existence of Particlelike Solutions to Nonlinear Field Theories*, J. Math. Phys. 7, 2066 (1966). G. Rosen, *Nonexistence of Localized Periodic Solutions to Nonlinear Field Theories*, J. Math. Phys. 7, 2071 (1966). G. Rosen, *Particlelike Solutions to Nonlinear Complex Scalar Field Theories with Positive-Definite Energy Densities*, J. Math. Phys. 9, 0996 (1968). G. Rosen, *Charged Particlelike Solutions to Nonlinear Complex Scalar Field Theories*, J. Math. Phys. 9, 0999 (1968). G. Pinski, *Interaction of Particlelike Solutions in a Classical Nonlinear Model Field Theory*, J. Math. Phys. 9, 1323 (1968). Derrick, Anderson, *Stability of Time-Dependent Particlelike Solutions in Nonlinear Field Theories. I*, J. Math. Phys. 11, 1336 (1970). Mario Soler, *Classical, Stable, Nonlinear Spinor Field with Positive Rest Energy*, PRD 1, 2766 (1970). Derrick, Anderson, *Stability of Time-Dependent Particlelike Solutions in Nonlinear Field Theories. II*, J. Math. Phys. 2, 945 (1971). A. F. Rañada, *Relativistic quantum mechanics of the hydrogen atom as the weak-field limit of a nonlinear theory*, Int. J. Theor. Phys. 16, 795 (1977). Luis García and Juan M. Usón, *A simple model for Poincaré self-stresses*, Found. Phys. 10, 137 (1980). Ph. Gueret and J. P. Vigier, *Nonlinear Klein Gordon equation carrying a nondispersive soliton like singularity*, Lett. Nuovo Cimento 35, 256 (1982). Ph. Gueret and J. P. Vigier, *Soliton model of Einstein's 'Nadelstrahlung' in real physical Maxwell waves*, Lett. Nuovo Cimento 35, 260 (1982). Ph. Gueret and J. P. Vigier, *Relativistic wave equations with quantum potential nonlinearity*, Lett. Nuovo Cimento 38, 125 (1983). F. I. Cooperstock, N. Rosen, *A nonlinear gauge-invariant field theory of leptons*, Int. J. Theor. Phys. 28, 423 (1989).

[36] Nonlinear static self fields:

V. Benci, D. Fortunato, *Solitary waves of the nonlinear Klein-Gordon equation coupled with the Maxwell equations,* Reviews in Mathematical Physics 4, 409 (2002). C. Sparber, P. Markowich, *Semiclassical Asymptotics for the Maxwell - Dirac System*,

arXiv:math-ph/0305009 (2003). Lubomir M. Kovachev, *Nonlinear Amplitude Maxwell-Dirac Equations. Optical Leptons*, arXiv:nlin/0304037 (2003). Antonio Azzollini, Alessio Pomponio, *Ground state solutions for the nonlinear Schrödinger-Maxwell equations with a singular potential,* arXiv:0706.1679 (2007). Antonio Azzollini, Alessio Pomponio, *Ground state solutions for the nonlinear Klein-Gordon-Maxwell equations*, arXiv:0810.1128 (2008). Eamonn Long, David Stuart, *Effective dynamics for solitons in the nonlinear Klein Gordon Maxwell system and the Lorentz force law*, Reviews in Mathematical Physics 21, 459 (2009). Vieri Benci, Donato Fortunato, *Existence of hylomorphic solitary waves in Klein-Gordon and in Klein-Gordon-Maxwell equations*, arXiv:0903.3508 (2009).

[37] Robert W. Boyd, *Nonlinear Optics*, Harcourt Brace & Company, 1992. Govind P. Agrawal, *Nonlinear Fiber Optics*, Academic Press, Incorporated, 1989.

[38] Electron self-energy:
J.R. Oppenheimer, *Note on the Theory of the Interaction of Field and Matter,* Phys. Rev. 35, 461 (1930). I. Waller, *Bemerkungen über die Rolle der Eigenenergie des Elektrons in der Quantentheorie der Strahlung*, Zeitschrift für Physik 62, 673 (1930). Werner Heisenberg, *Die Selbstenergie des Elektrons (The self-energy of the electron*), Zeitschrift für Physik 65, 004 (1930). V. Weisskopf, *Über die Selbstenergie des Elektrons (The self-energy of the electron*), Zeitschrift für Physik 89, 027 (1934). V. Weisskopf, *Berichtigung (Correction to the paper: The self-energy of the electron*), Zeitschrift für Physik 90, 817 (1934). W. H. Furry, A *Symmetry Theorem in the Positron Theory*, Phys. Rev. 51, 125 (1937). V. Weisskopf, *On the self-energy and the electromagnetic field of the electron,* Phys. Rev. 56, 0072 (1939). Sakata and Hara, *The self-energy of the electron and the mass difference of nucleons*, Prog. Theor. Phys. 2, 030 (1947). Jo Bovy, *The self-energy of the electron: a quintessential problem in the development of QED*, arXiv:physics/0608108 (2006).

[39] Renormalization:
E.C.C. Stückelberg, *An Unambiguous Method of Avoiding Divergence Difficulties in Quantum Theory*, Nature 153, 143 (1944). S. Tomonaga, *On a relativistically invariant formulation of the quantum theory of wave fields*, Prog. Theor. Phys. 1, 027 (1946). S. Tomonaga, KOBA, TATI, *On a Relativistically Invariant Formulation of the Quantum Theory of Wave Fields. II.: Case of Interacting Electromagnetic and Electron Fields*, Prog. Theor. Phys. 2, 101 (1947). S. Tomonaga, KOBA, TATI, *On a relativistic invariant formulation of the quantum theory of wave fields. III. Case of interacting electromagnetic and electron fields*, Prog. Theor. Phys. 2, 198 (1947). Richard Feynman, *Relativistic Cut-Off for Quantum Electrodynamics*, Phys. Rev. 74, 1430 (1948). J. Schwinger, *Quantum Electrodynamics. I. A Covariant Formulation*, Phys. Rev. 74, 1439 (1948). Kanesawa and S. Tomonaga, *On a relativistic invariant formulation of the quantum theory of wave fields. IV. Case of interacting electromagnetic and meson fields*, Prog. Theor. Phys. 3, 001 (1948). Kanesawa and S. Tomonaga, *On a relativistic invariant formulation of the quantum theory of wave fields. V. Case of interacting electromagnetic and meson fields*, Prog. Theor. Phys. 3, 101 (1948). Richard Feynman, *Space-time approach to non-relativistic quantum mechanics*, Rev. Mod. Phys. 20, 367 (1948). D. Rivier, *Une methode d'elimination des infinites en theorie des champs quantifies. Application au moment magnetique du neutron*, Helv. Phys. Acta 22, 265 (1949). F. Dyson, *The radiation theories of Tomonaga, Schwinger, and Feynman,* Phys. Rev. 75, 0486 (1949). J. Schwinger, *Quantum electrodynamics II,* Phys. Rev. 75, 0651 (1949). J. Schwinger, *On radiative corrections to electron scattering*, Phys. Rev. 75, 0898 (1949). F. Dyson, *The S-Matrix in quantum electrodynamics*, Phys. Rev. 75, 1736 (1949). Richard Feynman, *The theory of positrons*, Phys. Rev. 76, 0749 (1949). Richard Feynman, *Space-time approach to quantum electrodynamics*, Phys. Rev. 76, 0769 (1949). J. Schwinger, *Quantum electrodynamics, III: The electromagnetic properties of the electron-Radiative corrections to scatterin*g, Phys. Rev. 76, 0790 (1949). FUKUDA, MIYAMOTO, Tomonaga, *A Self-Consistent Subtraction Method in the Quantum Field Theory. II_1*, Prog. Theor. Phys. 4, 047 (1949). FUKUDA, MIYAMOTO, Tomonaga, *A Self-Consistent Subtraction Method in the Quantum Field Theory.II_2*, Prog. Theor. Phys. 4, 121 (1949). W. Pauli and F. Villars, *On the invariant regularization in relativistic quantum theory*, Rev. Mod. Phys. 21, 434 (1949). J.C. Ward, *An Identity in Quantum Electrodynamics*, Phys. Rev. 78, 182 (1950). G. C. Wick, *Evaluation of the collision matrix*, Phys. Rev. 80, 268 (1950). Richard Feynman, *Mathematical formulation of the quantum theory of electromagnetic interaction,* Phys. Rev. 80, 440 (1950). F. J. Dyson, *Heisenberg operators in quantum electrodynamics*. I., Phys. Rev. 82, 428 (1951). J. Schwinger, *On gauge invariance and vacuum polarization*, Phys. Rev. 82, 664 (1951). J. Schwinger*, The theory of quantized fields, I,* Phys. Rev. 82, 914 (1951). F. J. Dyson, *Heisenberg operators in quantum electrodynamics. II.,* Phys. Rev. 83, 0608 (1951). S. Gupta, *On the elimination of divergences from quantum electrodynamics*, Proc. Phys. Soc. A 64, 426 (1951). J. Schwinger, *On the Green's function of quantized fields, I*, Proceedings of the National Academy of Sciences USA 37, 452 (1951). G. Källen, *On the definition of the renormalization constants in quantum electrodynamics,* Helv. Phys. Acta 25, 417 (1952). J. Schwinger, *The theory of quantized fields, II,* Phys. Rev. 91, 0713 (1953).

[40] I.D. Feranchuk, *Nonperturbative Description Of The Mass And Charge Renormalization In Quantum Electrodynamics,* arXiv:hep-th/0309072 (2003); I.D.Feranchuk, S.I.Feranchuk, *Soliton-like excitation in large-alpha QED*, arXiv:math-ph/0605028 (2006). Ilya D. FERANCHUK and Sergey I. FERANCHUK, *Self-Localized Quasi-Particle Excitation in Quantum Electrodynamics and Its Physical Interpretation*, Symmetry, Integrability and Geometry: Methods and Applications 3, 117 (2007).

[41] Robert W. Boyd, *Nonlinear Optics*, Harcourt Brace & Company 1992, 2007.

[42] S. Weinberg, *The Quantum Theory of Fields: Foundations*, Cambridge University Press, 1995. S. Weinberg, *The Quantum Theory of Fields: Modern Applications*, Cambridge University Press, 1996.

[43] W. Greiner, J. Reinhardt, *Quantum Electrodynamics*, Springer, 3rd ed. 1992, page 48.

[44] R. Serber, *Linear Modifications in the Maxwell Field Equations*, Phys. Rev. 48, 049 (1935). E.A. Uehling, *Polarization Effects in the Positron Theory*, Phys. Rev. 48, 055 (1935).

[45] H. Euler and B. Kockel, *Über die Streuung von Licht an Licht nach der Diracschen Theorie*, Naturwissenschaften 23, 246 (1935). H. Euler, *Über die Streuung von Licht an Licht nach der Diracschen Theorie*, Ann. d. Physik 26, 398 (1936). W. Heisenberg and H. Euler, *Folgerungen aus der Diracschen Theorie des Positrons*, Zeitschrift für Physik 98, 714 (1936).

[46] C. Cohen-Tannoudji, J. Dupont-Roc and G. Grynberg, *Photons and Atoms: Introduction to Quantum Electrodynamics*, NY Wiley Interscience, 1989.

[47] J. D. Jackson, *Classical Electrodynamics*, John Wiley, New York, 1975, 1962.

[48] Marin Soljacic and Mordechai Segev, *Sellf-trapping of electromagnetic beams in vacuum supported by QED nonlinear effects*, Phys. Rev. A 62, 043817 (2000).

[49] Tomasz Radozycki, *Quantum effects in the evolution of vortices in the electromagnetic field*, Phys. Rev. E 69 066616 (2004), arXiv:quant-ph/0401115.

[50] Finn Ravndal, *Effective Field Theory of QED Vacuum Fluctuations*, Plenary talk at PASCOS-98, Boston, Massachusetts, USA, March 22-29, 1998, arXiv:hep-ph/9806338. Xinwei Kong, Finn Ravndal, *Radiative Corrections to the Casimir Energy*, Phys. Rev. Lett. 79, 545 (1997), arXiv:quant-ph/9703019.

[51] Ralph Schiller, *New Transition to Ray Optics*, Phys. Rev. 097, 1421 (1955). W. B. Joyce, *Classical-particle description of photons and phonons*, Phys. Rev. D 09, 3234 (1974). W. B. Joyce, *Radiation force and the classical mechanics of photons and phonons*, Am. J. Phys. 43, 245 (1975). Frank R. Tangherlini, *Particle approach to the Fresnel coefficients*, Phys. Rev. A 12, 0139 (1975). R. A. Young, *Thinking of the photon as a quantum-mechanical particle*, Am. J. Phys. 44, 1043 (1976). R. D. Prosser, *The interpretation of diffraction and interference in terms of energy flow*, Int. J. Theor. Phys. 15, 169 (1976). R. D. Prosser, *Quantum theory and the nature of interference*, Int. J. Theor. Phys. 15, 181 (1976). J. P. Wesley, *A resolution of the classical wave-particle problem*, Found. Phys. 14, 155 (1984). M. Davidovic, A. S. Sanz, D. Arsenovic, M. Bozic, S. Miret-Artes, *Trajectory Aspects of Electromagnetic Waves: A Prescription to Determine Photon Paths*, Phys. Scr. T135, 14009 (2009).

[52] Hans C. Ohanian, *What is spin?,* Am. J. Phys. 54, 0500 (1986).

[53] Richard A. Pakula, *Realistic model for radiation-matter interaction*, arXiv:quant-ph/0405055 (2004).

[54] Ralph Schiller, *New Transition to Ray Optics*, Phys. Rev. 097, 1421 (1955). W. B. Joyce, *Classical-particle description of photons and phonons*, Phys. Rev. D 09, 3234 (1974). W. B. Joyce, *Radiation force and the classical mechanics of photons and phonons*, Am. J. Phys. 43, 245 (1975). Frank R. Tangherlini, *Particle approach to the Fresnel coefficients*, Phys. Rev. A 12, 0139 (1975). R. A. Young, *Thinking of the photon as a quantum-mechanical particle*, Am. J. Phys. 44, 1043 (1976). R. D. Prosser, *The interpretation of diffraction and interference in terms of energy flow*, Int. J. Theor. Phys. 15, 169 (1976). R. D. Prosser, *Quantum theory and the nature of interference*, Int. J. Theor. Phys. 15, 181 (1976). J. P. Wesley, *A resolution of the classical wave-particle problem*, Found. Phys. 14, 155 (1984). M. Davidovic, A. S. Sanz, D. Arsenovic, M. Bozic, S. Miret-Artes, *Trajectory Aspects of Electromagnetic Waves: A Prescription to Determine Photon Paths*, Phys. Scr. T135, 14009 (2009).

[55] Xinwei Kong, Finn Ravndal, *Quantum Corrections to the QED Vacuum Energy*, Nucl.Phys. B526 627 (1998), arXiv:hep-ph/9803216. Finn Ravndal, Radiative *Corrections to the Stefan-Boltzmann Law*, arXiv:hep-ph/9709220 (1997).

[56] Walter Greiner, *Relativistic Quantum Mechanics, Wave Equations*, Springer Verlag 1981, 1997, Exercise 2.1, page 105.

[57] G. Breit, *The Effect of Retardation on the Interaction of Two Electrons*, Phys. Rev. 34, 553 (1929). G. Breit, *Dirac's Equation and the Spin-Spin Interactions of Two Electrons*, Phys. Rev. 39, 616 (1932).

[58] Cohen-Tannoudji, *Quantum Mechanics I & II*, John Wiley & Sons, 1973, 1992, page 789.

[59] J. Tersoff and David Bayer, *Quantum statistics for distinguishable particles*, Phys. Rev. Lett. 50, 0553 (1983).

[60] P.A.M. Dirac, *The Quantum Theory of the Emission and Absorption of Radiation*, Proc. Roy. Soc. London A 114, 243 (1927).

[61] Max Born and P. Jordan, *Zur Quantenmechanik*, Zeitschrift für Physik 34, 858 (1925).

[62] Werner Heisenberg and W. Pauli, *Zur Quantenelektrodynamik der Wellenfelder*, Zeitschrift für Physik 56, 001 (1929). Werner Heisenberg and W. Pauli, *Zur Quantenelektrodynamik der Wellenfelder II*, Zeitschrift für Physik 59, 168 (1930). Wolfgang Pauli and V. Weisskopf, *Über die Quantisierung der skalaren relativistischen Wellengleichung (The quantization of the scalar relativistic wave equation)*, Helv. Phys. Acta 7, 709 (1934).

[63] L. D. Landau and R. Peierls, *Quantenelektrodynamik im Konfigurationsraum*, Zeitschrift für Physik 62, 188 (1930).

[64] V. Fock, *Konfigurationstraum und zweite Quantelung*, Zeitschrift für Physik 75, 622 (1932).

[65] Richard Feynman, *Space-time approach to non-relativistic quantum mechanics*, Rev. Mod. Phys. 20, 367 (1948). Richard Feynman, *The theory of positrons*, Phys. Rev. 76, 0749 (1949). Richard Feynman, *Space-time approach to quantum electrodynamics*, Phys. Rev.76, 0769 (1949).

[66] S. Weinberg, *A Model of Leptons*, Phys. Rev. Lett. 19, 1264 (1967).

[67] Leslie C. Allen, Joseph H. Eberly, *Optical Resonance and Two-Level Atoms*, Dover Publisher, Wiley, New York, 1975.

[68] Rodney Loudon, *The Quantum Theory of Light*, Oxford University Press, Oxford, 1975, 1983.

[69] James Bjorken and Sidney Drell, *Relativistic Quantum Mechanics*, McGraw Hill, New York, 1965.

[70] W. Greiner, *Field Quantization,* Springer Verlag, 1996.

[71] Abrikosov, Gorkov, Dzyaloshinskii, *Quantum Field Theoretical Methods in Statistical Physics,* Pergamon Press , London New York, 2nd ed. 1965

[72] H. Nikolic, *Superluminal velocities and nonlocality in relativistic mechanics with scalar potential*, arXiv:1006.1986 (2010). H. Nikolic, *Relativistic Quantum Mechanics and Quantum Field Theory*, Chapter 8. of the book "Applied Bohmian Mechanics: From Nanoscale Systems to Cosmology", edited by X. Oriols and J. Mompart (Pan Stanford Publishing, 2012); ISBN: 978-981-4316-39-2, Copyright 2012 by Pan Stanford Publishing Pte. Ltd, arXiv:1205.1992 (2012). H. Nikolic, *Time and probability: From classical mechanics to relativistic Bohmian mechanics*, arXiv:1309.0400 (2013).

[73] A. O. Barut, *Electrodynamics and classical theory of fields and particles*, Dover Publications, Inc., New York, 1964, 1980.

[74] K. W. Chan, C. K. Law, and J. H. Eberly, *Localized Single-Photon Wave Functions in Free Space*, Phys. Rev. Lett. 88, 100402 (2002). K. W. Chan, C. K. Law, J. H. Eberly, *Quantum entanglement in photon-atom scattering*, Phys. Rev. A 68, 022110 (2003). M. V. Fedorov, M. A. Efremov, A. E. Kazakov, K. W. Chan, C. K. Law, J. H. Eberly, *Packet narrowing and quantum entanglement in photoionization and photodissociation*, Phys. Rev. A 69, 052117 (2004). M. V. Fedorov, M. A. Efremov, A. E. Kazakov, K. W. Chan, C. K. Law, and J. H. Eberly, *Spontaneous emission of a photon: wave packet structures and atom-photon entanglement*, Phys. Rev. A 72, 032110 (2005).

[75] Max Born and P. Jordan, *Zur Quantenmechanik*, Zeitschrift für Physik 34, 858 (1925). Edwin Schrödinger, *Der stetige Übergang von der Mikro- zur Makromechanik*, Naturwissenschaften 14, 664 (1926).

[76] P.A.M. Dirac, *Quantum mechanics and a preliminary investigation of the hydrogen atom*, Proc. Roy. Soc. London A110, 561 (1926). W. Pauli, *On the hydrogen spectrum from the standpoint of the new quantum mechanics*, Zeitschrift für Physik 36, 336 (1926). Edwin Schrödinger, *Quantizierung als Eigenwert Problem*, Ann. d. Physik 79, 361 (1926).

[77] P. W. Milonni, The Quantum Vacuum- An Introduction to Quantum Electrodynamics, Academic Press, Incorporated , San Diego, 1994.

[78] Mark Hillery, *An Introduction to the Quantum Theory of Nonlinear Optics*, Acta Physica Slovaca 59, 001 (2009). Chapter 9: Other Approaches. Also arXiv:quant-ph/0901.3439.

[79] P. W. Milloni, *Why spontaneous emission?*, Am. J. Phys. 52, 340 (1984).

[80] W. Greiner, J. Reinhardt, *Quantum Electrodynamics*, Springer, 3rd ed., 1992, page 48.

[81] Richard Feynman, *Space-time approach to non-relativistic quantum mechanics*, Rev. Mod. Phys. 20, 367 (1948). Richard Feynman, *The theory of positrons*, Phys. Rev. 76, 0749 (1949). Richard Feynman, *Space-time approach to quantum electrodynamics*, Phys. Rev.76, 0769 (1949).

[82] Julian Schwinger, *Particles and sources*, Phys. Rev. 152, 1219 (1966). Julian Schwinger, *Sources and electrodynamics*, Phys. Rev. 158, 1391 (1967).

[83] W.J. Lehr and J. L. Park, *A stochastic derivation of the Klein-Gordon equation*, J. Math. Phys. 18, 1235 (1977). N. Cufaro-Petroni and J. P. Vigier, *Markov process at the velocity of light: the Klein-Gordon statistic*, Int. J. Theor. Phys. 18, 807 (1979). J.P. Vigier , *Model of quantum statistics in terms of a fluid with irregular stochastic fluctuations propagating at the velocity of light: A derivation of Nelson's equations*, Lett. Nuovo Cimento 24, 265 (1979). C. Fenech, M. Moles, and J. P. Vigier, *Internal rotations of spinning particles*, Lett. Nuovo Cimento 24, 056 (1979). N. Cufaro-Petroni and J. P. Vigier, *Stochatsic derivation of the Dirac equation in terms of a fluid of spinning tops endowed with random fluctuations at the velocity of light*, Phys. Lett. A 81, 012 (1981). N. Cufaro-Petroni and J. P. Vigier*, Random motions at the velocity of light and relativistic quantum mechanics*, J. Phys. A: Math. Gen. Phys. 17, 599 (1984). A. Kyprianidis and D. Sardelis, *A H -theorem in the causal stochastic interpretation of quantum mechanics*, Lett. Nuovo Cimento 39, 337 (1984). N. Cufaro-Petroni, C. Dewdney, P. Holland, T. Kiprianidids and J. P. Vigier, *Realistic Physical Origin of the Quantum Observable Operator Algebra in the Frame of the Causal Stochastic Interpretation of Quantum Mechanics: The Relativistic Spin-Zero Case*, Phys. Rev. D 32, 1375 (1985). C. Dewdney, P. R. Holland, A. Kyprianidis, Z. Maric and J. P. Vigier, *Stochastic physical origin of the quantum operator algebra and phase space interpretation of the Hilbert space formalism: the relativistic spin zero case*, Phys. Lett. A 113, 359 (1986). A. Valentini, *Signal-locality, uncertainty, and the subquantum H-theorem. I.*, Phys. Lett. A 156, 005 (1991). A. Valentini, *Signal-locality, uncertainty, and the subquantum H-theorem. II.*, Phys. Lett. A 158, 001 (1991). Yuri Shtanovm, *Origin of quantum randomness in the pilot wave quantum mechanics*, arXiv:quant-ph/9705024 (1997).

[84] P. W. Milonni, *The Quantum Vacuum- An Introduction to Quantum Electrodynamics*, Academic Press, Incorporated , San Diego, 1994.

[85] T. A. Welton, *Some Observable Effects of the Quantum-Mechanical Fluctuations of the Electromagnetic Field,* Phys. Rev. 74, 1157 (1948).

[86] A. O. Barut and J. P. Dowling, *Self-field quantum electrodynamics: The two-level atom,* Phys. Rev. A, 41, 2284 (1990).

[87] V. Weisskopf , *Über die Selbstenergie des Elektrons (The self-energy of the electron)*, Zeitschrift für Physik 89, 027 (1934). V. Weisskopf, *Berichtigung (Correction to the paper: The self-energy of the electron)* , Zeitschrift für Physik 90, 817 (1934).

[88] A. H. Compton, *The Size and Shape of the Electron*, Phys. Rev. 14, 020 (1919). A. H. Compton, *The Size and Shape of the Electron*, Phys. Rev. 14, 247 (1919).

*Time dependence of the field energy densities surrounding sources: Application to scalar mesons near point sources and to electromagnetic fields near molecules*, Phys. Rev. A 36, 0475 (1987). E. A. Power and T. Thirunamachandran, *Quantum electrodynamics with nonrelativistic sources. IV. Poynting vector, energy densities, and other quadratic operators of the electromagnetic field,* Phys. Rev. A 45, 0054 (1992). A. Salam and T. Thirunamachandran, *Maxwell fields and Poynting vector in the proximity of a chiral molecule*, Phys. Rev. A 50, 4755 (1994). G Compagno, G M Palma, R Passante and F Persico, *Atoms dressed and partially dressed by the zero-point fluctuations of the electromagnetic field*, J. Phys. B: Atom. Molec. Phys. 28, 1105 (1995).

[112] P. Carbonaro, G. Compagno, and F. Persico, *Canonical dressing of atoms by intense radiation fields*, Phys. Lett. A 73, 097 (1979). G. Compagno and F. Persico, *Canonical transformatoin for single-atom resonance fluorescence: The strong-driving-field limit*, Phys. Rev. A 22, 2108 (1980). G. Compagno, R. Passante, and F. Persico, *The role of the cloud of virtual photons in the shift of the ground state energy of a hydrogen atom,* Phys. Lett. A 98, 0253 (1983). G. Compagno, S. Vivirito, and F. Persico, *Cloud of virtual photons surrounding a two-level atom driven by an external field,* Phys. Rev. A 46, 7303 (1992). G Compagno, G M Palma, R Passante and F Persico, *Atoms dressed and partially dressed by the zero-point fluctuations of the electromagnetic field*, J. Phys. B: Atom. Molec. Phys. 28, 1105 (1995). D. P. Craig and E. A. Power*, The asymptotic Casimir-Polder potential from second-order perturbation theory and its generalization for anisotropic polarizabilities*, Int. J. Quantum Chemistry 3, 903 (1969). E. A. Power*, Long-range retarded intermolecular forces*, Phys. Rev. A 10, 0756 (1974). E. A. Power and T. Thirunamachandran, *On the nature of the Hamiltonian for the interaction of radiation with atoms and molecules: (e/mc)p.A, -μ.E, and all that*, Am. J. Phys. 46, 370 (1978). R. Passante and E. A. Power, *The Lamb shift in non-relativistic quantum electrodynamics*, Phys. Lett. A 122, 014 (1987). G. Compagno and E. A. Power, *Alternative effective Hamiltonians in nonrelativistic quantum electrodynamics,* Phys. Rev. A 38, 4340 (1988). E. A. Power and T. Thirunamachandran, *Zero-point energy differences and many-body dispersion forces*, Phys. Rev. A 50, 3929 (1994).

[113] W. Nagourney, J. Sandberg, and H. Dehmelt, *Shelved optical electron amplifier: Observation of quantum jumps*, Phys. Rev. Lett. 56, 2797 (1986). Th. Sauter, W. Neuhauser, R. Blatt, and P. E. Toschek , *Observation of quantum jumps*, Phys. Rev. Lett. 57, 1696 (1986). J. C. Bergquist, R. G. Hulet, W. M. Itano, and D. J. Wineland, *Observation of quantum jumps in a single atom*, Phys. Rev. Lett. 57, 1699 (1986). E. Peik, G. Hollemann, and H. Walther, *Laser cooling and quantum jumps of a single indium ion,* Phys. Rev. A 49, 0402 (1994). O. Benson, G. Raithel, and H. Walther, *Quantum jumps of the micromaser field: Dynamic behavior close to phase transition points*, Phys. Rev. Lett. 72, 3506 (1994). J. von Zanthier, G. S. Agarwal, H. Walther, *Cavity-induced modifications of quantum jumps*, Phys. Rev. A 56, 2242 (1997).

[114] R. J. Cook and H. J. Kimble, *Possibility of direct observation of quantum jumps*, Phys. Rev. Lett. 54, 1023 (1985). J. Dalibard and C. Cohen-Tannoudji, *Single-atom laser spectroscopy. Looking for dark periods in fluorescent light*, Europhys. Lett. 1, 441 (1986). J. Javanainen, *Possibility of quantum jumps in a three-level system*, Phys. Rev. A 33, 2121 (1986). A. Schenzle, R. G. De Voe, and R. G. Brewer, *Possibility of quantum jumps*, Phys. Rev. A 33, 2127 (1986). D. T. Pegg, R. Loudon, and P. L. Knight, *Correlations in light emitted by three-level atoms*, Phys. Rev. A 33, 4085 (1986). A. Schenzle and R. G. Brewer, *Macroscopic quantum jumps in a single atom*, Phys. Rev. A 34, 3127 (1986). H. J. Kimble, R. J. Cook, and A. L. Wells, *Intermittent atomic fluorescence*, Phys. Rev. A 34, 3190 (1986).

M. B. Plenio and P. L. Knight, *The quantum jump approach to dissipative dynamics in quantum optics*, Rev. Mod. Phys. 70, 101 (1998). W. G. Teich and G. Mahler, *Stochastic dynamics of individual quantum systems: Stationary rate equations,* Phys. Rev. A 45, 3300 (1992). J. Dalibard, Y. Castin, and K. Molmer, *Wave-function approach to dissipative processes in quantum optics*, Phys. Rev. Lett. 68, 580 (1992).

[115] Richard A. Pakula, *Realistic model for radiation-matter interaction*, arXiv:quant-ph/0405055, (2004).

[116] Leslie C. Allen, Joseph H. Eberly, *Optical Resonance and Two-Level Atoms*, Dover Publisher, Wiley, New York, 1975. J. H. Shirley, *Solution of the Schrödinger equation with a Hamiltonian periodic in time*, Phys. Rev. 138, B0979 (1965). C. S. Chang and P. Stehle, *Quantum-Electrodynamical Theory of Atoms Interacting with High-Intensity Radiation Fields*, Phys. Rev. A 4, 0641 (1971). D. T. Pegg and G. W. Series, *Comment on "Quantum-Electrodynamical Theory of Atoms Interacting with High-Intensity Radiation Fields"*, Phys. Rev. A 7, 0371 (1973). F Ahmad and R K Bullough, *Analytical expressions for the Bloch-Siergert shift*, J. Phys. B: Atom. Molec. Phys. 7, L147 (1974). F. T. Hioe and J. H. Eberly, *Stimulated and spontaneous radiative frequency shifts of a two-level system*, Phys. Rev. A 11, 1358 (1975). Paul R. Berman and Jehuda Ziegler, *Generalized dressed-atom approach to atom-strong-field interactions—application to the theory of lasers and Bloch-Siegert shifts*, Phys. Rev. A 15, 2042 (1977). Guy Oliver, *Semiclassical and quantum treatments of the interaction between a polychromatic field and a multilevel quantum system coupled to a photon reservoir*, Phys. Rev. A 15, 2424 (1977). P. R. Berman and R. Salomaa, *Comparison between dressed-atom and bare-atom pictures in laser spectroscopy*, Phys. Rev. A 25, 2667 (1982). P. R. Berman, *Theory of fluorescence and probe absorption in the presence of a driving field using semiclassical dressed states*, Phys. Rev. A 53, 2627 (1996). J. C. Wells, I. Simbotin, M. Gavrila, *Physical Reality of Light-Induced Atomic States*, Phys. Rev. Lett. 80, 3479 (1998).

[117] Leslie C. Allen, Joseph H. Eberly, *Optical Resonance and Two-Level Atoms*, Dover Publisher, Wiley, New York, 1975

[118] C. Cohen-Tannoudji, J. Dupont-Roc and G. Grynberg, *Photons and Atoms: Introduction to Quantum Electrodynamics*, NY Wiley Interscience, 1989.

[119] C. Cohen-Tannoudji, J. Dupont-Roc and G. Grynberg, *Atom-Photon Interactions: Basic Processes and Applications* (Wiley Science Paperback Series), John Willey & Sons, New York, 1992.

[120] D. P. Craig T. Thirunamachandran*, Molecular Quantum Electrodynamics: An Introduction to Radiation-Molecule Interactions*, Dover Publications, Incorporated, 1998.

[121] P. Carbonaro, G. Compagno, and F. Persico, *Canonical dressing of atoms by intense radiation fields*, Phys. Lett. A 73, 097 (1979). G. Compagno and F. Persico, *Canonical transformatoin for single-atom resonance fluorescence: The strong-driving-field limit*, Phys. Rev. A 22, 2108 (1980).

[122] James Bjorken and Sidney Drell, *Relativistic Quantum Mechanics*, McGraw Hill, New York, 1965.

[123] M. Born and E. Wolf, *Principles of Optics, Chapter 12: Diffraction of light by ultrasonic waves*, Pergamon Press, 5th ed, Cambridge, Oxford, 1975.

[124] Arthur. H. Compton, *The Size and Shape of the Electron*, Phys. Rev. 14, 020 and 14, 247 (1919).

[125] Arthur H. Compton, *A Quantum Theory of the Scattering of X-rays by Light Elements*, Phys. Rev. 21, 483 (1923).

[126] G. I. Barenblatt, *Scaling, Self-similarity, and Intermediate Asymptotics: Dimensional Analysis and Intermediate Asymptotics*, Cambridge University Press, 1996

[127] W. Greiner, J Reinhardt, *Quantum Electrodynamics*, 3rd edition, , Springer Verlag, Berlin, New York, 2003.

[128] V. Weisskopf , *Über die Selbstenergie des Elektrons (The self-energy of the electron)*, Zeitschrift für Physik 89, 027 (1934).

[129] P. W. Milonni, *The Quantum Vacuum- An Introduction to Quantum Electrodynamics*, Academic Press, Incorporated , San Diego, 1994.